\documentclass{article}

\renewcommand{\theequation}{\arabic{section}.\arabic{equation}}

\usepackage{amssymb}
\usepackage{amsmath}
\usepackage{amsthm}
\usepackage{verbatim}

\input xy
\xyoption{all} \CompileMatrices

\newtheorem{theorem}{Theorem}[section]
\newtheorem{lemma}[theorem]{Lemma}
\newtheorem{proposition}[theorem]{Proposition}
\newtheorem{corollary}[theorem]{Corollary}

\theoremstyle{definition}
\newtheorem{definition}[theorem]{Definition}

\theoremstyle{definition}

\theoremstyle{remark}
\newtheorem{remark}[theorem]{Remark}

\theoremstyle{remark}
\newtheorem{example}[theorem]{Example}

\newcommand{\g}{\mathfrak{g}}
\newcommand{\fg}{\mathfrak{g}}
\newcommand{\n}{\mathfrak{n}}
\newcommand{\fn}{\mathfrak{n}}
\newcommand{\h}{\mathfrak{h}}
\newcommand{\fm}{\mathfrak{m}}
\newcommand{\fp}{\mathfrak{p}}
\newcommand{\fs}{\mathfrak{s}}
\newcommand{\lag}{\mathfrak{l}}
\newcommand{\m}{\mathfrak{m}}

\newcommand{\ssl}{\operatorname{sl}}

\newcommand{\ad}{\operatorname{ad}}

\newcommand{\id}{{1 \mskip -5mu {\rm I}}}

\newcommand{\vac}{|0\rangle}

\newcommand{\charge}{\mbox{charge}}
\newcommand{\Cur}{\mbox{Cur}}
\newcommand{\End}{\mbox{End}}

\newcommand{\Lie}{\mbox{Lie}}
\newcommand{\Res}{\mbox{Res}}

\newcommand{\Span}{\mbox{span}}
\newcommand{\Vir}{\mbox{Vir}}

\renewcommand{\deg}{\mbox{deg}}
\newcommand{\Ker}{\mbox{Ker}}
\newcommand{\im}{\mbox{Im}}

\newcommand{\Mod}{\text{Mod}}
\newcommand{\Ind}{\text{Ind}}

\newcommand{\sdim}{\mbox{sdim}}
\newcommand{\str}{\mbox{str}}

\newcommand{\A}{\mathcal{A}}
\newcommand{\B}{\mathcal{B}}

\newcommand{\F}{\mathcal{F}}
\newcommand{\I}{\mathcal{I}}

\renewcommand{\O}{\mathcal{O}}
\newcommand{\T}{\mathcal{T}}

\newcommand{\M}{\mathcal{M}}

\newcommand{\gr}{\text{gr}}
\newcommand{\grd}{d^\gr}

\newcommand{\tA}{\tilde{\A}}
\newcommand{\tI}{\tilde{\I}}
\newcommand{\ti}{{\tilde{\text{\i}}}}
\newcommand{\tSigma}{\tilde{\Sigma}}
\newcommand{\tS}{\tilde{S}}

\newcommand{\dz}{d^Z}

\newcommand{\ttt}{T}
\newcommand{\tf}{\text{fin}}

\newcommand{\st}[1]{\ensuremath{^{\scriptstyle \textrm{#1}}}}

\newcommand{\com}{{\rm com}}

\newcommand{\BI}{\text{BI}}
\newcommand{\LBI}{\text{LBI}}
\newcommand{\RBI}{\text{RBI}}

\newcommand{\cl}{\text{cl}}

\newcommand{\VPA}{\mathcal V}

    \newcommand{\cF}{\mathcal F}

\newcommand{\cW}{\mathcal W}

			\newcommand{\CC}{\mathbb{C}}

			\newcommand{\NN}{\mathbb{N}}

			\newcommand{\RR}{\mathbb{R}}

			\newcommand{\ZZ}{\mathbb{Z}}
\newcommand{\R}{\mathbb R}      
\newcommand{\Z}{\mathbb Z}
\newcommand{\C}{\mathbb C}

\newcommand{\romanparenlist}{% changes enumerate 1st level to (i)...(ix)
  \renewcommand{\theenumi}{\roman{enumi}}%
  \renewcommand{\labelenumi}{(\theenumi)}%
}
\newcommand{\alphaparenlist}{% changes enumerate 1st level to (a)...(z)
  \renewcommand{\theenumi}{\alph{enumi}}%
  \renewcommand{\labelenumi}{(\theenumi)}%
}
\newcounter{appendice}
\def\theappendice{{\normalsize\Alph{appendice}}} 
\newcommand{\appendice}[1]{
\refstepcounter{appendice}
\setcounter{equation}{0}
\def\theequation{\theappendice.\arabic{equation}} 
\setcounter{theorem}{0}
\def\thesection{{\normalsize\Alph{appendice}}}
\bigskip\bigskip\noindent 
\begin{center}
{\Large\textbf{Appendix.\ #1}}
\end{center}
\par\bigskip\nopagebreak}

\begin{document}

%%%%%%%%%%%%%%%%%%%%%%%%%%%%%%%%%%%%%%%%%%%%%%%%
%%%%%%%%%%% TITOLO %%%%%%%%%%%%%%%%%%%%%%%%%%%%

\begin{center}
{\Large {\bf
Finite vs affine W-algebras
}
}

\vfill

{\large
\begin{tabular}[t]{c}
$\mbox{Alberto De Sole}^{1}\ \ \ and
\phantom{mm}
\mbox{Victor G. Kac}^{2}$
\\
\end{tabular}
\par
}

\bigskip

{\small
\begin{tabular}[t]{ll}
{1} & {\it Department of Mathematics, Harvard University} \\
& 1 Oxford Street, Cambridge, MA 02138, USA \\
%& \qquad $\&$ \\
& $\&$\ \ {\it Istituto Nazionale di Alta Matematica} \\
&  Citt\'{a} Universitaria, 00185 Roma, Italy \\
& E-mail: {\tt desole@math.harvard.edu} \\
{2} & Department of Mathematics, MIT \\
& 77 Massachusetts Avenue, Cambridge, MA 02139, USA \\
& E-mail: {\tt kac@math.mit.edu}
\end{tabular}
}
\end{center}

\vfill

\begin{quotation}
  ``I am an old man, and I know that a definition  cannot be so complicated.''
\end{quotation}
\begin{center}
  I.M. Gelfand (after a talk on vertex algebras in his Rutgers seminar)
\end{center}

\vfill

\centerline{\bf Abstract} 
\smallskip

In Section \ref{sec:1} we review various equivalent definitions of a vertex algebra $V$.
The main novelty here is the definition in terms of an indefinite integral of the $\lambda$-bracket.
In Section \ref{sec:2} we construct, in the most general framework, the Zhu algebra 
$Zhu_{\Gamma}V$, an associative algebra which "controls" $\Gamma$-twisted representations
of the vertex algebra $V$ with a given Hamiltonian operator $H$.
An important special case of this construction is the $H$-twisted Zhu algebra $Zhu_H V$.
In Section \ref{sec:nonlin} we review the theory of non-linear Lie conformal algebras
(respectively non-linear Lie algebras). Their universal enveloping vertex algebras 
(resp. universal enveloping algebras) form an important class of freely generated vertex algebras
(resp. PBW generated associative algebras).
We also introduce the $H$-twisted Zhu non-linear Lie algebra $Zhu_H R$
of a non-linear Lie conformal algebra $R$ and we show that its universal enveloping algebra
is isomorphic to the $H$-twisted Zhu algebra of the universal enveloping vertex algebra of $R$.
After a discussion of the necessary cohomological material in Section \ref{sec:almost},
we review in Section \ref{sec:6} the construction and basic properties of affine and finite
$W$-algebras, obtained by the method of quantum Hamiltonian reduction.
Those are some of the most intensively studied examples of freely generated vertex algebras 
and PBW generated associative algebras.
Applying the machinery developed in Sections \ref{sec:nonlin} and \ref{sec:almost}, we then show
that the $H$-twisted Zhu algebra of an affine $W$-algebra is isomorphic to the
finite $W$-algebra, attached to the same data.
In Section \ref{sec:pva} we define the Zhu algebra of a Poisson vertex algebra,
and we discuss quasiclassical limits.

%\end{titlepage}
\vfill\eject

\setcounter{section}{-1}
%%%%%%%% SECTION %%%%%%%%%%%%%%%%%%%%%%%%%%%%%

%%%
\numberwithin{equation}{section}

\section{Introduction}
\label{sec:intro}
\subsection{ What is a vertex algebra?}
\label{sec:0.1} 

According to the general principles of quantum field theory, the
observables are quantum fields, which are operator valued
distributions on the space-time, satisfying certain axioms,
called Wightman's axioms.  In one sentence, a vertex algebra is a
chiral part of a $2$-dimensional conformally covariant quantum
field theory.

The data of a vertex algebra consist of the space of states $V$ 
(an arbitrary vector space), the vacuum vector $\vac \in V$,
the infinitesimal translation operator $T \in \End V$ and a
collection of $\End V$-valued quantum fields $\F$, 
subject to the axioms formulated below (which
are ``algebraic'' consequences of Wightman's axioms).

A quantum field $a(z)$ is a formal $\End V$-valued distribution, that is
an $\End V$-valued linear function on
the space of Laurent polynomials $\CC [z,z^{-1}]$ (the space of
test functions), which satisfies the condition that $a (z) v$ is a
Laurent series in $z$ for every $v \in V$.  We can
write $a(z)$ in the form 
\begin{equation}
\label{eq:0.1}
a(z) = \sum_{n \in \ZZ} a_{(n)} z^{-n-1}\, , \, \hbox{ where  } a_{(n)} \in  
\End V\, ,
\end{equation}
so that $a_{(n)}=\Res_z z^n a(z)$,  where, as usual, $\Res_z$ stands for the
coefficient of $z^{-1}$. Then the above condition means that $a_{(n)}v=0$
for $n$ sufficiently large.

The following are the four axioms of a vertex algebra $(V$,
$\vac$, $T$, $\F = \{ a^j(z) =\sum_{n\in \ZZ}
a^j_{(n)}z^{-n-1}\}_{j \in J})$: 
\begin{eqnarray*}
  \begin{array}{ll}
     \hbox{(vacuum)}   & T \vac = 0\, , \\[1ex]
     \hbox{(translation covariance)} & [T,a^j (z) ]
        = \partial _z a^j(z) \, ,\\[1ex]
     \hbox{(locality)} & (z-w)^{N_{jk}} [a^j(z) , a^k (w)]=0
        \hbox{ for some  } N_{jk}\in \ZZ_+ \, ,\\[1ex]
     \hbox{(completeness)} & \hbox{ vectors}\, \, 
       a^{j_1}_{(n_1)} \ldots a^{j_s}_{(n_s)} \vac \,\, \hbox {span}\, V\, .
 \end{array}
\end{eqnarray*}

This definition allows one to easily construct examples.  In
Section \ref{sec:1} we give several more equivalent definitions of  vertex
algebras, which allow us to work with them.  First, we show that,
if we enlarge $\F$ to the 
maximal collection $\bar{\F}$ of quantum fields for which the axioms
still hold, then the map $\bar{\F} \to V$, defined by
$a(z) \mapsto a_{(-1)} \vac$, is bijective.  We thus get the
state-field correspondence, defined as the inverse map $V\rightarrow \bar\F$

\begin{displaymath}
   V \ni a \mapsto Y (a,z) \in\bar\F\, ,
\end{displaymath}
which leads to the second definition of a vertex algebra.
We shall use the customary notation $Y(a,z)$ for the quantum
field, corresponding to the state $a \in V$, and the customary
name ``vertex operators'' for these quantum fields.

The state-field correspondence allows one to introduce bilinear
products on $V$ for each $n \in \ZZ$ by letting
\begin{displaymath}
  a_{(n)} b = \Res_z z^{n} Y (a,z)b\, .
\end{displaymath}
This leads to the third, the original Borcherds definition  of a
vertex algebra as a vector space $V$ with the vacuum vector $\vac$ 
and bilinear products $a_{(n)}b$ for each $n \in \ZZ$, satisfying a
simple vacuum identity and a quite complicated cubic identity,
called the Borcherds identity \cite{B}.  This identity is
somewhat similar to the Jacobi identity; the former is as
important for the theory of vertex algebras as the latter is for
the theory of Lie algebras.

A part of the structure of a vertex algebra is the
  $\lambda$-bracket, defined by
  \begin{displaymath}
    [a_{\lambda}b]=\Res_z e^{\lambda z} Y (a,z)b = \sum_{n\in\Z_+}\frac{\lambda^n}{n!}a_{(n)}b
       \quad (a,b \in V) \, ,
  \end{displaymath}
which satisfies axioms similar to the axioms of a Lie algebra.
This structure, extensively studied in the past few years, is
called a Lie conformal algebra \cite{K}.  The only additional operation
needed to get a vertex algebra structure is the $-1$\st{st}
product $a_{(-1)}b$, usually called the normally ordered
product.  The vacuum vector is the identity element for this
product, which is neither commutative nor associative.  However,
the normally ordered product is quasicommutative and quasiassociative, meaning
that the ``quantum corrections'' to the commutator and the
associator are certain explicit expressions in terms of the
$\lambda$-bracket.  Finally, the $\lambda$-bracket and the
normally ordered product are related by a ``non-commutative Wick
formula'', which is an analogue of the Leibniz rule, where the
``quantum correction'' is expressed in terms of a double
$\lambda$-bracket.  This is the fourth description of a vertex
algebra, given in \cite{BK}, similar to the definition of a
Poisson algebra.
On the other hand, this
definition is the most convenient one for the theory of quantum
integrable systems.

Remarkably, the fourth definition can be interpreted by saying that a vertex algebra is a Lie
conformal algebra whose $\lambda$-bracket can be "integrated".
This leads to a new, fifth definition of a vertex algebra, which is
discussed in Section \ref{sec:1.11}.

\subsection{ Non-linearities}
\label{sec:0.2} 

Some of the most important examples of vertex algebras, such as
affine, fermionic and Virasoro vertex algebras are the universal
enveloping vertex algebras of certain Lie conformal algebras.
However, the majority of affine $W$-algebras, which are central
to the present paper, cannot be obtained in this way due to the
``non-linearities'' in the commutation relations.  The first
example of this kind is Zamolodchikov's $W_3$ algebra \cite{Za}.
(The two generating quantum fields of the $W_3$ algebra
correspond to the basic invariants of the Weyl group of
$s\ell_3$, hence the name ``$W$-algebra''.)

In order to take into account the above mentioned
``non-linearities'', one needs the notion of a
non-linear Lie conformal algebra, introduced in \cite{DK}).  
This and the parallel, simpler notion of a non-linear Lie algebra, are
discussed in Section \ref{sec:nonlin}.  It follows from \cite{DK,KW1}
that all affine $W$-algebras can be obtained as universal
enveloping vertex algebras of certain non-linear Lie conformal algebras.
Likewise, all "finite" $W$-algebras are universal enveloping algebras
of certain non-linear Lie algebras.

\subsection{ Hamiltonian operators}
\label{sec:0.3}

We shall consider most of the time $H$-graded vertex
algebras.  A vertex algebra $V$ is called $H$-graded if one fixes
a Hamiltonian operator $H$, i.e. a diagonalizable operator on
$V$ such that
\begin{equation}
  \label{eq:0.2}
  [H ,Y (a,z)] =z \partial _z Y (a,z)+Y(Ha,z)\, .
\end{equation}
The main source of Hamiltonians is an energy-momentum field, that
is a vertex operator of the form $Y (\nu ,z) = \sum_{n\in\ZZ}
L_n z^{-n-2}$, with the following three properties:

\romanparenlist
\begin{enumerate}
\item %%i
the operators $L_n$ form the Virasoro algebra 
\begin{equation}
  \label{eq:0.3}
  [L_m , L_n]=(m-n) L_{m+n} + \frac{c}{12} (m^3-m)
      \delta_{m,-n} \, ,
\end{equation}
where $c\in\C$ is called the
central charge,

\item %%ii
  $L_{-1}=T $, 

\item  %%iii
  $L_0$ is a diagonalizable operator.
\end{enumerate}
Then $H=L_0$ is a Hamiltonian operator.

In connection to twisted representation theory of a vertex
algebra $V$ one also considers an abelian group $A$ of
automorphisms of the vertex algebra $V$, commuting with $H$.  It
turns out that $A$-twisted irreducible positive energy (with
respect to $H$) modules over $V$ are in a canonical one-to-one
correspondence with irreducible modules over an associative
algebra $Zhu_{H,A} V$, canonically associated to the triple
$(V,H,A)$.  In the case when the
group $A$ is trivial and all the eigenvalues of $H$ are integers
this construction was introduced by Zhu \cite{Z}.  Our
construction generalizes those of \cite{Z,KWa,DLM}, 
and the proof of the main theorem is closer to \cite{DLM}.
Especially important for this paper is the case when the grading defined by the action of $A$
is given by the eigenspaces of $e^{2\pi iH}$. In this case we denote $Zhu_{H,A}V$ by $Zhu_H V$,
and we call it the $H$-twisted Zhu algebra.

In order to define and study the properties of the Zhu algebra,
we introduce and study in Section \ref{sec:2} a certain deformation of
the vertex operators of $V$.  This deformed vertex algebra
satisfies all the axioms of a vertex algebra, except that the
translation covariance is replaced by a ``gauged'' translation
covariance with the potential $\hbar H /(1+\hbar z)$.  We show
that for this particular choice of the potential the deformed
vertex operators satisfy a very simple deformation of the
Borcherds identity.  This makes the exposition of the theory of
Zhu algebras very simple and transparent.

In Section \ref{sec:zhuenv} we introduce the (non-linear) Zhu Lie algebra $Zhu_H R$
of a non-linear Lie conformal algebra $R$ and we show that the Zhu algebra $Zhu_H V(R)$ 
of the universal enveloping vertex algebra $V(R)$ is the universal enveloping algebra
of $Zhu_H R$.
As an application, we show in Section \ref{sec:dirac}  that the $N=1$ extension of 
the Sugawara construction, given in \cite{KT}, is a "chiralization"
of the cubic Dirac operator, defined in \cite{AM,Ko1}, which immediately
implies all its properties.

In Section \ref{sec:almost} we introduce the notion of an almost linear
differential of a non-linear Lie algebra (resp. a non-linear Lie
conformal algebra) and show that, under certain conditions, the
cohomology of the universal enveloping algebra (resp. universal
enveloping vertex
algebra) equals the universal enveloping algebra (resp. universal
enveloping vertex
algebra) of the cohomology of the non-linear Lie algebra
(resp. Lie conformal algebra).

\subsection{ What is a $W$-algebra?}
\label{sec:0.4}

In the remainder of the paper we apply the developed machinery to
finite and affine $W$-algebras. They are universal enveloping algebras of certain 
non-linear Lie algebras and non-linear Lie conformal algebras respectively, obtained
by the method of quantum Hamiltonian reduction.

Let us first explain the classical Hamiltonian reduction in a
special situation that shall be used.  Let $\fg$ be a simple
finite-dimensional Lie algebra with a non-degenerate symmetric
invariant bilinear form $(.\, | \, .)$, and let $x \in \fg$ be an
$\ad$-diagonalizable element with eigenvalues in $\frac12 \Z$.  We have the
decomposition into a sum of subalgebras:  $\fg = \fg_- \oplus
\fg_0 \oplus \fg_+$, 
where $\fg_+$ (resp. $\fg_-$) is the sum of all
eigenspaces of $\ad x$ with positive (resp. negative) eigenvalues
and $\fg_0$ is the $0$\st{th} eigenspace.  Let $G_+$ be the
connected Lie group with the Lie algebra $\fg_+$.  Recall that
the algebra $S(\fg)$ of polynomial functions on $\fg^*$ carries
the classical Poisson algebra  structure, defined by $\{ a,b \} =
[a,b]$ if $a,b \in \fg$, and extended to $S(\fg)$ by the Leibniz
rule.  We shall identify $\fg^*$ with $\fg$ via the bilinear form
$(. \, | \, .)$.  The coadjoint action of $G_+$ on $\fg^*_+$ can
be conveniently thought of as the composition of its adjoint
action on $\fg_- \subset \fg$ with the projection $\pi_-$ of $\fg$ onto
$\fg_-$, the latter being canonically identified with $\fg^*_+$
via the invariant form $(. \, | \, .)$.  Let $f \in \fg_-$, let
$\O =G_+ \cdot f$ be the coadjoint orbit of $f$ in $\fg_-$,
and consider the $G_+$-invariant submanifold
$\tilde{\O}=\pi^{-1}_- (\O)$ of $\fg = \fg^*$.  The ideal $I$
consisting of functions from $S (\fg)$ which vanish on
$\tilde{\O}$ is not a Poisson ideal.  Nevertheless, it is easy to
check that the algebra $P=(S (\fg)/I)^{G_+}$ of $G_+$-invariants
(which can be thought of as the algebra of functions on
$\tilde{\O}/G_+$) carries a well-defined  Poisson bracket induced
from that on $S(\fg)$.  The Poisson algebra $P$ 
is called the Hamiltonian reduction of the
(Hamiltonian) action of $G_+$ on $\fg^*$.  (A more general setup
is a Hamiltonian action of a group $N$ on a Poisson manifold $X$
with the moment map $\mu : X\to \fn^*$.  Picking an $N$-orbit
$\O$ in $\fn^*$, one obtains a Poisson manifold $\mu^{-1} (\O)/N$.)

According to the general BRST idea, in order to quantize the
Poisson algebra $P$, one should try to represent it as a homology
of a complex $(C,d)$, which can be quantized.

This can be done, when the pair $(x,f)$ is \emph{good}, meaning
that $f\in\g_-$ is an eigenvector of $\ad x$ with eigenvalue $-1$, such
that its centralizer in $\fg$ lies in $\fg_-+\fg_0$.  The most
important examples of such pairs come from the $s\ell_2$-triples
$\{ e,h=2x,f \}$ in $\fg$.

One may replace $\fg_+$ by a slightly smaller
subalgebra $\fn_\fs = \fs+ \fg_{\geq 1}$, where $\fs$ is an arbitrary  subspace of
$\fg_{1/2}$ and $\fg_{\geq 1}$ is the sum of all eigenspaces of
$\ad x$ with eigenvalues $\geq 1$.  It is easy to see \cite{GG} that this
does not change the Poisson algebra $P$.

Let $N_\fs$ be the connected Lie group with the Lie algebra
$\fn_\fs$.  Then the orbit $N_\fs \cdot f$ of $f$ in $\fg^*_+ =
\fg_-$ is $\O_\fs =f + [f,\fs]$, hence its preimage in $\fg^* = \fg$
is $\tilde{\O}_\fs =f + [f,\fs]+\fg_0+\fg_+$.  This is defined by the
following equations in $\xi\in\g$:
\begin{displaymath}
 (\xi|a)=(f|a)\, , \quad a \in \fm_\fs\ ,
\end{displaymath}
where $\fm_\fs = ([f,\fs]^{\perp} \cap \fg_{1/2}) + \fg_{\geq 1}$, and $\perp$
stands for the orthocomplement in $\fg$ with respect to $(. \, | \, .)$.
Hence the ideal $I_\fs\subset S(\g)$ of all functions from $S(\g)$ which vanish on $\tilde\O_\fs$
is generated by
$\big\{a-(f|a)\ |\ a\in\fm_\fs\big\}$.
It follows that the coordinate ring $\CC [\tilde\O_\fs]=S(\g)/I_\fs$ can be
represented as the $0$\st{th} cohomology of the Koszul complex
\begin{displaymath}
  (S (\fg) \otimes \wedge (\fm_\fs) \, , \, d_K)\, , 
\end{displaymath}
where the degrees of elements of $\fg$ (resp. $\fm_\fs$) are equal to
$0$ (resp. $-1$), and 
\begin{displaymath}
  d_K (a) =0 \, , \quad d_K (\varphi_a)=a-(f|a)\, .
\end{displaymath}
Here $a \in \fg$ and $\varphi_a$ denotes the element of
$\fm_{\fs}$ corresponding to $a$, under the inclusion $\fm_\fs \subset \fg$.

The algebra $P$ is the algebra of invariants of the group $N_\fs$
acting on the algebra $\CC [\tilde\O_\fs]$, which is the $0$\st{th}
Lie algebra cohomology of the complex
\begin{displaymath}
  (\CC [\tilde\O_\fs] \otimes \wedge (\fn^*_\fs) \, , d_L)\, ,
\end{displaymath}
where $d_L$ is the standard Lie algebra cohomology differential of the
$\fn_\fs$-module $\CC [\tilde\O_\fs]$.  Therefore the algebra $P$ is
isomorphic to the $0$\st{th} cohomology of the complex \cite{KS}
\begin{equation}
  \label{eq:0.4}
  (S (\fg) \otimes \wedge (\fm_s \oplus \fn^*_\fs) \, , \, 
       d_K + d_L)\, .
\end{equation}

If $\fs$ is chosen in such a way that $\fm_\fs =\fn_\fs$, it is clear
how to quantize this complex.  Just replace $S(\fg)$ by the
universal enveloping algebra $U (\fg)$ and replace $\wedge
(\fn_\fs \oplus \fn^*_\fs)$ by the Clifford algebra on the space
$\fn_\fs \oplus \fn^*_\fs$ with the symmetric bilinear form defined
by the pairing between $\fn_\fs$ and $\fn^*_\fs$.  One can find such
$\fs$ by introducing the following non-degenerate skewsymmetric
bilinear form on $\fg_{1/2}$:
\begin{equation}
  \label{eq:0.5}
  \langle a,b \rangle = (f | [a,b])\, ,
\end{equation}
and taking $\fs$ to be any maximal isotropic subspace of
$\fg_{1/2}$.

Unfortunately in the Lie superalgebra case this may not work
(since the bilinear form (\ref{eq:0.5}) is symmetric on the odd
part of $\fg_{1/2}$).  The way out is to choose $\fs$ to be a
co-isotropic subspace of $\fg_{1/2}$ (the favorite choice is
$\fs =\fg_{1/2}$), so that $
\fs^{\perp}_{\langle \, , \, \rangle}\subset \fs$, hence
$\fm_{\fs}\subset \fn_{\fs}$, and the space $\fp=\fs /
\fs^{\perp}_{\langle \, . \, \rangle}$ carries the induced
non-degenerate bilinear form, also denoted by $\langle \, . \, ,
\, . \, \rangle$ .  One then replaces the complex (\ref{eq:0.4})
by a quasi-isomorphic complex
\begin{equation}
  \label{eq:0.6}
  (S (\fg) \otimes \wedge (\fn_{s}\oplus \fn^*_{\fs})
  \otimes S(\fp) \, , \, d  )
\end{equation}
where the degree of the elements of $\fp$ is $0$, and $d$ is a
suitable modification of the differential $d_K + d_L$.  This
complex is again easy to quantize by replacing, as before,
$S(\fg)$ by $U(\fg)$, $\wedge (\fn_\fs \oplus \fn^*_\fs)$ by the
corresponding Clifford algebra and $S(\fp)$ by the Weil algebra on
the space $\fp$ with the bilinear form  $\langle \, . \, ,
\, . \, \rangle$.  Finally, we write for the differential $d$ the
same formula as in the classical complex (\ref{eq:0.6}).

The cohomology of the quantization of the complex (\ref{eq:0.6})
is an associative algebra, called the finite $W$-algebra
associated to the triple $(\fg,x,f)$, and denoted by $W^\tf(\fg,x,f)$.
(In fact, this is the $0$\st{th} cohomology since the rest
vanishes.)

One associates to the same triple a family of vertex algebras
$W_k (\fg,x,f)$, called affine $W$-algebras,
depending on a complex parameter $k$, in a similar
fashion, replacing the associative algebras by the corresponding
vertex algebras.  Namely, $U(\fg)$  is replaced by the universal
affine vertex algebra $V^k(\fg)$, and Clifford and Weil algebras
are replaced by the vertex algebras of free superfermions
$F(\Pi \fn_{\fs} \oplus \Pi \fn_{\fs^*})$ and $F(\fp)$, where $\Pi$
stands for the change of parity.  The differential $d$ is
replaced by $d_{(0)}=\Res_z d(z)$, where the
vertex operator $d(z)$ is obtained by writing $d=\ad D$,
where $D$ is an element of the quantization of (\ref{eq:0.6}),
and replacing all factors in $D$ by the corresponding vertex
operators and the associative products by normally ordered products.

The first main result of the theory of $W$-algebras states that
the associative algebra $W^\tf(\fg,x,f)$ (resp. vertex algebra $W_k
(\fg,x,f)$) is freely generated in the sense of
Poincar\'e--Birkhoff--Witt by elements attached to a basis of the
centralizer of $f$ in $\fg$ (see \cite{KW1}).  Thus, according to
\cite{DK}, all of these $W$-algebras are universal enveloping algebras
(resp. enveloping vertex algebras) of non-linear Lie algebras
(resp. non-linear Lie conformal algebras).

\subsection{ Some background remarks}
\label{sec:0.5}

After the work of Zamolodchikov \cite{Za}, a number of papers on
affine $W$-algebras, appeared in
physics literature, mainly as ``extended conformal algebras'',
i.e. vertex algebra extensions of the Virasoro vertex algebra.  
A review
of the subject up to the early 1990s may be found in the
collection of a large number of reprints on $W$-algebras
\cite{BS}.  The most important work of this period is the work by
Feigin and Frenkel \cite{FF1,FF2}, where the approach
described in the previous subsection was introduced in the case of the
principal nilpotent element $f$.  Namely, they quantize the
classical Hamiltonian reduction, introduced by Drinfeld and
Sokolov \cite{DS}.  For example, if $\fg = s\ell_n$, the
Drinfeld--Sokolov classical reduction produces the Gelfand--Dikii
Poisson algebras, while their Feigin--Frenkel quantization
produces the Virasoro vertex algebra for $n=2$ and
Zamolodchikov's $W_3$ algebra for $n=3$.  Further important
contributions were made in \cite{BT,ST,FKW,FB}, and many other works. 
The framework described in (\ref{eq:0.4}) was
developed in \cite{KRW,KW1,KW2} and applied to
representation theory of superconformal algebras:  
it turned out that in the case of minimal nilpotent orbits
of small simple Lie superalgebras these $W$-algebras
are the N=1,2,3,4 and many other well-known superconformal algebras, 
which led to a further
progress in their representation theory. Of course, in the quasiclassical
limit of $W_k(\fg ,x,f)$ one recovers the Drinfeld-Sokolov
reduction and its generalizations.
For the most recent important development in representation
theory of $W$-algebras see \cite{A1,A2}.

The finite $W$-algebras were much less studied. 
Some of the few references in physics literature that we know of are \cite{BT1,DV,RS}.  
More recently there has been a revival of
interest in them in connection to geometry and representation theory of simple
finite-dimensional Lie algebras, see \cite{M,P1,P2,GG,BrK}.
The earliest work is \cite{Ko}, where it was shown that
the algebra $W^\tf(\fg,x,f)$ for the principal nilpotent $f$ is
canonically isomorphic to the center of $U(\fg)$.

It is proved in the Appendix, written jointly with A. D'Andrea, C. De Concini and R. Heluani, 
that the quantum Hamiltonian reduction 
definition of finite $W$-algebras adopted in the paper and the definition
via the Whittaker models which goes back to \cite{Ko} (see \cite{P1,GG})
are equivalent.
This result was independently proved in \cite{A2}.
In fact in \cite{A2} the result is stated only in the case of principal nilpotent $f$,
cf. \cite{KS}, but the proof there can be adapted to the general case.
The connection to the Whittaker models may be given as another reason for 
the name "W-algebras".

\subsection{The main results of the paper}
\label{sec:0.6}

The definition and structure theory of finite and affine
$W$-algebras are very similar, but,
as far as we know, the precise connection between
them has been unknown.  The main new result of the present paper
answers this question: the algebra $W^\tf (\fg,x,f)$ is the
$H$-twisted Zhu algebra $Zhu_{H} W_k (\fg,x,f)$, where $H=L_0$ 
comes from the energy-momentum field of the vertex algebra 
$W_k(\fg,x,f)$.
In the case of the principal nilpotent $f$ this
result has been independently obtained by Arakawa \cite{A2} (cf. \cite{FKW}).

In conclusion of the paper we introduce the notion of a Zhu algebra of a Poisson
vertex algebra and discuss its connection to the Zhu algebra of a vertex algebra
via the quasiclassical limit.
As a result, the quantum and classical affine $W$-algebras on the one hand,
and the finite $W$-algebra and the Poisson algebra of the Slodowy slice on the other hand,
are tied in one picture.

Though, for simplicity, we assumed above that $\fg$ is a Lie
algebra, we work in
the more general situation of a simple finite-dimensional Lie
superalgebra $\fg$ with a non-degenerate supersymmetric
invariant bilinear form, as in \cite{KRW,KW1,KW2}.  
This is important due to applications to representation theory  
of superconformal algebras.
Thus, throughout the paper we shall always talk 
about Lie superalgebras, Lie conformal superalgebras, associative
superalgebras, etc., unless otherwise stated. 
However, we shall often drop the prefix "super".
\medskip

We wish to thank Bojko Bakalov for many illuminating discussions.
In particular it was his suggestion to introduce $\hbar$ in Zhu's definition of the Zhu algebra,
which led us to the definition of $\hbar$-deformed vertex operators.
He also helped us with the proof of Theorem \ref{th:3.5}.

The main results of the paper were reported in June 2005 at the summer school in ESI, Vienna.
\medskip

All vector spaces, algebras and their tensor
products are considered over the field of complex numbers $\CC$,
unless otherwise stated.  We denote by $\RR$, $\R_+$, $\ZZ$, $\ZZ_+$ and
$\NN$ the sets of real numbers, non negative real numbers, integers,
non-negative integers and positive integers, respectively.

%\newpage
%%%%%%%% SECTION 1 %%%%%%%%%%%%%%%%%%%%%%%%%%%%%

\section{Five equivalent definitions of a vertex algebra}
\label{sec:1}
\subsection{Vector superspaces}
\label{sec:1.1}

Let $V = V_{\bar{0}} \oplus V_{\bar{1}}$ be a vector superspace,
i.e. a vector space, endowed with a decomposition in a direct
sum of subspaces $V_{\alpha}$, where $\alpha \in \ZZ /2\ZZ = \{
\bar{0}, \bar{1} \}$.  If $v \in V_{\alpha}$, one writes
$p(v)=\alpha$ and calls it the \emph{parity} of $v$.  One has the
associated vector superspace decomposition,
which makes $\End V$ an associative superalgebra
(i.e. a $\Z/2\Z$-graded associative algebra)
\begin{displaymath}
  \End V = (\End V)_{\bar{0}} \oplus (\End V)_{\bar{1}}\, ,
\end{displaymath}
where $(\End V)_{\alpha} = \{ a \in \End V\ |\ a V_{\beta} \subset
V_{\alpha + \beta} \}$.
One defines the bracket on an associative superalgebra $A$ by
\begin{equation}
  \label{eq:1.1}
  [a,b] = ab- p(a,b) ba \, ,
\end{equation}
which makes $A$ a Lie superalgebra.
Here and further we use the notation
\begin{equation}
  \label{eq:1.2}
  p(a,b) = (-1)^{p(a)p(b)} ,\,\, s(a)=(-1)^{p(a)}\, ,
\end{equation}
and, when a formula involves $p(a)$, we assume that $a$ is
homogeneous with respect to the $\ZZ /2\ZZ$-grading, and extend
this formula to arbitrary $a$ by linearity.

\subsection{The first definition and the Extension Theorem}
\label{sec:1.2} 

The first definition of a vertex algebra, given in the beginning
of the introduction in the case of a vector space $V$,
extends to the super case as follows. (We shall work in this generality.)
\begin{definition}\label{def1}
A {\itshape vertex algebra} is a quadruple $(V,\,\vac,\,T,\,\F)$,
where $V=V_{\bar{0}}\oplus V_{\bar{1}}$ is a vector superspace,
called the {\it space of states}, $\vac$ is an element of $V_{\bar{0}}$,
called the {\itshape vacuum vector}, $T$ is an even endomorphism of $V$,
called the {\itshape infinitesimal translation operator},
and $\F$ is a collection of $\End V$-valued quantum fields, i.e.
$$
\F\ =\ \big\{a^j(z)=\sum_{n\in\Z}a^j_{(n)}z^{-n-1}\big\}_{j\in J}\ ,
$$
where for each $j\in J$ all $a^j_{(n)}\in \End V$ have the same parity,
and for every $b\in V$ we have $a^j_{(n)}b=0$ for $n\gg0$.
The above data satisfy the following axioms:
\begin{enumerate}
\item vacuum: $T \vac = 0$,
\item translation covariance: $[T,a^j (z) ] = \partial _z a^j(z)$,
\item locality: $(z-w)^{N_{jk}} [a^j(z) , a^k (w)]=0$ for some $N_{jk}\in\Z_+$,
\item completeness: $V=\Span_\C\big\{a^{j_1}_{(n_1)} \ldots a^{j_s}_{(n_s)}\vac\big\}$.
\end{enumerate}
The commutator in the locality axiom is defined by (\ref{eq:1.1}).
\end{definition}

In order to pass to the second definition, we need the following
simple

\begin{lemma}
  \label{lem:1.1}

Let $V$ be a vector superspace and let $\Omega \in V_{\bar0},\, T\in(\End V)_{\bar0}$, 
be such that $T \Omega =0$.  Let $a(z) =\sum_{n \in \ZZ} a_{(n)} z^{-n-1}$ 
be a translation covariant $\End V$-valued quantum field, i.e. $[T,a(z)]=\partial_z a(z)$. 
Then $a_{(n)}\Omega =0$ for all $n \in \ZZ_+$.
\end{lemma}

\begin{proof}
Recall that $a_{(n)} \Omega =0 $ for all $n \geq N$, where $N$ is
a sufficiently large non-negative integer.  Choose minimal such
$N$; we have to show that $N=0$. By assumption we have for all $n \in \ZZ$:
\begin{equation}
  \label{eq:1.3}
  [T,a_{(n)}] =-n a_{(n-1)}\, .
\end{equation}
Applying both sides to $\Omega$ for $n=N$, we have:  $T a_{(N)}
\Omega = -N a_{(N-1)} \Omega$.  Hence, if $N>0$, we obtain
$a_{(N-1)}\Omega =0$, a contradiction.
\end{proof}

Let $(V,\vac,T,\F)$ be a vertex algebra.
Let $\widetilde{\F}$ be the set of all $\End V$-valued quantum
fields $a(z)$, which are translation covariant, i.e. $[T,a(z)] = \partial _z a(z)$.  
By Lemma \ref{lem:1.1}, we have
\begin{equation}
  \label{eq:1.4}
  a(z) \vac \in V [[z]]\ ,\ \  \hbox{  if  } a(z) \in \widetilde\F \, ,
\end{equation}
which allows us to define the linear map
  \begin{equation}
    \label{eq:1.5}
    s: \widetilde{\F} \to V \, , 
         \quad s(a(z)) = a(z) \vac |_{z=0}\, .
  \end{equation}

We next define the $n$\st{th} product of $\End V$-valued quantum fields, for $n\in\Z$:
\begin{equation}
  \label{eq:1.6}
  a(w)_{(n)} b(w) = \Res_z \big(i_{z,w} (z-w)^n a(z) b(w) -
      p(a,b) i_{w,z} (z-w)^n b (w) a(z)\big) \, ,
\end{equation}
where $i_{z,w}$ denotes the power series expansion
in the domain $|z| > |w|$.
\begin{lemma}\label{dongs}
\alphaparenlist
\begin{enumerate}
\item The $n$\st{th} product of two $\End V$-valued quantum fields 
is an $\End V$-valued quantum field.
\item If $a(z),\,b(z)$ are translation covariant $\End V$-valued quantum fields,
so are $\partial_z a(z)$ and $a(z)_{(n)}b(z)$ for every $n\in\Z$.
\item If $(a(z),b(z))$ is a local pair of $\End V$-valued quantum fields, 
so is $(\partial_z a(z),b(z))$.
\item If $(a(z),b(z),c(z))$ is a collection of pairwise local $\End V$-valued quantum fields,
then $(a(z)_{(n)}b(z),c(z))$ is a local pair for every $n\in\Z$.
\end{enumerate}
\end{lemma}
\begin{proof}
(a) is straightforward. (b) follows immediately by the definition (\ref{eq:1.6}) of $n$\st{th}
product of fields.
(c) is easy and (d) is known as Dong's Lemma.
Its proof can be found in \cite[Lemma 3.2]{K}.
\end{proof}
\begin{lemma}
  \label{lem:1.3}
\alphaparenlist
\begin{enumerate}
\item %%a
  If $a(z) , b(z) \in \widetilde{\F}$, $a(z) =\sum_{n \in \ZZ}
  a_{(n)}z^{-n-1}$, and $s (a(z)) = a,s (b(z))=b$, then
  $s (a(z)_{(n)} b(z)) = a_{(n)}b$ and $s(\partial_z a(z))=Ta$.

\item %%b
  If $a(z) \in \widetilde{\F}$, then $a(z) \vac =e^{Tz}a$, where $a=s(a(z))$.
\end{enumerate}
\end{lemma}
\begin{proof}
To prove the first part of (a), apply both
sides of (\ref{eq:1.6}) to $\vac$.  Since $a(z) \vac \in V [[z]]$ by (\ref{eq:1.4}) and 
$i_{w,z}(z-w)^n$ involves only non-negative powers of $z$ as well, the
second term of the RHS of (\ref{eq:1.6}) applied to $\vac$
vanishes.  Hence $(a(w)_{(n)} b(w)) \vac |_{w=0}=\Res_z z^n
a(z) b = a_{(n)}b$.  
The second part of (a) is immediate by
applying the translation covariance to $\vac$ and putting $z=0$.
Finally (b) is obtained by looking at the Taylor expansion of $a(z)\vac$,
and using the identities $\partial_z^na(z)\vac|_{z=0}=T^na$,
given by (a).
\end{proof}
Denote by
$\F^\prime$ the minimal subspace of the space of all $\End V$-valued quantum
fields, containing $I_V$ and $\F$, which is $\partial_z$-invariant
and closed under all $n$\st{th} products.
We can construct $\F^\prime$ as follows.
Define by induction an increasing sequence $\F_k,\, k\geq0$,
of subspaces of the space of $\End V$-valued quantum fields,
by letting $\F_0=\Span_\C\{I_V,\F\}$, and 
$\F_{k+1}=\F_k+\big(\sum_{i\geq0}\partial_z^i\F_k\big)+\big(\sum_{n\in\Z}{\F_k}_{(n)}\F_k\big)$.
Then
$\F^\prime=\bigcup_{k\geq0}\F_k$.
We also let $\bar\F$ be the collection of all $\End V$-valued quantum fields which 
are translation covariant and local to all elements of $\F$.
By Lemma \ref{dongs} we have the following inclusions:
\begin{equation}\label{aster}
\F\ \subset\ \F^\prime\ \subset\ \bar\F\ \subset\ \widetilde\F\ .
\end{equation}
\begin{theorem}\label{th:1.2}
Let $(V, \vac ,T ,\F)$ be a vertex algebra. Then
\begin{enumerate}
\item $(V,\vac, T , \bar{\F})$ is also a vertex algebra.
\item The map $s: \bar{\F} \to V$, defined by (\ref{eq:1.5}), is bijective.
\item $\F^\prime=\bar\F$.
\end{enumerate}
\end{theorem}
\begin{proof}
Lemma \ref{dongs} guarantees that $\F^\prime$ is a collection of translation covariant
pairwise local fields,
and that all pairs of quantum fields $(a(z),b(z))$, 
with $a(z)\in\F^\prime$ and $b(z)\in\bar\F$, are local.
In particular $(V,\vac,T,\F^\prime)$ is a vertex algebra.
Hence, by
Lemma~\ref{lem:1.3}(a) and the completeness axiom,
the map $s: \F^\prime \to V$ is surjective.
Let now $a(z) \in \bar{\F}$ be such that $s(a(z))=0$.  
In particular $a(z)\vac=0$ by Lemma \ref{lem:1.3}(b).
Since $s:\,\F^\prime\to V$ is surjective, for each $b \in V$ there exists $b(z) \in \F^\prime$
such that $s(b(z))=b$.  We have, by locality of the pair $(a(z),b(z))$,
\begin{displaymath}
  (z-w)^N a(z) b (w)\vac = p(a,b) (z-w)^N b(w)a(z)\vac=0 \, .
\end{displaymath}
Letting $w=0$ and canceling $z^N$, we get $a(z) b=0$ for every $b\in V$, 
hence $a(z)=0$.  
Thus, the map $s:\,\bar\F \rightarrow V$ is injective.
Since $s:\,\F^\prime\rightarrow V$ is surjective and,
by (\ref{aster}), $\F^\prime\subset\bar\F$, we immediately get that $\F^\prime=\bar\F$
and that $s:\bar\F\rightarrow V$ is bijective.
\end{proof}

\begin{remark}
  \label{rem:1.4}
Theorem~\ref{th:1.2} implies the Existence Theorem~4.5 from
\cite{K} with the assumption (iii) and part of assumption~(ii)
removed.  Lemma~\ref{lem:1.1} was pointed out by Nikolov and
Todorov (this property is an axiom in \cite{K}), see also
Remark~4.4b in \cite{K}.  
\end{remark}

Lemma \ref{lem:1.3}(a), and Theorem~\ref{th:1.2} and its proof
imply the following

\begin{corollary}
  \label{cor:1.5}
Let $(V, \vac ,T,\F)$ be a vertex algebra.  Then for each $a \in
V$ there exists a unique $\End V$-valued quantum field $Y(a,z)$
which is translation covariant and local with all quantum fields from $\F$,
and such that $Y(a,z)\vac|_{z=0}=a$.  Furthermore, one has:
\begin{eqnarray}
  \label{eq:1.9}
  Y (\vac ,z) &=& I_V \, , \\
  \label{eq:1.10}
  Y (Ta ,z) &= & \partial_z Y(a,z)\, , \\
  \label{eq:1.11}
  Y (a_{(n)}b,z) &=& Y(a,z)_{(n)} Y (b,z)\, .
\end{eqnarray}
\end{corollary}

\subsection{The second definition of a vertex algebra} 
\label{sec:1.4} 

We arrive at the second, equivalent
definition of a vertex algebra, which was given in \cite{K}. 
\begin{definition}\label{def2}
A {\itshape vertex algebra} is a triple $(V, \vac ,Y)$, where 
$V=V_{\bar{0}}\oplus V_{\bar{1}}$ is  a vector superspace (the space of
states), $\vac \in V_{\bar{0}}$ is a vector (vacuum vector), and
$Y$ is a parity preserving linear map from $V$ to the space of
$\End V$-valued quantum fields:
 $ Y(a) = Y(a,z) = \sum_{n \in \ZZ} a_{(n)} z^{-n-1}$ ,
called the \emph{state-field correspondence}.  The infinitesimal
translation operator $T$ is encoded in the state-field
correspondence via the formula
\begin{equation}
  \label{eq:1.12}
  Ta = a_{(-2)} \vac \, .
\end{equation}
The axioms are as follows,
\begin{enumerate}
\item vacuum: $T \vac =0,\ Y(a,z) \vac |_{z=0}=a$,
\item translation covariance: $[T, Y (a,z)] =\partial_z Y (a,z)$,
\item locality: $(z-w)^N [Y(a,z), Y (b,w)]= 0$ for some $N \in \ZZ_+$.
\end{enumerate}
\end{definition}

Notice that the definition (\ref{eq:1.12}) of the infinitesimal translation operator
follows from (\ref{eq:1.10}) by applying both sides to $\vac$ and putting $z=0$.
We thus have proved already that the first definition implies the
second.  The converse is obvious, just define $T$ by
(\ref{eq:1.12}) and let $\F = \{ Y (a,z)\}_{a \in V}$.  The
completeness axiom is automatic since, by the vacuum axiom,
\begin{equation}
  \label{eq:1.13}
  a=a_{(-1)} \vac \quad \hbox{for  }a \in V \, .
\end{equation}
Note also that we automatically have (\ref{eq:1.9}), (\ref{eq:1.10}) and (\ref{eq:1.11}) 
for all $a,b \in V$.

\subsection{Borcherds identity and the third definition of a vertex algebra}
\label{sec:1.5} 

Formula~(\ref{eq:1.11}), called the $n$\st{th} product identity,
leads to the most important formula of the theory of vertex algebras,
called the Borcherds identity.  In order to derive it from
(\ref{eq:1.11}), recall the simple decomposition formula, which
holds for any local formal distribution $a(z,w)=\sum_{m,n \in
  \ZZ} a_{(m,n)} z^{-m-1} w^{-n-1}$ in $z$ and $w$, where by
locality one means that $(z-w)^N a(z,w)=0$ for some $N \in \ZZ_+$ 
(see \cite[Corollary~2.2]{K}):
\begin{equation}
  \label{eq:1.14}
  a(z,w) =\sum_{j \in \ZZ_+} c^j(w) \partial_w^j 
     \delta (z-w)/j!\, ,
\end{equation}
where the sum is finite (actually $j<N$), the quantum fields $c^j(w)$ are given by
\begin{equation}
  \label{eq:1.15}
  c^j(w) =\Res_z (z-w)^j a(z,w)\, ,
\end{equation}
and $\delta (z-w)$ is the formal delta function, defined by
\begin{equation}
  \label{eq:1.16}
  \delta (z-w) =i_{z,w} \frac{1}{z-w}-i_{w,z} \frac{1}{z-w}
      = z^{-1} \sum_{n \in \ZZ} \left( \frac{w}{z}\right)^n\, .
\end{equation}

The Borcherds identity reads:
\begin{eqnarray}
  \label{eq:1.17}
  Y (a,z) Y (b,w) i_{z,w} (z-w)^n-p(a,b) Y(b,w) Y(a,z) i_{w,z} (z-w)^n\\
\nonumber
  = \sum_{j \in \ZZ_+} Y (a_{(n+j)}b,w) 
     \partial_w^j \delta (z-w)/j! \, ,
\end{eqnarray}
for all $n \in \ZZ$ and $a,b \in V$.  Note that the sum on the
right is finite since $a_{(N)}b=0$ for $N \gg 0$, because
$Y(a,z)$ is a quantum field.

In order to prove (\ref{eq:1.17}), note that the LHS is local,
since, multiplied by $(z-w)^N$, it is equal to
\begin{equation}
  \label{eq:1.18}
  Y(a,z) Y(b,w) i_{z,w} (z-w)^{n+N} -p(a,b) Y(b,w) Y(a,b)
     i_{w,z} (z-w)^{n+N}\, , 
\end{equation}
which is $[Y (a,z), Y (b,w)] (z-w)^{n+N}$ if $n+N \geq 0$.
We thus apply the decomposition formula (\ref{eq:1.14})
to the LHS.
By (\ref{eq:1.15}), (\ref{eq:1.16}) and (\ref{eq:1.18}), we have $c^j (w) =
Y(a,w)_{(n+j)} Y(b,w)$. We then apply (\ref{eq:1.11}) to obtain (\ref{eq:1.17}).

A special case of (\ref{eq:1.17}), for $n=0$, is the commutator
formula $(a,b \in V)$:
\begin{equation}
  \label{eq:1.19}
  [Y (a,z), Y (b,w)]=\sum_{j \in \ZZ_+} Y(a_{(j)}b,w)
     \partial_w^{j} \delta (z-w)/j!  \, .
\end{equation}

We thus arrive at the third definition of a vertex algebra:
\begin{definition}\label{def3}
A {\itshape vertex algebra} is a triple 
$(V,\vac ,Y)$, where $V$ and $\vac$ are as in
Section \ref{sec:1.4} and $Y$ is an injective state-field
correspondence $a \mapsto Y(a,z)$, such that (\ref{eq:1.9}) and
the Borcherds identity (\ref{eq:1.17}) hold.
\end{definition}

In order to prove that this definition implies that first one, and
for further use, we shall digress now to the important notion of
the normally ordered product of $\End V$-valued quantum fields
$a(z)= \sum_{n \in \ZZ} a_{(n)}z^{-n-1}$ and $b(z)$.  Let
\begin{displaymath}
  a(z)_+ = \sum_{n<0} a_{(n)} z^{-n-1} \, , \quad 
        a(z)_- =  \sum_{n \geq 0} a_{(n)}z^{-n-1}\, , 
\end{displaymath}
and define the \emph{normally ordered product} by 
\begin{equation}
  \label{eq:1.20}
   :  a(z) b(z)  : = a (z)_+ b(z) + p(a,b) b(z) a(z)_- \, .
\end{equation}
Using the formal Cauchy formulas
\begin{displaymath}
  \Res_z a(z) i_{z,w} \frac{1}{z-w}=a(w)_+ \, , \quad
     \Res_z a(z) i_{w,z} \frac{1}{z-w} =-a(w)_- \, , 
\end{displaymath}
and the observation that $\partial_z (a(z)_{\pm})=
(\partial_ z a (z))_{\pm}$, we deduce from (\ref{eq:1.6})
for $n \in \ZZ_+$:
\begin{equation}
  \label{eq:1.21}
  a(w)_{(-n-1)} b(w) = : (\partial_w^n 
       a(w))    b(w)  :/n! \, .
\end{equation}

Returning to the third definition, note that commutator formula
(\ref{eq:1.19}) implies locality of the pair $Y(a,z)$, $Y(b,z)$, since
$a_{(j)}b =0$ for $j \gg 0$ and $(z-w)^{j+1}\partial_w^j \delta (z-w)=0$.  
Next, notice that the
$n$\st{th} product formula (\ref{eq:1.11}) follows from the
Borcherds identity by taking $\Res_z$ of both sides.  Letting
$b=\vac$ in (\ref{eq:1.11}) and using (\ref{eq:1.9}) and
(\ref{eq:1.6}), (\ref{eq:1.21}), we obtain:
\begin{eqnarray}
  \label{eq:1.22}
  Y (a_{(n)} \vac, z) =0\ ,\ \ \hbox{  for  } n \in \ZZ_+\, ,\\
  \label{eq:1.23}
Y (a_{(-n-1)} \vac ,z) = \partial_z ^n
    Y (a,z)/n!\ ,\ \  \hbox{  for  } n \in \ZZ_+ \, .
\end{eqnarray}
It follows by injectivity of the state-field correspondence that
\begin{equation}
  \label{eq:1.24}
  a_{(n)} \vac = \delta_{n,-1}a\ ,\ \ \hbox{  for  } n \geq -1\, .
\end{equation}
The latter formula for $n=-1$ implies the completeness axiom of
the first definition.

Next, define $T \in \End V$ by (\ref{eq:1.12}).  Then $T \vac =0$
due to (\ref{eq:1.9}), which is the vacuum axiom of the first
definition.  Letting $n=1$ in (\ref{eq:1.23}), we obtain
(\ref{eq:1.10}), which implies that $Y(T^n a,z) =
\partial_z ^n Y (a,z)$.  Applying both sides of the
latter formula to $\vac$ and letting $z=0$ (which is possible due
to (\ref{eq:1.14})), we obtain
\begin{equation}
  \label{eq:1.25}
  Y(a,z) \vac = e^{zT}a \, .
\end{equation}
Applying both sides of the $n$\st{th} product identity to
$\vac$ and using (\ref{eq:1.25}), we obtain:
\begin{displaymath}
  e^{zT} (a_{(n)}b) = (Y (a,z)_{(n)} Y (b,z))\vac \, .
\end{displaymath}
Differentiating both sides by $z$ and letting $z=0$, we obtain,
using (\ref{eq:1.10}) and (\ref{eq:1.24}):
\begin{equation}
  \label{eq:1.26}
  T (a_{(n)} b) = (Ta)_{(n)} b + a_{(n)} Tb \, .
\end{equation}
Finally, the equivalent form (\ref{eq:1.3}) of the translation
covariance axiom follows from (\ref{eq:1.26}) and the formula
\begin{equation}
\label{eq:1.27}
  (Ta)_{(n)} = -na_{(n-1)}\, ,
\end{equation}
which is an equivalent form of
(\ref{eq:1.10}).  This completes the proof of the equivalence of
the third and the first definitions.

The vertex operators $Y(a,z)=\sum_{n \in \ZZ} a_{(n)}z^{-n-1}$
allow one to define $n$\st{th} bilinear products on $V$ for
each $n \in \ZZ$ by the obvious formula $a_{(n)}b =$ the operator
$a_{(n)}$ applied to vector $b$.  Recall that we have (see
(\ref{eq:1.9}) and (\ref{eq:1.24})):
\begin{equation}
  \label{eq:1.28}
  \vac_{(n)} a=\delta_{n,-1} a \hbox{  for  } n \in \ZZ\, , \quad
  a_{(n)} \vac =\delta_{n,-1} a \hbox{  for  } n \geq -1 .
\end{equation}

Applying both sides of identity (\ref{eq:1.17}) to  $c\in V$ and
comparing coefficients of $z^{-m-1} w^{-k-1}$, we obtain the
original form of the Borcherds identity:
\begin{eqnarray}
  \label{eq:1.29}
  \sum_{j \in \ZZ_+} (-1)^j \binom{n}{j} \Big(a_{(m+n-j)}
  (b_{(k+j)}c)-(-1)^n p(a,b)b_{(n+k-j)} (a_{(m+j)}c)\Big)\\[-2ex]
\nonumber
= \sum_{j \in \ZZ_+} \binom{m}{j} (a_{(n+j)}b)_{(m+k-j)} c
\end{eqnarray}
for all $a,b, c \in V$, $m,n,k \in \ZZ$.

It is easy to see that the third definition of the vertex algebra
is essentially the same as the original Borcherds definition as a
vector (super)space $V$ with a vacuum vector $\vac$ and
$\CC$-bilinear products $a_{(n)}b$ for each $n \in \ZZ$, such
that (\ref{eq:1.28}) and (\ref{eq:1.29}) hold.

From this point of view all the standard algebra notions of a
homomorphism and isomorphism, subalgebras and ideals, carry over
to vertex algebras in the obvious way.  Note that the right
ideals are automatically $T$-invariant due to (\ref{eq:1.21}).
Also, right ideals and $T$-invariant left ideals are
automatically two-sided ideals, due to (\ref{eq:1.12}) and the simple skewsymmetry relation
(see \cite{FHL}, \cite[(4.2.1)]{K}):
\begin{equation}
  \label{eq:1.30}
  Y (a,z)b = p(a,b) e^{zT} Y(b,-z)a \, .
\end{equation}

\vspace{-2ex}
\subsection{Lie conformal algebras and the fourth definition of a vertex algebra}
\label{sec:1.7} 
For the fourth definition of a vertex algebra we need the
definition of  a Lie conformal algebra \cite{K,D'K}.

\begin{definition}
  \label{def:1.6}
A {\itshape Lie conformal superalgebra} is a $\ZZ /2\ZZ$-graded $\CC
[T]$-module $R=R_{\bar{0}} \oplus R_{\bar{1}}$, endowed with a
parity preserving $\CC$-bilinear $\lambda$-bracket $R \otimes R \to \CC [\lambda]
\otimes R$, denoted by $[a_{\lambda}b]$, such that the following
axioms hold:

\vspace*{-1ex}
\begin{table*}[h]
  \begin{tabular}{ll}
(sesquilinearity) & $[T a_{\lambda}b] = -\lambda [a_{\lambda} b]$, 
     $T[a_{\lambda}b] = [Ta_{\lambda}b] + [a_{\lambda}Tb]$,\\%[1ex]
(skewsymmetry)  & $[b_{\lambda}a] =-p(a,b) 
         [a_{-\lambda -T}b]$,\\%[1ex]
(Jacobi identity) & $[a_{\lambda}[b_{\mu}c]] -
    p(a,b) [b_{\lambda} [a_{\mu}c]] =
    [[a_{\lambda}b]_{\lambda + \mu}c]$.
  \end{tabular}
\end{table*}
\end{definition}

One writes $[a_{\lambda}b] = \sum_{n \in \ZZ_+}
\frac{\lambda^n}{n!} (a_{(n)}b)$, where the sum is finite, and
the bilinear products $a_{(n)}b$ are called the $n$\st{th}
products of $R$.  The expression $[a_{-\lambda-T}b]$ in the
skewsymmetry relation is interpreted as $\sum_{j \in \ZZ_+}
\frac{(-\lambda -T)^n}{n!} (a_{(n)}b)$.  Of course, $\lambda$ and
$\mu$ are (even) indeterminates.
Thus, the skewsymmetry relation is equivalent to the following identities ($n\in\Z_+$):
\begin{equation}\label{n-skew}
b_{(n)}a\ =\ p(a,b)\sum_{j\in\Z_+}(-1)^{n+j+1}T^{j}(a_{(n+j)}b)/j!\ .
\end{equation}
Similarly, the Jacobi identity is equivalent to the following identities ($m,n\in\Z_+$):
\begin{equation}\label{n-jac}
a_{(m)}(b_{(n)}c)-p(a,b)b_{(n)}(a_{(m)}c)\ =\ \sum_{i\in\Z_+}\binom{m}{i} (a_{(i)}b)_{(m+n-i)}c\ .
\end{equation}

Define the \emph{formal Fourier transform} by 
\begin{displaymath}
    F^{\lambda}_z a(z) = \Res_z e^{\lambda z}a(z) \, .
\end{displaymath}
It is a linear map from the space of $U$-valued formal
distributions to $U[[\lambda]]$, and it has the following
properties, which are immediate to check:
\begin{eqnarray}
  \label{eq:1.31}
  F^{\lambda}_z \partial_z a(z) 
       &=& -\lambda F^{\lambda}_z a(z)  \, \\
 \label{eq:1.32}
 F^{\lambda}_z (e^{zT} a(z)) &=&
 F^{\lambda+T}_z a(z) \hbox{  if  }a (z) \in U((z))\, , \\
 \label{eq:1.33}
   F^{\lambda}_z a(-z) &=& -F^{-\lambda}_z a (z)\, , \\
 \label{eq:1.34}
  F^{\lambda}_z  \partial^n_w \delta (z-w)
      &=& e^{\lambda w}  \lambda^n \, .
\end{eqnarray}

Note that (\ref{n-skew}) for all $n\in\Z$ is an equivalent form of (\ref{eq:1.30}),
and (\ref{n-jac})  for all $m,n\in\Z$ is an equivalent form of (\ref{eq:1.19}).
This is not surprising in view of the following proposition \cite{K}.
\begin{proposition}
  \label{prop:1.7}

Define a $\lambda$-bracket on a vertex algebra $V$ by
\begin{displaymath}
  [a_{\lambda}b] = F^{\lambda}_z Y (a,z)b \, .
\end{displaymath}
Then the $\CC [T]$-module $V$ with this $\lambda$-bracket is a
Lie conformal algebra.
\end{proposition}

\begin{proof}
  
The sesquilinearity relations follow from
(\ref{eq:1.27}) and (\ref{eq:1.26}) 
for $n \in \ZZ_+$.  Applying $F^{\lambda}_z$ to
both sides of (\ref{eq:1.30}) and using (\ref{eq:1.32}) and
(\ref{eq:1.33}), we get the skewsymmetry relation.
In order to prove the Jacobi identity, apply
$F^{\lambda}_z$ to the commutator formula (\ref{eq:1.19}),
applied to $c$, and use (\ref{eq:1.34}) to obtain:
\begin{equation}
  \label{eq:1.35}
  [a_{\lambda} Y (b,w)c] = p(a,b) Y (b,w) [a_{\lambda}c]
     + e^{\lambda w} Y([a_{\lambda}b], w)c \, .
\end{equation}
Applying $F^{\mu}_w$ to both sides of this formula, we get
the Jacobi identity.

\end{proof}

Next, define the \emph{normally ordered product} on $V$ by
\begin{displaymath}
  : ab : = a_{(-1)}b \quad (=\Res_z z^{-1} Y (a,z)b) 
\end{displaymath}
and recall that $a_{(0)}=\Res_z Y (a,z)$.  The following
proposition is an immediate corollary of identity (\ref{eq:1.35}).

\begin{proposition}
  \label{prop:1.8}
\alphaparenlist
\begin{enumerate}
\item %%a
The operator $a_{(0)}$ is a derivation of all $n$\st{th} products of $V$, and
it commutes with $T$. Consequently, $Ker\, a_{(0)}$ is a subalgebra of the 
vertex algebra $V$, and its intersection with $Im \, a_{(0)}$ is an ideal of
$Ker\, a_{(0)}$.

\item %%b
One has the following identity for all $a,b,c \in V$:
\begin{equation}
  \label{eq:1.36}
  [a_{\lambda}:bc:] = : [a_{\lambda}b]c : + p(a,b) :
  b[a_{\lambda}c]:  + \int^{\lambda}_0 [[a_{\lambda}b]_{\mu}c]\,
  d\mu \, . 
\end{equation}
\end{enumerate}
\end{proposition}

\begin{proof}
The first (resp. second) part of (a) follows from (\ref{eq:1.35})
(resp. the second sesquilinearity relation) by letting $\lambda
=0$.  (b) is obtained from (\ref{eq:1.35}) by comparing
coefficients of $w^0$.
\end{proof}

Formula (\ref{eq:1.36}) is called in \cite{K} the
\emph{non-commutative Wick formula}.  It is extremely useful in
calculations of the commutators of normally ordered products of
quantum fields.

Skewsymmetry of the $\lambda$-bracket and (\ref{eq:1.36}) imply
the right non-commutative Wick formula \cite{BK}:
\begin{equation}
  \label{eq:1.37}
  [:a b:_{\lambda} c] =
  : (e^{T\partial_{\lambda}}a)
     [b_{\lambda}c]: + p (a,b) 
  : (e^{T\partial_{\lambda}}b)
     [a_{\lambda}c]: + p(a,b) \int^{\lambda}_0
     [b_{\mu} [a_{\lambda -\mu}c]]\, d\mu \, .
\end{equation}

The normally ordered product is not commutative, however, taking
the coefficient of $z^0$ of the skewsymmetry relation
(\ref{eq:1.30}), we obtain the ``quasicommutativity'', which can
be written in the following form:
\begin{equation}
  \label{eq:1.38}
  : ab : -p(a,b) : ba : = \int^0_{-T} [a_{\lambda}b]\,
  d\lambda\, .
\end{equation}
Likewise, this product is not associative, but, taking the
coefficient of $z^0$ of the $-1$\st{st} product  identity,
i.e. (\ref{eq:1.11}) for $n=-1$, we obtain the
``quasi-associativity'', which can be written in the following
form:
\begin{equation}
  \label{eq:1.39}
  :: ab :c: -:a: bc:: =: (\int^T_0 \, d \lambda a)
     [b_{\lambda}c]: + p(a,b) : (\int^T_0 \, d\lambda b)
     [a_{\lambda}c]: \, .
\end{equation}
The integrals in formulas (\ref{eq:1.38}) and (\ref{eq:1.39}) are
interpreted as follows:  expand the $\lambda$-bracket and put the powers
of $\lambda$ on the left, under the sign of integral, then take
the definite integral by the usual rules inside the parenthesis.
For example we have $:(\int^T_0\, d\lambda a) [b_{\lambda}c]:\ =\
:(\int^T_0 \sum_{j \geq 0} \frac{\lambda^j}{j!} 
      d\lambda a) (b_{(j)}c):\ =\ \sum_{j \geq 0} : 
      \frac{T^{j+1}a}{(j+1)!} (b_{(j)}c):\
      =\sum_{j \geq 0} a_{(-j-2)} (b_{(j)}c)$.

We thus arrive at the fourth definition of a vertex algebra \cite{BK}.
\begin{definition}\label{def4}
A vertex algebra is a quintuple $(V, \vac, T , 
[ \, . \, {}_{\lambda} \, . \, ], :\,:)$, where
\romanparenlist
\begin{enumerate}
\item %%i
$(V,T, [\, .\, {}_{\lambda} \, . \, ])$ is a Lie conformal
superalgebra,
\item %%ii
  $(V,\vac ,T , : \, :)$ is a unital differential (i.e. $T$ is a
  derivation) superalgebra, satisfying (\ref{eq:1.38}) and
  (\ref{eq:1.39}),
\item %%iii
  the $\lambda$-bracket $ [\, .\, {}_{\lambda} \, . \, ]$ and the
  product $: \, :$ are related by the non-commutative Wick
  formula (\ref{eq:1.36}).
\end{enumerate}
\end{definition}

We have shown that the fourth definition follows from the
previous ones.  A proof of the fact that the fourth definition
implies the first one can be found in \cite{BK}.

\begin{remark}\label{weak}
It is proved in \cite{BK} that one can replace the quasi-associativity axiom (\ref{eq:1.39})
with a weaker version of it, namely the condition that the associator $:(:ab:)c:-:a(:bc:):$
is symmetric in $a$ and $b$, or, equivalently,
\begin{equation}\label{eq:1.39-weak}
:a(:bc:):-p(a,b):b(:ac:):\ =\ :(:ab:)c:-p(a,b):(:ba:)c:\ .
\end{equation}
\end{remark}

\subsection{Conformal weight}
\label{sec:1.8} 

Let $(V,\vac ,T,Y)$ be a vertex algebra.  Recall that a
diagonalizable operator $H$ on the vertex superspace $V$ is
called a \emph{Hamiltonian operator} if equation~(\ref{eq:0.2})
holds; the eigenspace decomposition for $H$:
\begin{equation}
  \label{eq:1.40}
  V= \oplus_{\Delta \in \CC} V [\Delta]
\end{equation}
is called the $H$-\emph{grading} of $V$.  If $a$ is an
eigenvector of $H$, its eigenvalue is called the \emph{conformal
  weight} of $a$ and is denoted by $\Delta_a$ or $\Delta(a)$.

\begin{proposition}
  \label{prop:1.10}
A decomposition (\ref{eq:1.40}) is an $H$-grading of  $V$ if and only if
either of the following two equivalent properties holds:
\romanparenlist
\begin{enumerate}
\item %%i
If for $a \in V [\Delta]$ we write $Y (a,z)=\sum_{a \in \ZZ -\Delta} a_n
z^{-n-\Delta}$, then
\begin{equation}
  \label{eq:1.41}
  [H, a_n] = -na_n \, .
\end{equation}
\item %%ii
  If $a \in V [\Delta_a]$, $b \in V [\Delta_b]$, $n \in \ZZ$,
  then 
  \begin{equation}
    \label{eq:1.42}
    \Delta_{a_{(n)}b}  = \Delta_a + \Delta_b -n-1 \, .
  \end{equation}
\end{enumerate}
\end{proposition}
\begin{proof}
It is immediate to see that (\ref{eq:0.2}) and (\ref{eq:1.41})
are equivalent.  
We have, by definition,
\begin{equation}
 \label{eq:1.44}
  a_{(n)}=a_{n-\Delta_a +1}\ ,\ \  n\in\Z\ ,\ \ 
  a_{m}=a_{(m+\Delta_a -1)}\ ,\ \  m\in\Z-\Delta_a\ .
\end{equation}
Hence (\ref{eq:1.42}) is equivalent to
$$
\Delta_{a_mb}\ =\ \Delta_{a_{(m+\Delta_a-1)}b}\ =\ \Delta_b-m\ ,\ \ m\in\Z-\Delta_a\ ,
$$
which is in turn equivalent to (\ref{eq:1.41}), by applying both sides to $b$.
\end{proof}
\begin{remark}
  \label{rem:1.11}
Notice that, using (\ref{eq:1.12}) and (\ref{eq:1.13}), we get from (\ref{eq:1.42}) that
\begin{equation}
  \label{eq:1.43}
  \Delta_{\vac} = 0\, , \quad \Delta_{Ta} = \Delta_a +1\, .
\end{equation}
Formulas  (\ref{eq:1.42}) and  (\ref{eq:1.43}) show that $e^{2\pi
iH}$ is an automorphism of the vertex algebra $V$. (However, the $H$-grading
is not a vertex algebra grading.)
\end{remark}

In the most important case, when the Hamiltonian operator comes from an
energy-momentum field $Y(L,z)$, i.e. $H=L_0$ (see
Subsection \ref{sec:0.3}), its eigenvector $a \in V$ has
conformal weight $\Delta$ if and only if
\begin{equation}
  \label{eq:1.45}
  [L_{\lambda}a] = (T+\Delta \lambda) a + O(\lambda^2)\, .
\end{equation}
In the case when $[L_{\lambda}a] = (T+\Delta \lambda)a$, the
vector $a$ and the corresponding vertex operator $Y(a,z)$ are
called \emph{primary}.  This is equivalent to $L_n a =
\delta_{n,0}\Delta a$ for all $n \in \ZZ_+$.

Taking the formal Fourier transform of both sides of (\ref{eq:0.2}) applied to $b$,
we obtain the definition of a Hamiltonian operator on a Lie conformal algebra $R$,
as a diagonalizable operator $H$ on $R$ such that
\begin{equation}\label{29set}
[Ha\ _\lambda\ b]+[a\ _\lambda\ Hb]\ = \ (H+\lambda\partial_\lambda+I_R)[a\ _\lambda\ b]\ .
\end{equation}
The most important example comes from a conformal derivation $L$ of $R$,
such that $L_{(0)}=T$, and $L_{(1)}=H$ is diagonalizable (cf. (\ref{eq:1.45})).

\subsection{Universal enveloping vertex algebras of Lie conformal algebras}
\label{sec:1.9} 

A fairly general and very important class of vertex algebras is
obtained from regular formal distribution Lie algebras as
their universal enveloping vertex algebras \cite{K}.

Recall that a formal distribution Lie (super)algebra is a pair
$(\fg ,\F)$ where $\fg$ is a Lie (super)algebra and $\F$ is a
local family of $\fg$-valued formal distributions $\{ a^j (z)
=\sum_{n \in \ZZ} a^j_{(n)}z^{-n-1}\}_{j \in J}$, whose
coefficients $a^j_{(n)}$ span $\fg$. 
Recall that locality means
that $(z-w)^{N_{ij}}[a^i(z), a^j (w)]=0$ for some $N_{ij}\in\Z_+$ and all $i,j \in J$. 
The formal distribution Lie superalgebra $(\g,\F)$
is called regular if there exists a derivation $T$ of the
algebra $\fg$, such that
\begin{equation}
  \label{eq:1.46}
  Ta^j(z) = \partial_z a^j (z) \, , \quad j \in J \, .
\end{equation}
Due to Dong's lemma (see e.g. \cite{K}, Lemma~3.2), the minimal family
$\bar{\F}$ of $\g$-valued formal distributions
containing $\F$, which is $\partial_z$-invariant and closed under all
products (cf.~(\ref{eq:1.6}))
\begin{displaymath}
  a(w)_{(j)} b(w) = \Res_z (z-w)^j [a(z), b(w)]\, ,
     \quad j \in \ZZ_+ \, ,
\end{displaymath}
is again local.  Hence, by (\ref{eq:1.14}) and (\ref{eq:1.15}) we
have for each pair $a(z), b(w) \in \bar{\F}$:
\begin{equation}
  \label{eq:1.47}
  [a(z), b(w)] =\sum_{j \in \ZZ_+} (a(w)_{(j)}b(w))
    \partial_w ^{j} \delta (z-w)/j! \, ,
\end{equation}
where the sum is finite by locality.
In terms of coefficients, equation (\ref{eq:1.47}) can be written as
follows:
\begin{equation}
  \label{eq:1.48}
  [a_{(m)}, b_{(n)}] = \sum_{j \in \ZZ_+} \binom{m}{j}
     (a_{(j)} b)_{(m+n-j) \, ,}
\end{equation}
where $(a_{(j)}b)_{(n)}=\Res_w w^n a(w)_{(j)}b(w)$.

Denote by $\fg_-\subset\g$ the span of all $a_{(m)}$ with $a(z) \in
\bar{\F}$, $m \in \ZZ_+$.  It follows from (\ref{eq:1.48}) that
$\fg_-$ is a subalgebra of the Lie superalgebra $\fg$ (it is
called the \emph{annihilation subalgebra}) and, moreover, since
(\ref{eq:1.46}) for $a(z)$ means that $T (a_{(n)}) =-na_{(n-1)}$,
$\fg_-$ is a $T$-invariant subalgebra.  Let $V \equiv V (\fg , \F) =
U(\fg)/U (\fg)\fg_-$, and let $\vac$ denote the image of $1 \in U (\fg)$ in
$V$.  Extending $T$ to a derivation of $U(\fg)$, we get the induced
endomorphism of $V$, which we again denote by $T$.  Finally, each
$\fg$-valued formal distribution $a^j(z) \in \F$ induces an $\End
V$-valued distribution, which we again denote by $a^j (z)$, via
the representation given by
the left multiplication, so that (\ref{eq:1.47}) still holds.
Hence $\F = \{ a^j (z)\}_{j \in \ZZ}$ is again a local system.
Moreover, the $a^j (z)$ are quantum fields.  Indeed, for each
monomial $v=a^{j_1}_{(n_1)} \ldots a^{i_s}_{(n_s)} \vac \in V$
one easily checks by induction on $s$, using (\ref{eq:1.48}),
that $a^j_{(n)}v=0$ for $n \gg 0$.  Thus, all axioms of the first
definition of a vertex algebra hold, and we obtain

\begin{theorem}
  \label{th:1.12}
$(V(\fg ,\F), \vac ,T,\F)$ is a vertex algebra.
\end{theorem}

The vertex algebra $V(\fg ,\F)$ is called \emph{the universal
enveloping vertex algebra} of the regular formal distribution Lie
algebra $(\fg ,\F)$.

\begin{remark}
  \label{rem:1.13}
Note that $\bar{\F}$ is a $T$-module.  Define a $\lambda$-bracket
on $\bar{\F}$ by (cf.~(\ref{eq:1.47})):
$ [a_{\lambda} b] = \sum_{j \in \ZZ_+} \frac{\lambda^j}{j!}
    (a_{(j)}b)$.
It is not difficult to show (see \cite{K}) that we thus define on
$\bar{\F}$ a structure of a Lie conformal algebra.
\end{remark}

A special case of the above construction is the universal
enveloping vertex algebra $V(R)$ of a Lie conformal algebra
$(R,T, [\, . \, {}_{\lambda} \, . \, ])$.  First we construct the
\emph{maximal} formal distribution Lie algebra $(\Lie R , R)$,
associated to $R$, by letting \cite{K}:
\begin{displaymath}
  \Lie R = R[t,t^{-1}] / (T + \partial_t) R[t,t^{-1}]
\end{displaymath}
with the bracket given by (\ref{eq:1.48}), where $a_{(n)}$ stands
for the image of $at^n$ in $\Lie R$, and the family of $\Lie
R$-valued formal distributions
$\{ a(z) = \sum_{n \in \ZZ} a_{(n)} z^{-n-1}\}_{a \in R}$.
Then $(\Lie R ,R)$ is regular since $T (=-\partial_t)$ on 
$\Lie R$ satisfies (\ref{eq:1.46}), and the image
of $R[t]$ in $\Lie R$ is the annihilation
subalgebra.

Then $V(R) = V(\Lie R ,R)$ is called the \emph{universal enveloping
vertex algebra} of the Lie conformal algebra $R$.

It is easy to show that the map $a \mapsto at^{-1}\vac$
defines a Lie conformal algebra embedding: $R \hookrightarrow V(R)$.
Moreover, it is easy to see that in $V(R)$:
\begin{equation}
  \label{eq:1.49}
  : aB : =aB \,\hbox{  if  } a \in R \, , \quad
  B \in V(R) \, ,
\end{equation}
where the product on the right is the product of $U (\Lie R)$.

\begin{definition}
  \label{def:1.14}
A set of elements $\B=\big\{a_i\,|\,i\in\I\big\}$
\emph{strongly generates} a vertex algebra $V$ if
the monomials $: a_{j_1} a_{j_2} \ldots a_{j_s}:$ span $V$. 
Here and further the normally ordered product of $s>2$ (resp. $s=0$) elements 
is understood from right to left (resp. is $\vac$). 
An ordered set of elements $\B=\big\{a_i\,|\,i\in\I\big\}$
\emph{freely generates} $V$ if the monomials 
$:a_{j_1}\dots a_{j_s}:$ with $j_r \leq j_{r+1}$, 
and $j_r < j_{r+1}$ when $p(a_{j_r}) = \bar{1}$, $1
\leq r<s$, form a basis (over $\CC$) of $V$.
\end{definition}

\begin{proposition}
    \label{prop:1.15}
\alphaparenlist
\begin{enumerate}
\item %%a
The inclusion $R \hookrightarrow V (R)$ satisfies the 
universality property , similar to that of 
the inclusion of a Lie algebra in
its universal enveloping algebra.
\item %%b
Any ordered basis (over $\CC$) of $R$ freely generates $V(R)$ .
\end{enumerate}
\end{proposition}

\begin{proof}
It follows from the corresponding properties of the universal
enveloping algebras of Lie (super)algebras.
For example, in order to prove (b), we use the decomposition
$$
\Lie R\ =\ Rt^{-1}\oplus(\Lie R)_-\ ,
$$
as direct sum of subspaces, which follows by the identity $at^{-n-1}=(T^na/n!)t^{-1}$ in $(\Lie R)_-$.
Hence $R(\simeq Rt^{-1})$ freely generates $V(R)\simeq U(Rt^{-1})$ by the PBW Theorem
for ordinary Lie algebras.
\end{proof}

An equivalent construction of $V(R)$, purely in terms of the Lie
conformal algebra $R$, is as follows \cite{BK,GMS}.
Denote by $R_L$ the Lie (super)algebra structure on $R$, defined
by (cf.~(\ref{eq:1.38})) $[a,b]=\int^0_{-T} [a_{\lambda}b]\, d
\lambda$, and let $U(R_L)$ be its universal enveloping
(super)algebra.  Then $U(R_L)$ carries a unique structure of a
vertex algebra with $\vac =1$, with $T$, extended from $R_L$ to
$U(R_L)$ as a derivation, the $\lambda$-bracket defined on $R_L$
as on $R$, the normally ordered product defined by
(cf.~(\ref{eq:1.49}))
\begin{displaymath}
  : aB : =aB \hbox{  if  } a \in R_L \, , 
      \quad  B \in U (R_L)\, , 
\end{displaymath}
and the $\lambda$-bracket
 and normally ordered product extended to $U (R_L)$ by
 (\ref{eq:1.36}), (\ref{eq:1.37}), (\ref{eq:1.38}),
 (\ref{eq:1.39}).
The vertex algebra isomorphism $U(R_L)
\overset{\sim}{\rightarrow} V(R)$
 is induced by the map $a \mapsto a_{(-1)} \vac$, $a \in R_L$.

\begin{remark}
It is easy to see that a Hamiltonian operator on a Lie conformal algebra $R$, 
defined by (\ref{29set}), extends uniquely to a Hamiltonian operator of the vertex algebra $V(R)$.
\end{remark}

\subsection{Three examples of vertex algebras and energy-\\momentum fields}
\label{sec:1.10} 

The most important examples of regular formal distribution Lie
superalgebras and the associated universal enveloping vertex
algebras are the following three.

\subsubsection*{Universal affine vertex algebra}

Let $\fg = \fg_{\bar{0}} + \fg_{\bar{1}}$ be a simple
finite-dimensional Lie superalgebra with a non-degenerate
supersymmetric invariant bilinear form $(\, . \, | \, . \, )$.
Recall that ``supersymmetric'' means that the form  $(\, . \, |
\, . \, )$ is symmetric (resp. skewsymmetric) on $\fg_{\bar{0}}$
(resp. $\fg_{\bar{1}}$) and    $(\fg_{\bar{0}}
|\fg_{\bar{1}})=0$, and ``invariant'' means that $([a,b]|c)=
(a|[b,c])$.

The \emph{Kac--Moody affinization} is the formal distribution Lie
superalgebra $(\hat{\fg}, \F)$, where $\hat{\fg} = \fg
[t,t^{-1}] \oplus\CC K$ is the Lie superalgebra with parity given by 
$p(t) =p(K) =\bar{0}$, and Lie bracket $(a,b \in \fg, \, \, m,n \in \ZZ)$
\begin{equation}
  \label{eq:1.50}
  [at^m , bt^n]= [a,b]t^{m+n} +m (a|b) \delta_{m,-n}K\, , 
       \quad [K,\hat{\fg}]=0 \, ,
\end{equation}
and $\F$ is the following collection of $\hat\g$-valued formal distributions:
$$
\F = \Big\{ a(z) =\sum_n (at^n) z^{-n-1}\ \Big|\ a\in \fg\Big\} \cup \big\{K \big\}\ .
$$
Notice that the collection $\F$ is local since an equivalent form of
(\ref{eq:1.50}) is
\begin{equation}
  \label{eq:1.51}
   [a(z) ,b(w)] = [a,b](w) \delta (z-w)+ (a|b) K
      \partial_w \delta (z-w)\, , \quad
        [a(z),K]=0 \, .
\end{equation}
The pair $(\hat{\fg}, \F)$ is regular with $T=-\frac{d}{dt}$.
Hence we have the universal enveloping vertex algebra
$V(\hat{\fg}, \F)$.

Given $k \in \CC$, the subspace $(K-k)V(\hat{\fg}, \F)$ is an ideal of $V
(\hat{\fg},\F)$.  The vertex algebra
\begin{displaymath}
  V^k (\fg) = V (\hat{\fg}, \F) /(K-k)V (\hat{\fg},F)
\end{displaymath}
is called the \emph{universal affine} vertex algebra of
\emph{level} $k$.
Note that the  Lie conformal algebra associated to $(\hat{\fg},\F)$ is 
\begin{displaymath}
  \Cur \fg = (\CC [T] \otimes \fg) \oplus \CC K \ ,\ \  [a_{\lambda}b]
     = [a,b] + \lambda (a|b)K\ ,\ \ [a_{\lambda}K]=0=[K_\lambda K]\ ,
\end{displaymath}
for  $a,b \in 1 \otimes \fg$.
It is called the current Lie conformal algebra.  It is easy
to see that
\begin{equation}
  \label{eq:1.52}
  V(\Cur \fg) = V(\hat{\fg}, \F)\, .
\end{equation}

\subsubsection*{Femionic vertex algebra}

Let $A=A_{\bar{0}} + A_{\bar{1}}$ be a vector superspace
with a non-degenerate skew-supersymmetric bilinear form
$\langle \, . \, , \, .\, \rangle$.  Recall that
``skew-supersymmetric'' means supersymmetric if the parity of $A$
is reversed.
The \emph{Clifford affinization} is the formal distribution Lie
superalgebra $(\hat{A} , \F)$, where
\begin{displaymath}
  \hat{A} = A [t,t^{-1}] \oplus \CC K \, , \quad
  \F = \Big\{ a(z) = \sum_{n\in\Z} (at^n)z^{-n-1}\ \Big|\ a \in A\Big\}
       \cup \big\{ K \big\} \, , 
\end{displaymath}
and the bracket on $\hat{A}$ is defined by
\begin{equation}
  \label{eq:1.53}
  [at^m , bt^n] = \langle a,b \rangle \delta_{m, -n-1}K\, , 
  \quad [K,\hat{A}]=0 \, , 
\end{equation}
or, equivalently, by
\begin{equation}
  \label{eq:1.54}
  [a(z),b(w)] = \langle a,b \rangle \delta (z-w)K\, , 
     \quad [a(z) ,K]=0 \, .
\end{equation}
For the same reason as above, $(\hat{A}, \F)$ is a regular formal
distribution Lie superalgebra, hence we obtain the universal
enveloping vertex algebra $V(\hat{A},\F)$.  The vertex algebra
\begin{displaymath}
  F(A) = V (\hat{A},\F)/(K-1)V(\hat{A},\F)
\end{displaymath}
is called the \emph{fermionic} vertex algebra, attached to
$(A,\langle\, . \, , \, . \, \rangle )$.
\begin{remark}
It is important to point out that the properties of $V^k(\fg)$
depend essentially on $k$ (for example it is simple for generic
$k$, but not simple for a certain infinite set of values of $k$),
whereas all $F^k(A)$ are isomorphic (and simple) if $k \neq 0$,
that is why we put $k=1$.
\end{remark}
The \emph{fermionic} Lie conformal algebra,  associated to $(\hat{A},\F)$, is
\begin{displaymath}
  (\CC [T] \otimes A) \oplus \CC K\ , \ \ [a_{\lambda}b]
     =\langle a,b \rangle K \ , \ \ [a_{\lambda}K]=0=[K_\lambda K]\ ,
\end{displaymath}
for $a,b \in 1\otimes A$,
and we have an isomorphism, similar to (\ref{eq:1.52}).

\subsubsection*{Universal Virasoro vertex algebra}

This example is constructed starting with the formal distribution
Lie superalgbera $(\Vir,\F)$, where $Vir=\big(\bigoplus_{n\in\Z}\C L_n\big)\oplus\C C$
is the Virasoro Lie algebra, with Lie bracket as in (\ref{eq:0.3}) with $c$ replaced by $C$,
and $\F = \{ L(z) =\sum_{n} L_n
z^{-n-2} \, , \, C \}$.  Locality holds since (\ref{eq:0.3}) is
equivalent to
\begin{equation}
  \label{eq:1.55}
  [L(z), L(w)] =\partial_w L(w) \delta (z-w) +2L (w)
     \partial_w \delta (z-w) +\frac{C}{12} 
     \partial_w ^3 \delta (z-w)\, ,
\end{equation}
and the pair $(\Vir, \F)$ is regular with $T=L_{-1}$.  The
vertex algebra $(c \in \CC)$
\begin{displaymath}
  V^c = V (\Vir ,\F) /(C-c) V (\Vir, \F)
\end{displaymath}
is called the universal Virasoro vertex algebra with central
charge $c$.

Note that the associated Lie conformal algebra, called the
Virasoro Lie conformal algebra, is
\begin{equation}
  \label{eq:1.56}
  (\CC [T] \otimes L) \oplus \CC C\ , \ \ 
  [L_{\lambda} L] = (T+2\lambda) L+\frac{\lambda^3}{12}C\ , \ \  
    [L_{\lambda}C]=0=[C_\lambda C] \, , 
\end{equation}
and an isomorphism similar to (\ref{eq:1.52}) holds.

\subsubsection*{Energy-momentum fields}

We conclude this section by constructing energy-momentum
fields $Y (L,z)$ for each of the above examples, so that
$H=L_0=L_{(1)}$ is a Hamiltonian operator (cf. Subsection \ref{sec:0.3}).

For $V^k (\fg)$ it is the well-known Sugawara construction (defined for $k\neq-h^{\vee}$)
\begin{equation}
  \label{eq:1.57}
  L^{\fg} = \frac{1}{2(k+h^{\vee})} \sum_{i\in I} : a^ia_i : \, ,
\end{equation}
where $\{ a_i \}_{i\in I}$ and $\{ a^i\}_{i\in I}$ is a pair of dual bases of
$\fg$, i.e. $(a_i | a^j) =\delta_{ij}$, we identify $a \in \fg$
with $(at^{-1}) \vac \in V^k (\fg)$, and $h^{\vee}$ is the
dual Coxeter number, i.e.
the eigenvalue of the Casimir operator $\sum_i a^ia_i$ on $\fg$ divided by 2.
Then $Y(L^{\fg},z)=\sum_n L^{\fg}_n z^{-n-2}$ is an energy-momentum
field with central charge
\begin{equation}
  \label{eq:1.58}
  c^{\fg}_k = \frac{k\, \sdim \fg}{k+h^{\vee}}\, , \hbox{  where  }
     \sdim  \fg = \dim \fg_{\bar{0}} - \dim \fg_{\bar{1}}\, .
\end{equation}
Moreover, all $a(z)$,  where $a\in \fg$, are
primary of conformal weight $1$.  This is easy to prove using the
non-commutative Wick formula (\ref{eq:1.36}).

\begin{remark}
The vertex algebra $V^k(\g)$ can be constructed as above for any Lie superalgbera $\g$
and a supersymmetric invariant bilinear form $(.|.)$.
Also, the Hamiltonian operator $H=L_0^\g$ can be constructed in this generality
for any $k$ by making use of the derivation $-t\frac{d}{dt}$ of $\hat\g$.
More generally, if $h$ is a diagonalizable derivation of $\g$, then a Hamiltonian operator
can be constructed by using the derivation $-t\frac d{dt}-\ad h$ of $\hat\g$.
In this case the field $a(z)$ has conformal weight $1-j$ if $[h,a]=ja,\ j\in\C$.
Note also that all derivations $-t^{n+1}\frac d{dt}$ of $\hat\g$ with $n\geq-1$
extend to the endomorphisms of the space $V^k(\g)$, which form the
"annihilation subalgebra" $\sum_{n\geq-1}\C L_n$ of the Virasoro algebra,
and one has:
$$
[L_n,Y(a,z)]\ =\ \sum_{j\geq-1}\binom{n+1}{j+1}Y(L_ja,z)z^{n-j}\ ,\ \ n\geq-1\ .
$$
\end{remark}

For $F(A)$ we let
\begin{equation}
  \label{eq:1.59}
  L^A = \frac{1}{2} \sum_{i\in I} : (T \Phi^i)\Phi_i: \, , 
\end{equation}
where $\{ \Phi_i \}_{i\in I}$ and $\{ \Phi^i \}_{i\in I}$ is a pair of dual bases
of $A$, i.e. $\langle \Phi_i , \Phi^i \rangle = \delta_{ij}$,
and we identify $\Phi \in A$ with $(\Phi t^{-1}) \vac$.  Then $Y
(L^A ,z)$ is an energy-momentum field with central charge
\begin{equation}
  \label{eq:1.60}
  c^A = - \frac{1}{2} \sdim A \, .
\end{equation}
All the fields $\Phi (z)$, where $\Phi \in A$, are primary of
conformal weight $1/2$.

Next, we consider the following important modification of this
construction.  Suppose that $A=A_+ \oplus A_-$, where both $A_+$
and $A_-$ are isotropic subspaces with respect to the bilinear
form $\langle \, . \, , \, .\, \rangle$.  Choose a basis $\{
  \varphi_i \}$ of  $A_+$ and the dual basis $\{ \varphi^i \}$ of
  $A_-$, so that $\langle \varphi_i , \varphi^j \rangle
  =\delta_{ij}$.  Then we can define a family of energy-momentum
  fields, parameterized by a collection $\vec{m} = (m_i)$ of
  complex numbers:
  \begin{equation}
    \label{eq:1.61}
    L^{A,\vec{m}}=-\sum_i m_i : \varphi^i (T \varphi_i) :
       + \sum_i (1-m_i) : (T\varphi^i) \varphi_i : \, .
  \end{equation}
The central charge of $Y (L^{A,\vec{m}},z)$ is equal 
\begin{equation}
  \label{eq:1.62}
  c^A_{\vec{m}} = \sum_i s(\varphi_i)
     (12 m^2_i -12m_i +2)\, ,
\end{equation}
and all the fields $\varphi^i (z)$ and $\varphi_i (z)$ are
primary of conformal weight $m_i$ and $1-m_i$, respectively.

Note that the vertex algebra $F (A=A_+ \oplus A_-)$ has an
important $\ZZ$-grading,
\begin{equation}
  \label{eq:1.63}
  F(A) = \oplus_{m \in \ZZ} F_m (A) \, ,
\end{equation}
called the \emph{charge decomposition}, which is defined by
\begin{equation}
  \label{eq:1.64}
  \charge \,\vac =0 \, , \quad
  \charge \,\varphi^i (z) =1\, ,\quad
  \charge \,\varphi_i (z) =-1\, .
\end{equation}

Finally, for the Virasoro vertex algebra $V^c$ the
energy-momentum field is, of course, $L(z) = Y(L_{-2} \vac, z)$,
with central charge~$c$.

%%%
\subsection{The indefinite integral definition of a vertex algebra}
\label{sec:1.11}

In this section we introduce a new, fifth definition of vertex algebras.
In some sense it just consists in a change of notation in the
fourth definition discussed in Section \ref{sec:1.7},
but it is much more elegant, as it encodes all axioms, such as
formulas (\ref{eq:1.36}), (\ref{eq:1.37}), (\ref{eq:1.38}) and (\ref{eq:1.39}),
in the axioms of a Lie conformal algebra (see Definition \ref{def:1.6})
written in "integral form".

In a nutshell, the fifth definition says that a vertex algebra is a Lie conformal algebra
in which the $\lambda$-bracket can be "integrated".
More precisely we have
\begin{definition}\label{def5}
A vertex algebra is a  $\ZZ /2\ZZ$-graded $\CC[T]$-module $V$,
endowed with an even element $\vac\in V_{\bar{0}}$ 
and an {\itshape integral} $\lambda$-{\itshape bracket},
namely a linear map $V \otimes V \to \CC [\lambda] \otimes V$, 
denoted by $\int^\lambda dx[a\ _x\ b]$, such that 
the following axioms hold:
\begin{description}
\item (unity)
$$
\int^\lambda dx[\vac\ _x\ a]=\int^\lambda dx[a\ _x\ \vac]=a\ ,
$$
\item (sesquilinearity)
\begin{equation}\label{ax1}
\int^\lambda\!\! dx\, [T a\ _x\ b] = -\int^\lambda\!\! dx\,x[a\ _x\ b]\ ,\ \  
\int^\lambda\!\! dx\,[a\ _x\ Tb] = \int^\lambda\!\!dx\, (T+x)[a\ _x\ b]\ ,
\end{equation}
\item (skewsymmetry)
\begin{equation}\label{ax2}
\int^\lambda\!\! dx\,[b\ _x\ a] =-p(a,b)\int^\lambda\!\! dx\,[a\ _{-x -T}\ b]\ ,
\end{equation}
\item (Jacobi identity)
\begin{equation}\label{ax3}
\int^\lambda\!\! dx\!\! \int^\mu\!\! dy \Big\{
 [a\,_x[b\,_y\,c]] - p(a,b) [b\,_y[a\,_x\,c]] - [[a\,_x\,b]_{x+y}\,c] \Big\} = 0\ . 
\end{equation}
\end{description}
\end{definition}

Before showing that Definition \ref{def5} is equivalent to Definition \ref{def4},
some words of explanation are needed.
In particular, not all expressions above are immediately defined in terms of the integral
$\lambda$-bracket $\int^\lambda dx[a\ _x\ b]$,
but in some sense they are "canonically" defined, since there is a unique way
to make standard "integral manipulations", such as change of variables and change of order
of integration, to rewrite them as expressions which make sense, i.e. which are well defined
only in terms of the integral $\lambda$-bracket.
In all manipulations it is convenient to think of $\int^\lambda\!\! dx[a\ _x\ b]$ as being a definite
integral with lower bound $-\infty$.
Moreover, not all relations above are in fact axioms: some of them should be considered
as defining relations for some new notation.
This is best understood if we write the integral $\lambda$-bracket as
$\int^\lambda dx[a\ _x\ b]=I_\lambda(a,b)$,
and we try to write all relations above in terms of $I_\lambda$ only.

The first sesquilinearity relation (\ref{ax1}) includes a piece of notation,
\begin{equation}\label{sesq-1.1}
\int^\lambda dx\, x[a\ _x\ b]\ :=\ -I_\lambda(Ta,b)\ ,
\end{equation}
and an axiom,
\begin{equation}\label{sesq-1.2}
\frac{d}{d\lambda}I_\lambda(Ta,b)\ =\ -\lambda\frac{d}{d\lambda}I_\lambda(a,b)\ .
\end{equation}
The latter follows by the "fundamental Theorem of calculus",
which is encoded in the integral notation.
Using (\ref{sesq-1.1}), the second sesquilinearity relation (\ref{ax1}) is equivalent to
\begin{equation}\label{sesq-2}
TI_\lambda(a,b)\ =\ I_\lambda(Ta,b)+I_\lambda(a,Tb)\ .
\end{equation}
Notice that here we are implicitly assuming that the action of $T$ commutes with taking integrals.
The right hand side of the skewsymmetry relation (\ref{ax2}) is defined by a
change of variable $y=-x-T$:
$$
\int^\lambda\!\! dx\,[a\ _{-x-T}\ b]\ =\ -\int^{-\lambda-T}\!\! dy\,[a\ _y\ b]\ =\ -I_{-\lambda-T}(a,b)\ .
$$
Hence, in terms of $I_\lambda$ axiom (\ref{ax2}) is equivalent to
\begin{equation}\label{skew}
I_\lambda(b,a)\ =\ p(a,b)I_{-\lambda-T}(a,b)\ .
\end{equation}
We are left to consider the Jacobi identity.
The first two terms are quite clear. For example, the first term should be understood as
$$
\int^\lambda\!\! dx {[a\ _x\ \Big(\int^\mu\!\! dy[b\ _y\ c]\Big)]}\ =\ I_\lambda(a,I_\mu(b,c))\ ,
$$
and similarly the second term, obtained by exchanging $\lambda$ with $\mu$ and $a$ with $b$.
Less clear is the last term in (\ref{ax3}). For this, we first need to perform a change of variable
$z=x+y$, which gives
\begin{equation}\label{double}
-\int^\lambda\!\! dx\int^{\mu+x}\!\!\! dz\,[[a\ _x\ b]_z\ c]\ .
\end{equation}
Since we want to perform {\itshape first} the integral in $x$ (corresponding to the inner bracket)
and {\itshape then} the integral in $z$, we now need to exchange the order of integration.
For this we have to be careful (as we usually are, when exchanging order of integration in multiple
integrals), and, as explained above, we think of $\int^\lambda\!\! dx$ as $\int_{-\infty}^\lambda dx$.
With this in mind, (\ref{double}) is an integral on the following region of the $(x,z)$-plane:
$\Big\{(x,z)\ \big|\ x\leq\lambda,\ z\leq\mu+x\Big\}$,
which can be equivalently written as
$\Big\{(x,z)\ \big|\ z\leq\lambda+\mu,\ z-\mu\leq x\leq\lambda \Big\}$.
Hence, after exchanging the order of integration, (\ref{double}) becomes
$$
-\int^{\lambda+\mu}\!\!\!dz\,[\Big(\int_{z-\mu}^\lambda\!\! dx\,[a\ _x\ b]\Big)\ _z\ c]\ .
$$
In the inner integral we finally replace $z$  by $-T$, thanks to (\ref{sesq-1.1}),
and now it is easy to rewrite the above expression in terms of the function $I_\lambda$:
$$
-I_{\lambda+\mu}\big(I_\lambda(a,b)-I_{-\mu-T}(a,b),c\big)\ .
$$
Thus, the Jacobi identity is equivalent to the following equation:
\begin{equation}\label{jac1}
I_\lambda\big(a,I_\mu(b,c)\big)-p(a,b)I_\mu\big(b,I_\lambda(a,c)\big)\ =\ 
I_{\lambda+\mu}\big(I_\lambda(a,b)-I_{-\mu-T}(a,b),c\big)\ .
\end{equation}

To go from Definition \ref{def5} to Definition \ref{def4} we put
\begin{equation}\label{corresp}
:ab:\ = \int^0\!\! dx\,[a\ _x\ b] = I_0(a,b)\  ,\ \ 
[a\ _\lambda\ b] = \frac{d}{d\lambda}\int^\lambda\!\! dx\,[a\ _x\ b] = \frac{d}{d\lambda}I_\lambda(a,b)\ ,
\end{equation}
and conversely, to go from Definition \ref{def4} to Definition \ref{def5} we let
$$
\int^\lambda\!\! dx\,[a\ _x\ b]\ =\ :ab:+\int_0^\lambda\!\! dx\,[a\ _x\ b]\ .
$$
We then only need to check that the above axioms, written in terms of the normally ordered
product and the $\lambda$-bracket defined by (\ref{corresp}),
correspond to the axioms of vertex algebra as stated in Definition \ref{def4}.

It is immediate to check that the sesquilinearity relations (\ref{sesq-1.2}) and (\ref{sesq-2}) 
are equivalent both to the sesquilinearity of the $\lambda$-bracket $[a\ _\lambda\ b]$,
and to the fact that $T$ is a derivation of the normally ordered product $:ab:$.
Similarly, the skewsymmetry axiom (\ref{skew}) gives both the skewsymmetry
of the $\lambda$-bracktet $[a\ _\lambda\ b]$ and the skewsymmetry relation  (\ref{eq:1.38})
for the normally ordered product $:ab:$.
We are left to consider the Jacobi identity  (\ref{jac1}).
It gives three different conditions.
If we look at the constant term in $\lambda$ and $\mu$, namely we put $\lambda=\mu=0$
in (\ref{jac1}), we get
$$
:a(:bc:):-p(a,b):b(:ac:):\ =\ :\big(\int_{-T}^0\!\! d\lambda\, [a\ _\lambda\ b]\big)c:\ ,
$$
which, in view of (\ref{eq:1.38}), is equivalent to the "weak" quasi-associativity (\ref{eq:1.39-weak}).
If instead we put $\mu=0$ in (\ref{jac1}) and we take the derivative of both sides with
respect to $\lambda$, we get the left Wick formula (\ref{eq:1.36}).
Notice that (\ref{jac1}) is unchanged if we exchange $\lambda$ with $\mu$ and $a$ with $b$.
Hence we do not need to look separately at the constant term in $\lambda$.
Finally, if we take the derivatives with respect to both $\lambda$ and $\mu$ of both sides
of (\ref{ax3}) (equivalent to (\ref{jac1}), we get the Jacobi identity for 
the $\lambda$-bracket $[a\ _\lambda\ b]$.

The above arguments, together with Remark \ref{weak}, clearly prove
that Definition \ref{def4} and Definition \ref{def5} are indeed equivalent.

%\newpage
%%%%%%%% SECTION %%%%%%%%%%%%%%%%%%%%%%%%%%%%%

\section{The Zhu algebra and representation theory}\label{sec:2}
\setcounter{equation}{0}

%%%

\subsection{$\hbar$-deformation of  vertex operators}
\label{sec:2.1}

Let $V$ be a vertex algebra with the vacuum vector $\vac$, the
infinitesimal translation operator~$T$ and the vertex operators
$Y(a,z)=\sum_{n \in \Z} a_{(n)} z^{-n-1}$ $(a \in V)$.  
Fix a Hamiltonian operator $H$ on $V$ (see Subsection \ref{sec:1.8}), 
and let (\ref{eq:1.40})
be the eigenspace decomposition for $H$. 

Define the $\hbar$\emph{-deformation} of the vertex operator $Y(a,z)$ as follows:
\begin{equation}
  \label{eq:2.1}
  Y_{\hbar}(a,z) = (1+\hbar z)^{\Delta_a} Y(a,z) =
     \sum_{n\in \Z} a_{(n,\hbar)} z^{-n-1}\, .
\end{equation}
Here and further, as before, when a formula involves 
$\Delta_a$ 
it is assumed that $a$ is an eigenvector of $H$ with eigenvalue 
$\Delta_a$ , and the formula is
extended to arbitrary $a$ by linearity.  
We have, obviously, the following
formula for $\hbar$-deformed $n$\st{th} products:
\begin{equation}
  \label{eq:2.2}
  a_{(n,\hbar)} b = \sum_{j \in \Z_+} \binom{\Delta_a}{j}
     \hbar^j a_{(n+j)} b \, .
\end{equation}

We view $\hbar$ as an indeterminate, and we expand
$(1+\hbar z)^{\Delta} =\sum_{j \in \Z_+} \binom{\Delta}{j} \,
\hbar^jz^j$, so that $Y_{\hbar} (a,z) \in (\End V) [[\hbar]]
[[z,z^{-1}]]$, i.e. $a_{(n,\hbar)} \in (\End V) [[\hbar]]$ is a
formal power series in $\hbar$ (with coefficients in $\End V$).
However, $a_{(n,\hbar)}b$ is a polynomial in $\hbar$ (with
coefficients in $V$), so that we can specialize $\hbar$ to any
value in $\C$.  Note also that the space $(\End V)
[[\hbar]][[z,z^{-1}]]$ is also a module over the algebra $\C
[[\hbar z]]$, which allows us to expand binomials $(1+\hbar z)^{\Delta}$ 
into formal power series.

The $0$\st{th} product is often viewed as a bracket on $V$.  We
introduce the following $\hbar$-deformation of this bracket
(which is different from $a_{(0,\hbar)}b$):
\begin{eqnarray}
    \label{eq:2.3}
    [a,b]_{\hbar} =
\sum_{j \in \Z_+}(-\hbar)^j a_{(j,\hbar)}b\, .
\end{eqnarray}

Note that, by (\ref{eq:2.1}),
\begin{displaymath}
    [a,b]_{\hbar} 
=Res_z(1+\hbar z)^{-1}Y_{\hbar}(a,z)b
=Res_z(1+\hbar z)^{\Delta_a-1}Y(a,z)b \, ,
\end{displaymath}
hence we obtain
\begin{equation}
  \label{eq:2.4}
  [a,b]_{\hbar} =\sum_{j \in \Z_+} \binom{\Delta_a-1}{j}\hbar^j a_{(j)}b \, .
\end{equation}
The following is a simple nice relation between the $(-1,\hbar)$-product (\ref{eq:2.2})
and the $\hbar$-bracket (\ref{eq:2.4}), which can be checked directly:
\begin{equation}\label{new}
a_{(-1,\hbar)}b\ =\ :ab:+\int_0^\hbar dx [H(a),b]_x\ .
\end{equation}

\renewcommand{\theenumi}{\alph{enumi}}
\renewcommand{\labelenumi}{(\theenumi)}
\begin{lemma}\label{lem-sesq}
\begin{enumerate}
% a
\item The following sesquilinearity properties of the $\hbar$-deformed
vertex operators hold
\begin{eqnarray}
  \label{eq:2.5}
  Y_{\hbar}(Ta,z) &=& (1+\hbar z) [T,Y_{\hbar}(a,z)]\\
\nonumber
     &=& \big((1+\hbar z) \partial_z -\hbar \Delta_a\big) Y_{\hbar}(a,z)\, .
\end{eqnarray}
% b
\item The $(n,\hbar)$-product satisfies the following 
$\hbar$-deformation of the sesquilinearity equations for the usual $n$\st{th} 
product:
\begin{eqnarray}
&& ((T+\hbar(n+1+H))a)_{(n,\hbar)}b\ =\ -na_{(n-1,\hbar)}b\ , \label{2.6-eq}\\
&& T(a_{(n,\hbar)}b)\ =\ \sum_{k\in\Z_+}(-\hbar)^k (Ta)_{(n+k,\hbar)}b+a_{(n,\hbar)}
(Tb)\ .\label{2.7-eq}
\end{eqnarray}
In particular,
\begin{equation}\label{eq:2.8}
a_{(-2,\hbar)}b\ =\ \big((T+\hbar H)a\big)_{(-1,\hbar)}b\ .
\end{equation}
% c
\item The $\hbar$-bracket satisfies the relations:
\begin{eqnarray}
[(T+\hbar H)a,b]_\hbar &=& 0\ , \label{2.9-eq}\\
(T+\hbar H)[a,b]_\hbar &=&[a,(T+\hbar H)b]_\hbar\ .\label{2.10-eq}
\end{eqnarray}
% d
\item The following identity holds for $k \in \ZZ_+$,
\begin{equation}\label{2.11-eq}
a_{(-k-1,\hbar)}b\ =\ \sum_{j=0}^k \binom{\Delta_a}{k-j}\hbar^{k-j} 
(T^{(j)}a)_{(-1,\hbar)}b\ ,
\end{equation}
or, equivalently, in terms of $\hbar$-deformed vertex operators:
\begin{displaymath}
Y_{\hbar}(a,z)_+ = (1+\hbar z)^{\Delta_a} (e^{zT}a)_{(-1,\hbar)}.
\end{displaymath}
\end{enumerate}
\end{lemma}
\renewcommand{\theenumi}{\arabic{enumi}}
\renewcommand{\labelenumi}{\theenumi.}
\begin{proof}
Equations (\ref{eq:2.5}) follow by the definition (\ref{eq:2.1}) of the $\hbar$-deformed vertex operator
and the translation covariance axiom for vertex algebras.
By looking at the coefficients of $z^{-n-1}$ in the second identity in (\ref{eq:2.5}) we get (\ref{2.6-eq}),
while if we first divide both sides of the first identity in (\ref{eq:2.5}) by $1+\hbar z$ and then
look at the coefficients of $z^{-n-1}$ we get (\ref{2.7-eq}).
Equations (\ref{2.9-eq}) and (\ref{2.10-eq}) are easily checked using the 
formula 
(\ref{eq:2.4}) for the $\hbar$-bracket.
We are left to prove part (d). From (\ref{2.6-eq}) we have
$$
a_{(-k-1 ,\hbar)}b\ =\ 
\frac 1k (Ta)_{(-k,\hbar)}b
+ \hbar \frac{\Delta_a-k+1}{k} a_{(-k,\hbar)}b\ .
$$
Equation (\ref{2.11-eq}) can be easily proved by induction using the above 
relation.
\end{proof}

\begin{proposition}
  \label{prop:2.2}
One has the following $\hbar$-deformed $n$\st{th} product identity
for $\hbar$-deformed vertex operators:
\begin{displaymath}
  (1+\hbar w)^{n+1} Y_{\hbar}(a_{(n,\hbar)}b,w) =
     Y_{\hbar}(a,w)_{(n)} Y_{\hbar} (b,w)\, .
\end{displaymath}
\end{proposition}

\begin{proof}
By (\ref{eq:2.2}) and (\ref{eq:1.42}), the LHS of the
$\hbar$-deformed $n$\st{th} product identity 
is equal to
\begin{displaymath}
  \sum_{j \in \Z_+} \binom{\Delta_a}{j} \hbar ^j
     (1+\hbar w)^{\Delta_a+\Delta_b-j} Y (a_{(n+j)}b,w)\, .
\end{displaymath}
By the $n$\st{th} product identity for usual vertex operators,
$Y(a_{(n+j)}b,w) = \Res_z i_{z,w}$ $(z-w)^{n+j}
Y(a,z)Y(b,w)-p(a,b)\Res_z i_{w,z} (z-w)^{n+j} Y (b,w)Y (a,z)$. 
Hence the LHS of the $\hbar$-deformed $n$\st{th} product identity is equal to
\begin{eqnarray*}
& \Res_z \sum_{j \in \Z_+}
\binom{\Delta_a}{j}\hbar^j (z-w)^j (1+\hbar w)^{\Delta_a+\Delta_b-j}
\Big\{i_{z,w} (z-w)^n Y (a,z) Y(b,w) \\
& -p(a,b) i_{w,z} (z-w)^n Y (b,w) Y(a,z)\Big\}
\end{eqnarray*}
But 
$\sum_{j \in \Z_+} \binom{\Delta_a}{j} \hbar^j (z-w)^j
(1+\hbar w)^{\Delta_a+\Delta_b-j}=(1+\hbar z)^{\Delta_a}(1+\hbar w)^{\Delta_b}$,
by the binomial formula,
which gives the RHS of the $\hbar$-deformed $n$\st{th} product identity .
\end{proof}

\begin{theorem}
  \label{th:2.3}
One has the following identity for the $\hbar$-deformed vertex
operators ($\hbar$-deformed Borcherds identity) for arbitrary $n
\in \Z$:
\begin{eqnarray*}
  i_{z,w}(z-w)^n Y_{\hbar} (a,z)Y_{\hbar} (b,w)-p(a,b)i_{w,z}(z-w)^n
  Y_{\hbar} (b,w) Y_{\hbar}(a,z)\\
  = \sum_{j \in \Z_+} (1+\hbar w)^{n+j+1}Y_{\hbar} (a_{(n+j,\hbar)}
     b,w) \partial^j_w \delta(z-w)/j! \, .
\end{eqnarray*}

\end{theorem}

\begin{proof}
It is the same as the proof of the usual Borcherds identity,
given in Section \ref{sec:1.5}, by making use of the $\hbar$-deformed
$n$\st{th} product identity, 
instead of the $n$\st{th} product identity.
\end{proof}

Comparing the coefficients of $z^{-m-1}$ of both sides of the
$\hbar$-deformed Borcherds identity applied to $c \in V$, we
obtain:
\begin{eqnarray}
  \label{eq:2.12}
&\displaystyle{
 \sum_{j \in \Z_+}\!\! (-1)^j\!\! \binom{n}{j}\!\!\Big(\!a_{(m+n-j,\hbar)} 
 \big(Y_{\hbar}(b,w)c\big)w^j\!
    -\!p(a,b)(-1)^n  Y_{\hbar}(b,w)\big(a_{(m+j,\hbar)}c\big)w^{n-j}\!\Big) 
    }\nonumber\\
&\displaystyle{
 =\ \sum_{j\in \Z_+} (1+\hbar w)^{n+j+1} \binom{m}{j} Y_{\hbar}(a_{(n+j,\hbar)}b,w)cw^{m-j}\ .}
\end{eqnarray}
Note that, remarkably, the $\hbar$-deformation does not
change the $-1$\st{st} product identity:
\begin{displaymath}
  : Y_{\hbar} (a,w) Y_{\hbar} (b,w):\
      =\ Y_{\hbar} (a_{(-1,\hbar)}b,w) \, .
\end{displaymath}
The $\hbar$-deformed commutator formula
\begin{equation}
  \label{eq:2.10}
  [Y_{\hbar}(a,z),Y_{\hbar}(b,w)] =\sum_{j \in \Z_+}
     (1+\hbar w)^{j+1}Y_{\hbar} (a_{(j,\hbar)} b,w)
     \partial^j_w \delta (z-w)/j!
\end{equation}
is a special case of the $\hbar$-deformed Borcherds identity for $n=0$.

We want to rewrite the Borcherds identity (\ref{eq:2.12}) in terms of the 
$(n,\hbar)$-products.
We can do it in two ways:
either we compare the coefficients of 
 $w^{-k-1}$ in both sides of (\ref{eq:2.12}),
%\begin{eqnarray}\label{eq:2.14}
%&\displaystyle{
%\sum_{j \in \Z_+}\!\!  \binom{n}{j}\! (-1)^j\!
%\Big(\!
% a_{(m+n-j,\hbar)} \big( b_{(k+j,\hbar)}c \big)
% -(-1)^n p(a,b)  b_{(k+n-j,\hbar)} \big(a_{(m+j,\hbar)}c\big)
%\!\Big) 
%}\nonumber\\
%&\displaystyle{
%=\ \sum_{i,j\in \Z_+} \binom{m}{j} \binom{n+j+1}{i} \hbar^i (a_{(n+j,\hbar)}b)_{(k+m-j+i,\hbar)}c\ . }
% \end{eqnarray}
or, for the form it will be used,
we first multiply both sides of (\ref{eq:2.12}) by 
$(1+\hbar w)^{-n-1}$
and then compare the coefficients of $w^{-k-1}$. We then get
\begin{eqnarray}\label{eq:2.15}
&\displaystyle{
\sum_{i,j \in \Z_+} \binom{n}{j} \binom{-n-1}{i} (-1)^j \hbar^i
\Big(
a_{(m+n-j,\hbar)} \big( b_{(k+j+i,\hbar)}c \big) 
}\nonumber\\
&\displaystyle{
 -(-1)^n p(a,b) b_{(k+n-j+i,\hbar)} \big(a_{(m+j,\hbar)}c\big)
\Big) 
}\\
&\displaystyle{
=\ \sum_{i,j\in \Z_+} \binom{m}{j} \binom{j}{i} \hbar^i (a_{(n+j,\hbar)}b)_{(k+m-j+i,\hbar)}c\ . }\nonumber
\end{eqnarray}

We now state some results about $\hbar$-deformed vertex operators which 
will be useful in the
following sections.
\renewcommand{\theenumi}{\alph{enumi}}
\renewcommand{\labelenumi}{(\theenumi)}
\begin{lemma}
\begin{enumerate}
% a
\item One has for 
$n,k\in\Z_+$:
\begin{eqnarray}\label{eq:2.16}
\big(a_{(n,\hbar)}b\big)_{(k,\hbar)}c
&=& \displaystyle{
\sum_{i,j \in \Z_+} \binom{n}{j} \binom{-n-1}{i} (-1)^j \hbar^i
\Big(
a_{(n-j,\hbar)} \big( b_{(k+j+i,\hbar)}c \big) 
}\nonumber\\
&-& \displaystyle{
(-1)^n p(a,b)  b_{(k+n-j+i,\hbar)} \big(a_{(j,\hbar)}c\big)
\Big) \ .}
\end{eqnarray}
% b
\item The $(-1,\hbar)$-product satisfies the following 
"quasi-associativity" relation
\begin{eqnarray}\label{eq:2.17}
& \displaystyle{
\big(a_{(-1,\hbar)}b\big)_{(-1,\hbar)}c - a_{(-1,\hbar)}\big(b_{(-1,\hbar)}c\big)
} \\
& \displaystyle{
= \sum_{j \in \Z_+}
\Big(
a_{(-j-2,\hbar)} \big( b_{(j,\hbar)}c \big) 
+ p(a,b)  b_{(-j-2,\hbar)} \big(a_{(j,\hbar)}c\big)
\Big)\ .\nonumber }
\end{eqnarray}
% c
\item The following identity holds:
\begin{eqnarray}\label{eq:2.18}
& \displaystyle{
\big(a_{(-2,\hbar)}b\big)_{(-1,\hbar)}c = a_{(-2,\hbar)}\big(b_{(-1,\hbar)}c\big)
} \\
& \displaystyle{
+ \sum_{j \in \Z_+}
\Big(
(Ta)_{(-j-2,\hbar)} \big( b_{(j,\hbar)}c \big) 
+ p(a,b)  b_{(-j-2,\hbar)} \big((Ta)_{(j,\hbar)}c\big)
\Big) 
\nonumber } \\
& \displaystyle{
+ \hbar\Delta_a \sum_{j \in \Z_+}
\Big(
a_{(-j-2,\hbar)} \big( b_{(j,\hbar)}c \big) 
+ p(a,b)  b_{(-j-2,\hbar)} \big(a_{(j,\hbar)}c\big)
\Big)\ . \nonumber }
\end{eqnarray}
\end{enumerate}
\end{lemma}
\renewcommand{\theenumi}{\arabic{enumi}}
\renewcommand{\labelenumi}{\theenumi.}
\begin{proof}
Equation (\ref{eq:2.16}) is a special case of (\ref{eq:2.15}), with $m=0$.
Equation (\ref{eq:2.17}) is obtained by putting $n=k=-1$ in (\ref{eq:2.16}).
Finally, (\ref{eq:2.18}) follows immediately by using (\ref{eq:2.8})
in the LHS and applying (\ref{eq:2.17}).
\end{proof}

\begin{lemma}\label{lem:2.5}
The following identity holds (where $T^{(n)}$ stands for $T^n/n!$):
\begin{eqnarray}\label{eq:2.19}
&& a_{(-1,\hbar)} b- p(a,b) b_{(-1,\hbar)}a-\hbar [a,b]_{\hbar} \\
&& \displaystyle{
= \sum_{n\in\Z_+}\sum_{l=0}^n \binom{\Delta_b}{n-l} \hbar^{n-l} (-1)^n
\Big(
T^{(l+1)}(a_{(n)}b)-\hbar^{l+1}\binom{-\Delta_{a_{(n)}b}}{l+1} a_{(n)}b
\Big)\ .
}\nonumber
\end{eqnarray}
\end{lemma}
\begin{proof}
Recall the skewsymmetry relation in a vertex algebra (see (\ref{eq:1.30})):
$$
p(a,b)b_{(n)}a\ =\ \sum_{i\in\Z_+}(-1)^{n+i+1} T^{(i)}(a_{(n+i)}b)\ .
$$
Using this, we get, by the definition of the $(-1,\hbar)$-product,
\begin{eqnarray}\label{14sett_1}
&& \!\!\!\!\!\!\!\!\!\!\!\!\!\!\!\!\!\!\!\!\!
p(a,b)b_{(-1,\hbar)}a
 \ =\ \sum_{k,i\in\Z_+} \binom{\Delta_b}{k} \hbar^k (-1)^{k+i} T^{(i)}(a_{(k+i-1)}b) \nonumber\\
&& \ =\ \sum_{n\in\Z_+}\sum_{l=0}^n \binom{\Delta_b}{n-l} \hbar^{n-l} (-1)^{n+1} T^{(l+1)}(a_{(n)}b) \\
&&\ \ \ +\sum_{k\in\Z_+} \binom{\Delta_b}{k} (-\hbar)^k a_{(k-1)}b\ . \nonumber
\end{eqnarray}
The last identity is obtained by separating the summand with $i=0$ from the rest, and, in the latter,
by performing the change of variables $l=i-1,\ n=k+i-1=k+l$.
We can add and subtract the appropriate quantity to rewrite the first term in the RHS
of (\ref{14sett_1}) as
\begin{eqnarray}\label{14sett_2}
&\displaystyle{
\sum_{n\in\Z_+}\sum_{l=0}^n \binom{\Delta_b}{n-l} \hbar^{n-l} (-1)^{n+1} 
\Big(T^{(l+1)}(a_{(n)}b) -\hbar^{l+1} \binom{-\Delta_{a_{(n)}b}}{l+1} a_{(n)}b\Big) 
}\nonumber\\
&\displaystyle{
+\sum_{n\in\Z_+}\sum_{l=0}^n \binom{\Delta_b}{n-l} \binom{-\Delta_{a_{(n)}b}}{l+1} 
(-\hbar)^{n+1} a_{(n)}b\ .
}
\end{eqnarray}
We now perform the change of variables $k=n+1,\ i=l+1$ in the last term of (\ref{14sett_2})
to rewrite it as
$$
\sum_{k\in\Z_+}\sum_{i=1}^k \binom{\Delta_b}{k-i} \binom{-\Delta_{a_{(k-1)}b}}{i} 
(-\hbar)^{k} a_{(k-1)}b\ ,
$$
which, combined to the last term in the RHS of (\ref{14sett_1}), gives
\begin{equation*}
\sum_{k\in\Z_+} \binom{\Delta_a-1}{k} \hbar^k a_{(k-1)}b\ .
\end{equation*}
For this we used the obvious combinatorial identity ($n\in\Z_+$):
\begin{equation}  \label{eq:2.20}    
\binom{x+y}{n} = \sum^n_{j=0}\binom{x}{j}\binom{y}{n-j}
\end{equation}
with $x=-\Delta_a-\Delta_b+k$ and $y=\Delta_b$.
The above discussion gives
\begin{eqnarray*}
&\displaystyle{
p(a,b)b_{(-1,\hbar)}a
\ =\ \sum_{n\in\Z_+}\sum_{l=0}^n \binom{\Delta_b}{n-l} \hbar^{n-l} (-1)^{n+1}
\Big(
T^{(l+1)}(a_{(n)}b) 
}\\
&\displaystyle{
-\hbar^{l+1}\binom{-\Delta_{a_{(n)}b}}{l+1} a_{(n)}b
\Big) 
 +\sum_{k\in\Z_+} \binom{\Delta_a-1}{k} \hbar^k a_{(k-1)}b\ .
}
\end{eqnarray*}
Equation (\ref{eq:2.19}) now follows by the definitions of the $(-1,\hbar)$-product
and $\hbar$-bracket, and by the relation
$\binom{\Delta_a}{k}-\binom{\Delta_a-1}{k}-\binom{\Delta_a-1}{k-1} = 0$.
\end{proof}
\begin{lemma}\label{lem:2.6}
Taking the $(n,\hbar)$-bracket is a derivation of the $\hbar$-product:
\begin{equation}\label{14sett_4}
[a,b_{(n,\hbar)}c]_\hbar\ =\ \big([a,b]_\hbar\big)_{(n,\hbar)}c
+p(a,b)b_{(n,\hbar)}\big([a,c]_\hbar\big)\ ,
\end{equation}
or, equivalently, in terms of the $\hbar$-deformed vertex operators:
$$
[a,Y_\hbar(b,w)c]_\hbar\ =\ Y_\hbar([a,b]_\hbar,w)c + 
p(a,b)Y_\hbar(b,w)[a,c]_\hbar\ .
$$
\end{lemma}
\begin{proof} %%%INSERT I
Take identity (\ref{eq:2.12}) for $n=0$, multiply both sides by
$(-\hbar)^m$ and sum over $m \in \ZZ_+$.  Then the LHS of the
obtained identity is
\begin{displaymath}
  [a,Y_{\hbar} (b,w)c]_{\hbar}- p(a,b) Y_{\hbar}(b,w)
     [a,c]_{\hbar} \, ,
\end{displaymath}
while the RHS becomes $Y_{\hbar} ([a,b]_{\hbar}, w)c$, after
changing the order of summation by $m$ and $j$, and using the
expansion in $\CC [[x]]$ (for $x=-\hbar w$):
$\frac{x^j}{(1-x)^{j+1}} = \sum^{\infty}_{m=0} \binom{m}{j}x^m$.
\end{proof}

Let $J_{\hbar} \subset V[\hbar]$ be the $\C[\hbar]$-span of all the elements
$a_{(-2,\hbar)} b$  $(a,b \in V)$.
By (\ref{eq:1.28}) and (\ref{eq:2.8}) with $b=\vac$, we have $(T+\hbar H)a\in J_\hbar$ 
for all $a\in V$. Hence, more generally, we obtain by induction on $k$, using (\ref{eq:1.43}):
\begin{equation}\label{eq:2.21}
T^{(k)}a\ \equiv\ \hbar^k\binom{-\Delta_a}{k} a\ \ \mod J_\hbar .
\end{equation}
\renewcommand{\theenumi}{\alph{enumi}}
\renewcommand{\labelenumi}{(\theenumi)}
\begin{theorem} \label{th:2.7}
  \begin{enumerate}
% a
\item $J_{\hbar}$ contains all the elements
    $a_{(-j,\hbar)}b$, where $j \geq 2$ and $a,b \in V$.
% b
\item $(T+\hbar H) V \subset J_{\hbar}$.
% c
\item $J_{\hbar}$ is a two-sided ideal with respect to the
  product
 ~$a _{(-1,\hbar)}b$.
% d
\item $J_{\hbar}$ is a two-sided ideal with respect to the
 bracket $\hbar[a,b]_{\hbar}$.
% e
\item $\vac$ is the identity element for the product
   ~$a _{(-1,\hbar)}b$.
% f
\item $a_{(-1,\hbar)}b-p(a,b)b_{(-1,\hbar)}a-\hbar[a,b]_\hbar \in J_\hbar$.
% h
\item $(a _{(-1,\hbar)}b) _{(-1,\hbar)}c -a_{(-1,\hbar)}(b_{(-1,\hbar)}c)  
\in J_{\hbar}$.
\end{enumerate}
\end{theorem}
\renewcommand{\theenumi}{\arabic{enumi}}
\renewcommand{\labelenumi}{\theenumi.}
\begin{proof}
Claim (a) follows from (\ref{2.6-eq}); 
(b) is a special case of (\ref{eq:2.21}) for $k=1$;
(e) follows from (\ref{eq:2.2}) and the properties (\ref{eq:1.28})
 of the vacuum; 
(f) follows from Lemma \ref{lem:2.5} and (\ref{eq:2.21});
(g) follows from equation (\ref{eq:2.17}) and (a).
Equation (\ref{eq:2.18}), together with (a), implies 
$(a_{(-2,\hbar)}b)_{(-1,\hbar)}c\ \in\ J_\hbar$,
hence $J_\hbar$ is a right ideal with respect to the $(-1,\hbar)$-product.
Moreover, equation (\ref{14sett_4}) with $n=-2$ implies that
$[a,b_{(-2,\hbar)}c]_\hbar\ \in\ J_\hbar$,
hence $J_\hbar$ is a left ideal with respect to the $\hbar$-bracket.
Therefore, due to (f), $J_\hbar$ is also a left ideal with respect to the 
$(-1,\hbar)$-product, proving (c). Now (d) follows from (c) and (f).
\end{proof}

By Theorem \ref{th:2.7}(c), (e) and (g), we have a unital associative superalgebra over $\C[\hbar]$:
$$
Zhu_{\hbar,H}V\ :=\ V[\hbar]/J_\hbar\ ,
$$
with the product induced by the $(-1,\hbar)$-product on $V[\hbar]$,
and the image of $\vac$ is the identity element.
Furthermore, if we denote by $J_{\hbar=1}$ the $\C$-span of all emenets
$\{a_{(-2,\hbar=1)}b\ |\ a,b\in V\}$, then $V/J_{\hbar=1}$ is a unital associative algebra over $\C$
(cf. \cite{Z}), which will play an important role in the second part of the paper.
\begin{definition}\label{hzhu}
The $H$-{\itshape twisted Zhu algebra} of a vertex algebra $V$
with a Hamiltonian $H$ is
$$
Zhu_{H}V\ :=\ V/J_{\hbar=1}\ .
$$
\end{definition}
\begin{remark}
  \label{rem:1.1}
The $\hbar$-deformation of a vertex algebra is a special case
of the following notion of a gauged vertex algebra.  In addition
to the usual data $(V,\vac, T, Y (a,z)$) of a vertex algebra one
is also given an $\End V$-valued power series
$A(z)=\sum^{\infty}_{n=0}A_n z^n$, such that $A(z)\vac =0$.  
The vacuum and locality axioms are
the same as that for a vertex algebra, but the translation
covariance axiom is changed:
\begin{displaymath}
  [T,Y(a,z)] = \partial_z Y (a,z)-Y(A(z)a,z)\, .
\end{displaymath}
(For convergence one needs that for each $k \in \Z$,
$(A_ja)_{(k+j)} b=0$ as $j \gg 0$.)  The $\hbar$-deformed vertex
algebra corresponds to the choice (cf.~(\ref{eq:2.5})):
\begin{displaymath}
  A(z) = \hbar H /(1+\hbar z)\, .
\end{displaymath}
Note that vertex operators of a gauged vertex algebra can be obtained from
the vertex operators $X(a,z)$ of an ordinary vertex algebra by letting $Y(a,z)
=X(B(z)a,z)$, where B(z) is the unique formal power series solution of the
equation $B'(z)=A(z)B(z)$, $B(0)=I$. One then has $B(z)\vac =\vac$ .
\end{remark}

%%%

\subsection{($\hbar ,\Gamma$)-deformation of vertex operators and 
$Zhu_{\hbar ,\Gamma}V$}
  \label{sec:2.2}

Let $\Gamma$ be an additive subgroup of $\R$, containing $\Z$.
For $\gamma \in \R$ we shall denote the coset $\gamma +\Z$ by
$\bar{\gamma}$. 
\begin{definition}\label{def-gamma}
A vertex algebra $V$ with a Hamiltonian operator $H$ is said to 
be $\Gamma/\Z$-{\itshape graded}
if we have a decomposition
\begin{equation}
\label{eq:2.22}
  V\ =\ \bigoplus_{\bar{\gamma} \in \Gamma /\Z}
{}^{\bar{\gamma}}V
\end{equation}
which is a vertex algebra grading, compatible
with the Hamiltonian operator $H$, i.e.
$$
{}^{\bar{\alpha}} V_{(n)} {}^{\bar{\beta}}V
\subset {}^{\bar{\alpha} +\bar{\beta}} V\ ,\ \ 
H({}^{\bar{\gamma}}V) \subset {}^{\bar{\gamma}}V\ .
$$
In particular, $\vac \in {}^{\bar{0}}V$.
We shall assume for simplicity, that the eigenvalues of $H$ are
real.
\end{definition}

Given a vector $a \in V$ of conformal weight $\Delta_a$ and 
degree $\bar{\gamma}_a$
(i.e. $a \in {}^{\bar{\gamma}_a}V$), denote by $\epsilon_a$ the maximal
non-positive real number in the coset $\bar{\gamma}_a
-\bar{\Delta}_a$. This number has the following properties:
\begin{equation}
  \label{eq:2.23}
  \epsilon_{\vac}=0 \, , \quad \epsilon_{Ta}=\epsilon_a \, , \quad
     \epsilon_{a_{(n)}b} = \epsilon_a + \epsilon_b +\chi (a,b)\, ,
\end{equation}
where $\chi (a,b)=1$ or $0$, depending on whether $\epsilon_a
+\epsilon_b \leq-1$ or not. 
For example, $\Delta_{\vac}=\epsilon_{\vac}=0$, so that
$\chi (a,\vac) = \chi (\vac ,a)=0$.

Let $\gamma_a = \Delta_a
+\epsilon_a$ .  Then (\ref{eq:1.43}),(\ref{eq:1.42}) and (\ref{eq:2.23}) imply
\begin{equation}
  \label{eq:2.24}
  \gamma_{\vac} = 0 \, , \quad \gamma_{Ta} = \gamma_a +1 \, , \quad
     \gamma_{a_{(n)}b}= \gamma_a + \gamma_b +\chi (a,b) -n-1\,.
\end{equation}
We denote by $H'$ the operator on $V$, defined by
$H'(a)=\gamma_a a$.

\begin{example}\label{ex-triv}
If $\Gamma=\Z$, the decomposition (\ref{eq:2.22}) is trivial: $V={}^{\bar{0}}V$.
In this case, for $a\in V,\ \gamma_a$ is the largest integer less than or equal to $\Delta_a$
and $\epsilon_a=0$ if and only if $\Delta_a$ is an integer.
\end{example}
\begin{example}\label{ex-delta}
A case of special interest for the present paper is when the $\Gamma/\Z$-grading (\ref{eq:2.22}) 
is induced by $H$, namely
\begin{equation}\label{delta-twist}
{}^{\bar{\gamma}}V =\ \bigoplus_{\Delta \in \bar{\gamma}}V[\Delta] ,
\end{equation}
where $\Gamma$ is the additive subgroup of $\RR$, generated by 1 and the eigenvalues 
of $H$. In this case 
$\epsilon_a=0,\ \gamma_a=\Delta_a$ for all $a\in V$, and $\chi(a,b)=0$ 
for all $a,b\in V$.
\end{example}

The $(\hbar , \Gamma)$\emph{-deformation} of a vertex operator
$Y(a,z)$ is defined as follows:
\begin{displaymath}
  Y_{\hbar ,\Gamma}(a,z) = (1+\hbar z)^{\gamma_a}
     Y(a,z)=\sum_{n\in \Z}a_{(n,\hbar ,\Gamma)}z^{n-1}\, .
\end{displaymath}
We have:
\begin{equation}
  \label{eq:2.26}
  a_{(n,\hbar ,\Gamma)} b = \sum_{j \in \Z_+}
  \binom{\gamma_a}{j}\hbar^j a_{(n+j)}b \, .
\end{equation}
The same proof as that of (\ref{2.6-eq}) gives
\begin{equation}
  \label{eq:2.27}
  -na_{(n-1,\hbar ,\Gamma)}b =
      ((T+\hbar  H')a)_{(n,\hbar,\Gamma)}b
         + \hbar (n+1)a_{(n,\hbar ,\Gamma)}b\, .
\end{equation}
Define the $(\hbar,\Gamma)$-deformed bracket in the same way as (\ref{eq:2.3}):
\begin{displaymath}
  [a,b]_{\hbar ,\Gamma} =\sum_{j \in \Z_+}(-\hbar)^j
     a_{(j,\hbar,\Gamma)}b\, ,
\end{displaymath}
and in the same way as (\ref{eq:2.4}), we show that
\begin{equation}
  \label{eq:2.28}
  [a,b]_{\hbar,\Gamma} =\sum_{j \in \Z_+} \binom{\gamma_a-1}{j}
     \hbar^j a_{(j)}b \, .
\end{equation}

The same proof as that of Proposition~\ref{prop:2.2}, using
(\ref{eq:2.24}) in place of (\ref{eq:1.42}), gives the $(\hbar
,\Gamma)$-deformed $n$\st{th}-product identity:
\begin{equation}
  \label{eq:2.29}
  (1+\hbar w)^{n+1-\chi (a,b)} Y_{\hbar ,\Gamma}
     (a_{(n,\hbar ,\Gamma)}b,w) =Y_{\hbar ,\Gamma} (a,w)_{(n)}
        Y_{\hbar ,\Gamma}(b,w)\, .
\end{equation}
The same proof as that of Theorem~\ref{th:2.3} gives now the
$(\hbar ,\Gamma)$-deformed  Borcherds identity:
\begin{eqnarray}
  \label{eq:2.30}
&\displaystyle{
i_{z,w}(z-w)^n Y_{\hbar ,\Gamma} (a,z)Y_{\hbar ,\Gamma}
     (b,w)-p(a,b) i_{w,z} (z-w)^n Y_{\hbar ,\Gamma} (b,w)
       Y_{\hbar ,\Gamma} (a,z) 
}\nonumber\\
&\displaystyle{
= \sum_{j \in \Z_+} (1+\hbar w)^{n+j+1-\chi (a,b)}
     Y_{\hbar, \Gamma} (a_{(n+j,\hbar, \Gamma)}b,w) \partial^j_w \delta (z-w)/j!\, .
}
\end{eqnarray}

Denote by $V_{\Gamma}$ (resp. $V_{\hbar ,\Gamma}$)
the $\C$-span (resp. $\C [\hbar]$-span in $V[\hbar]$) of all elements
$a \in V$ such that $\epsilon_a =0$, and denote by $J_{\hbar
  ,\Gamma}$ the $\C [\hbar]$-span of all the elements $a_{(-2+\chi (a,b) ,
  \hbar ,\Gamma)}b$, where $a,b \in V$ and $\epsilon_a
+\epsilon_b \in \Z$.  By (\ref{eq:2.23}), $V_{\hbar ,\Gamma}$ is  closed
under the $(-1,\hbar,\Gamma)$-product, which coincides on this subspace
with the $(-1,\hbar)$-product, and $J_{\hbar ,\Gamma}
\subset V_{\hbar, \Gamma}$.

For example, if $\Gamma=\Z$, then we have $V_\Z=\big\{a\in V\ | 
\Delta_a\in \Z\big\}$. On the other hand,
if we consider the $\Gamma/\ZZ$-grading of $V$ induced by $H$, as in 
Example \ref{ex-delta}, then $V_\Gamma=V$ and $J_{\hbar,\Gamma}=J_\hbar$.
In this case the $(n,\hbar,\Gamma)$-product coincides with the $(n,\hbar)$-
product (\ref{eq:2.2}).

The same proof as that of Theorem~\ref{th:2.7}, gives the
following
\begin{theorem}
  \label{th:2.12}
\renewcommand{\theenumi}{\alph{enumi}}
\renewcommand{\labelenumi}{(\theenumi)}
  \begin{enumerate}
  \item %%a
$J_{\hbar ,\Gamma}$ contains all the elements $a_{(-j+\chi
  (a,b),\hbar ,\Gamma)}b$, where $j \geq 2$ and $a,b \in V$,
$\epsilon_a +\epsilon_b \in \Z$.
\item %%b
$(T + \hbar H)V_{\hbar ,\Gamma} \subset J_{\hbar ,\Gamma}$.
\item %%c
$J_{\hbar,\Gamma}$ is a two-sided ideal of $V_{\hbar ,\Gamma}$ with
respect to the $(-1,\hbar ,\Gamma)$-product.
\item %%d
$J_{\hbar,\Gamma}$ is a two-sided ideal of $V_{\hbar ,\Gamma}$ with
respect to the bracket $\hbar[a,b]_{\hbar ,\Gamma}$.
\item %%e
$\vac $ is the identity element for the $(-1,\hbar,\Gamma)$-product.
\item %%f
$a_{(-1,\hbar ,\Gamma)}b-p(a,b) b_{(-1,\hbar,\Gamma)}a \equiv
(1-\chi (a,b)) \hbar [a,b]_{\hbar ,\Gamma} \mod J_{\hbar
  ,\Gamma}$, $a,b \in V$.
\item %%g
The $(-1,\hbar ,\Gamma)$-product is associative  on $V_{\hbar ,\Gamma} 
\mod J_{\hbar,  \Gamma}$.
 \end{enumerate}
\end{theorem}

Define the following algebra over $\C [\hbar]$:
\begin{displaymath}
  Zhu_{\hbar ,\Gamma}V  = V_{\hbar ,\Gamma }/ J_{\hbar,\Gamma}
\end{displaymath}
with the algebra structure induced  by the $(-1,\hbar,\Gamma)$-product 
on $V_{\hbar,\Gamma}$.
By Theorem \ref{th:2.12}, this product is well defined, and it is associative. 
Moreover, by the same theorem, the bracket 
$\hbar[a,b]_{\hbar,\Gamma}$ induces a bracket on $Zhu_{\hbar,\Gamma}V$, 
satisfying the relation
\begin{equation}
  \label{eq:2.31}
  a_{(-1,\hbar,\Gamma)}b -p(a,b)b_{(-1,\hbar,\Gamma)}a\ =\ \hbar [a,b]_{\hbar,\Gamma}\ .
\end{equation}
Furthermore, the image of $\vac$ in $Zhu_{\hbar ,\Gamma}V$, which we 
denote by $1$,
is the identity element for the associative product of $Zhu_{\hbar ,\Gamma}V$.
Also, 
% it is clear that $V_{\hbar ,\Gamma}$ and $J_{\hbar ,\Gamma}$ are $H$-invariant
% and $H=H'$ on $V_{\hbar ,\Gamma}$, hence, 
by Theorem \ref{th:2.12}(b),
% they are $T$-invariant, and, keeping for these operators acting on
% $V_{\hbar ,\Gamma}/J_{\hbar ,\Gamma}$ the same notations, 
we have:
\begin{equation}
  \label{eq:2.32}
  T+\hbar H =0 \hbox{  on  }Zhu_{\hbar,\Gamma}V\, .
\end{equation}

We can specialize $\hbar = c \in \CC$, i.e. we let
\begin{displaymath}
Zhu_{\hbar =c ,\Gamma}V 
\,=\, Zhu_{\hbar,\Gamma}V\,/\,(\hbar-c)Zhu_{\hbar,\Gamma}V
\,=\, V_{\Gamma }/ J_{\hbar =c,\Gamma}\ .
\end{displaymath}
Note that the algebras $Zhu_{\hbar =c,\Gamma}V$ (over $\C$) are isomorphic
for all $c \in \C \backslash \{ 0 \}$.
This can be seen by rescaling $\sigma_c(a)=c^{\Delta_a}a$ and the obvious formula
\begin{equation}\label{isozhu}
\sigma_c(a_{(n,\hbar=c,\Gamma)}b)\ =\ c^{-n-1}\sigma_c(a)_{(n,\hbar=1,\Gamma)}
\sigma_c(b)\ ,
\end{equation}
so that $\sigma_c$ induces an isomorphism $\sigma_c:\ Zhu_{\hbar=c,\Gamma}V
\stackrel{\sim}{\longrightarrow} Zhu_{\hbar=1,\Gamma}V$.

We define the ($H,\Gamma$)-{\itshape twisted Zhu algebra} by 
putting $\hbar=1$:
$$
Zhu_{\Gamma} V\ :=\ Zhu_{\hbar=1 ,\Gamma}V\ .
$$
(Of course, $Zhu_{\Gamma} V$ depends on $H$ as well,
as does $Zhu_{\hbar,\Gamma}$, but we omit $H$ to simplify notation.)
Special cases of $Zhu_\Gamma V$ appeared in \cite{Z,KWa,DLM}.
\begin{example}
The (non-$\Gamma$-twisted) Zhu algebra (considered in \cite{Z,KWa}
in the most important cases) is defined by putting $\hbar=1$ and taking 
$\Gamma=\Z$, as in Example \ref{ex-triv}, i.e.
$$
Zhu_{\Z} V\ =\ V_\Z/J_{\hbar=1,\Z}\ .
$$
\end{example}
\begin{example}\label{ex2}
Take the $\Gamma/\ZZ$-grading of $V$ induced by $H$, as in
Example \ref{ex-delta}. 
Then $V_{\Gamma}=V$ and the $(n,\hbar,\Gamma)$-product
on $V$ coincides with the $(n,\hbar)$-product (\ref{eq:2.2}).
Let
\begin{equation}\label{star-pr-br}
a*_nb=a_{(n,\hbar=1)}b \ ,\ \  [a_*b]=[a,b]_{\hbar=1}
\end{equation}
in order to simplify notation. Then  
 we obtain the $H$-twisted Zhu algebra
$Zhu_H V\ =\ V/J_{\hbar=1}$ introduced in Definition \ref{hzhu}.
Recall that its product is induced by $a*_{-1}b$, and
$J_{\hbar=1}=V *_{-2}V$ is the span of all elements 
$a*_{-2}b$, with $a,b\in V$.
Note that, in view of (\ref{eq:2.8}), $J_{\hbar=1}$ is the two-sided ideal
of the algebra $(V,*_{-1})$, generated by $(T+H)V$.
\end{example}

Later we shall need the following relation, which is obtained from
(\ref{2.11-eq}) by specializing $\hbar =1$ and using identity
(\ref{eq:2.20}) with $y=-x=\Delta_a$:
\begin{equation}
\label{eq:2.33}
a*_{-k-1}b\ =\ \sum_{j=1}^k \binom{\Delta_a}{k-j}\Big((T^{(j)}a)*_{-1}b-
\binom{-\Delta_a}{j}a*_{-1}b\Big)\ .
\end{equation}

\begin{remark}
A vertex algebra $V$ with a Hamiltonian operator $H$ is called a {\itshape chiralization}
of the unital associative algebra $Zhu_H V$.
An interesting open question is which unital associative algebras $U$ admit a chiralization.
For example, the affine vertex algebra $V^k(\g)$ is a chiralization of $U(\g)$
for any Lie superalgebra $\g$.
Theorem \ref{fine} of the present paper says that the affine $W$-algebra $W_k(\g,x)$
is a chiralization of the finite $W$-algebra $W^\tf(\g,x)$.
Note that any commutative $U$ admits a chiralization $(V=U,\vac=1,Y(a,z)=L_a)$ with $H=0$.
Another very interesting example of a chiralization is the so called Kac--Todorov 
construction \cite{KT} of the Neveu--Schwarz vertex algebra, which is a chiralization
of the non-commutative Weil complex \cite{AM},
the corresponding coset construction of \cite{KT} being the chiralization of the cubic
Dirac operator of \cite{Ko1}.
We shall discuss this in more detail in Section \ref{sec:dirac}.
\end{remark}

The eigenspace decomposition of $V$ with respect to $H$
induces an increasing filtration of $V$ with respect to the
product $a_{(-1,\hbar =c,\Gamma)}b$ $(p \in \R)$:
\begin{equation}\label{H-filtr}
F^p V = \oplus_{\Delta \leq p} V[\Delta]\ ,
  \ \ (F^pV)_{(-1,\hbar=c,\Gamma)}(F^qV) \subset F^{p+q}V\ ,
\end{equation}
which descends to a filtration of the algebra $Zhu_{\hbar=c,\Gamma}$ by subspaces
\begin{equation}\label{240}
F^p Zhu_{\hbar =c,\Gamma}V
\,=\, F^p V_\Gamma/\big(F^pV_\Gamma\cap J_{\hbar=c,\Gamma}\big)\ .
\end{equation}
\begin{proposition}\label{old:2.17}
\renewcommand{\theenumi}{\alph{enumi}}
\renewcommand{\labelenumi}{(\theenumi)}
\begin{enumerate}
% a
\item (cf. \cite{Z}) The algebra
$Zhu_{\hbar=0,\Gamma}\,=\,V_\Gamma / V_\Gamma\cap(V_{(-2+\chi)}V)$
has a Poisson superalgebra structure,
with commutative associative product induced by the $-1\st{st}$ product of $V_\Gamma$,
and Lie bracket induced by the $0\st{th}$ product of $V_\Gamma$.
Here $V_{(-2+\chi)}V$ denotes the $\C$-span of all elements $a_{(-2+\chi(a,b))}b$
with $a,b\in V$.
Moreover the action of $H$ on $V_\Gamma$ induces an algebra grading 
on $Zhu_{\hbar=0,\Gamma} V$.
% b
\item For every $c\in\C$, the algebra $\gr\, Zhu_{\hbar,\Gamma}V$,
associated graded to filtration (\ref{240}),
is a Poisson superalgebra, with associative commutative product
$$
\gr^{\Delta_1}Zhu_{\hbar=c,\Gamma}V\times \gr^{\Delta_2}Zhu_{\hbar=c,\Gamma}V
\,\longrightarrow\, \gr^{\Delta_1+\Delta_2}Zhu_{\hbar=c,\Gamma}V\ ,
$$
induced by the $(-1,\hbar=c,\Gamma)$-product on $V$,
and with Lie bracket
$$
\gr^{\Delta_1}Zhu_{\hbar=c,\Gamma}V\times \gr^{\Delta_2}Zhu_{\hbar=c,\Gamma}V
\,\longrightarrow\, \gr^{\Delta_1+\Delta_2-1}Zhu_{\hbar=c,\Gamma}V\ ,
$$
induced by the $(\hbar=c,\Gamma)$-deformed bracket (\ref{eq:2.28}) on $V$.
% c
\item There is a canonical surjective homomorphism of graded Poisson algebras
\begin{equation}\label{surj}
Zhu_{\hbar=0,\Gamma}V\ \twoheadrightarrow\ \gr\, Zhu_{\hbar=c,\Gamma}V\ .
\end{equation}
In particular, $\dim Zhu_{\hbar=c,\Gamma}V\leq \dim Zhu_{\hbar=0,\Gamma}V$.
\end{enumerate}
\renewcommand{\theenumi}{\arabic{enumi}}
\renewcommand{\labelenumi}{\theenumi.}
\end{proposition}
\begin{proof}
% a
Notice that, for $\hbar=0$, the $(-1,\hbar,\Gamma)$-product of $V_\Gamma$
is equal to the $-1\st{st}$ product,
and the $(\hbar,\Gamma)$-bracket on $V_\Gamma$ is equal to the $0\st{th}$ product,
and it is easy to see from the fourth definition of a vertex algebra that
they induce a Poisson algebra structure 
on $V_\Gamma/V_\Gamma\cap(V_{(-2+\chi)}V)$.
To conclude the proof of (a) we just notice that $V_\Gamma\cap(V_{(-2+\chi)}V)$
is invariant under the action of $H$,
hence the quotient space $V_\Gamma/V_\Gamma\cap(V_{(-2+\chi)}V)$ has an
induced action of $H$.
More explicitly, the eigenspace of $H$ on $V_\Gamma\cap(V_{(-2+\chi)}V)$
with eigenvalue $\Delta$ is, by (\ref{eq:1.42}),
\begin{eqnarray}\label{oct30}
V_\Gamma\cap(V_{(-2+\chi)}V) &=& 
\sum_{\substack{
\Delta_1+\Delta_2+1=\Delta \\ \bar{\gamma_1},\bar{\gamma_2}|\epsilon_1+\epsilon_2=0
}}
V^{\bar{\gamma_1}}[\Delta_1]_{(-2)}V^{\bar{\gamma_2}}[\Delta_2] \\
&+& \sum_{\substack{
\Delta_1+\Delta_2=\Delta \\ \bar{\gamma_1},\bar{\gamma_2}|\epsilon_1+\epsilon_2=-1
}}
V^{\bar{\gamma_1}}[\Delta_1]_{(-1)}V^{\bar{\gamma_2}}[\Delta_2]\ . \nonumber
\end{eqnarray}
% b
To prove (b), just notice that, by Theorem \ref{th:2.12}(f),
$ab-ba\equiv c[a,b]_{\hbar=c,\Gamma}\ \ \mod J_{\hbar=c,\Gamma}$,
for $a,b\in V_\Gamma$, and that, by definition of the $(\hbar=c,\Gamma)$-bracket,
we have
$$
[F^pV,F^qV]_{\hbar=c,\Gamma}\,\subset\, F^{p+q-1}V\ .
$$
% c
We are left to prove (c).
By definition we have, for $c\in\C$,
\begin{eqnarray*}
\gr^\Delta Zhu_{\hbar=c,\Gamma}V &=&
\frac{F^\Delta V_\Gamma \big/ (F^\Delta V_\Gamma \cap J_{\hbar=c,\Gamma})}
{F^{\Delta_-}V_\Gamma \big/ (F^{\Delta_-} V_\Gamma \cap J_{\hbar=c,\Gamma})} \\
&\simeq&
\frac{F^\Delta V_\Gamma}
{\big(F^{\Delta_-}V_\Gamma\big) + \big(F^\Delta V_\Gamma \cap J_{\hbar=c,\Gamma}\big)}\ ,
\end{eqnarray*}
and in particular, for $\hbar=0$,
$$
\gr^\Delta Zhu_{\hbar=0,\Gamma}V\,\simeq\,
\frac{F^\Delta V_\Gamma}
{\big(F^{\Delta_-}V_\Gamma\big) + \big(F^\Delta V_\Gamma \cap (V_{(-2+\chi)}V)\big)}\ ,
$$
where $F^{\Delta_-}V_\Gamma=\bigcup_{\Delta^\prime<\Delta}F^{\Delta^\prime}V_\Gamma$.
On the other hand, by the definition of the $(n,\hbar,\Gamma)$-products
and by (\ref{oct30}), we have
$$
(F^\Delta V_\Gamma) \cap (V_{(-2+\chi)}V)\,\subset\,
(F^{\Delta_-}V_\Gamma) + (F^\Delta V_\Gamma \cap J_{\hbar=c,\Gamma})\ ,
$$
which induces a quotient map 
$\gr^\Delta Zhu_{\hbar=0,\Gamma}V\twoheadrightarrow \gr^\Delta Zhu_{\hbar=c,\Gamma}V$.
Statement (c) now follows by the fact that, by (a), 
$\gr\, Zhu_{\hbar=0,\Gamma}V\simeq Zhu_{\hbar=0,\Gamma}V$.
\end{proof}
\begin{example}\label{stupid}
We will consider in Section \ref{sec:zhuenv} (see Proposition \ref{last})
a large class of vertex algebras (i.e. pregraded, freely generated by a free
$\C[T]$-module) for which the map (\ref{surj}) is actually an isomorphism.
We give here an example of a vertex algebra $V$, with a Hamiltonian operator $H$,
for which the linear map (\ref{surj}) fails to be an isomorphism.
Let $V=\bigoplus_{n\in\Z_+}V[n]$ be a unital, associative, commutative, graded algebra,
with $V[0]=\C1$.
Then $V$ is a vertex algebra with $\vac=1,\,T=0$ and $a_{(n)}b=\delta_{n,-1}ab$.
It has a Hamiltonian operator, acting as $nI_{V[n]}$ on the subspace $V[n]$.
In this case $V_{(-2)}V=0$, so that
$Zhu_{\hbar=0,H}V\ =\ V/V_{(-2)}V\ \simeq\ V$.
On the other hand, since $a_{(-2,\hbar)}\vac=\hbar H(a)$, we have
$J_{\hbar=1}=\bigoplus_{n\geq1}V[n]$. Hence
$Zhu_{\hbar=1,H}V=V/J_{\hbar=1}\simeq\C1$.
Thus the map (\ref{surj}) has non-zero kernel, unless $V=V[0]$.
\end{example}

%%%

\subsection{$\Gamma$-twisted positive energy modules and 
representations of the algebra $Zhu_{\Gamma}V$}
\label{sec:2.3}

\begin{definition}\label{modV}
A $\Gamma$-\emph{twisted module} over a $\Gamma/\Z$-graded vertex 
algebra $V$ is a vector superspace $M$ and a parity preserving
linear map from $V$ to the space of $\End M$-valued 
$\Gamma$-{\itshape twisted quantum fields} 
$a \mapsto Y^M (a,z) = \sum_{m \in \bar{\gamma}_a} a^M_{(m)}
z^{-m-1}$ (i.e. $a^M_{(m)} \in \End M$ and $a^M_{(m)} v=0$ for
each $v \in M$ and $m \gg 0$), such that the following properties hold:
\begin{gather}
  \vac _{(n)}^M = \delta_{n,-1} I_M \, ,\tag{M1}\\
    \sum_{j \in \Z_+} \binom{m}{j} (a_{(n+j)}b)^M_{(m+k-j)}v\tag{M2}\\
    = \sum_{j \in \Z_+} (-1)^j \binom{n}{j} (a^M_{(m+n-j)}
       b^M_{(k+j)} -p(a,b)(-1)^n b^M_{(k+n-j)}a^M_{(m+j)}) v\, ,\notag
  \end{gather}
where $a \in {}^{\bar{\gamma}_a}V$, $m \in \bar{\gamma}_a$, $n \in \Z$, $k \in \bar{\gamma}_b$.
\end{definition}

Formula (M2) is called the $\Gamma$-twisted Borcherds identity.  In the
case when $\Gamma =\Z$, i.e. the grading (\ref{eq:2.1}) is
trivial, $M$ is called a (non-twisted) $V$-module.  A special
case of this is $M=V$ with $Y^M (a,z) = Y (a,z)$, when (M2)
turns into the Borcherds identity (\ref{eq:1.29}).
Note that an equivalent form of (M2) is 
\begin{eqnarray}
\label{eq:2.34}
&\displaystyle{
i_{z,w} (z-w)^n Y^M (a,z)Y^M (b,w) -p(a,b) i_{w,z}(z-w)^n
Y^M (b,w) Y^M (a,z)
}\nonumber\\
&\displaystyle{
= \sum_{j \in \Z_+} Y^M (a_{(n+j)} b,w) \partial^j_w
\delta_{\bar{\gamma}_a}(z-w)/j! \, ,
}\end{eqnarray}
where $\delta_{\bar{\gamma}}(z-w) = z^{-1} \sum_{n \in \bar{\gamma}}
(\frac{w}{z})^n$ for a $\Z$-coset $\bar{\gamma}$.

\begin{remark}
For $\Gamma$-twisted quantum fields
$a(z)=\sum_{m\in\bar{\gamma}_a}a_{(m)}z^{-m-1}$,
the usual notion (\ref{eq:1.6}) of $n\st{th}$ product of fields
is not quite right, since, for example, it is identically zero if $\gamma_a\notin\Z$.
So we introduce the $\Gamma$-{\itshape deformed} $n\st{th}$ {\itshape product}
of $\Gamma$-twisted quantum fields ($n\in\Z$):
\begin{equation}
  \label{g-nprod}
  a(w)_{(n,\Gamma)} b(w) = \Res_z (z/w)^{\gamma_a}\Big( a(z) b(w) i_{z,w}-
      p(a,b)  b (w) a(z)i_{w,z} \Big)(z-w)^n \, .
\end{equation}
Notice that the above formula depends on the choice of the representative $\gamma_a$
in the coset $\bar{\gamma}_a=\gamma_a+\Z$.
Moreover (\ref{g-nprod}) reduces to (\ref{eq:1.6}) if $\gamma_a=0$.
If we multiply both sides of (\ref{eq:2.34}) by $(z/w)^{\gamma_a}$
and take residues in $z$, we get
\begin{equation}\label{g-nform}
Y^M(a_{(n,\hbar=\frac 1w,\Gamma)}b,w)\ =\ Y^M(a,w)_{(n,\Gamma)}Y^M(b,w)\ .
\end{equation}
To get (\ref{g-nform}) we used the definition (\ref{eq:2.26}) of $(\hbar,\Gamma)$-deformed
$n\st{th}$ products in $V$,
and the definition (\ref{g-nprod}) of $\Gamma$-deformed
$n\st{th}$ product of $\Gamma$-twisted $\End M$-valued quantum fields.
Equation (\ref{g-nform}) is a generalization of the $n\st{th}$ product identity (\ref{eq:1.11})
to arbitrary $\Gamma$-twisted $V$-modules $M$,
and it reduces to it for $M=V$, the adjoint representation
(or, more generally, for a non-twisted $V$-module $M$).
It has the following interpretation:
the space of $\Gamma$-twisted $\End M$-valued quantum fields
$\F^M=\{a(x)=Y^M(a,x)\ |\ a\in V\}$
together with the $\Gamma$-deformed $n\st{th}$ product (\ref{g-nprod}),
form an "$(\hbar=\frac 1x,\Gamma)$-deformed vertex algebra",
namely the $(\hbar,\Gamma)$-deformed Borcherd's identity (\ref{eq:2.30}) 
holds for $\hbar=\frac 1x$ and for any
two elements $a(x),\, b(x)\in \F^M$.
This is the first indication that $(\hbar,\Gamma)$-deformed vertex operators
are related to $\Gamma$-twisted
representations of the vertex algebra $V$
(as it will become clear from Theorem \ref{th:3.5}).
\end{remark}

From now on we fix a Hamiltonian operator $H$ on $V$
with the eigenspace decomposition (\ref{eq:1.40}). As before, we shall assume 
that the eigenvalues of $H$ are real.
If $a \in V[\Delta_a]$, one often writes (cf. Proposition \ref{prop:1.10})
$Y^M (a,z) = \sum_{n \in \bar{\epsilon}_a } a^M_n z^{-n-\Delta_a}$,
so that
\begin{equation}
  \label{eq:2.35}
  a^M_{(n)} = a^M_{n-\Delta_a +1}\, , n\in\bar{\gamma}_a\ \ ,
  \qquad a^M_{n} = a^M_{(n+\Delta_a -1)}\, ,\  n\in\bar{\epsilon}_a\ .
\end{equation}

After making the change of indices (\ref{eq:2.35}), (M1) becomes
$\vac_n =\delta_{n,0}$, and the twisted Borcherds identity (M2)
becomes $\R$-graded:
\begin{eqnarray}
  \label{eq:2.36}
  \sum_{j \in \Z_+} \binom{m+\Delta_a -1}{j}
     (a_{(n+j)} b)^M_{m+k}\\
\nonumber
= \sum_{j \in \Z_+} (-1)^j \binom{n}{j}
(a^M_{m+n-j} b^M_{k+j-n}- p(a,b) (-1)^n b^M_{k-j} a^M_{m+j})\, ,
\end{eqnarray}
where $m \in \bar{\epsilon}_a$, $k \in \bar{\epsilon}_b$, $n \in
\Z$.  Letting in this identity $b=\vac$, $n=-2$ and $k=0$, we
obtain:
\begin{equation}
  \label{eq:2.37}
  (Ta)^M_m = -(m+\Delta_a) a^M_m \, .
\end{equation}
Letting $n=0$ in (\ref{eq:2.36}), we obtain the commutator
formula:
\begin{equation}
  \label{eq:2.38}
  [a^M_m , b^M_k] = \sum_{j \in \Z_+} \binom{m+\Delta_a-1}{j}
     (a_{(j)}b)^M_{m+k}\, .
\end{equation}

\begin{definition}\label{pe}
The $\Gamma$-twisted $V$-module $M$ is called a \emph{positive
  energy} ($\Gamma$-twisted) $V$-module if $M$ has an
$\R$-grading $M =\oplus_{{j \geq 0}} M_j$ such that
\begin{equation}
  \label{eq:2.39}
  a^M_n M_j \subset M_{j -n}\, .
\end{equation}
The subspace ~$M_0$ is called the
\emph{minimal energy subspace}.  Then, of course,
\begin{equation}
  \label{eq:2.40}
  a^M_n M_0 = 0 \hbox{  for  } n>0 \hbox{  and  }a^M_0 M_0
  \subset M_0 \, .
\end{equation}
\end{definition}

Letting $m=\epsilon_a +1$ and $k=\epsilon_b
-1 + \chi (a,b)$ in (\ref{eq:2.36}), we obtain:
\begin{eqnarray}
  \label{eq:2.41}
  \sum_{j \in \Z_+} \binom{\gamma_a}{j}
      (a_{(n+j)}b)^M_{\epsilon_a +\epsilon_b +\chi (a,b)} v= 
     \sum_{j \in \Z_+} (-1)^j\binom{n}{j}\\
\nonumber
\times    ( a^M_{n-j+1+\epsilon_a}b^M_{j-n-1+\chi (a,b) +\epsilon_b}
        -p(a,b) (-1)^n b^M_{\epsilon_b+\chi (a,b)-j-1}
      a^M_{j+1+\epsilon_a}) v\, .
\end{eqnarray}
Recall from Section \ref{sec:2.2} the definition of the 
$(H,\Gamma)$-twisted Zhu algebra 
$Zhu_{\Gamma}V$.
We are going to establish a correspondence between positive energy 
$\Gamma$-twisted
$V$-modules and $Zhu_{\Gamma}V$ -modules, generalizing the construction
of \cite {Z,KWa,DLM}. For this we need a few lemmas.

\begin{lemma}
  \label{lem:3.1}
Let $M$ be a $\Gamma$-twisted positive energy $V$-module.
Then the map
\begin{equation}
\label{eq:2.42}
  a \mapsto a^M_0 |_{M_0} \, , \qquad a \in V_{\Gamma} \, ,
\end{equation}
defines a linear map $\varphi : V_{\Gamma} \mapsto \End M_0$
which is zero on $J_{\hbar =1}$ and induces a homomorphism
$Zhu_{\Gamma}V \to \End M_0$.

\end{lemma}

\begin{proof}
Let $a,b \in V$ be such that $\epsilon_a +\epsilon_b \in \Z$, i.e.
$\epsilon_a +\epsilon_b =0$
or $-1$, hence $\epsilon_a +\epsilon_b +\chi (a,b) =0$.
If $n \leq -1$, and $v \in M_0$, then (\ref{eq:2.41}) becomes, due
to (\ref{eq:2.26}) and (\ref{eq:2.40}),
\begin{displaymath}
 %%%% \label{eq:3.9}
  (a_{(n,\hbar=1,\Gamma)}b)^M_0 v = \sum_{j \in \Z_+} \binom{\gamma_a}{j}
(a_{(n+j)}b)^M_0 v
     = a^M_{n+1+\epsilon_a} b^M_{-n-1+\chi (a,b)+\epsilon_b} v\, .
\end{displaymath}
The RHS of this equality is zero if $n \leq -2$, or if $n=-1$ and
$\chi (a,b) =1$.  Thus, $\varphi (J_{\hbar =1})=0$.  Next, if $a,b \in
V_{\Gamma}$, i.e. $\epsilon_a=\epsilon_b=0$, and $n=-1$,
this formula becomes, due to (\ref{eq:2.40}):
\begin{displaymath}
  (a_{(-1,\hbar =1,\Gamma)}b)^M_0 v=a^M_0 b^M_0 v \, , \quad
      v \in M_0 \, .
\end{displaymath}
The lemma is proved.
\end{proof}

The above lemma defines a functor from the category $\Mod_V^{\Gamma,+}$
of $\Gamma$-twisted positive  energy $V$-modules, to the category 
$\Mod_{Zhu_{\Gamma} V}$
of modules over\break $Zhu_{\Gamma} V$:
$$
\Mod_V^{\Gamma,+}\ \rightarrow\ \Mod_{Zhu_{\Gamma} V}\ \ ,\qquad M\ 
\mapsto\ M_0\ .
$$
We next want to define a functor in the opposite direction. 
For this we will need the following "Uniqueness Lemma".

\begin{lemma}
  \label{lem:3.2}
Let $M$ be a vector superspace and consider a family
$\cF = \{ a^j (z) =\sum_{n \in\gamma_j+\Z} a^j_{(n)} z^{-n-1}\}_{j \in J}$
of $\Gamma$-twisted $\End M$-valued quantum
fields, i.e. $a^j_{(n)} \in \End M$ and for each $j \in J$, $v \in
M$, $a^j_{(n)}v=0$ for $n \gg 0$.  Assume that all pairs $\{ a^i (z),
a^j(z) \}$ are local.
Let $M_0$ be a generating subspace of $M$,
i.e. the span of all vectors $a^{j_1}_{(n_1)} \ldots a^{j_s}_{(n_s)}
M_0$ is $M$.  Let $b(z)$ be another $\Gamma$-twisted $\End M$-valued field
such that all pairs $(b(z), a^j(z))$ are local and $b(z)M_0 =0$.  Then $b(z)=0$.
\end{lemma}

\begin{proof}
Since, by assumption, $M_0$ generates the whole space $M$, 
we just have to prove that
\begin{equation}
  \label{eq:2.43}
  b(z) a^{j_1}(w_1) \ldots a^{j_s}(w_s)v=0\ ,
\end{equation}
for any $j_1 ,\ldots ,j_s \in J$ and $v \in M_0$.
On the other hand, by locality of
the pairs $(b(z), a^{j_k}(z))$, we have for some $N_i \in \Z_+$ ,
\begin{equation}
  \label{eq:2.44}
  \prod^s_{i=1} (z-w_i)^{N_i} b(z) a^{j_1} (w_1)\ldots
     a^{j_s} (w_s) v =0 \, ,
\end{equation}
since we can move $b(z)$ to the right, and $b(z)v=0$ by assumption.
But the LHS of the above
expression is a Laurent series in $w_s$, i.e.~the exponents in
$w_s$ that occur are bounded below.  Hence we can cancel
$(z-w_s)^{N_s}$ in (\ref{eq:2.44}).  Multiplying after that both
sides of (\ref{eq:2.44}) by $(w_{s-1}-w_s)^N$ for some $N \in
\Z_+$ , we can permute the factors $a^{j_{s-1}}(w_{s-1})$ and
$a^{j_s}(w_s)$ to get
\begin{displaymath}
  (w_{s-1}-w_s)^N \prod^{s-1}_{i=1} (z-w_i)^{N_i}
    b(z) a^{j_1}(w_1)\ldots a^{j_{s-2}}(w_{s-2})
      a^{j_s}(w_s) a^{j_{s-1}}(w_{s-1}) v=0 \, .
\end{displaymath}
Since the LHS is a Laurent series in $w_{s-1}$, we can cancel the product\\
$(w_{s-1}-w_s)^N (z-w_{s-1})^{N_{s-1}}$.  After finitely many such
steps we obtain (\ref{eq:2.43}) with the $a^{j_i}(w_i)$'s in
reversed order.
\end{proof}

Recall that any vertex algebra $V$ is, in particular, a Lie
conformal superalgebra, and to the latter one canonically
associates the (maximal formal distribution) Lie superalgebra
$L(V)$ \cite{K}.  Given a Hamiltonian operator $H$ on $V$ and a 
$\Gamma /\Z$-grading
(\ref{eq:2.22}), one likewise associates to $V$ an
$\R$-graded Lie superalgebra $L_{H,\Gamma}(V)$ as follows.

\begin{definition}
The $(H,\Gamma)$-\emph{twisted formal distribution Lie superalgebra}  
$L_{H,\Gamma}(V)$ is,
as a vector superspace, the quotient space of the
vector superspace with the basis
$\{a_n\ |\ a \in {}^{\bar{\gamma}}V[\Delta],\,n \in \bar{\epsilon}_a\}$, 
by the  span of the vectors $(\lambda a +\mu b)_n -\lambda a_n -\mu b_n$ and
(cf.~(\ref{eq:2.37})):
\begin{equation}
  \label{eq:2.45}
  (Ta)_n + (n+\Delta_a)a_n
\end{equation}
where $\lambda ,\mu \in \C$, $a,b \in V$, $n
\in \bar{\epsilon}_a$.  The Lie bracket on $L_{H,\Gamma}(V)$ is defined by
(cf.~(\ref{eq:2.38}))
\begin{equation}
  \label{eq:2.46}
  [a_m,b_k] = \sum_{j \in \Z_+} \binom{m+\Delta_a-1}{j}
     (a_{(j)} b)_{m+k}\, ,
\end{equation}
where $m \in \bar{\epsilon}_a,k\in\bar{\epsilon}_b$.  
It is easy to see that (\ref{eq:2.46})
defines a structure of a Lie superalgebra on $L_{H,\Gamma}(V)$. 
Moreover $L_{H,\Gamma}(V)$ is
$\R$-graded by $\deg\, a_m =m \in \R$:
\begin{equation}
  \label{eq:2.47}
  L_{H,\Gamma} (V) =\oplus_{j \in \R} L_j \, .
\end{equation}
The corresponding family $\F$ of $\Gamma$-twisted formal distributions consists of ($a\in V$):
\begin{equation}\label{nov3}
Y_\Gamma(a,z)\ =\ \sum_{m\in\bar{\epsilon}_a}a_m z^{-m-\Delta_a}\in
L_{H,\Gamma}(V)[[z^{\pm1}]]z^{\gamma_a}\ .
\end{equation}
\end{definition}
Let $U$ be the universal enveloping superalgebra of
$L_{H,\Gamma}(V)$, and let $U =\oplus_{j \in \R} U_j$ be the
grading induced by (\ref{eq:2.47}).  Consider the following
fundamental system of neighborhoods of $0$ in $U:\{ UF_N \}_{N \geq 0}
$, where $F_N=\oplus_{j\geq N}L_j$.  
Denote by $U^{\com}$ the completion of the algebra
$U$ in this topology and by $\bar{U}_j$ the closure of $U_j$ in
$U^{com}$.  Then $\bar{U} := \oplus_{j \in \R}$ $\bar{U}_j$ is a
subalgebra of the algebra $U^{\com}$.

Given $a,b \in V$ (homogeneous in $H$- and $\Gamma /\Z$-grading),
$m \in \bar{\epsilon}_a$, $k \in
\bar{\epsilon}_b$ and $n \in \Z$, introduce the following element
of $\bar{U}_{m+k}$ (cf.~(\ref{eq:2.36})):
\begin{eqnarray}\label{BI}
  \BI (a,b;m,k,n) = \sum_{j \in \Z_+}\binom{m+\Delta_a-1}{j}
     (a_{(n+j)}b)_{m+k}\\
     -\sum_{j \in \Z_+} (-1)^j \binom{n}{j}
      (a_{m+n-j} b_{k+j-n} -p(a,b)(-1)^n b_{k-j}a_{m+j})\, .\nonumber
\end{eqnarray}
Let $\A$ denote the factor algebra of the associative
superalgebra $\bar{U}$ by the $2$-sided ideal generated by all
elements $\BI(a,b;m,k,n)$, and the elements $\vac_n - \delta_{n,0}
1 (n \in \Z)$.
\begin{remark}
Since relations (\ref{eq:2.36}) imply the
relations (\ref{eq:2.37}) and (\ref{eq:2.38}), it follows from the
definition, that a $\Gamma$-twisted $V$-module $M$ is the same as a module
over the associative superalgebra $\A$, such that $a_n v =0$ for
any given $a \in V$ and $v \in M$ for $n \gg 0$.
\end{remark}

\begin{lemma}\label{lem:3.4}
\alphaparenlist
\begin{enumerate}
\item Given $a,b,c\in V,\ m\in\bar{\epsilon}_a,k\in\bar{\epsilon}_b, s\in\bar{\epsilon}_c,n\in\Z$,
we have
\begin{eqnarray}\label{al0}
[c_s\, ,\, \BI(a,b;m,k,n)] &=&
\sum_{j\in\Z_+}\binom{s+\Delta_c-1}{j}\big(\BI(c_{(j)}a,b;m+s,k,n) \nonumber\\
&+& p(a,c)\BI(a,c_{(j)}b,m,k+s,n)\big)\ .
\end{eqnarray}
\item Let $B$ denote the (graded) subspace of the algebra $\bar{U}$,
spanned by all the elements $\BI(a,b; m,k,n)$, with $a,b\in V,
m\in\bar{\epsilon}_a,k\in\bar{\epsilon}_b,n\in\Z$.  Then
$[c_s , B] \subset B$,
for any $c \in V$ and $s \in \bar{\epsilon}_c$.
\end{enumerate}
\renewcommand{\theenumi}{\arabic{enumi}}%
\renewcommand{\labelenumi}{\theenumi.}%
\end{lemma}
\begin{proof}
We denote by $\LBI(a,b;m,k,n)$ and $\RBI(a,b;m,k,n)$ respectively the first
and the second sum
in the RHS of (\ref{BI}).
By the definition (\ref{eq:2.46}) of the Lie bracket in $L_{H,\Gamma}(V)$, 
we have
\begin{eqnarray}\label{al1}
&& \lefteqn{\hspace{-.5in}[c_s\, ,\, \LBI(a,b;m,k,n)]\ =\
\sum_{j\in\Z_+}\binom{m+\Delta_a-1}{j}[c_s\, ,\, (a_{(n+j)}b)_{m+k}]} \\
&&\qquad =\
\sum_{i,j\in\Z_+}\binom{m+\Delta_a-1}{j}\binom{s+\Delta_c-1}{i}(c_{(i)}(a_{(n+j)}b))_{m+k+s}\ .
\nonumber
\end{eqnarray}
By the commutator formula for the vertex algebra $V$ we have
\begin{equation}\label{al2}
c_{(i)}(a_{(n+j)}b)\ =\ \sum_{l\in\Z_+}\binom{i}{l}(c_{(l)}a)_{(n+i+j-l)}b+p(a,c)
a_{(n+j)}(c_{(i)}b)\ .
\end{equation}
Now we use the following combinatorial identity, which is a generalization 
of (\ref{eq:2.20}) ($l,r\in\Z_+$):
\begin{equation}\label{al3}
\sum_{j=0}^r\binom{x}{j}\!\!\!\binom{y}{r+l-j}\!\!\!\binom{r+l-j}{l}
\!\!=\!\! \binom{y}{l}\!\!\!\binom{x+y-l}{r}\ .
\end{equation}
By (\ref{al2}), and (\ref{al3}) with $x=m+\Delta_a-1$,
$y=s+\Delta_c-1$, and using the change of the summation variable
$r=i+j-l$, we can rewrite the RHS of (\ref{al1}) as
\begin{eqnarray*}
&& \sum_{l,r\in\Z_+}
\binom{s+\Delta_c-1}{l}\!\!\!\binom{m+s+\Delta_a+\Delta_c-l-2}{r}
((c_{(l)}a)_{(n+r)}b)_{m+k+s} \\
&&\qquad
+\sum_{i,j\in\Z_+}\binom{m+\Delta_a-1}{j}\binom{s+\Delta_c-1}{i}
p(a,c)(a_{(n+j)}(c_{(i)}b))_{m+k+s}\ .
\end{eqnarray*}
By the definition of $LBI(a,b;m,k,n)$ we thus get
\begin{eqnarray*}
[c_s\, ,\, \LBI(a,b;m,k,n)] &=&
\sum_{j\in\Z_+}\binom{s+\Delta_c-1}{j}\big(\LBI(c_{(j)}a,b;m+s,k,n) \nonumber\\
&+& p(a,c)\LBI(a,c_{(j)}b,m,k+s,n)\big)\ .
\end{eqnarray*}
An easier computation shows that the same formula holds with $\RBI$ 
instead of $\LBI$,
thus proving (\ref{al0}). The second part of the lemma is now obvious.
\end{proof}

For $n\in\Z$ we denote by $\BI(a,b;z,w;n)$ the generating series of all 
elements (\ref{BI}), namely (see (\ref{nov3}))
\begin{eqnarray}\label{al4}
& \displaystyle{
\BI(a,b;z,w;n)\ =\ \sum_{m\in\bar{\epsilon}_a,k\in\bar{\epsilon}_b}z^{-m-\Delta_a}w^{-k-\Delta_b+n}
\BI(a,b;m,k,n)
}\nonumber\\
& \displaystyle{
=\ \sum_{j\in\Z_+}Y_\Gamma(a_{(n+j)}b,w)\partial_w^{j}\delta_{\bar{\gamma}_a}(z-w)/j!
}\\
& \displaystyle{
-\Big(Y_\Gamma(a,z)Y_\Gamma(b,w)i_{z,w}
-p(a,b)Y_\Gamma(b,w)Y_\Gamma(a,z)i_{w,z}\Big)(z-w)^n\ .
}\nonumber
\end{eqnarray}
The second equality follows by a straightforward computation. 
Notice that, since (\ref{eq:2.38}) is equivalent to (\ref{eq:2.34})
with $n=0$, the $\Gamma$-twisted fields $Y_\Gamma(a,z),\ a\in V$, are
pairwise local formal distributions, i.e. one has
$(z-w)^N[Y_\Gamma(a,z),Y_\Gamma(b,w)]$ $=0$ for an integer $N\gg0$.
By comparing the coefficients of $z^{-m-\Delta_a}$ in both sides of
equation (\ref{al4}) we get, by a straightforward computation,
\begin{eqnarray}\label{al5}
& \displaystyle{
\sum_{k\in\bar{\epsilon}_b}\!w^{\!-k-\Delta_b+n}
\BI(a,b;m,k,n)
=\Res_{z-w}Y_\Gamma(Y(a,z-w)b,w)i_{w,z-w} F(z,w)
}\nonumber\\
& \displaystyle{
-\Res_z \Big(Y_\Gamma(a,z)Y_\Gamma(b,w)i_{z,w}
-p(a,b)Y_\Gamma(b,w)Y_\Gamma(a,z)i_{w,z}\Big)F(z,w)\ ,
}
\end{eqnarray}
where $F(z,w)=z^{m+\Delta_a-1}(z-w)^n$, and 
as before, $i_{z,w}$ denotes expansion in the domain
$|z|>|w|$. This formula is analogous to
yet another form of Borcherds identity, see \cite{FHL}, \cite[Sec4.8]{K}. 
Note that
if we replace $F(z,w)$ by $w^lF(z,w)$, the RHS of
(\ref{al5}) changes by an overall factor $w^l$. Hence the subspace
$B$ of $\bar{U}$ is spanned by the coefficients of all powers of $w$
in the RHS of (\ref{al5}), where $F(z,w)$ is
$z^{\gamma_a}w^r,\ r\in\R$, times an arbitrary rational function in
$z$ and $w$ with poles at $z=0,w=0,z=w$. 

For every $a\in V$,
homogeneous with respect to $H$- and $\Gamma/\Z$-grading, fix an
element $m_a\in\bar{\epsilon}_a$ (later we will choose
$m_a=\epsilon_a+1$).
We define the \emph{normally ordered product} of $\Gamma$-twisted 
quantum fields as follows (see \ref{g-nprod}):
\begin{eqnarray}\label{nop}
:Y_\Gamma(a,z)Y_\Gamma(b,z):\ &=& \Res_z
(z/w)^{m_a+\Delta_a-1}\Big(Y_\Gamma(a,z)Y_\Gamma(b,w)i_{z,w} \nonumber\\
&-& p(a,b)Y_\Gamma(b,w)Y_\Gamma(a,z)i_{w,z}\Big)(z-w)^{-1}\ .
\end{eqnarray}
By computing the residue in $z$, we can rewrite the above definition
in the more familiar form
$$
:Y_\Gamma(a,z)Y_\Gamma(b,z):\ \ =\ Y_\Gamma(a,z)_+Y_\Gamma(b,z)
+p(a,b)Y_\Gamma(b,z)Y_\Gamma(a,z)_-\ ,
$$
where
$$
Y_\Gamma(a,z)_+\ =\!\!\! \sum_{m\in\bar{\epsilon}_a,m<m_a}\!\!\! a_m z^{-m-\Delta_a}
\ \ ,\qquad
Y_\Gamma(a,z)_-\ =\!\!\! \sum_{m\in\bar{\epsilon}_a,m\geq m_a}\!\!\! 
a_m z^{-m-\Delta_a}\ .
$$
(Notice that the above definition depends on the choice of 
$m_a\in\bar{\epsilon}_a$.)
If we put $n=-1$ and $m=m_a$ in (\ref{al5}), we then get, after
computing the residue in $z-w$,
\begin{eqnarray}\label{al6}
&& \displaystyle{ \BI_1(a,b;w)\ :=\
\sum_{k\in\bar{\epsilon}_b}w^{-k-\Delta_b-m_a-\Delta_a}
\BI(a,b;m_a,k,-1) }\\&& \displaystyle{ \quad =\
\sum_{j\in\Z_+}\binom{m_a+\Delta_a-1}{j}Y_\Gamma(a_{(j-1)}b,w)w^{-j}
-\ :Y_\Gamma(a,w)Y_\Gamma(b,w):\ . }\nonumber
\end{eqnarray}

\begin{lemma}
  \label{lem:3.5}
The subspace $B\subset\bar{U}$ is spanned by all elements of the
form $\BI(a,b;m_a,k,-1)$, with $a,b\in V$ and $k\in\bar{\epsilon}_b$.
\end{lemma}
\begin{proof}
By the above observations $B$ is spanned by the coefficients of all powers of $w$
in the RHS of (\ref{al5}), with $F(z,w)=z^\alpha w^\beta(z-w)^\gamma,\
\alpha\in\bar{\epsilon}_a,\beta,\gamma\in\Z$.
Notice that every such function $F(z,w)$ can be written, by partial fraction decomposition,
as finite linear combination over $\C[w^{\pm1}]$ of monomials $z^p$, with $p\in\bar{\gamma}_a$,
and $z^{m_a+\Delta_a-1}(z-w)^q$, with integer $q\leq-1$.
But the RHS of (\ref{al5}) with $F(z,w)=z^p$ is identically zero in $\bar{U}$.
This follows by the commutator formula (\ref{eq:2.46}) for elements of 
$L_{H,\Gamma}(V)\subset\bar{U}$.
Moreover it is easy to show using (\ref{eq:2.45}) that the RHS of (\ref{al5})
with $F(z,w)=z^{m_a+\Delta_a-1}(z-w)^{-n-1},\ n\geq0$, is a finite linear combination
over $\C[w^{\pm1}]$ of expressions as in the RHS of (\ref{al5})
with $a$ replaced by $a,Ta,\dots,T^na$, and $F(z,w)=z^{m_a+\Delta_a-1}(z-w)^{-1}$.
It immediately follows by the above observations that the subspace $B\subset\bar{U}$
is spanned by the coefficients of all powers of $w$ in the RHS of (\ref{al5})
with $F(z,w)=z^{m_a+\Delta_a-1}(z-w)^{-1}$.
The claim follows immediately.
\end{proof}

Consider the subalgebras $L_0$, $L_{\geq}=\oplus_{j \geq 0}L_j$,
$L_>=\oplus_{j>0} L_j$ and $L_{\leq} =\oplus_{j \leq 0}L_j$ of
the Lie superalgebra $L_{H,\Gamma}(V)$.  Note that,
\begin{displaymath}
L_0 = V_{\Gamma} /(T+H)  V_{\Gamma} \, ,
\end{displaymath}
with the Lie bracket (cf.~(\ref{eq:2.46}))
\begin{displaymath}
  [a_0,b_0] =\sum_{j \in \Z_+} \binom{\Delta_a-1}{j}
       (a_{(j)} b)_0 \, .
\end{displaymath}
Due to Theorem~\ref{th:2.12}(c) and (f), and formula
(\ref{eq:2.4}), we have a canonical surjective Lie superalgebra
homomorphism
\begin{equation}
  \label{eq:2.57}
  \psi :L_0 \to Zhu_{\Gamma}V \, ,
\end{equation}
induced by the map $a_0 \mapsto a$. Here the associative superalgebra
$Zhu_{\Gamma}V$ is viewed as a Lie superalgebra with the usual Lie
superbracket.

Now let $M_0$ be a module over the (associative) superalgebra
$Zhu_{\Gamma}V$.  Then, in particular, it is a module over
$Zhu_{\Gamma}V$ viewed as a Lie superalgebra, hence, using the
homomorphism $\psi$ defined by (\ref{eq:2.57}), $M_0$ is an
$L_0$-module.  Extend it to a $L_{\geq}$-module by letting $L_j
M_0 =0$ for $j >0$ and consider the induced module $\tilde{V}
(M_0):=\Ind^L_{L_\geq0} M_0$.  The reversed grading of $L$ $(j
\mapsto -j)$ induces on $\tilde{V} (M_0)$ an $\R_+$-grading,
which makes it a "positive energy" $L_{H,\Gamma}(V)$-module,
in analogy with Definition \ref{pe}:
\begin{displaymath}
  \tilde{V} (M_0) = \oplus_{j_{\geq 0}} \tilde{M}_j \ \ , \qquad
    \tilde{M}_0 =M_0 \ \ ,\quad  L_i M_j \subset M_{j-i} \ .
\end{displaymath}
Denote by $V(M_0)$ the quotient of the $L_{H,\Gamma}(V)$-module
$\tilde{V} (M_0)$ by the maximal graded submodule intersecting
$M_0$ trivially.  It is again a positive energy
$L_{H,\Gamma}(V)$-module:
\begin{displaymath}
  V(M_0)=\oplus_{j_{\geq 0} } M_j \, .
\end{displaymath}
Of course, $\tilde{V}(M_0)$ and $V(M_0)$ are $U$-modules and, by the "positive energy" condition,
they are also $\bar{U}$-modules.

\begin{lemma}\label{op}
Let $M_0$ be a module over the associative algebra $Zhu_{\Gamma} V$.
Then $V(M_0)$ is naturally a $\Gamma$-twisted positive energy module over the vertex algebra $V$.
\end{lemma}

\begin{proof}
By the definition of a $\Gamma$-twisted positive energy module over $V$,
it suffices to prove that
\begin{equation}
  \label{eq:2.58}
  B\, V (M_0) = 0 \ ,
\end{equation}
where $B\subset \bar{U}$ is the subspace spanned by the elements $\BI(a,b;m,k,n)$ in (\ref{BI}).
For that consider the left submodule $M^\prime=\bar{U} B\, M_0$ of the
$\bar{U}$-module $\tilde{V}(M_0)$.  It is obviously a graded submodule:
$M^\prime =\oplus_{j_{\geq 0}} M^\prime_j$.
We want to show that $M^\prime$ intersects $M_0$ trivially, so that it maps to zero in $V(M_0)$.
Due to the PBW theorem, $M^\prime$ is
the span of elements of the form
\begin{displaymath}
  a^1_{-m_1} \ldots a^s_{-m_s} \,\, b^1_{n_1} \ldots b^t_{n_t} \,\, B(M_0),
\end{displaymath}
where $a^i_{-m_i} \in L_{\leq}$, i.e. $m_i\geq0$, and $b^i_{n_i} \in L_>$, i.e. $n_i>0$.
Since by Lemma~\ref{lem:3.4}, $[b_i,B]\subset B$, and since $b_n M_0 =0$ for $n>0$,
we conclude that $M^\prime$ is the span of elements of the form
\begin{equation}
  \label{eq:2.59}
  a^1_{-m_1} \ldots a^s_{-m_s} \,\, B (M_0)\, , \hbox{  where  }
     a^i_{-m_i} \in L_{\leq} \, .
\end{equation}
Obviously $B=\oplus_{j \in \R} B_j$ is a graded subspace of $U$ and
$B_jM_0=0$ for $j>0$.
Hence the zero degree component of $M^\prime$ is
$$
M^\prime\cap M_0\ =\ U(L_0)B_0(M_0)\ .
$$
We need to show that this is zero, namely that $B_0(M_0)=0$.
By Lemma~\ref{lem:3.5}, $B_0$ is spanned by the elements
$\BI(a,b;\epsilon_a+1, -\epsilon_a -1,-1)$, with $a,b\in V$ such that $\epsilon_a+\epsilon_b\in\Z$.
Hence $B_0(M_0)$ is spanned by (cf. (\ref{eq:2.26}))
\begin{eqnarray}\label{fin}
&& \BI(a,b;\epsilon_a+1,-\epsilon_a-1,-1)(M_0) =
 \big((a_{(-1,\hbar=1,\Gamma)}b\big)_0 \\
&& \qquad \qquad -\ \sum_{j \in \Z_+} \big(a_{\epsilon_{a-j}}
     b_{-\epsilon_a+j} +p(a,b) b_{-\epsilon_a-1-j}
     a_{\epsilon_a+j+1})\big) M_0\, .\nonumber
\end{eqnarray}
But if $\epsilon_a \neq 0$, then $a _{(-1,\hbar=1,\Gamma)} b \in J_{\hbar =1}$,
hence $(a_{(-1,\hbar=1,\Gamma)}b)_0 \, M_0 =0$, since $M_0$ is by assumption
a $Zhu_{\Gamma} V$-module. Moreover, if $\epsilon_a\neq 0$ all summands 
of the sum in the RHS of (\ref{fin}) annihilate
$M_0$ since $-\epsilon_a >0$.
Finally, if $\epsilon_a=0$ then $a,b \in V_{\Gamma}$ and the RHS of (\ref{fin})
becomes $((a_{(-1,\hbar=1,\Gamma)}b)_0 - a_0b_0) \, M_0$,
which is again zero since, by assumption, $M_0$ is a $Zhu_{\Gamma} V$-module.

So far we proved  that $M^\prime$ is
a submodule of $\tilde{V}(M_0)$ intersecting $M_0$ trivially, 
hence in $V(M_0)$ we have $BM_0 =0$,
or equivalently $\BI_1(a,b;z)M_0=0$, where $\BI_1(a,b;z)$ was defined in (\ref{al6}).
We remarked above that all $\End V(M_0)$-valued fields
$Y_\Gamma(a,z)=\sum_{n\in\bar{\epsilon}_a} a_n z^{-n-\Delta_a}$ are
pairwise local.
Moreover the same argument used to prove Dong's Lemma (see e.g.
\cite[Lemma 3.2]{K}), can be used to show that the $\Gamma$-twisted
fields $Y_\Gamma(a,z)$ are local with the normally ordered products
$:Y_\Gamma(b,z)Y_\Gamma(c,z):$ defined in (\ref{nop}), and hence
with the formal distributions $\BI_1(a,b;z)$ in (\ref{al6}).
Then by Lemma~\ref{lem:3.2}, $\BI_1 (a,b;z) V(M_0)=0$.  But by
Lemma~\ref{lem:3.5}, this implies (\ref{eq:2.58}).
\end{proof}

Lemma \ref{lem:3.1} defines a "restriction" functor $M\mapsto M_0$
from the category $\Mod^{\Gamma,+}_V$ of $\Gamma$-twisted positive energy 
$V$-modules
to the category $\Mod_{Zhu_{\Gamma} V}$ of modules over $Zhu_{\Gamma} V$.
Lemma \ref{op} defines an "induction" functor in the opposite direction,
from the category $\Mod_{Zhu_{\Gamma}V}$ to the category $\Mod^{\Gamma,+}_V$,
given by $M_0\mapsto V(M_0)$.
We next want to understand in what sense the above functors are inverse
of each other,
thus establishing an equivalence of categories.

\begin{definition}\label{ai}
We shall call an $\R_+$-graded module $M= \oplus_{j_{\geq
    0}}M_j$ over an $\R$-graded Lie superalgebra $L$ or over a
vertex algebra $V$, an \emph{almost irreducible module} if $M_0$
generates the module $M$ and $M$ contains no graded submodules
intersecting $M_0$ trivially.  For example, the
$L_{H,\Gamma}(V)$-module $V(M_0)$ is almost irreducible.
\end{definition}

It is clear that the restriction functor $M \mapsto M_0$ from the
category of almost irreducible $L_{H,\Gamma}(V)$-modules to the
category of $L_0$-modules and the induction functor $M_0 \mapsto
V(M_0)$ are inverse to each other.  Moreover, these functors
establish a bijective correspondence between irreducibles of
these categories, since clearly an $L_{H,\Gamma}(V)$-module
$V(M_0)$ is irreducible if and only if the $L_0$ -module $M_0$ is
irreducible.

We can now state the main theorem of this section, which is an immediate 
consequence
of the above results and Definition \ref{ai}.
\begin{theorem}
  \label{th:3.5}
The restriction functor $M \mapsto M_0$ is a functor from the
category of positive energy $\Gamma$-twisted $V$-modules to the category
of $Zhu_{\Gamma}V$-modules, which is inverse to the induction functor 
$M_0\mapsto V(M_0)$,
on the full subcategory of almost irreducible $\Gamma$-twisted $V$-modules.
In particular,  these functors
establish a bijective correspondence between irreducible positive
energy $\Gamma$-twisted $V$-modules and irreducible $Zhu_{\Gamma}V$-modules.
\end{theorem}

%\newpage
%%%%%%%% SECTION %%%%%%%%%%%%%%%%%%%%%%%%%%%%%
\section{Non-linear Lie algebras and non-linear Lie conformal algebras}\label{sec:nonlin}
\setcounter{equation}{0}

We review here the notions of non-linear Lie superalgebras and 
non-linear Lie conformal superalgebras, introduced in \cite{DK}.

In this section, $\Gamma_+$ is a discrete, additively closed subset of $\R_+$
containing $0$.
We let $\Gamma_+^\prime=\Gamma_+\backslash\{0\}$,
and for $\zeta\in\Gamma_+^\prime$ we denote by $\zeta_-$ the largest element
of $\Gamma_+$ strictly smaller than $\zeta$. 

%%%
\subsection{Non-linear Lie superalgebras}\label{sec-nonlinlie}

% Definition

Let $\g$ be a vector superspace, and let $\T(\g)$ denote the tensor superalgebra over $\g$.
We assume that $\g$ is endowed with a $\Gamma_+^\prime $-grading
$$
\g=\bigoplus_{\zeta\in\Gamma_+^\prime }\g_\zeta\ ,
$$
and we extend it to a grading 
$\T(\g)=\bigoplus_{\zeta\in\Gamma_+}\T(\g)[\zeta]$ in the obvious way,
namely, writing $\zeta(A)=\zeta$ when $A\in\T(\g)[\zeta]$,
we require that $\zeta(1)=0$ and $\zeta(A\otimes B)=\zeta(A)+\zeta(B)$.
We have the induced increasing $\Gamma_+$-filtration by subspaces
$$
\T_\zeta(\g)\ =\ \bigoplus_{\zeta^\prime\leq\zeta}\T(\g)[\zeta^\prime]\ .
$$
If $\g$ is endowed with a linear map
$$
[\ ,\ ]\ :\ \ \g\otimes\g\ \longrightarrow\ \T(\g)\ ,
$$
we extend it to $\T(\g)$ by the Leibniz rule:
\begin{eqnarray*}
{[A,B\otimes C]} &=& [A,B]\otimes C+p(A,B)B\otimes [A,C]\ ,\\
{[A\otimes B, C]} &=& A\otimes [B,C]+p(B,C) [A,C]\otimes B\ .
\end{eqnarray*}
Let, for $\zeta\in\Gamma_+$,
$$
\M_\zeta(\g)=\Span
\bigg\{
        \begin{array}{c}
         A\otimes (b\otimes c-p(b,c)c\otimes b\\
         -[b,c]) \otimes D
         \end{array}
\left|\begin{array}{c}
 b,c\in \g,A,D\in\T(\g),\\
 \zeta(A\otimes b\otimes c\otimes D))\leq \zeta
 \end{array}\right\}\ ,
$$
and let $\M(\g)=\cup_\zeta \M_\zeta(\g)$. Note that $\M(\g)$ is the two sided ideal 
of the tensor algebra $\T(\g)$ generated by elements $a\otimes b-p(a,b)b\otimes a-[a,b]$, 
where $a,b\in\g$.
\begin{definition}\label{g}
A  {\itshape non-linear Lie superalgebra} $\g$ is a $\Gamma_+^\prime $-graded vector superspace
with a parity preserving linear map $[\ ,\ ]:\ \g\otimes\g\rightarrow\T(\g)$ satisfying the following
properties ($a,b,c\in\g$):
\begin{enumerate}
\item grading condition: $[\g_{\zeta_1},\g_{\zeta_2}]\ \subset\ \T_{(\zeta_1+\zeta_2)_-}(\g)$,
\item skewsymmetry: $[a,b]=-p(a,b)[b,a]$,
\item Jacobi identity: 
$$
[a,[b,c]]-p(a,b)[b,[a,c]] -[[a,b],c]\in\M_{(\zeta(a)+\zeta(b)+\zeta(c))_-}(\g)\ .
$$
\end{enumerate}
\end{definition}

% main theorem 1

The associative superalgebra $U(\g)=\T(\g)/\M(\g)$ is called the universal enveloping
algebra of the non-linear Lie superalgebra $\g$.
\begin{definition}\label{32}
Let $U$ be an associative superalgebra and let $\B=\{E_i\}_{i\in\I}$
be an ordered set of elements of $U$, compatible with the parity.
We shall say that $\B$ is a set of {\it PBW generators} of $U$ 
(and that $U$ is {\itshape PBW generated} by $\B$)
if the collection of {\itshape ordered monomials}
$$
S\ =\ \Big\{ E_{i_1}\dots E_{i_s}\ \Big|\ i_i\leq\dots\leq i_s 
\text{ and } i_k<i_{k+1} \text{ if } p(i_k)=\bar 1\Big\}\ .
$$
is a basis of $U$.
If $\g$ is a $\Gamma_+^\prime$-graded subspace of $U$ and any basis of $\g$,
compatible with parity and $\Gamma_+^\prime$-grading, PBW generates $U$,
we shall say that $U$ is {\itshape PBW generated by} $\g$.
\end{definition}
It is not difficult to show by the usual method (see e.g. \cite{J})
that the PBW theorem holds for the universal enveloping superalgebra $U(\g)$
of a non-linear Lie superalgebra $\g$:
\begin{theorem}\label{gmain1}
Let $\g$ be a non-linear Lie superalgebra.
Then $U(\g)$ is PBW generated by $\g$.
\end{theorem}

% main theorem 2

In order to state a converse theorem, we need the following
\begin{definition}\label{predef}
Let $U$ be an associative superalgebra, generated by a $\Gamma_+^\prime$-graded
subspace $\g=\bigoplus_{\zeta\in\Gamma_+^\prime}\g_\zeta$.
Define an increasing filtration of $U$ induced by the $\Gamma_+^\prime$-grading of $\g$,
i.e. the filtration by the subspaces
\begin{equation}\label{35}
U_\zeta\ =\ \sum_{\zeta_1+\dots+\zeta_s\leq\zeta} \g_{\zeta_1}\dots\g_{\zeta_s}\ .
\end{equation}
We shall say that $U$ is {\itshape pregraded by} $\Gamma_+$ if
\begin{equation}\label{pregrad}
[\g_{\zeta_1},\g_{\zeta_2}]^U\ \subset\ U_{(\zeta_1+\zeta_2)_-}\ ,
\end{equation}
where $[\,,\,]^U$ denotes the commutator in the associative superalgebra $U$.
\end{definition}
Let $U$ be an associative superalgebra and 
let $\g=\bigoplus_{\zeta\in\Gamma_+^\prime}\g_\zeta$
be a subspace of $U$.
We shall assume that $U$ is PBW generated by an ordered basis $\B=\big\{E_i\,|\,i\in\I\big\}$
of $\g$ compatible with parity and $\Gamma_+^\prime$-grading.
Moreover we assume that $U$ is pregraded by $\Gamma_+$.
We have $\I=\bigsqcup_\zeta\I_\zeta$,
where $\zeta(E_i)=\zeta$ for $i\in\I_\zeta$.
It is convenient to introduce the following notation
for $\vec{j}=(j_1,\dots,j_s)\in\I^s$  with $j_k\in\I_{\zeta_k}$:
\begin{eqnarray}\label{notat}
&\displaystyle{\!\!\!\!\!\!\!\!\!\!\!\!\!\!\!\!\!\!\!\!\!\!\!\!\!\!\!\!\!\!\!\!\!\!\!\!\!\!\!\!\!\!\!\!\!\!\!\!\!\!\!\!\!\!\!\!\!
\!\!\!\!\!\!\!\!\!\!\!\!\!\!\!\!\!\!\!\!
\zeta(\vec{j})\ =\ \zeta_1+\cdots+\zeta_s\ ,
}\\
&\displaystyle{
\#(\vec{j}) = \#\left\{(p,q)\ \left|\ 
        \begin{array}{c}
        1\leq p<q\leq s \text{ and either } 
         j_p>j_q\ ,\\
        \text{or } j_p=j_q \text{ and } p(E_{j_p})=\bar{1}
        \end{array}\right.\right\}\ .\nonumber
        }
\end{eqnarray}
Let
\begin{eqnarray*}
S_\zeta &=& \left\{E_{i_1}\dots E_{i_s}\ \big|\ \#(\vec{i})=0,\ \zeta(\vec{i})\leq\zeta\right\}\ ,
\quad S\ =\ \bigcup_{\zeta\in\Gamma_+} S_\zeta\ ; \\
\tilde{S}_\zeta &=& \left\{E_{i_1}\otimes\dots\otimes E_{i_s}\ \big|\ 
\#(\vec{i})=0,\ \zeta(\vec{i})\leq\zeta\right\}\ ,
\quad \tilde{S}\ =\ \bigcup_{\zeta\in\Gamma_+}\tilde{S}_\zeta\ .
\end{eqnarray*}
If $A=E_{j_1}\otimes\cdots\otimes E_{j_s}$, we will also let sometimes
$\zeta(A)=\zeta(\vec{j})$ and $\#(A)=\#(\vec{j})$.

Let $\pi$ be the natural quotient map
$$
\pi\ :\ \T(\g)\ \twoheadrightarrow\ U\ ,
$$
given by
$$
\pi(a_1\otimes\cdots\otimes a_n)\ =\ a_1\cdots a_n\ ,\ \ \text{ for } a_1,\dots,a_n \in\g\ ,
$$
and let $\rho$ be the embedding
$$
\rho\ :\ U\ \hookrightarrow\ \T(\g)\ ,
$$
given by 
$\rho(E_{i_1}\cdots E_{i_s})\ =\ E_{i_1}\otimes\cdots\otimes E_{i_s}$ for every element 
from $S$.
Note that $\pi\circ\rho=I_U$.
The surjective map $\pi$ induces the $\Gamma_+$-filtration on $U$,
given by the subspaces (\ref{35}), namely
$U_\zeta=\pi(\T_\zeta(\g))$.
On the other hand, since by assumption $U$ is PBW generated by $\B$,
the injective map $\rho$ defines
a $\Gamma_+$-grading $U=\bigoplus_\zeta U[\zeta]$ (as a vector space, not as an algebra),
by the condition $\rho(U[\zeta])\subset \T(\g)[\zeta]$, i.e.
\begin{equation}\label{gradU}
U[\zeta]\ =\ 
\Span_\C \big(S_\zeta\backslash S_{\zeta_-}\big)\ .
\end{equation}
Moreover, thanks to the embedding $\rho:\ U\rightarrow\T(\g)$, we can define a linear map
$$
[\ ,\ ]\ :\ \ \g\otimes\g\ \longrightarrow\ \T(\g)\ ,
$$
given by
\begin{equation}\label{L}
[a,b]\ =\ \rho(ab-p(a,b)ba)\ ,
\end{equation}
and we extend it to a linear map $[\ ,\ ]:\ \T(\g)\otimes\T(\g)\rightarrow\T(\g)$, 
by the Leibniz rule, as explained above.
% lemma1
\begin{lemma}\label{l1}
\alphaparenlist
\begin{enumerate}
\item The $\Gamma_+$-filtration (\ref{35})
is induced by the $\Gamma_+$-grading (\ref{gradU}):
$$
U_\zeta\ =\ \bigoplus_{\zeta^\prime\leq\zeta}U[\zeta^\prime]\ .
$$
\item We have
$$
\T_\zeta(\g) \ =\ \Span_\C\tilde{S}_\zeta+\M_\zeta(\g)\ .
$$
\end{enumerate}
\end{lemma}
\begin{proof}
It is clear by definition that $\bigoplus_{\zeta^\prime\leq\zeta}U[\zeta^\prime]\subset U_\zeta$.
In order to prove the reverse inclusion,
consider an arbitrary (not necessarily ordered) monomial
$A=E_{j_1}\dots E_{j_s}$,
and assume that $\zeta(\vec{j})=\zeta$
(recall notation (\ref{notat})).
We want to prove, by induction on $(\zeta(\vec{j}),\#(\vec{j}))$,
that $A\in\bigoplus_{\zeta^\prime\leq\zeta}U[\zeta^\prime]$.
If $\#(\vec{j})=0$, we have $A\in U[\zeta]$, so there is nothing to prove. 
Let $\#(\vec{j})\geq1$ and let $k$ be such that
either $j_k>j_{k+1}$, or $j_k=j_{k+1}$ and $p(E_{j_k})=\bar{1}$.
In the first case we have
\begin{eqnarray*}
A
&=& p(E_{j_k},E_{j_{k+1}}) E_{j_1}\dots 
E_{j_{k+1}}E_{j_k}\cdots E_{j_s} \\
&+& E_{j_1}\dots[E_{j_k},E_{j_{k+1}}]\cdots E_{j_s}\ =\ B+C\ .
\end{eqnarray*}
Obviously $\zeta(B)=\zeta(\vec{j})$ and $\#(B)=\#(\vec{j})-1$,
so by the inductive assumption $B\in \bigoplus_{\zeta^\prime\leq\zeta}U[\zeta^\prime]$.
Moreover, by the grading condition
(\ref{pregrad}), we have $C\in U_{\zeta_-}$, hence,
by the inductive assumption, $C\in \bigoplus_{\zeta^\prime\leq\zeta}U[\zeta^\prime]$.
Suppose now $j_k=j_{k+1}$ and $p(E_{j_k})=\bar{1}$. 
In this case we have
$$
A\ =\ \frac{1}{2}E_{j_1}\dots[E_{j_k},E_{j_{k+1}}]\cdots E_{j_s}\ ,
$$
so that, by (\ref{pregrad}), $A\in U_{\zeta_-}$,
and, by inductive assumption, $A\in \bigoplus_{\zeta^\prime\leq\zeta}U[\zeta^\prime]$.
This proves part (a).
In order to prove part (b) 
we need to show that, given an arbitrary (not necessarily ordered) monomial
$A=E_{j_1}\otimes\dots\otimes E_{j_n}$, with $\zeta(\vec{j})=\zeta$,
we can decompose it as
\begin{equation}\label{dec}
A\ =\ P+M\ ,
\end{equation}
where $P\in\Span_\C\tilde{S}_\zeta$ and $M\in\M_\zeta(\g)$.
The argument is very similar to that used in part (a), by induction 
on the pair $(\zeta(\vec{j}),\#(\vec{j}))$.
For example, if $\#(\vec{j}))\geq1$ and $k$ is such that $j_k>j_{k+1}$,
we have, by definition of $\M_\zeta(\g)$,
\begin{eqnarray}\label{29sett}
A &\equiv& p(E_{j_k},E_{j_{k+1}}) E_{j_1}\otimes\dots
E_{j_{k+1}}\otimes E_{j_{k}}\dots\otimes E_{j_s} \\
&+& E_{j_1}\otimes\dots\otimes [E_{j_{k}},E_{j_{k+1}}]\otimes\dots\otimes E_{j_s}
\quad \mod\M_{\zeta}(\g) \ .\nonumber
\end{eqnarray}
By the grading condition (\ref{pregrad}) and part (a), we have
$[E_{j_{k}},E_{j_{k+1}}]^U\in U_{(\zeta_k+\zeta_{k+1})_-}
=\bigoplus_{\zeta^\prime<\zeta_k+\zeta_{k+1}}U[\zeta^\prime]$.
This implies, by the definition (\ref{gradU}) of the grading of $U$, that
$[E_{j_{k}},E_{j_{k+1}}]=\rho([E_{j_{k}},E_{j_{k+1}}]^U)\in 
\bigoplus_{\zeta^\prime<\zeta_k+\zeta_{k+1}}\T(\g)[\zeta^\prime]
=\T_{(\zeta_k+\zeta_{k+1})_-}(\g)$.
Therefore the second term in the RHS of (\ref{29sett}) belongs to $\T_{\zeta_-}(\g)$.
We thus use induction to conclude that both terms in the RHS of (\ref{29sett})
decompose as in (\ref{dec}).
\end{proof}
\begin{theorem}\label{gmain2}
Let $U$ be an associative algebra,
let $\g$ be a $\Gamma_+^\prime$-graded subspace of $U$,
and let $\B$ be a basis of $\g$ compatible with parity and $\Gamma_+^\prime$-grading.
Suppose that $U$ is PBW generated by $\B$, and that it is pregraded by $\Gamma_+$.
Then the linear map
$[\ ,\ ]\ :\ \ \g\otimes \g\ \longrightarrow\ \T(\g)$,
defined by (\ref{L}), defines a structure of a non-linear Lie superalgebra on $\g$,
such that $U$ is canonically isomorphic 
to the universal enveloping algebra $U(\g)$.
\end{theorem}
\begin{proof}
% preliminaries
First, let us list some simple facts about $U$ and $\T(\g)$:
\renewcommand{\theenumi}{\alph{enumi}}
\renewcommand{\labelenumi}{(\theenumi)}
\begin{enumerate}
\item $[U_{\zeta_1},U_{\zeta_2}]^U\subset U_{(\zeta_1+\zeta_2)_-}$,
where, as before, $[\,,\,]^U$ denotes the commutator in $U$.
\item $[\T_{\zeta_1}(\g),\T_{\zeta_2}(\g)]\subset \T_{(\zeta_1+\zeta_2)_-}(\g)$,
where here $[\,,\,]$ denotes the "non-linear" bracket of $\g$, given by (\ref{L}),
extended to $\T(\g)$ by the Leibniz rule.
\item $\pi([A,B])=[\pi(A),\pi(B)]^U$, for $A,B\in\T(\g)$.
\end{enumerate}
\renewcommand{\theenumi}{\arabic{enumi}}
\renewcommand{\labelenumi}{\theenumi.}
By (\ref{pregrad}) we have $[\g_{\zeta_1},\g_{\zeta_2}]^U\subset U_{(\zeta_1+\zeta_2)_-}$. 
Hence (a) follows by applying the Leibniz rule to the commutator of arbitrary monomials
and by the definition (\ref{35}) of the filtration $U_\zeta$.
By Lemma \ref{l1}(a) we also have
$$
[\g_{\zeta_1},\g_{\zeta_2}]\subset \rho(U_{(\zeta_1+\zeta_2)_-})
=\rho\Big(\!\!\! \bigoplus_{\zeta^\prime<\zeta_1+\zeta_2}U[\zeta^\prime]\Big)
\subset\!\!\!\! \bigoplus_{\zeta^\prime<\zeta_1+\zeta_2}\T(\g)[\zeta^\prime]
=\T_{(\zeta_1+\zeta_2)_-}(\g)\ .
$$
(b) then follows since, by definition, $[\ ,\ ]$ is defined on $\T(\g)$ by the Leibniz rule.
(c) is obviously true when $A,B\in\g$ , since $\pi\circ\rho=\id|_U$.
Hence (c) holds for arbitrary monomials $A,B\in\T(\g)$, 
since both the left and the right hand side satisfy the Leibniz rule.

We can now prove the theorem.
We need to show that $[\ ,\ ]$ satisfies the axioms of a non-linear Lie superalgebra 
(see Definition \ref{g}).
The grading condition is a special case of (b).
Skewsymmetry is obvious.
It remains to prove the Jacobi identity.
We denote, for $a,b,c\in\g$
$$
J(a,b,c)\ =\ [a,[b,c]]-p(a,b)[b,[a,c]]-[[a,b],c]\ \in\ \T(\g)\ ,
$$
and we need to show that $J(a,b,c)\in\M_{\zeta_-}(\g)$, where $\zeta=\zeta(a)+\zeta(b)+\zeta(c)$.
By (b) we have $J(a,b,c)\in\T_{\zeta_-}(\g)$.
Hence by Lemma \ref{l1}(b), we can decompose
$J=P+M$, with $P\in\Span_\C\tilde{S}_{\zeta_-}$ and $M\in\M_{\zeta_-}(\g)$.
By (c) we have $\pi(J(a,b,c))=[a,[b,c]^U]^U-p(a,b)[b,[a,c]^U]^U-[[a,b]^U,c]^U=0$, 
since obviously the Jacobi identity holds on $U$.
On the other hand $\pi(M)=0$, by definition of $\M_\zeta(\g)$.
We thus have $\pi(P)=0$.
Since, by assumption, $U$ is PBW generated by $\B$,
we have $\pi:\ \Span_\C\tilde{S}\stackrel{\sim}{\longrightarrow} U$.
Hence $P=0$,
and $J(a,b,c)=M\in\M_{\zeta_-}(\g)$, as required.
\end{proof}

%%%
\subsection{Non-linear Lie conformal superalgebras}

In this section, following \cite{DK}, we recall the definitions and results in the
(more difficult) "confrmal" case, similar to those in the non-linear Lie algebra case,
discussed in the previous section.
In particular we show that the category of non-linear Lie conformal algebras is
equivalent to that of pregraded freely generated vertex algebras.
\medskip

Let $R$ be a $\C[T]$-module together with a $\Gamma_+^\prime$-grading by
$\C[T]$-submodules:
$$
R\ =\ \bigoplus_{\zeta\in\Gamma_+^\prime}R_\zeta\ .
$$
We extend the action of $T$ to be a derivation of the tensor algebra $\T(R)$
of $R$, viewed as a vector superspace, 
and we extend the $\Gamma_+$-grading to 
$\T(R)=\bigoplus_{\zeta\in\Gamma_+}\T(R)[\zeta]$
in the obvious way.
We also let $\T_\zeta(R)=\bigoplus_{\zeta^\prime\leq\zeta}\T(R)[\zeta^\prime]$
be the induced increasing filtration.

We say that $R$ is a {\itshape non-linear conformal algebra}
if it is endowed with a linear map
$$
[\ _\lambda\ ]\ :\ \ R\otimes R\ \longrightarrow\ \C[\lambda]\otimes\T(R)\ ,
$$
called the $\lambda$-{\itshape bracket}, satisfying the following conditions:
\begin{enumerate}
\item grading condition:
$[R_{\zeta_1}\ _\lambda\ R_{\zeta_2}]\ \subset\ \C[\lambda]\otimes\T_{(\zeta_1+\zeta_2)_-}(R)$,
\item sesquilinearity:
${[Ta\ _\lambda\ b]} = -\lambda{[a\ _\lambda\ b]}\ , \ \ 
{[a\ _\lambda\ Tb]} = (\lambda+T){[a\ _\lambda\ b]}$.
\end{enumerate}

It is not hard to show, by induction on $\zeta\in\Gamma_+$, that the $\lambda$-bracket
can be extended to the whole tensor algebra $\T(R)$, by using the axioms in the
fourth definition of a vertex algebra (see Section \ref{sec:1.7}):
\begin{lemma}
\cite{DK} If $R$ is a non-linear conformal algebra,
there is a unique way to define linear maps
\begin{eqnarray*}
\vphantom{\Big(}
\T(R)\otimes \T(R) &\longrightarrow& \T(R) \ , \ \ \ \ \ \ \ \ \ \ A\otimes B\mapsto\ \ :AB:\ ,\\
\vphantom{\Big(}
\T(R)\otimes \T(R)
&\longrightarrow& \C[\lambda]\otimes\T(R) \ ,\ \ A\otimes B\mapsto\ [A\ _\lambda\ B]\ ,
\end{eqnarray*}
such that $[a _\lambda\ b]$, with $a,b\in R$, coincides with the $\lambda$-bracket on $R$,
$:aB:=a\otimes B$ if $a\in R$ and $B\in\T(R)$, and the following relations hold, with 
$a,b\in R,\ A,B,C\in \T(R)$:
\begin{eqnarray*}
:1A: &=& :A1:\ =\ A\ ,  \ \ {[1\ _\lambda\ A]}\ =\ {[A\ _\lambda\ 1]}\ =\ 0\ ,\\
:(a\otimes B)C: &=& :a(:BC:):
 +:\Big(\int_0^Td\lambda\ a\Big) [B\ _\lambda\ C]: \nonumber\\
&+& p(a,B):\Big(\int_0^Td\lambda\ B\Big)[a\ _\lambda\ C]:\ , \\
{[a\ _\lambda\ (b\otimes C)]}
&=& :{[a\ _\lambda\ b]}C:+p(a,b):b{[a\ _\lambda\ C]}: \\
&+& \int_0^\lambda d\mu\  {[[a\ _\lambda\ b]\,_\mu\ C]} \ ,  \\
{[(a\otimes B)\ _\lambda\ C]}
&=& :\Big(e^{T\partial_\lambda}a\Big){[B\ _\lambda\ C]}:
+ p(a,B):\Big(e^{T\partial_\lambda}B\Big){[a\ _\lambda\ C]}: \\
&+& p(a,B)\int_0^\lambda d\mu\  {[B\ _\mu[a\ _{\lambda-\mu}\ C]]} \ . 
\end{eqnarray*}
The above linear maps satisfy the following grading conditions
$$
:\T_{\zeta_1}(R)\T_{\zeta_2}(R):\ \subset\ \T_{\zeta_1+\zeta_2}(R)\ ,\ \ 
[\T_{\zeta_1}(R)\ _\lambda\ \T_{\zeta_2}(R)]\ \subset\ \C[\lambda]\T_{(\zeta_1+\zeta_2)_-}(R)\ .
$$
\end{lemma}
For $\zeta\in\Gamma_+$ we let $\M_\zeta(R)$ be the subspace of $\T(R)$ spanned by the 
elements
\begin{equation}\label{MR}
A\otimes \Big((b\otimes c\otimes D-p(b,c)c\otimes b\otimes D 
-:\Big(\int_{-T}^0 d\lambda\ [b\ _\lambda\ c]\Big)D:\Big)\ ,
\end{equation}
with $b,c\in R,\ A,D\in\T(R)$, such that $A\otimes b\otimes c\otimes D\in\T_\zeta(R)$.
We also let
$\M(R)=\bigcup_{\zeta\in\Gamma_+}\M_\zeta(R)$.
This is the analogue of the space $\M(\g)$ defined in the previous section, but now
$\M(R)$ is only a left ideal of $\T(R)$, not a two-sided ideal.
In particular the quotient space $\T(R)/\M(R)$ is not an associative algebra.
\begin{definition}\label{non-lin-conf}
A {\itshape non-linear Lie conformal algebra} is a 
non-linear conformal algebra (defined above)
satisfying the following additional conditions:
\begin{enumerate}
\setcounter{enumi}{+2}
\item skewsymmetry:
$[a\ _\lambda\ b] = -p(a,b)[b\ _{-\lambda-T}\ a]$ for $a,b\in R$,
\item Jacobi identity:
$$
[a\ _\lambda[b\ _\mu\ c]]-p(a,b)[b\ _\mu[a\ _\lambda\ c]] 
- [[a\ _\lambda\ b]\,_{\lambda+\mu}\ c] 
\in \C[\lambda,\mu]\otimes\M_{\zeta}(R)\ ,
$$
where $a,b,c\in R$ and $\zeta=(\zeta(a)+\zeta(b)+\zeta(c))_-$.
\end{enumerate}
\end{definition}
\begin{example}\label{newex}
Let $R$ be a Lie conformal algebra with $\lambda$-bracket $[.\,_\lambda\,.]$
and let $\hat R=R\oplus \C C$ be its central extension, where $C$ is a central element,
$TC=0$ and the $\lambda$-bracket of $a,b\in R\subset\hat R$ is
$[a\ _\lambda\ b]\widehat{\phantom{a}}=[a\ _\lambda\ b]+\varphi(a,b)C$, with $\varphi(a,b)\in\C$.
Then we can construct a non-linear Lie conformal algebra structure on $R$ by
$$
[a\ _\lambda\ b]\widehat{\phantom{a}}=[a\ _\lambda\ b]+\varphi(a,b)\ ,
$$
and the trivial grading $R=R_1$.
Similarly, a central extension of a Lie algebra $\g$ can be viewed as a non-linear Lie algebra
structure on $\g$.
\end{example}
The following theorem, is a "conformal" analogue of the PBW Theorem \ref{gmain1} for non-linear
Lie algebras.
Its proof is much more difficult (see \cite{DK} for details).
\begin{theorem}\label{Rmain1}
\cite{DK} Let $R$ be a non-linear Lie conformal algebra.
Then there is a canonical structure of a vertex algebra on the space
$V(R)=\T(R)/\M(R)$, called the {\itshape universal enveloping vertex algebra of $R$}, 
such that the vacuum vector is $\vac=1$ mod$\M(R)$,
and such that the infinitesimal translation operator, the normally ordered product
and the $\lambda$-bracket on $V(R)$ are induced by $T,\ :\ \ :$ and $[\ _\lambda\ ]$
respectively.
Moreover the canonical map $\pi:\,\T(R)\rightarrow V(R)$ induces a $\C[T]$-module 
embedding $R\subset V(R)$ and $V(R)$ is freely generated by $R$,
namely, any ordered basis $\B$ of $R$, compatible with parity and $\Gamma_+^\prime$-grading,
freely generates $V(R)$
(in the sense of Definition \ref{def:1.14}).
\end{theorem}
\medskip

We next want to state the converse of Theorem \ref{Rmain1}, i.e. the "conformal analogue"
of Theorem \ref{gmain2} of the previous section.

\begin{definition}\label{pregr}
Let $V$ be a vertex algebra, strongly generated by a free $\C[T]$-submodule
$R=\C[T]\otimes\g$,
where $\g\ =\ \bigoplus_{\zeta\in\Gamma_+^\prime}\g_\zeta$.
Define an increasing $\Gamma_+$-filtration by subspaces $V_\zeta$ of $V$,
induced by the $\Gamma_+^\prime$-grading of 
$R=\bigoplus_{\zeta\in\Gamma_+^\prime}\big(\C[T]\otimes\g_\zeta\big)$,
where we let $\deg T=0$. Namely
\begin{equation}\label{ind-filtr}
V_\zeta\ =\ \sum_{\substack{k_1,\dots,k_s\in\Z_+\\ \zeta_1+\dots+\zeta_s\leq\zeta}} 
:(T^{k_1}\g_{\zeta_1})\dots(T^{k_s}\g_{\zeta_s}):\ .
\end{equation}
We shall say that
the vertex algebra $V$ is
{\itshape pregraded} by $\Gamma_+$ if
the $\lambda$-bracket on $V$ satisfies the following {\itshape grading condition}
\begin{equation}\label{eq-pregr}
[\g_{\zeta_1}\ _\lambda\ \g_{\zeta_2}]\ \subset\ \C[\lambda]\otimes V_{(\zeta_1+\zeta_2)_-}\ .
\end{equation}
\end{definition}
\begin{example}\label{pregr_ex}
Suppose the vertex algebra $V$ is freely generated by $\C[T]\otimes\g$ and it
has a Hamiltonian operator $H$ with eigenvalues in $\Gamma_+$,
such that $H\g\subset\g$ and $\g\cap V[0]=0$.
Then the corresponding eigenspace decomposition 
$\g=\bigoplus_{\Delta\in \text{Spec} H}\g[\Delta]$
makes $V$ a {\itshape pregraded} vertex algebra. 
Notice that in this case the $\Gamma_+$-filtration $V_\Delta$ defined in (\ref{ind-filtr})
is not induced by the $H$-eigenspace decomposition of $V$.
Indeed the filtered space $V_\Delta$ is preserved by $T$, while we know that
$T$ increases the conformal weight by 1.
On the other hand we clearly have
$$
\bigoplus_{\Delta^\prime\leq\Delta}V[\Delta^\prime]\ \subset\ V_\Delta\ .
$$
From this we get
$$
[\g[\Delta_1]\ _\lambda\ \g[\Delta_2]]
\ \subset\ \C[\lambda]\otimes \Big(\!\!\!\!\!\!
\bigoplus_{\Delta\leq\Delta_1+\Delta_2-1}V[\Delta]\Big)
\ \subset\  \C[\lambda]\otimes V_{\Delta_1+\Delta_2-1}\ ,
$$
which proves, as we wanted, that $V$ is pregraded by $\Gamma_+$.
\end{example}
It is not hard to show, by induction, that the grading condition (\ref{eq-pregr})
can be extended to the whole vertex algebra $V$, namely we have
\begin{lemma}\label{grlem}
\cite{DK} If $V$ is pregraded by $\Gamma_+$, then we have, for $\zeta_1,\zeta_2\in\Gamma_+$,
$$
:V_{\zeta_1}V_{\zeta_2}:\ \subset\ V_{\zeta_1+\zeta_2}\ \ ,\qquad
[V_{\zeta_1}\ _\lambda\ V_{\zeta_2}] \subset \C[\lambda]\otimes V_{(\zeta_1+\zeta_2)_-}\ .
$$
\end{lemma}
The proof of the following result is similar, though somewhat more complicated,
than the proof of Theorem \ref{gmain2} in the previous section,
and it can be found in \cite{DK}.
\begin{theorem}\label{Rmain2}
\cite{DK} Let $V$ be a vertex algebra.
Let $R=\bigoplus_{\zeta\in\Gamma_+^\prime}\big(\C[T]\otimes\g_\zeta\big)$
be a $\Gamma_+^\prime$-graded free $\C[T]$-submodule.
Assume that there exists a basis of $R$, compatible with 
parity and $\Gamma_+^\prime$-grading of $R$,
which freely generates $V$.
Moreover, assume that $V$ is
pregraded by $\Gamma_+$.
Then $R$ admits a structure of a non-linear Lie conformal algebra
$$
[\ _\lambda\ ]\ :\ \ R\otimes R\ \longrightarrow\ \C[\lambda]\otimes\T(R)\ ,
$$
compatible with the $\lambda$-bracket of $V$, in the sense that
$\pi([a\ _\lambda\ b]^R)\ =\ [a\ _\lambda\ b]^V$ for $a,b\in R$.
We are writing upper indices $R$ and $V$ to distinguish the $\lambda$-bracket of $V$
from the non-linear $\lambda$-bracket of $R$,
and, as before, $\pi$ denotes the quotient map $\T(R)\twoheadrightarrow V(R)$.
Moreover $V$ is canonically isomorphic to the universal enveloping vertex algebra $V(R)$.
\end{theorem}
\begin{remark}
A {\itshape homomorphism} of $\Gamma_+^\prime$-graded
non-linear Lie conformal algebras $R$ and $R^\prime$ is a $\C[T]$-module
homomorphism $\phi:\ R\rightarrow R^\prime$ preserving $\Z/2\Z$ and $\Gamma_+^\prime$-gradings,
such that the induced map $\T(R)\rightarrow \T(R^\prime)$ satisfies the following properties:
\begin{eqnarray*}
&& \phi(:AB:)-:\phi(A)\phi(B):\ \in\ \M_{\zeta(A)+\zeta(B)}(R^\prime)\ ,\\
&& \phi([A\ _\lambda\ B])-[\phi(A)\ _\lambda\ \phi(B)]\ \in\ 
\C[\lambda]\otimes\M_{(\zeta(A)+\zeta(B))_-}(R^\prime)\ .
\end{eqnarray*}
It is not hard to see, using Theorems \ref{Rmain1} and \ref{Rmain2},
that the category of $\Gamma_+^\prime$-graded free, as $\C[T]$-modules,
non-linear Lie conformal algebras is equivalent to the category of $\Gamma_+$-pregraded
vertex algebras, freely generated by a free $\C[T]$-module.
In particular $R$ and $R^\prime$ are isomorphic non-linear Lie conformal algebras
if and only if the corresponding universal enveloping vertex algebras $V(R)$ and $V(R^\prime)$
are isomorphic vertex algebras.
Of course, similar facts hold in the non-linear Lie algebra case as well.
\end{remark}

%%% addition 2: observation on non-lin Lie conformal algebras

\begin{remark}\label{rem:3.16}
Let $R$ be a non-linear conformal algebra which satisfies the skewsymmetry condition 3
of Definition \ref{non-lin-conf}.
The necessary and sufficient condition for $R$ to be a non-linear Lie conformal algebra,
i.e. to satisfy condition 4 of Definition \ref{non-lin-conf}, is as follows.
Choose an ordered basis $\B$ of $R$, compatible with $\Z/2\Z$ and $\Gamma_+^\prime$-grading.
Write the LHS of the Jacobi identity $J(a,b,c;\lambda,\mu)$ of three elements $a,b,c\in\B$,
as an element of $\T(R)[\lambda,\mu]$ in terms of linear combinations
of products of $a^i$'s.
Using skewsymmetry and quasiassociativity rules, rewrite $J(a,b,c;\lambda,\mu)$
as a sum of an element of $\M(R)$, which we ignore, and a
linear combination of ordered monomials.
Then all coefficients of this linear combination should be zero.
It immediately follows from Definition \ref{non-lin-conf} and Theorem \ref{Rmain2} that this condition
is respectively sufficient and necessary.
If $R$ is freely generated, as a $\C[T]$-module, by a set $\bar\B$,
we can take $\B=\{T^{(k)}a_i\ |\ k\in\Z_+,\,a_i\in\bar\B\}$.
Then it suffices to check the above condition for $J(a,b,c;\lambda,\mu)$
for all triples of elements of $\bar\B$.
Similar remarks hold in the ordinary algebra case.
\end{remark}

%%%

\subsection{The $H$-twisted Zhu algebra  of a  
non-linear Lie conformal algebra}
\label{sec:zhuenv}

Let $V$ be a vertex algebra with a Hamiltonian operator $H$.
Let $R\subset V$ be a free $\C[T]$-submodule
$R=\C[T]\otimes \g$, where $\fg$ is an $H$-invariant 
subspace of $V$.
As usual, we let $V[\Delta]$ be the eigenspace of $H$ with eigenvalue $\Delta$,
and we denote $\g[\Delta]=\g\cap V[\Delta]$.
% pregr
We also fix a $\Gamma_+^\prime$-grading of $\g$
by $H$-invariant subspaces:
\begin{equation}\label{g+gr}
\g\ =\ \bigoplus_{\zeta\in\Gamma_+^\prime}\g_\zeta\ ,
\end{equation}
which we extend to a $\Gamma_+^\prime$-grading of $R$ by letting $\deg(T)=0$.
We assume that there exists an ordered basis of $R$,
compatible with parity and $\Gamma_+^\prime$-grading,
which freely generates $V$.
Moreover, we assume that $V$ is pregraded by $\Gamma_+$, according 
to Definition \ref{pregr}.

By Theorems \ref{Rmain1} and \ref{Rmain2}, the above assumptions are 
equivalent to say that $R=\C[T]\g$ admits a structure of non-linear Lie 
conformal algebra, and that $V$ is isomorphic to its universal enveloping 
vertex algebra $V(R)$.

Thanks to Lemma \ref{grlem}, condition (\ref{eq-pregr}) immediately 
implies the following stronger grading conditions:
\begin{equation}\label{4.2-gr-cond}
:V_{\zeta_1}V_{\zeta_2}:\ \subset\ V_{\zeta_1+\zeta_2}\ \, , TV_\zeta\ 
\subset\ V_\zeta\ \, ,
[V_{\zeta_1}\ _\lambda\ V_{\zeta_2}]\ \subset\ \C[\lambda]\otimes 
V_{(\zeta_1+\zeta_2)_-}\ .
\end{equation}
Recall the definitions  of the $*_n$-products and the $*$-bracket on $V$, 
given in Example \ref{ex2}.
We immediately get from (\ref{4.2-gr-cond})
the following further grading 
conditions, for $k\in\Z_+$:
\begin{eqnarray}\label{4.21-gr-cond}
& \vphantom{\Big(}
V_{\zeta_1} *_{-k-1} V_{\zeta_2}\ \subset\ V_{\zeta_1+\zeta_2}\ \ ,
\qquad V_{\zeta_1} *_{k} V_{\zeta_2}\ \subset\ V_{(\zeta_1+\zeta_2)_-}\ ,
\nonumber\\
& \vphantom{\Big(}
[V_{\zeta_1}\ _*\ V_{\zeta_2}]\ \subset\ V_{(\zeta_1+\zeta_2)_-}\ .
\end{eqnarray}

\begin{remark}
If the Hamiltonian operator $H$ has eigenvalues in $\Gamma_+^\prime $,
then we can just take $\g_\zeta=\g[\zeta]$, and the grading condition 
(\ref{eq-pregr})
is satisfied.
Notice that, in this case, the $\Gamma_+$-filtration $V_{\zeta}$ is not induced 
by the $H$-grading, as 
the subspaces $V_{\zeta}$ are preserved by $T$, while $T$ increases 
the conformal weight by $1$.
On the other hand $\bigoplus_{\Delta\leq\zeta} V[\Delta] \subset V_{\zeta}$ ,
from which condition (\ref{eq-pregr}) follows immediately.
\end{remark}

We fix an ordered basis $\bar{\B}$ of $\g$, compatible with 
the $\Gamma_+^\prime $-grading (\ref{g+gr}) and the $H$-grading:
$$
\bar{\B}\ =\ \big\{e_\alpha\ \big|\ \ \alpha\in\bar{\I}\big\}\ .
$$
We let $\zeta_\alpha=\zeta$ if $e_\alpha\in\g_\zeta$, 
and $\Delta_\alpha=\Delta$ if $e_\alpha\in\g[\Delta]$.
Moreover we let $\bar{\I}_\zeta=\{\alpha\ |\ \zeta_\alpha=\zeta\}$
and $\bar{\I}[\Delta]=\{\alpha\ |\ \Delta_\alpha=\Delta\}$, and
we extend $\bar{\B}$ to the following ordered basis of $R=\C[T]\g$:
$$
\B\ =\ \Big\{e_i=T^{(k)}e_\alpha\ \big|\ \ i=(\alpha,k)\in \bar{\I}\times\Z_+
=: \I\Big\}\ .
$$
As before, $T^{(k)}=T^k/k!$.
Recalling that $TV_\zeta\subset V_\zeta$ and $TV[\Delta]\subset V[\Delta+1]$, 
we let
$\I_\zeta=\bar{\I}_\zeta\times\Z_+$ and $\I[\Delta]=\bigsqcup_{k\in\Z_+}
\bar{\I}[\Delta-k]\times\{k\}$.
By Theorem \ref{Rmain1}, $V$ is freely generated by $R$, hence
the collection of ordered monomials
$$
S\ =\ \big\{:e_{i_1}\dots e_{i_s}:\ \big|\ \#(\vec{i})=0\big\}\ .
$$
is a basis of $V$.
Here and further, for $\vec{j}=(j_1,\dots,j_s)\in\I^s$, 
with $j_l=(\alpha_l,k_l),\ \alpha_l\in\bar{\I}_{\zeta_l}\cap\bar{\I}[\Delta_l],
\ k_l\in\Z_+$, we denote $\zeta(\vec{j})$ and $\#(\vec{j})$ as in (\ref{notat}),
and we let
$$
\Delta(\vec{j}) = \Delta_1+k_1+\cdots+\Delta_s+k_s\ .
$$
It is clear by definition that, for $\vec{j}=(j_1,\dots,j_s)\in\I^s$,
we have $:e_{j_1}\dots e_{j_s}:\ \in V_{\zeta(\vec{j})}\cap V[\Delta(\vec{j})]$.
We define the sets
\begin{eqnarray*}
S_\zeta &=& \Big\{:e_{i_1}\cdots e_{i_s}:\ \Big|\ 
        \zeta(\vec{i})\leq\zeta\ ,\ \ \#(\vec{i})=0\Big\}\ , \\
S^*_\zeta &=& \Big\{e_{i_1}*_{-1}\cdots *_{-1}e_{i_s}\ \Big|\ 
        \zeta(\vec{i})\leq\zeta\ ,\ \ \#(\vec{i})=0\Big\}\ ,
\end{eqnarray*}
so that the basis $S$ can be written as $S=\bigcup_\zeta S_\zeta$, and similarly we define
$S^*=\bigcup_\zeta S^*_\zeta$.
In the above formula, the $*_{-1}$-product of more than two elements is 
defined, as for the normally ordered product, by taking products from right to left:
$e_{i_1}*_{-1}\cdots *_{-1}e_{i_s}=e_{i_1}*_{-1}(e_{i_2}*_{-1}(\cdots *_{-1}e_{i_s}))$.
Recall the filtration $V_\zeta$ introduced above; we also define spaces $V^*_\zeta$
in a similar way, with $*_{-1}$-products instead of normally ordered products. In other words,
\begin{eqnarray*}
V_\zeta &=& \Span_\C\Big\{:e_{j_1}\cdots e_{j_s}:\ \Big|\ 
        \zeta(\vec{j})\leq\zeta\Big\}\ , \\
V^*_\zeta &=& \Span_\C\Big\{e_{j_1}*_{-1}\cdots *_{-1}e_{j_s}\ \Big|\ 
        \zeta(\vec{j})\leq\zeta\Big\}\ .
\end{eqnarray*}
\renewcommand{\theenumi}{\alph{enumi}}
\renewcommand{\labelenumi}{(\theenumi)}
\begin{lemma}\label{lem1}
\begin{enumerate}
% a
\item For every $\vec{j}=(j_1,\dots,j_s)$, we have
$$
e_{j_1}*_{-1}\cdots *_{-1}e_{j_s}\ -\ :e_{j_1}\cdots e_{j_s}:\ \ \in\ 
V_{\zeta(\vec{j})_-}\ .
$$
% b
\item The following spaces are equal:
$$
V_\zeta\ =\ \Span_\C S_\zeta\ =\ V^*_\zeta\ =\ \Span_\C S^*_\zeta\ .
$$
% c
\item $S^*$ is a basis of $V$.
\end{enumerate}
\end{lemma}
\renewcommand{\theenumi}{\arabic{enumi}}
\renewcommand{\labelenumi}{\theenumi.}
\begin{proof}
We prove (a) by induction on $\zeta(\vec{j})$. For $s=1$ there is nothing to prove,
so let $s\geq2$. By inductive assumption we have
$$
e_{j_2}*_{-1}\cdots *_{-1}e_{j_s}\ =\ :e_{j_2}\cdots e_{j_s}:\ +\ A\ ,
$$
where $A\in V_{(\zeta(\vec{j})-\zeta(e_{j_1}))_-}$.
Hence, by definition of the $*_{-1}$-product, we have
\begin{eqnarray*}
e_{j_1}*_{-1}\cdots *_{-1}e_{j_s} 
&=& :e_{j_1}e_{j_2}\cdots e_{j_s}:\ +\  e_{j_1}*_{-1}A \\
&+& \sum_{n\geq0}
        \left(\begin{array}{c} \!\!\! \Delta(e_{j_1})\!\!\! \\ \!\!\! n+1 \!\!\! \end{array}\right)
        e_{j_1}\ _{(n)}\ (:e_{j_2}\cdots e_{j_s}:)\ .
\end{eqnarray*}
The grading conditions (\ref{4.2-gr-cond}) guarantee that
$e_{j_1}\ _{(n)}\ (:e_{j_2}\cdots e_{j_s}:)\in V_{\zeta(\vec{j})_-}$, 
and similarly conditions (\ref{4.21-gr-cond}) guarantee that 
$e_{j_1}*_{-1}A\in V_{\zeta(\vec{j})_-}$.
This proves (a).

We now prove, by induction on $\zeta$, that $V^*_\zeta=V_\zeta$.
Let $j=(j_1,\dots,j_s)$ be such that $\zeta(\vec{j})=\zeta$.
By (a) we have
\begin{equation}\label{29jul}
e_{j_1}*_{-1}\cdots *_{-1}e_{j_s}\ -\ :e_{j_1}\cdots e_{j_s}:\ =\ A\ \in\ V_{\zeta_-}\ .
\end{equation}
Now we can read equation (\ref{29jul}) in two ways.
On one hand, since
$:e_{j_1}\cdots e_{j_s}:\ \in V_\zeta$,
and $A\in V_{\zeta_-}\subset V_\zeta$,
we get $e_{j_1}*_{-1}\cdots *_{-1}e_{j_s}\in V_\zeta$.
Hence $V^*_\zeta\subset V_\zeta$.
On the other hand, we have
$e_{j_1}*_{-1}\cdots *_{-1}e_{j_s}\in V^*_\zeta$,
and, by inductive assumption,
$A\in V_{\zeta_-}=V^*_{\zeta_-}\subset V^*_\zeta$,
so that $:e_{j_1}\cdots e_{j_s}:\ \in V^*_\zeta$.
Hence $V_\zeta\subset V^*_\zeta$.

To show that $V_\zeta=\Span_\C S_\zeta$, we use the standard argument.
First notice that, obviously, $\Span_\C S_\zeta\subset V_\zeta$.
So we only need to show that, given an arbitrary (non necessarily ordered)
monomial $:e_{j_1}\cdots e_{j_s}:\ \in V_\zeta,\ \zeta=\zeta(\vec{j})$,
we can write it as a linear combination of ordered monomials 
$:e_{i_1}\cdots e_{i_s}:\ \in S_\zeta$.
We will prove the statement by induction on the pair $(\zeta(\vec{j}),\#(\vec{j}))$ 
ordered lexicographically.
If $\#(\vec{j})=0$, we have $:e_{j_1}\cdots e_{j_s}:\ \in S_\zeta$, so there is nothing to prove.
Suppose then $\#(\vec{j})\geq1$, and let $p$
be such that $j_p>j_{p+1}$, or $j_p=j_{p+1}$ and $p(e_{j_p})=\bar{1}$.
In the first case we have:
\begin{eqnarray*}
:e_{j_1}\cdots e_{j_s}:
 &=& p(e_{j_p},e_{j_{p+1}}):e_{j_1}\cdots e_{j_{p+1}}e_{j_p}\cdots e_{j_s}: \\
&+& :e_{j_1}\cdots \left(\int_{-T}^0 d\lambda[e_{j_{p}}\ _\lambda\ e_{j_{p+1}}]\right)\dots e_{j_s}:
\ \ =\ B+C\ .
\end{eqnarray*}
Obviously $\zeta(B)=\zeta(\vec{j})$ 
and $\#(B)=\#(\vec{j})-1$,
so $B\in \Span_\C S_\zeta$ by inductive assumption.
Moreover, it immediately follows by the grading conditions (\ref{4.2-gr-cond}) 
that $C\in V_{\zeta_-}$,
hence $C\in \Span_\C S_\zeta$,  again by inductive assumption.
Consider now the second case, namely $j_p=j_{p+1}$ and $p(e_{j_p})=\bar{1}$.
We have
$$
:e_{j_1}\cdots e_{j_s}:
\ =\ \frac{1}{2}
 :e_{j_1}\cdots \left(\int_{-T}^0 d\lambda[e_{j_{p}}\ _\lambda\ e_{j_{p+1}}]\right)\dots e_{j_s}:\ .
$$
As before, the RHS belongs to $\Span_\C S_\zeta$ thanks to 
the grading conditions (\ref{4.2-gr-cond}) and inductive assumption.

To conclude the proof of (b), we are left to show that $V^*_\zeta=\Span_\C S^*_\zeta$.
We could proceed in the same way as above.
Instead we will prove, equivalently, that $\Span_\C S^*_\zeta=\Span_\C S_\zeta$.
The proof is again by induction on $\zeta$.
Let $\vec{i}=(i_1,\dots,i_s)$ be such that $\zeta(\vec{i})=\zeta$ and $\#(\vec{i})=0$.
By (a) we have
\begin{equation}\label{29jul-1}
e_{i_1}*_{-1}\cdots *_{-1}e_{i_s}\ -\ :e_{i_1}\cdots e_{i_s}:\ =\ A\ \in\ V_{\zeta_-}\ .
\end{equation}
By the previous result and inductive assumption we have
$A\in V_{\zeta_-}=\Span_\C S_{\zeta_-}=\Span_\C S^*_{\zeta_-}$.
Hence it follows by (\ref{29jul-1}) that
both $e_{i_1}*_{-1}\cdots *_{-1}e_{i_s}\in \Span_\C S_\zeta$
and $:e_{i_1}\cdots e_{i_s}:\in \Span_\C S^*_\zeta$.
This clearly implies that $\Span_\C S^*_\zeta=\Span_\C S_\zeta$.

We are left to prove (c).
By (b)  we only need to show that
elements of $S^*$ are linearly independent.
Suppose by contradiction that there is a non-trivial linear dependence,
\begin{equation}\label{29jul-2}
\sum_{\vec{i}}c_{\vec{i}}\ e_{i_1}*_{-1}\cdots *_{-1}e_{i_s}\ =\ 0\ .
\end{equation}
Let $\zeta$ be the largest value of $\zeta(\vec{i})$ for which $c_{\vec{i}}\neq0$.
Thanks to (a), we can rewrite (\ref{29jul-2}) as
\begin{equation}\label{29jul-3}
\sum_{\vec{i}\ |\ \zeta(\vec{i})=\zeta}c_{\vec{i}}\ :e_{i_1}\cdots e_{i_s}:\ +\ A\ =\ 0\ ,
\end{equation}
where $A\in V_{\zeta_-}=\Span_\C S_{\zeta_-}$. 
On the other hand, by assumption, $S$ is a basis of $V$. 
Hence (\ref{29jul-3}) implies that $c_{\vec{i}}=0$ for all $\vec{i}$ such that $\zeta(\vec{i})=\zeta$,
which is a contradiction.
\end{proof}

For $\zeta\in\Gamma_+$, we let
$$
\Sigma_\zeta = \left\{
        \begin{array}{c}
        (T^{(h_1)}e_{\alpha_1})*_{-1}\cdots *_{-1}(T^{(h_s)}e_{\alpha_s}) \\
        -\binom{-\Delta_1}{h_1}\cdots\binom{-\Delta_s}{h_s}
        e_{\alpha_1}\!*_{-1}\!\cdots\! *_{-1}\!e_{\alpha_s}
        \end{array}
        \left|
        \begin{array}{c}
        \zeta(\vec{i})\leq\zeta,\ \#(\vec{i})=0,\sum_i h_i>0\\
        \text{where } 
        \vec{i}=(i_1,\dots,i_s),\\
        i_l\!\!\!=\!\!\!(\!\alpha_l,h_l\!),\, \alpha_l\!\in \bar{\I}[\Delta_l],\, h_l\!\in\Z_+
        \end{array}
        \right.\right\}\ ,
$$
and
$$
W_\zeta = \Span_\C\left\{
        \begin{array}{c}
        (T^{(k_1)}e_{\beta_1})*_{-1}\cdots *_{-1}(T^{(k_s)}e_{\beta_s}) \\
        -\binom{-\Delta_1}{k_1}\cdots\binom{-\Delta_s}{k_s}
        e_{\beta_1}\!*_{-1}\!\cdots\! *_{-1}\!e_{\beta_s}
        \end{array}
        \left|
        \!\!\begin{array}{c}
        \!\!\zeta(\vec{j})\leq\zeta, \text{ where} \\ 
        \vec{j}\!=\!(\!j_1\dots j_s\!),\,k_l\in\Z_+\\
        j_l\!=\!(\!\beta_l\!,\!k_l\!),\ \beta_l\!\in\!\bar{\I}[\Delta_l]
        \end{array}
        \right.\right\}\ .
$$
Moreover we let $\Sigma=\bigcup_\zeta \Sigma_\zeta$ and $W=\bigcup_\zeta W_\zeta$. 
Obviously $\Span_\C \Sigma_\zeta\subset W_\zeta$. We will prove that the converse is also true.

\begin{lemma}\label{trivw}
If $e=T^{(k_1)}e_{\beta_1},\ \beta_1\in\bar{\I}_{\zeta_1}\cap\bar{\I}[\Delta_1],\ k_1\in\Z_+$,
and $A\in W_\zeta$, then
$$
e*_{-1}A\ \in\ W_{\zeta_1+\zeta}\ .
$$
\end{lemma}
\begin{proof}
Let
$$
A = (T^{(k_2)}e_{\beta_2})*_{-1}\cdots *_{-1}(T^{(k_s)}e_{\beta_s}) \\
        -\binom{-\Delta_2}{k_2}\cdots\binom{-\Delta_s}{k_s}
        e_{\beta_2}\!*_{-1}\!\cdots\! *_{-1}\!e_{\beta_s}\ .
$$
The lemma then immediately follows by the following obvious identity,
obtained by adding and subtracting terms:
\begin{eqnarray*}
e*_{-1}A 
&=& \Bigg((T^{(k_1)}e_{\beta_1})*_{-1}(T^{(k_2)}e_{\beta_2})*_{-1}
        \cdots *_{-1}(T^{(k_s)}e_{\beta_s}) \\
&-& \left(\begin{array}{c}\!\!\!\! -\Delta_1\!\!\!\!\\\!\!\!\! k_1 \!\!\!\!\end{array}\right)
        \!\!\left(\begin{array}{c}\!\!\!\! -\Delta_2\!\!\!\!\\\!\!\!\! k_2 \!\!\!\!\end{array}\right)\!\!
        \cdots \!\!\left(\begin{array}{c}\!\!\!\! -\Delta_s\!\!\!\!\\\!\!\!\! k_s \!\!\!\!\end{array}\right)\!\!
        e_{\beta_1}*_{-1}e_{\beta_2}\!*_{-1}\!\cdots\! *_{-1}\!e_{\beta_s}
        \Bigg) \\
&-&  \!\!\left(\begin{array}{c}\!\!\!\! -\Delta_2\!\!\!\!\\\!\!\!\! k_2 \!\!\!\!\end{array}\right)\!\!
        \cdots \!\!\left(\begin{array}{c}\!\!\!\! -\Delta_s\!\!\!\!\\\!\!\!\! k_s \!\!\!\!\end{array}\right)\!\!
        \!\!\left(\!\!T^{(k_1)}\!e_{\beta_1}\!
                -\!\left(\begin{array}{c}\!\!\!\! -\Delta_1\!\!\!\!\\\!\!\!\! k_1 \!\!\!\!\end{array}\right)
                \!\!e_{\beta_1}
        \!\!\right)\!\!*_{-1}\!e_{\beta_2}\!*_{-1}\!\cdots\! *_{-1}\!e_{\beta_s}\ .
\end{eqnarray*}
\end{proof}

Recall from Section \ref{sec:2} that we denote by $J_{\hbar=1}=V*_{-2}V$ the 
span of elements $a*_{-2}b$ with $a,b\in V$.
We also denote, for $\zeta\in\Gamma_+$,
$$
\big(V*_{-2} V\big)_\zeta\ =\ 
\sum_{\substack{\zeta_1,\zeta_2\in\Gamma_+\\\zeta_1+\zeta_2\leq\zeta}} 
V_{\zeta_1} *_{-2} V_{\zeta_2}\ ,
$$
so that $V*_{-2}V=\bigcup_{\zeta\in\Gamma_+} \big(V*_{-2} V\big)_\zeta$.
We have the following
\renewcommand{\theenumi}{\alph{enumi}}
\renewcommand{\labelenumi}{(\theenumi)}
\begin{lemma}\label{lem-4.9}
\begin{enumerate}
% a
\item For $A\in V_{\zeta_1},\ B\in V_{\zeta_2}$, we have
\begin{eqnarray}
&&\displaystyle{\!\!\!\!\!\!\!\!\!\!\!\!\!\!\!\!\!\!\!\!\!\!\!\!
 (T^{(k)}A)*_{-1}B-\binom{\Delta(A)}{k} A*_{-1}B\ \in\ V_{\zeta_1}*_{-2}V_{\zeta_2}\ , 
 }\label{prelim}\\
&&\displaystyle{\!\!\!\!\!\!\!\!\!\!\!\!\!\!\!\!\!\!\!\!\!\!\!\!
 [(T^{(h)}A) _* (T^{(k)} B)]
 }\nonumber\\
 &&\displaystyle{\!\!\!\!\!\!\!\!\!\!\!\!\!\!\!\!\!\!\!\!\!\!\!\!
        -\!\!\binom{-\Delta(A)}{h}\!\!\binom{-\Delta(B)}{k}\! [A _* B]
        \ \in\ (V\! *_{-2}\!V)_{(\zeta_1+\zeta_2)_-} \label{prelim2}\ .
        }
\end{eqnarray}
% b
\item For $A\in V_{\zeta_1},\ B\in V_{\zeta_2},\ C\in V_{\zeta_3}$, we have
\begin{eqnarray}
\vphantom{\Big(}
& A*_{-k-1}B\ \in\ V_{\zeta_1}*_{-2}V_{\zeta_2}\ ,\ \ \forall k\geq1\ , \label{-2eq8}\\
\vphantom{\Big(}
& (A*_{-1}B)*_{-1}C-A*_{-1}(B*_{-1}C) \in (V*_{-2}V)_{(\zeta_1+\zeta_2+\zeta_3)_-}
        \ , \label{-2eq13}\\
\vphantom{\Big(}
& A\!*_{-1}\!B-p(A,B)B\!*_{-1}\!A-[A\ _*\ B] \in (V\!*_{-2}\!V)_{(\zeta_1+\zeta_2)_-}
        \ .\label{-2eq14}
\end{eqnarray}
\item The following relations hold
\begin{eqnarray}
\vphantom{\Big(}
&& V_{\zeta_1}*_{-1}(V_{\zeta_2}*_{-2}V_{\zeta_3})\ \subset\ (V*_{-2}V)_{\zeta_1+\zeta_2+\zeta_3}
        \ ,\label{-2eq9}\\
\vphantom{\Big(}
&& (V_{\zeta_1}*_{-2}V_{\zeta_2})*_{-1}V_{\zeta_3}
        \ \subset\ (V*_{-2}V)_{\zeta_1+\zeta_2+\zeta_3}\ , \label{-2eq10}\\
\vphantom{\Big(}
&& {[V_{\zeta_1}\ _*\ (V_{\zeta_2}*_{-2}V_{\zeta_3})]}\ \subset\ (V*_{-2}V)_{(\zeta_1+\zeta_2+\zeta_3)_-}
        \ ,\label{-2eq11}\\
\vphantom{\Big(}
&& [(V_{\zeta_1}*_{-2}V_{\zeta_2})\ _*\ V_{\zeta_3}]
        \ \subset\ (V*_{-2}V)_{(\zeta_1+\zeta_2+\zeta_3)_-}
        \ . \label{-2eq12}
\end{eqnarray}
\end{enumerate}
\end{lemma}
\renewcommand{\theenumi}{\arabic{enumi}}
\renewcommand{\labelenumi}{\theenumi.}
\begin{proof}
We want to prove (\ref{prelim}) by induction on $k$. 
For $k=1$ it follows immediately by equation (\ref{eq:2.8}).
For $k\geq2$ we have
\begin{eqnarray*}
&& (T^{(k)}A)*_{-1}B
        -\binom{-\Delta(A)}{k} A*_{-1}B \\
&&\ =\ \frac{1}{k}\left(
        (T^{(k-1)}T A)*_{-1}B
        -\binom{-\Delta(A)-1}{k-1} (T A)*_{-1}B
        \right) \\
&&\qquad
+\ \frac{1}{k} \binom{-\Delta(A)-1}{k-1} A*_{-2}B\ .
\end{eqnarray*}
The above identity is obtained by adding and subtracting
$\frac1k \binom{-\Delta(A)-1}{k}(TA)*_{-1}~B$, and using (\ref{eq:2.8}) with $\hbar=1$.
Notice that, by definition, the filtered space $V_\zeta$ is invariant by the action of $T$.
Hence, by inductive assumption, both terms on the RHS side belong 
to $V_{\zeta_1}*_{-2}V_{\zeta_2}$, as wanted.
By (\ref{2.9-eq}) we have
$$
[(T^{(h)}A) _* B]
\ =\  -\left(\begin{array}{c}\!\!-\Delta(A)\!\!\\\!\! h \!\!\end{array}\right)[A _* B]\ ,
$$
so it suffices to prove (\ref{prelim2}) for $h=0$.
We will prove it by induction on $k$.
For $k=1$ we have, by (\ref{2.10-eq}),
$$
 [A _* (T B)]+\Delta(B)[A _* B]
\ =\ (T+H)[A _* B]\ =\ [A\ _*\ B]*_{-2}\vac\ ,
$$
and the RHS belongs to
$V_{(\zeta_1+\zeta_2)_-} *_{-2}V_0\subset(V*_{-2}V)_{(\zeta_1+\zeta_2)_-}$.
For $k\geq2$ we have, by adding and subtracting terms,
\begin{eqnarray*}
&& [A _* (T^{(k)} B)]
 -\left(\begin{array}{c}\!\!\! -\Delta(B) \!\!\!\\\!\!\! k \!\!\!\end{array}\right)[A\ _*\ B] \\
&&\ \ =\ \frac{1}{k}\left(
         [A _* (T^{(k-1)} T B)]
         -\left(\begin{array}{c}\!\!\! -\Delta(B)-1 \!\!\!\\\!\!\! k-1 \!\!\!\end{array}\right)[A\ _*\ T B]
         \right) \\
&&\ \ +\ \frac{1}{k}\left(\begin{array}{c}\!\!\! -\Delta(B)-1 \!\!\!\\\!\!\! k-1 \!\!\!\end{array}\right)\Big(
        {[A _* (T B)]}+\Delta(B)[A\ _*\ B]\Big)\ .
\end{eqnarray*}
By inductive assumption, both terms on the RHS belong 
to $(V*_{-2}V)_{(\zeta_1+\zeta_2)_-}$, thus proving (\ref{prelim2}).

We can now use the results of the previous section to deduce the relations in part (b).
Equation (\ref{-2eq8}) follows immediately by  (\ref{eq:2.33}) and (\ref{prelim}),
equation(\ref{-2eq13}) follows by (\ref{eq:2.17}), (\ref{4.21-gr-cond}) and (\ref{-2eq8}),
equation (\ref{-2eq14}) follows by  (\ref{eq:2.19}), (\ref{4.2-gr-cond}) and (\ref{prelim}) with $B=\vac$.

We are left to prove part (c).
Condition (\ref{-2eq10}) follows by (\ref{eq:2.18}), (\ref{4.21-gr-cond}) and (\ref{-2eq8}).
Condition (\ref{-2eq11}) follows by Lemma \ref{lem:2.6} and (\ref{4.21-gr-cond}).
For (\ref{-2eq9}) we have, by adding and subtracting terms,
\begin{eqnarray*}
\vphantom{\Big(}
& A*_{-1}(B*_{-2}C)\ =\ 
p(A,B)p(A,C)
(B*_{-2}C)*_{-1}A \\
\vphantom{\Big(}
& \ +\  [A\ _*\ (B*_{-2}C)]\ 
+\ \Big(A*_{-1}(B*_{-2}C) \\
\vphantom{\Big(}
&\ \  -
p(A,B)p(A,C)
(B*_{-2}C)*_{-1}A -[A\ _*\ (B*_{-2}C)] 
 \Big)\ ,
\end{eqnarray*}
so that (\ref{-2eq9}) follows by (\ref{-2eq10}), (\ref{-2eq11}) and (\ref{-2eq14}).
Finally,  with a similar argument, based on adding and subtracting the
appropriate terms to $[(A*_{-2}B)_*C]$, it is not hard to show that 
(\ref{-2eq12}) follows by (\ref{-2eq11}) and (\ref{-2eq14}).
\end{proof}

\begin{lemma}\label{lem2}
(a) The following spaces are equal:
$$
\Span_\C \Sigma_\zeta\ =\ W_\zeta\ =\ (V*_{-2}V)_\zeta\ .
$$
In particular 
$$
\Span_\C \Sigma\ =\ W\ =\ V*_{-2}V\ .
$$
\noindent (b) $\Sigma$ is a basis of $V*_{-2}V$.
\end{lemma}
\begin{proof}
Let $\vec{j}=(j_1,\dots,j_s)$ be such that 
$j_l=(\beta_l,k_l),\ \beta_l\in\bar{\I}_{\zeta_l}\cap \bar{\I}[\Delta_l],\ k_l\in\Z_+$,
and $\zeta(\vec{j})=\zeta_1+\cdots\zeta_s=\zeta$. We then have,
by adding and subtracting terms,
\begin{eqnarray*}
& (T^{(k_1)}e_{\beta_1})*_{-1}\cdots *_{-1}(T^{(k_s)}e_{\beta_s}) 
        -\left(\begin{array}{c}\!\! -\Delta_1\!\!\\\!\! k_1 \!\!\end{array}\right)
        \cdots \left(\begin{array}{c}\!\! -\Delta_s\!\!\\\!\! k_s \!\!\end{array}\right)
        e_{\beta_1}*_{-1}\cdots *_{-1}e_{\beta_s} \\
& =\ 
\left(
        T^{(k_1)}e_{\beta_1}
        -\left(\begin{array}{c}\!\! -\Delta_1\!\!\\\!\! k_1 \!\!\end{array}\right)e_{\beta_1}
        \right)*_{-1}\Bigg(
        (T^{(k_2)}e_{\beta_2})*_{-1}\cdots *_{-1}(T^{(k_s)}e_{\beta_s}) 
        \Bigg) \\
& +\ 
\left(\begin{array}{c}\!\! -\Delta_1\!\!\\\!\! k_1 \!\!\end{array}\right)
        e_{\beta_1}*_{-1}\Bigg(
        (T^{(k_2)}e_{\beta_2})*_{-1}\cdots *_{-1}(T^{(k_s)}e_{\beta_s}) \\
& -\ \left(\begin{array}{c}\!\! -\Delta_2\!\!\\\!\! k_2 \!\!\end{array}\right)
        \cdots \left(\begin{array}{c}\!\! -\Delta_s\!\!\\\!\! k_s \!\!\end{array}\right)
        e_{\beta_2}*_{-1}\cdots *_{-1}e_{\beta_s}
        \Bigg)\ .
\end{eqnarray*}
The first term in the RHS belongs to $V_{\zeta_1}*_{-2}V_{\zeta-\zeta_1}$ 
by (\ref{prelim}), while the second term in the RHS belongs to $(V*_{-2}V)_\zeta$
by induction on $s$ and (\ref{-2eq9}).
This proves the inclusion $W_\zeta\subset (V*_{-2}V)_\zeta$.

We next prove the reverse inclusion: $(V*_{-2}V)_\zeta\subset W_\zeta$.
Recall that, by Lemma \ref{lem1}(b), $V_\zeta$ is spanned 
by elements $e_{j_1}*_{-1}\cdots *_{-1}e_{j_s}$ with $\zeta(\vec{j})\leq\zeta$.
Let $\vec{i}=(i_1,\dots,i_s)$, with 
$i_l=(\alpha_l,h_l),\ \alpha_l\in\bar{\I}_{\zeta^1_l}\cap\bar{\I}[\Delta^1_l],\ h_l\in\Z_+$,
and $\vec{j}=(j_1,\dots,j_t)$, with 
$j_l=(\beta_l,k_l),\ \beta_l\in\bar{\I}_{\zeta^2_l}\cap\bar{\I}[\Delta^2_l],\ k_l\in\Z_+$,
be such that $\zeta(\vec{i})=\zeta^1_1+\cdots+\zeta^1_s=\zeta_1$,
$\zeta(\vec{j})=\zeta^2_1+\cdots+\zeta^2_t=\zeta_2$, and $\zeta_1+\zeta_2=\zeta$.
We need to prove, by induction on $\zeta$, that
$$
(e_{i_1}*_{-1}\cdots *_{-1}e_{i_s})*_{-2}(e_{j_1}*_{-1}\cdots *_{-1}e_{j_t})\ \in\ W_\zeta\ .
$$
We will consider separately the two cases $s=1$ and $s\geq2$.
For $s=1$ we have
\begin{eqnarray*}
&& e_{i_1}*_{-2}(e_{j_1}*_{-1}\cdots *_{-1}e_{j_t})
\ =\ 
((T+\Delta) e_{i_1})*_{-1}e_{j_1}*_{-1}\cdots *_{-1}e_{j_t} \\
&&\ =\ 
(h_1+1)\Bigg(
        (T^{(h_1+1)} e_{\alpha_1})*_{-1}(T^{(k_1)}e_{\beta_1})*_{-1}
                \cdots *_{-1}(T^{(k_t)}e_{\beta_t}) \\
&&\ \ \ \ -\left(\begin{array}{c}\!\!\!-\Delta^1_1 \!\!\!\\\!\!\! h_1+1 \!\!\!\end{array}\right)
        \left(\begin{array}{c}\!\!\!-\Delta^2_1 \!\!\!\\\!\!\! k_1 \!\!\!\end{array}\right)
        \cdots\left(\begin{array}{c}\!\!\!-\Delta^2_t \!\!\!\\\!\!\! k_t \!\!\!\end{array}\right)
        e_{\alpha_1}*_{-1}e_{\beta_1}*_{-1}\cdots *_{-1}e_{\beta_t}
        \Bigg) \\
&&\ +\ 
(\Delta^1_1+h_1)\Bigg(
        (T^{(h_1)} e_{\alpha_1})*_{-1}(T^{(k_1)}e_{\beta_1})*_{-1}
                \cdots *_{-1}(T^{(k_t)}e_{\beta_t}) \\
&&\ \ \ \ -\left(\begin{array}{c}\!\!\!-\Delta^1_1 \!\!\!\\\!\!\! h_1 \!\!\!\end{array}\right)
        \left(\begin{array}{c}\!\!\!-\Delta^2_1 \!\!\!\\\!\!\! k_1 \!\!\!\end{array}\right)
        \cdots\left(\begin{array}{c}\!\!\!-\Delta^2_t \!\!\!\\\!\!\! k_t \!\!\!\end{array}\right)
        e_{\alpha_1}*_{-1}e_{\beta_1}*_{-1}\cdots *_{-1}e_{\beta_t}
        \Bigg)\ \in\ W_\zeta\ .
\end{eqnarray*}
The last equality is obtained by adding and subtracting the same term and using
the obvious identity 
$(h_1+1)\binom{-\Delta^1_1}{h_1+1}=-(\Delta^1_1+h_1)\binom{-\Delta^1_1}{h_1}$.
For $s\geq2$ we use equation (\ref{eq:2.16}) with 
$a=e_{i_1},\,b=e_{i_2}*_{-2}\cdots *_{-1}e_{i_s},\,c=e_{j_1}*_{-1}\cdots *_{-1}e_{j_t},\,
n=-1$ and $k=-2$. We get
\begin{eqnarray*}
\vphantom{\Big(}
&& (e_{i_1}*_{-1}\cdots *_{-1}e_{i_s})*_{-2}(e_{j_1}*_{-1}\cdots *_{-1}e_{j_t}) \\
\vphantom{\Big(}
&&\ =\ 
e_{i_1}*_{-1}((e_{i_2}*_{-1}\cdots *_{-1}e_{i_s})*_{-2}(e_{j_1}*_{-1}\cdots *_{-1}e_{j_t})) \\
\vphantom{\Big(}
&&\ +
e_{i_1}*_{-2}((e_{i_2}*_{-1}\cdots *_{-1}e_{i_s})*_{-1}(e_{j_1}*_{-1}\cdots *_{-1}e_{j_t})) \\
&&\ +
\sum_{j\geq0}\Bigg(
e_{i_1}*_{-j-3}((e_{i_2}*_{-1}\cdots *_{-1}e_{i_s})*_{j}(e_{j_1}*_{-1}\cdots *_{-1}e_{j_t})) \\
&&\ \ \ \pm
        (e_{i_2}*_{-1}\cdots *_{-1}e_{i_s})*_{-j-3}(e_{i_1}*_{j}(e_{j_1}*_{-1}\cdots *_{-1}e_{j_t}))
\Bigg)\ ,
\end{eqnarray*}
where $\pm=p(e_{i_1}e_{i_2})\cdots p(e_{i_1},e_{i_s})$.
The first term in the RHS belongs to $W_\zeta$ by Lemma \ref{trivw} and 
by induction.
Moreover, by the above result (for $s=1$), also the second term of the RHS 
belongs to $W_\zeta$.
Finally, by (\ref{4.21-gr-cond}) and (\ref{-2eq8}), the last terms of the RHS 
belong
to $(V*_{-2}V)_{\zeta_-}$, so it belongs 
to $W_\zeta$ by induction on $\zeta$.

To conclude part (a) we are left to prove that $W_\zeta\subset\Span_\C \Sigma_\zeta$.
We will use the usual argument.
Let $\vec{j}=(j_1,\dots,j_s)$ be an arbitrary (not necessarily ordered) $s$-tuple
with $j_l=(\beta_l,k_l),\,\beta_l\in{\bar I}_{\zeta_l}[\Delta_l],\,k_l\in\Z_+$
and such that $\zeta(\vec{j})=\zeta$. Consider the corresponding element
\begin{eqnarray*}
A &=& (T^{(k_1)}e_{\beta_1})*_{-1}\cdots *_{-1}(T^{(k_s)}e_{\beta_s}) \\
&-& \left(\begin{array}{c}\!\! -\Delta_1\!\!\\\!\! k_1 \!\!\end{array}\right)
        \cdots \left(\begin{array}{c}\!\! -\Delta_s\!\!\\\!\! k_s \!\!\end{array}\right)
        e_{\beta_1}*_{-1}\cdots *_{-1}e_{\beta_s}\ \in\ W_\zeta\ .
\end{eqnarray*}
We need to prove, by induction on the pair $(\zeta(\vec{j}),\#(\vec{j}))$,
that we can write $A$ as a linear combination of "ordered" elements from $\Sigma_\zeta$.
If $\#(\vec{j})=0$, we have $A\in \Sigma_\zeta$, so there is nothing to prove.
Then, suppose that $\#(\vec{j})\geq1$, and we consider the case in which there is
some $p\in\{1,\dots,s\}$  such that $j_p>j_{p+1}$.
The case $j_p=j_{p+1},\ p(e_{j_p})=\bar{1}$, can be studied in the same way.
By adding and subtracting the appropriate terms, we can rewrite $A$ as
the sum of the following ten elements:
\begin{eqnarray*}
A_1 
&=&
e_{j_1}*_{-1} \cdots *_{-1}
        \Bigg(
        e_{j_p}*_{-1}e_{j_{p+1}}*_{-1}\cdots*_{-1}e_{\beta_s} \\
&-& \Big(e_{j_p}*_{-1}e_{j_{p+1}}\Big)*_{-1}\cdots*_{-1}e_{j_s}
        \Bigg)\ , \\
A_2
&=& - \prod_{l=1}^s\!\left(\begin{array}{c}\!\!\!\! -\Delta_l \!\!\!\!\\\!\! k_l \!\!\end{array}\right)
        e_{\beta_1}*_{-1}\cdots *_{-1}
        \Bigg(
        e_{\beta_p}*_{-1}e_{\beta_{p+1}}*_{-1}\cdots*_{-1}e_{\beta_s} \\
&-& \Big(e_{\beta_p}*_{-1}e_{\beta_{p+1}}\Big)*_{-1}\cdots*_{-1}e_{\beta_s}
        \Bigg)\ , %\\
\end{eqnarray*}
\begin{eqnarray*}
A_3 
&=&
e_{j_1}*_{-1} \cdots *_{-1}
        \Bigg(
        \!\Big(\!e_{j_p}\!*_{-1}\!e_{j_{p+1}}\!\Big)\!*_{-1}\cdots*_{-1}e_{j_s} \\
&\mp& 
        \Big(e_{j_{p+1}}*_{-1}e_{j_p}\Big)*_{-1}\cdots*_{-1}e_{j_s} 
        - [e_{j_p}\ _*\ e_{j_{p+1}}]*_{-1}\cdots*_{-1}e_{j_s}\!
        \Bigg)\ , \\
A_4
&=& - \prod_{l=1}^s\!\left(\begin{array}{c}\!\!\!\! -\Delta_l \!\!\!\!\\\!\! k_l \!\!\end{array}\right)
        e_{\beta_1}*_{-1} \cdots *_{-1}
        \Bigg(
        \Big(e_{\beta_p}*_{-1}e_{\beta_{p+1}}\Big)*_{-1}\cdots*_{-1}e_{\beta_s} \\
&\mp& 
        \!\Big(\!e_{\beta_{p+1}}\!*_{-1}\!e_{\beta_p}\!\Big)\!*_{-1}\cdots*_{-1}e_{\beta_s} 
        - [e_{\beta_p}\ _*\ e_{\beta_{p+1}}]*_{-1}\cdots*_{-1}e_{\beta_s}\!\!
        \Bigg)\ , \\
A_5
&=& \pm
        e_{j_1}*_{-1} \cdots *_{-1}
        \Bigg(
        \Big(e_{j_{p+1}}*_{-1}e_{j_p}\Big)*_{-1}\cdots*_{-1}e_{j_s} \\
&-& e_{j_{p+1}}*_{-1}e_{j_p}*_{-1}\cdots*_{-1}e_{j_s} 
        \Bigg)\ , \\
A_6
&=& \mp
        \prod_{l=1}^s\!\left(\begin{array}{c}\!\!\!\! -\Delta_l \!\!\!\!\\\!\! k_l \!\!\end{array}\right)
        e_{\beta_1}*_{-1} \cdots *_{-1}
        \Bigg(
        \Big(e_{\beta_{p+1}}*_{-1}e_{\beta_p}\Big)*_{-1}\cdots*_{-1}e_{\beta_s} \\
&-&     e_{\beta_{p+1}}*_{-1}e_{\beta_p}*_{-1}\cdots*_{-1}e_{\beta_s}
        \Bigg)\ , \\
A_7
&=& e_{j_1}*_{-1} \cdots *_{-1}
        \Bigg(
        [e_{j_p}\ _*\ e_{j_{p+1}}]*_{-1}e_{j_{p+2}}*_{-1}\cdots*_{-1}e_{j_s} \\
&-& \left(\begin{array}{c}\!\!\!\! -\Delta_p \!\!\!\!\\\!\! k_p \!\!\end{array}\right)
        \left(\begin{array}{c}\!\!\!\! -\Delta_{p+1} \!\!\!\!\\\!\! k_{p+1} \!\!\end{array}\right)
        [e_{\beta_p}\ _*\ e_{\beta_{p+1}}]*_{-1}e_{j_{p+2}}*_{-1}\cdots*_{-1}e_{j_s}
        \Bigg)\ , \\
A_8
&=& \left(\begin{array}{c}\!\!\!\! -\Delta_p \!\!\!\!\\\!\! k_p \!\!\end{array}\right)
        \left(\begin{array}{c}\!\!\!\! -\Delta_{p+1} \!\!\!\!\\\!\! k_{p+1} \!\!\end{array}\right)
        e_{j_1}*_{-1} \cdots *_{-1}e_{j_{p-1}}*_{-1} \\
&&      \Bigg(
        [e_{\beta_p}\ _*\ e_{\beta_{p+1}}]*_{-1}
        (T^{(k_{p+2})}e_{\beta_{p+2}})*_{-1}\cdots*_{-1}
        (T^{(k_s)}e_{\beta_s}) \\
&&      -\prod_{l=p+2}^s\!\left(\begin{array}{c}\!\!\!\! -\Delta_l \!\!\!\!\\\!\! k_l \!\!\end{array}\right)
        [e_{\beta_p}\ _*\ e_{\beta_{p+1}}]*_{-1}
        e_{\beta_{p+2}}*_{-1}\cdots*_{-1}e_{\beta_s}
        \Bigg)\ , \\
A_9
&=& \prod_{l=p}^s\!\left(\begin{array}{c}\!\!\!\! -\Delta_l \!\!\!\!\\\!\! k_l \!\!\end{array}\right)
        (T^{(k_1)}e_{\beta_1})*_{-1} \cdots *_{-1}(T^{(k_{p-1})}e_{\beta_{p-1}})*_{-1} \\
&&\ \   \Big([e_{\beta_p}\ _*\ e_{\beta_{p+1}}]*_{-1}
        e_{\beta_{p+2}}*_{-1}\cdots*_{-1}e_{\beta_s}\Big) \\
&-&  \prod_{l=1}^s\!\left(\begin{array}{c}\!\!\!\! -\Delta_l \!\!\!\!\\\!\! k_l \!\!\end{array}\right)
        e_{\beta_1}*_{-1} \cdots *_{-1} e_{\beta_{p-1}}*_{-1} \\
&&\ \   \Big([e_{\beta_p}\ _*\ e_{\beta_{p+1}}]*_{-1}e_{\beta_{p+2}}*_{-1}\cdots*_{-1}e_{\beta_s}\Big)\ , \\
A_{10}
&=& \pm
        (\!T^{(k_1)}e_{\beta_1}\!)\!*_{-1}\! \cdots\! (\!T^{(k_{p+1})}e_{\beta_{p+1}}\!)\!*_{-1}
        \!(\!T^{(k_p)}e_{\beta_p}\!)\!\cdots\!*_{-1}\!(\!T^{(k_s)}e_{\beta_s}\!) \\
&\mp&
        \prod_{l=1}^s\!\left(\begin{array}{c}\!\!\!\! -\Delta_l \!\!\!\!\\\!\! k_l \!\!\end{array}\right)
        e_{\beta_1}*_{-1} \cdots *_{-1}e_{\beta_{p+1}}*_{-1}e_{\beta_p}*_{-1}\cdots*_{-1}e_{\beta_s}\ .
\end{eqnarray*}
Here $\pm=p(e_{j_p},e_{j_{p+1}})$.
All terms $A_1,\dots,A_{10}$ belong to $\Span_\C \Sigma_\zeta$ by inductive 
assumption.
More precisely,
$A_1,\ A_2,\ A_5$ and $A_6$ 
belong to $(V*_{-2}V)_{\zeta_-}=W_{\zeta_-}$,
thanks to (\ref{-2eq13}) and (\ref{-2eq9}), 
hence by induction they belong to $\Span_\C \Sigma_{\zeta_-}\subset\Span_\C T_\zeta$.
Similarly, $A_3$ and $A_4$ belong to $(V*_{-2}V)_{\zeta_-}$,
thanks to (\ref{-2eq14}), (\ref{-2eq9}) and (\ref{-2eq10}).
$A_7$ belongs to $(V*_{-2}V)_{\zeta_-}$,
thanks to (\ref{prelim2}), (\ref{-2eq9}) and (\ref{-2eq10}).
$A_8$ belongs to $(V*_{-2}V)_{\zeta_-}$,
thanks to the equality $W_\zeta=(V*_{-2}V)_\zeta$ and conditions (\ref{4.21-gr-cond}) and (\ref{-2eq9}).
$A_9$ belongs to $(V*_{-2}V)_{\zeta_-}$,
by a simple induction argument based on (\ref{prelim}).
Finally, $A_{10}$ is such that $\zeta((j_1,\dots,j_{p+1},j_p,\dots,j_s))=
\zeta(\vec{j})$,
but $\#((j_1,\dots,j_{p+1},j_p,\dots,j_s))=\#(\vec{j})-1$,
hence by inductive assumption it belongs to $\Span_\C \Sigma_\zeta$.
This completes the proof of (a).

To prove (b) we only have to show that elements of $\Sigma$ are linearly independent.
But this follows immediately by the definition of $\Sigma$ and Lemma \ref{lem1}(c).
\end{proof}

\begin{corollary}\label{r-2}
One has
$$
R\ \cap\  (V*_{-2}V)\ =\ (T+H)R\ .
$$
\end{corollary}
\begin{proof}
The inclusion $(T+H)R\subset R\cap (V*_{-2}V)$ is obvious, 
thanks to equation (\ref{eq:2.8}).
On the other hand, by Lemma \ref{lem2}, the set
$$
\left\{T^{(k)}e_\alpha
-\left(\begin{array}{c}\!\!\! -\Delta_{\alpha} \!\!\!\\\!\!\! k \!\!\!\end{array}\right)e_\alpha\ \Big|\ 
\alpha\in\hat{\I}_\Delta,\ k\geq0\right\}\ ,
$$
is a basis for $R\cap (V*_{-2}V)$
and each such basis element is obviously in $(T+H)R$.
\end{proof}

Recall from Example \ref{ex2} that the $H$-twisted Zhu algebra of $V$ is,
by definition, the space $Zhu_H V= V/ (V*_{-2}V)$,
with the associative product induced by the $*_{-1}$-product on $V$.
We denote by $Zhu_HR$ the image of $R\subset V$ in $Zhu_HV$ via 
the natural quotient map $\pi_Z:\ V\twoheadrightarrow Zhu_HV=V/(V*_{-2}V)$.
Moreover, thanks to Corollary \ref{r-2}, we have the following identifications:
 $Zhu_HR=R/(T+H)R\simeq\g$.
If $\overline{e}_\alpha$ is the image of the basis element $e_\alpha\in\g$
in $Zhu_HR$,  we let, by abuse of notation,
$$
\overline{\B}\ =\ \{\overline{e}_\alpha\ |\ \ \alpha\in\bar{\I}\}\ .
$$
\begin{lemma}\label{lem3}
The associative algebra $Zhu_HV$ is PBW generated by $\bar\B$.
\end{lemma}
\begin{proof}
It is an immediate corollary of Lemma \ref{lem1}(c) and Lemma \ref{lem2}(b).
\end{proof}

Notice that
$Zhu_HR$ has a natural $\Gamma_+^\prime $-grading
induced by the isomorphism 
$Zhu_HR\simeq\g=\bigoplus_{\zeta\in\Gamma_+^\prime }\g_\zeta$.
\begin{lemma}\label{lem4}
The Zhu algebra $Zhu_HV$ is pregraded by $\Gamma_+$.
\end{lemma}
\begin{proof}
Recall the definition of the $\Gamma_+$-filtration of $Zhu_HV$
induced by the $\Gamma_+^\prime $-grading of $Zhu_HR$.
It is given by
$$
(Zhu_HV)_\zeta\ =\ \Span_\C\Big\{\overline{e}_{\beta_1}\dots\overline{e}_{\beta_s}\ \Big|
        \beta_k\in\bar{\I}_{\zeta_k}\ ,\ \  
        \zeta_1+\cdots+\zeta_s\leq\zeta
        \Big\}\ .
$$
In particular, by Lemma \ref{lem1}(b), we have $(Zhu_HV)_\zeta=\pi_Z(V_\zeta)$.
We then have, by (\ref{4.21-gr-cond}),
\begin{eqnarray*}
[(Zhu_HR)_{\zeta_1},(Zhu_HR)_{\zeta_2}]
\ =\  \pi_Z([\g_{\zeta_1}\ _*\ \g_{\zeta_2}])
\\
\ \subset\ \pi_Z(V_{(\zeta_1+\zeta_2)_-})
\ =\ (Zhu_HV)_{(\zeta_1+\zeta_2)_-}\ .
\end{eqnarray*}
According to Definition \ref{predef}, this implies that $V$ is pregraded by 
$\Gamma_+$.
\end{proof}

Lemma \ref{lem3} and Lemma \ref{lem4} guarantee that all the hypotheses of
Theorem \ref{gmain2} are satisfied. In other words $Zhu_HR$ has a structure 
of non-linear 
Lie algebra and $Zhu_HV\simeq U(Zhu_HR)$.

We can summarize all the above results in the following
\begin{theorem}\label{zhuenvel}
Let $V$ be an $H$-graded vertex algebra
and let $R\subset V$ be a free
$\C[T]$-module  $R=\C[T]\otimes \g$.
Assume that $\g$ has a an $H$-invariant $\Gamma_+^\prime $-grading,
$\g=\bigoplus_{\zeta\in\Gamma_+^\prime }\g_\zeta$,
which we extend to $R$ by letting $\deg(T)=0$.
Assume that there is a basis of $R$, compatible with parity and $\Gamma_+^\prime$-grading
which freely generates $V$ (according to Definition \ref{def:1.14}).
Moreover we assume that $V$ is pregraded by $\Gamma_+$, i.e. (\ref{eq-pregr}) holds.
Then  the space
$$
Zhu_HR\ =\ R/(T+H)R\ \simeq\ \g\ ,
$$
is naturally endowed with the following structure of a non-linear Lie algebra as follows.
Let $\{e_\alpha\ |\ \alpha\in\bar{\I}\}$ be an ordered basis of $\g$
compatible with parity, $\Gamma_+^\prime$ and $H$-gradings of $\g$.
Then the non-linear Lie bracket,
$$
[\ ,\ ]\ :\ \ Zhu_HR\otimes Zhu_HR\ \longrightarrow\ \T(Zhu_HR)\ ,
$$
is given by
$$
[\overline{a},\overline{b}]\ =\ \rho(\pi_Z([a\ _*\ b])\ ,
$$
where $a,b\in R$, $\bar{a},\bar{b}$ are the corresponding elements in 
$Zhu_HR$,
$\pi_Z$ denotes the natural quotient map $\pi_Z:\ V\twoheadrightarrow 
Zhu_HV=V/(V*_{-2}V)$
and $\rho$ is the embedding $\rho:\ Zhu_HV\hookrightarrow\T(Zhu_HR)$,
defined by
$\rho(\overline{e}_{\alpha_1}\cdots \overline{e}_{\alpha_s})
\ =\ \overline{e}_{\alpha_1}\otimes\cdots \otimes\overline{e}_{\alpha_s}$,
for all $\vec{\alpha}=(\alpha_1,\dots,\alpha_s)$ such that $\#(\vec{\alpha})=0$.
Moreover the universal enveloping associative algebra of $Zhu_HR$
is naturally isomorphic to the $H$-twisted Zhu algebra of $V$:
$$
Zhu_HV\ \simeq\ U(Zhu_HR)\ .
$$
\end{theorem}
\begin{corollary}\label{corzhuenv}
Let $R=\C[T]\g$ be a non-linear Lie conformal algebra, freely generated as a $\C[T]$-module,
and suppose that $\g$ is
endowed with a $\Gamma_+^\prime$-grading which induces a Hamiltonian operator 
$H$ on the universal enveloping vertex algebra $V(R)$.
Then the space $Zhu_HR=R/(T+H)R\simeq\g$
is naturally endowed with a structure of a non-linear Lie algebra,
so that
$$
Zhu_HV(R)\ \simeq\ U(Zhu_HR)\ .
$$
The non-linear Lie bracket on $Zhu_HR$ is "compatible" with the $*$-bracket
on $R$ in the sense that the following diagram commutes:
$$
\UseTips
\xymatrix{
R\otimes R \ar[d]_{\pi_Z} & \ar[r]^{[\ _*\ ]} & & V(R) \ar[d]^{\pi_Z} \\
Zhu_HR\otimes Zhu_HR & \ar[r]^{[\ ,\ ]} & & Zhu_HV(R)
}
$$
\end{corollary}
\begin{definition}\label{zhulie}
Under the assumptions of Theorem \ref{zhuenvel} or Corollary \ref{corzhuenv},
we will call $Zhu_HR$ the $H$-{\itshape twisted Zhu non-linear Lie 
algebra} associated to $R$.
\end{definition}
\begin{remark}
It can be useful for many applications to have an explicit way to compute the map
$\pi_Z :\ V\rightarrow Zhu_H V$.
Suppose $V\simeq V(R)$ is the universal enveloping vertex algebra 
of a non-linear Lie conformal algebra $R=\C[T]\g$,
so that $Zhu_H V\simeq U(Zhu_H R)$.
As before, for $a\in\g$, we denote by $\bar{a}$ the same element thought as an element
of $Zhu_H R\simeq\g$.
Recall the defining relation of $\pi_Z$, i.e.
$$
\pi_Z(A*_{-1}B)\ =\ \pi_Z(A)\pi_Z(B)\ .
$$
We have the following identity, for $a\in\g,\,B\in V$,
\begin{eqnarray}\label{sept15}
&\displaystyle{
\pi_Z(:(T^{(k)}a)B:)\ =\
\binom{-\Delta_a}{k}\bar{a}\pi_Z(B)
-\int_0^1d\hbar \pi_Z\big([H(T^{(k)}a),B]_\hbar\big) 
}\\
&\displaystyle{
= \binom{-\Delta_a}{k}\Big(
\bar{a}\pi_Z(B)
-(\Delta_a+k)\sum_{j\in\Z_+}\binom{\Delta_a-1}{j}\frac{1}{k+j+1}\pi_Z(a_{(j)}B)
\Big)\ .
}\nonumber
\end{eqnarray}
In the proof of the first identity we use (\ref{new}),
while the second equality follows from the definition (\ref{eq:2.4}) of the $\hbar$-bracket.
Equation (\ref{sept15}) can be used inductively 
to compute the image via $\pi_Z$ of an arbitrary element of $V$.
\end{remark}
\begin{corollary}
Let $R=\C[T]\g$ be a non-linear Lie conformal algebra as in Corollary \ref{corzhuenv}.
Then $Zhu_H V(R)$ is an associative algebra generated by $\g$
with defining relations ($a,b\in\g$)
$$
ab-p(a,b)ba\ =\ \sum_{j\in\Z_+}\binom{\Delta_a-1}{j}\pi_Z(a_{(j)}b)\ ,
$$
where the RHS should be computed with the help of relations (\ref{sept15}).
\end{corollary}
\begin{proposition}\label{last}
Suppose $V$ is a vertex algebra which satisfies all assumptions of Theorem \ref{zhuenvel},
and let $\Gamma$ be as in Example \ref{ex-delta}.
Then, for every $c\in\C$, the map (\ref{surj}) is an isomorphism of Poisson algebras:
$Zhu_{\hbar=0,H}V\ \simeq\ \gr\,Zhu_{\hbar=c,H}V$.
\end{proposition}
\begin{proof}
As in (\ref{H-filtr}), we denote by $F^\Delta V$ the increasing filtration of $V$ induced by the
conformal weight.
Recall that, by definition, $J_{\hbar=1}=V*_{-2}V$.
It follows by Lemma \ref{lem2} that
\begin{equation}\label{albilate}
F^\Delta\cap J_{\hbar=1}\,\subset\,
\sum_{\Delta_1+\Delta_2+1\leq\Delta} V[\Delta_1]*_{-2}V[\Delta_2]\,\subset\,
F^{\Delta_-} V+F^\Delta V\cap(V_{(-2)}V)\ .
\end{equation}
The statement then follows from the proof of Proposition \ref{old:2.17}(c) and
the fact that, by (\ref{isozhu}), 
$Zhu_{\hbar=c,H}V\simeq Zhu_{\hbar=1,H}V$ for arbitrary $c\neq0$.
\end{proof}
\begin{remark}
We may consider, more generally, the $(H,\Gamma)$-twisted Zhu algebra $Zhu_{\Gamma}V$
defined in Section \ref{sec:2.2}, for an arbitrary $\Gamma$-grading
$V=\bigoplus_{\bar\gamma \in\Gamma/\Z}{}^{\bar{\gamma}}V$ of $V$
(not necessarily induced by the action of $H$).
Recall that we denoted by $V_\Gamma$ the $\C$-span of all elements $a\in V$ such that
$\Delta_a\in\bar{\gamma}_a$.
It is a vertex subalgebra of $V$, and we clearly have, by definition, 
$J_\hbar(V_\Gamma)\subset J_{\hbar,\Gamma}(V)$.
Hence we have, in general, a surjective associative algebra homomorphism
$$
Zhu_H V_\Gamma=V_\Gamma/J_{\hbar=1}(V_\Gamma)\ 
\twoheadrightarrow\ V_\Gamma/J_{\hbar=1,\Gamma}(V)=Zhu_{\Gamma} V\ .
$$
In particular $Zhu_{\Gamma} V$ will not be, in general, a PBW generated
algebra, even if $V\simeq V(R)$ for some non-linear Lie conformal algebra $R$.
It does not make sense then to talk about the non-linear  
Lie algebra $Zhu_{\Gamma} R$.
\end{remark}

%%% addition 1: subsection on cubic Dirac operator

\subsection{Application: the (non-commutative) Weil algebra and 
the cubic Dirac operator}\label{sec:dirac}

Let $\g$ be a finite-dimensional simple or abelian Lie algebra
with a non-degenerate symmetric invariant bilinear form $(\cdot\,|\,\cdot)$.
Here, for simplicity, we assume $\g$ is an honest, i.e. purely even, Lie algebra,
and we denote by $\bar\g$ the space $\g$ with reversed parity, i.e. purely odd.
Consider the following non-linear Lie conformal superalgebra
$$
R\,=\,\C[T]\otimes\g+\C[T]\otimes\overline\g\ ,
$$
where the $\lambda$-brackets are ($a,b\in\g,\,\overline a,\overline b\in\overline \g$):
\begin{eqnarray}\label{l-gener}
[a\,_\lambda\,b] &=& [a,b]+\lambda(k+h^{\vee})(a|b)\ ,\nonumber\\ 
{[\overline a\,_\lambda\,\overline b]} &=& (k+h^{\vee})(a|b)\ ,\\ 
{[a\,_\lambda\,\overline b]} &=& [\overline a\,_\lambda\,b]\ =\ \overline{[a,b]}\ .\nonumber
\end{eqnarray}
Let $\{a^i\}$ be an orthonormal basis of $\g$.
Introduce the following fields of $V(R)$ (see \cite{KT}):
\begin{eqnarray}\label{GL}
G &=& \frac{1}{k+h^{\vee}}\left(
\sum_i:a^i\overline{a}^i:+\frac{1}{3(k+h^{\vee})}\sum_{i,j}:\overline{[a^i,a^j]}\overline{a}^i\overline{a}^j:
\right)\ , \nonumber\\
L &=& \frac{1}{2(k+h^{\vee})}\left(
\sum_i:a^i a^i:+%\frac{k+3h^{\vee}}{k+h^{\vee}}
\sum_i:(T\overline{a}^i)\overline{a}^i: \right.\\
&&\left. +\frac{1}{k+h^{\vee}}\sum_{i,j}:\overline{a}^i[a^i,a^j]\overline{a}^j:
\right)\ .\nonumber
\end{eqnarray}
The fields $G$ and $L$ satisfy the $\lambda$-brackets of the Neveu-Schwarz Lie conformal 
superalgebra with central charge
$$
c\ =\ \frac{k\dim\g}{k+h^{\vee}}+\frac{\dim\g}2\ ,
$$
i.e. $L$ is an energy-momentum field with central charge $c$, $G$ is an odd,
primary with respect to $L$, field of conformal weight $3/2$, and
\begin{equation}\label{gg}
[G\,_\lambda\,G]\ =\ 2L+\frac{\lambda^2}{3}c\ .
\end{equation}
Moreover, the fields $a\in\g$ (respectively $\overline a\in\overline\g$) are primary
with respect to $L$ of conformal weight 1 (resp. 1/2), and
$[a\,_\lambda\,G]=\lambda\overline a,\,[\overline a\,_\lambda\,,G]=a$ 
for $a\in\g,\,\overline a\in\overline\g$.

Let $H=L_{(1)}$ be the Hamiltonian operator on $V(R)$,
let $\cW$ be the $H$-twisted Zhu algebra of $V(R)$
and let $\pi_Z:V(R)\rightarrow \cW$ be the quotient map.
We put $k+h^{\vee}=1$ (which can be achieved by rescaling the form $(\cdot\,|\,\cdot)$,
provided that $k\neq-h^{\vee}$).
By Corollary \ref{corzhuenv}, $\cW$ is the universal enveloping superalgebra of the non-linear Lie
superalgebra $\g+\overline\g$ with non-linear Lie brackets, extending that on $\g$ by
($a,b\in\g,\,\overline a,\overline b\in\overline\g$):
\begin{equation}\label{lie-gener}
{[a,\overline b]}\,=\,[\overline a,b]\,=\,\overline{[a,b]}\ ,\ \ 
{[\overline a,\overline b]} = (a|b)\ .
\end{equation}
Let $D=\pi_Z(G)$ and $C=\pi_Z(L)$. Then
\begin{eqnarray*}
D &=&
\sum_i a^i\overline{a}^i+\frac{1}{3}\sum_{i,j} \overline{[a^i,a^j]}\overline{a}^i\overline{a}^j
\ , \\
C &=& \frac{1}{2}\sum_i a^i a^i
+\frac{1}{2}\sum_{i,j}[a^i,a^j]\overline{a}^i\overline{a}^j
+\frac{1+2h^{\vee}}{16}\dim\g
\ .
\end{eqnarray*}
Equation (\ref{gg}) implies, by looking at images via $\pi_Z$,
\begin{equation}\label{dd}
D^2\ =\ C+\left(\frac{h^{\vee}}{24}-\frac{1}{16}\right)\dim\g\ ,
\end{equation}
while from the other $\lambda$-bracket relations above we get
$[D,a]=0,\,[D,\overline a]=a,\,C$ is central.
Furthermore, one has the following Lie algebra embedding $a\mapsto\widehat a$
of $\g$ in $\cW$, whose image commutes with $\bar\g$:
$$
\widehat a\ =\ a-\frac12\sum_i\overline{[a,a^i]}\overline{a}^i\ .
$$
Let $\widehat C=\sum_i\widehat{a}^i\widehat{a}^i$ be the corresponding Casimir operator.
Then
$$
\widehat{C}\ =\ 2C-\frac18\dim\g\ ,\ \ 
D^2=\frac12\widehat C+\frac{h^{\vee}}{24}\dim\g
\ =\ \frac12\big(\widehat C+(\rho|\rho)\big)\ ,
$$
where $\rho$ is half of the sum of positive roots of $\g$.
The latter relation was derived in \cite{AM,Ko1}, where the element $D$ was called
the cubic Dirac operator, by a direct computation.

Let $\h$ be a subalgebra of $\g$, such that the restriction of $(\cdot\,|\,\cdot)$ to $\h$
is non-degenerate, and let $\h=\bigoplus_s\h_s$ be the decomposition of $\h$
in a direct sum of simple and abelian components.
For each $\h_s$ constructs the elements $G_s$ and $L_s$ as above,
and let $G^\h=\sum_sG_s,\,L^\h=\sum_sL_s$.
Then each pair $G_s,\,L_s$ (resp. $G^\h,\,L^\h$)
satisfies the relations of the Neveu-Schwarz algebra with central charge $c_s$
(resp. $c^\h=\sum_sc_s$),
and we can form the coset Neveu-Schwarz algebra $G-G^\h,\,L-L^\h$
with central charge $c-c^\h$,
which commutes with all fields $a\in\h,\,\overline a\in\overline\h$ (cf. \cite{KT}).
Applying $\pi_Z$, we obtain the relative cubic Dirac operator $D-D^\h$, introduced in \cite{Ko1},
and the relation, derived there
$$
(D-D^\h)^2\ =\ \frac12\Big(\widehat C-\widehat{C}^\h+(\rho|\rho)-(\rho_\h|\rho_\h)\Big)\ .
$$

Starting from the vertex algebra $V^{QFT}=V(R)$, we can construct all "fundamental
objects" in diagram (\ref{maxi}), as explained in Section \ref{sec:pva}.
The "classical analogue" of $V^{QFT}$, namely its quasiclassical limit $V^{clFT}$,
is the following Poisson vertex algebra.
As commutative associative unital superalgebra, $V^{clFT}=S(R)$,
where $R=\C[T]\otimes\g+\C[T]\otimes\bar\g$ as above,
and the $\lambda$--bracket $\{\cdot\,_\lambda\,\cdot\}$ on $S(R)$ is defined 
by extending formulas (\ref{l-gener}) to $S(R)$ by sesquilinearity and 
left and right commutative Wick formulas.
Consider the elements $G^{cl}$ and $L^{cl}$ of the Poisson vertex algebra $S(R)$, 
given by the same formulas (\ref{GL}) as $G$ and $L$ respectively, 
with the normally ordered product signs $:\ \ :$ removed. 
Then $L^{cl}$ is a Virasoro element with zero central charge,
$a,\,\bar a,\, G^{cl}$ are primary with respect to $L^{cl}$ with conformal
weights 1,1/2 and 3/2 respectively and
we have (cf (\ref{gg}):
$$
\{G^{cl}\,_\lambda\,\bar a\}\ =\ a\ ,\ \ 
\{G^{cl}\,_\lambda\,a\}\ =\ (T+\lambda)\bar a\ ,\ \ 
\{G^{cl}\,_\lambda\,G^{cl}\}\ =\ 2L^{cl}\ .
$$ 

The "quantum mechanical" analogue $V^{QM}$ of $V(R)$
is given by the $H$-twisted Zhu algebra.
As explained above, it is $V^{QM}=U(r)$, where $r=\g+\bar\g$
is the non-linear Lie superalgebra with Lie bracket given in (\ref{lie-gener}).
This algebra is also referred to as the {\itshape non-commutative Weil algebra} associated to $\g$,
\cite{AM}.
The "classical mechanical" analogue $V^{clM}$ of $V(R)$ is the quasiclassical limit of $Zhu_H V$,
or, equivalently, the $H$-twisted Zhu algebra of $V^{clFT}$, as defined in Section \ref{sec:pva}.
It is the commutative associative unital superalgebra $S(r)=S(\g)\otimes\Lambda(\bar\g)$
together with the Poisson bracket defined by extending formulas (\ref{lie-gener}) to $S(r)$
via the Leibniz rule.
This algebra is commonly known as the Weil algebra $W_\g$ associated to $\g$, \cite{AM}.
It is not hard to check that, in this case, the "quantum" and "classical" Zhu algebras 
at $\hbar=0$, namely the bottom left and right corners in diagram (\ref{maxi}),
are both isomorphic to the Weil algebra $W_\g=V^{clM}$.
Applying the canonical map $\pi_Z^{cl}: V^{clFT}\rightarrow Zhu_H V^{clFT}=V^{clM}\simeq S(r)$ 
(see Section \ref{sec:pva}) to $G^{cl}$ and $L^{cl}$, we obtain the "classical analogues" of the
elements $D$ and $C$ in $V^{QM}$.
As explained in Section \ref{sec:pva}, Theorem \ref{P-2.7}, the associative product in $V^{clM}$
is induced by the associative product in $V^{clFT}$, while the Poisson bracket is induced
by the bracket (\ref{P-hbr}) with $\hbar=1$.
Hence we get, for $k+h^{\vee}=1$,
\begin{eqnarray*}
D^{cl} &=& 
\sum_i a^i\overline{a}^i+\frac{1}{3}\sum_{i,j} \overline{[a^i,a^j]}\overline{a}^i\overline{a}^j
\ , \nonumber\\
C^{cl} &=& \frac{1}{2}
\sum_i a^i a^i +
\frac{1}{2}\sum_{i,j} \overline{a}^i[a^i,a^j]\overline{a}^j\ .
\nonumber
\end{eqnarray*}
With respect to the Poisson bracket, $C^{cl}$ is central,
$\{D^{cl},a\}=0,\, \{D^{cl},\bar a\}=a$ and $\{D^{cl},D^{cl}\}=2C^{cl}$.
Notice that $\ad D^{cl}$ coincides with the Weil differential of the Weil algebra $W_\g$
in its Koszul form.

%%%%%%%% SECTION %%%%%%%%%%%%%%%%%%%%%%%%%%%%%

\section{Almost linear differentials in non-linear Lie algebras and Lie conformal algebras}
\label{sec:almost}
\setcounter{equation}{0}

Throughout this section we let $\Gamma$ be a discrete additive subgroup of $\R$ 
containing $\Z$, i.e. $\Gamma=\frac{1}{N}\Z$ for some positive integer $N$, 
and we let $\epsilon=\frac{1}{N}$.
We also let $\Gamma_+$ (respectively $\Gamma_+^\prime$) be the set of
non-negative (resp. positive) elements of $\Gamma$.

By a {\itshape differential} $d$ on a vector superspace $U$ we mean an odd endomorphism 
of $U$ such that $d^2=0$.
Moreover, if $U$ is endowed with one or more products (or non-linear products), like a Lie algebra 
or a Lie conformal algebra, we assume that $d$ is an odd derivation of these products,
and in the Lie conformal algebra case we also assume that $d$ commutes with $T$.
For example, in the non-linear Lie algebra case we require that, for $a,b\in U$,
$d[a,b]=[da,b]+s(a)[a,db]\in\T(U)$.

%%%
\subsection{Review of homological algebra:
good filtered complexes}

Let $(U,d)$ be a filtered complex, namely $U$ is a $\Z_+$-graded vector superspace
$U=\bigoplus_{n\geq0}U^n$, with a compatible decreasing $\Gamma$-filtration 
$\{F^pU,\ p\in\Gamma\}$,
i.e. $F^pU=\bigoplus_{n\geq0}F^pU^n$, where 
$F^pU^n=F^pU\cap U^n$,
and $d$ is a  differential of $U$ of degree 1
preserving the filtration: $d(F^pU^n)\subset F^pU^{n+1}$. 

The cohomology $H(U,d)$ of this complex is again a
$\Z_+$-graded and $\Gamma$-filtered space, with
$$
F^pH^n(U,d)\ =\ \Ker(d:F^pU^n\rightarrow F^pU^{n+1})/(\im d\cap
F^pU^n)\ .
$$
Its associated graded $\gr\,H(U,d)$ is then the $\Gamma$-bigraded space
with
\begin{eqnarray*}
\gr^{pq}H(U,d) &=& F^pH^{p+q}(U,d)/F^{p+\epsilon}H^{p+q}(U,d) \\
&\simeq& \frac{F^pU^{p+q}\cap\Ker d}{\big(F^{p+\epsilon}U^{p+q}\cap\Ker d\big) +
\big(F^pU^{p+q}\cap\im d\big)}\ .
\end{eqnarray*}

The associated graded complex $(\gr\,U,\grd)$ is the $\Gamma$-bigraded
space $\gr\,U=\bigoplus_{p,q\in\Gamma} \gr^{pq}U$, where
$\gr^{pq}U = F^pU^{p+q}/F^{p+\epsilon}U^{p+q}$, together with the induced
differential $\grd$. 
Its cohomology $H(\gr\,U,\grd)$ is then the
$\Gamma$-bigraded space given by
\begin{eqnarray*}
H^{pq}(\gr\,U,\grd) &=&
\displaystyle{\frac{\Ker(\grd:\gr^{pq}U\rightarrow
\gr^{p,q+1}U)}{\im(\grd:\gr^{p,q-1}U\rightarrow \gr^{p,q}U)}} \\
&\simeq& \displaystyle{\frac{F^pU^{p+q}\cap
d^{-1}(F^{p+\epsilon}U^{p+q+1})}{F^{p+\epsilon}U^{p+q}+d(F^pU^{p+q-1})}
\ .}
\end{eqnarray*}
\begin{definition}\label{def-good}
The complex $(U,d)$ and the differential $d$ are said to be {\itshape good} if
$H^{pq}(\gr\,U,\grd)=0$ for all $p,q\in\Gamma$ such that $p+q\neq0$.
\end{definition}
\begin{lemma}\label{homolalg}
Assume $U$ is {\itshape locally finite}, namely for every $n\geq0$
we have $F^pU^n=0$ for $p\gg0$. 
Then if $(U,d)$ is a good complex we have
$$
\gr^{pq}H(U,d) \simeq H^{pq}(\gr\,U,\grd)\ ,\ \ \forall
p,q\in\Gamma\ .
$$
More precisely, we have the following identities ($p\in\Gamma$),
\begin{eqnarray}
\vphantom{\Big(}  F^pU^n\cap\Ker d &=&
F^pU^n\cap\im d\ ,\ \
\text{ for } n\neq0\ , \label{hom1}\\
\vphantom{\Big(} F^pU^0\cap d^{-1}(F^{p+\epsilon}U^1) &=&
F^{p+\epsilon}U^0+F^pU^0\cap\Ker d\ .\label{hom2}
\end{eqnarray}
\end{lemma}
\begin{proof}
The statement of the lemma is standard in the theory of
spectral sequences. For completeness, we will provide an
elementary proof of it. By assumption we have
$$
F^pU^{n}\cap d^{-1}(F^{p+\epsilon}U^{n+1})=F^{p+\epsilon}U^{n}+d(F^pU^{n-1})\ ,
\ \ \forall n\neq0\ .
$$
Suppose first $n\neq0$. In this case we have
\begin{eqnarray}
\vphantom{\Big(} F^pU^n\cap\Ker d &=&
 F^pU^n\cap d^{-1}(F^{p+\epsilon}U^{n+1})\cap\Ker d \nonumber\\
\vphantom{\Big(} &=& F^{p+\epsilon}U^n\cap\Ker d+d(F^pU^{n-1}) \label{hom3}\\
\vphantom{\Big(} &\subset& F^{p+\epsilon}U^n\cap\Ker d+F^pU^n\cap\im d\
.\nonumber
\end{eqnarray}
The reverse inclusion is obvious. Hence we have
$$
F^pU^n\cap\Ker d\ =\ F^{p+\epsilon}U^n\cap\Ker d+F^pU^n\cap\im d\ .
$$
We can now iterate the above result $N$ times to get
$$
F^pU^n\cap\Ker d\ =\ F^{p+N\epsilon}U^n\cap\Ker d+F^pU^n\cap\im d\ .
$$
Equation (\ref{hom1}) then follows by the locally finiteness assumption.
We are left to prove (\ref{hom2}). It follows by (\ref{hom3}) with
$n=1$ that
\begin{eqnarray*}
\vphantom{\Big(} d(F^pU^0)\cap F^{p+\epsilon}U^1 &=& d(F^pU^0)\cap F^{p+\epsilon}U^1\cap\Ker d\\
\vphantom{\Big(} &=& d(F^{p+\epsilon}U^0)+d(F^pU^0)\cap
F^{p+2\epsilon}U^1\ .
\end{eqnarray*}
We can now repeat the same argument for the second term in the
RHS to get, by induction,
$$
d(F^pU^0)\cap F^{p+\epsilon}U^1\ =\ d(F^{p+\epsilon}U^0)+d(F^pU^0)\cap
F^{p+r\epsilon}U^1\ ,\ \ \forall r\geq1\ .
$$
By locally finiteness, this implies
$$
d(F^pU^0)\cap F^{p+\epsilon}U^1\ =\ d(F^{p+\epsilon}U^0)\ ,
$$
which is equivalent to (\ref{hom2}).
\end{proof}

%%%%%%%%%%%%%%%%%%%%%%%%%%%%%%%
\subsection{Lie algebras and Lie conformal algebras with a differential}

We now review some basic facts about non-linear Lie algebras and non-linear Lie conformal
algebras with a "linear" differential.
In particular we show that taking cohomolgy commutes with taking the universal enveloping algebra.
In the next section we will see how to generalize these results to the case of an almost linear
differential.
First, we recall here the well know K\"{u}nneth Lemma:
\begin{lemma}[K\"{u}nneth Lemma]
Let $V_1,V_2$ be vector superspace with differentials $d_i:\
V_i\rightarrow V_i,\ i=1,2$, and define the differential
$d$ on $V=V_1\otimes V_2$
by $d(a\otimes b)=d(a)\otimes b+s(a)a\otimes d(b)$.
Then there is a canonical vector space isomorphism
$$
H(V,d)\ \simeq\ H(V_1,d_1)\otimes H(V_2,d_2)\ .
$$
\end{lemma}
As usual, we denote by $\T(V)$ and $S(V)$ respectively the tensor algebra and 
the symmetric algebra of  the vector superspace $V$. We have
\begin{lemma}\label{symlinear}
Let $V$ be a vector superspace with a differential $d$. 
Extend $d$ to a derivation of $\T(V)$ and $S(V)$.
Then
$$
H(\T(V),d)\simeq \T(H(V,d))\ ,\ \ 
H(S(V),d)\simeq S(H(V,d))\ .
$$
\end{lemma}
\begin{proof}
The statement immediately follows by the K\"{u}nneth Lemma and the fact that the action
of the symmetric group $S_n$ on $V^{\otimes n}$ commutes with the action of $d$.
\end{proof}

We can think at $V$ in the previous lemma as an abelian Lie algebra.
The following result is then a generalization of Lemma \ref{symlinear}
to the case in which the space $V$ has a structure of a non-linear
Lie algebra.
\begin{theorem}\label{glinear}
Let $\g$ be a non-linear Lie superalgebra,
together with a differential  $d:\g\rightarrow \g$ preserving the 
$\Gamma_+^\prime $-grading, i.e. $d(\g_\zeta)\subset\g_\zeta$ 
for $\zeta\in\Gamma_+^\prime $.
Consider the space $H(\g,d)$ with induced $\Gamma_+^\prime $-grading.
The non-linear Lie bracket of $\g$ induces a non-linear product
$$
H(\g,d)\otimes H(\g,d)\ \rightarrow\ \T(H(\g,d))\ ,
$$
and we assume that this defines a structure of a non-linear Lie superalgebra on $H(\g,d)$.
Then there is a canonical associative algebra isomorphism
$$
H(U(\g),d)\ \simeq\ U(H(\g,d))\ .
$$
\end{theorem}
\begin{proof}
First, let us describe the non-linear product
$H(\g,d)\otimes H(\g,d)\ \stackrel{[\ ,\ ]}{\rightarrow}\ \T(H(\g,d))$.
By Lemma \ref{symlinear} we have an isomorphism $H(\T(\g),d)\simeq\T(H(\g,d))$.
Then $[\ ,\ ]$ is defined by the following commutative diagram
$$
\UseTips
\xymatrix{
H(\g,d)\otimes H(\g,d) & \ar[r]^{[\ ,\ ]} & & {\T}(H(\g,d))=H({\T}(\g),d) \\
\text{Ker}(d:\g\rightarrow\g)^{\otimes2}  \ar@{>>}[u] 
& \ar[r] & & \text{Ker}(d: {\T}(\g)\rightarrow {\T}(\g)) \ar@{>>}[u]
}
$$
where the map on the bottom line is given by the non-linear Lie bracket on $\g$.
By assumption this makes $H(\g,d)$ into a non-linear Lie superalgebra,
hence its universal enveloping algebra is well defined.

We are left to show that $U(H(\g,d))\simeq H(U(\g),d)$.
We will use the same argument as in \cite[Lemma 3.2]{KW1}.
The $\Gamma_+^\prime$-grading of $\g$ induces increasing $\Gamma_+$-filtration
$\{U_\zeta(\g),\ \zeta\in\Gamma_+\}$ of $U(\g)$,
and a $\Gamma_+$-grading of the symmetric algebra
$S(\g)=\bigoplus_{\gamma\in\Gamma_+}S(\g)[\zeta]$.
By the grading condition of the non-linear Lie bracket, we have an exact sequence
$$
0\ \rightarrow\ U_{\zeta-\epsilon}(\g)\ \rightarrow\ U_\zeta(\g)\ \rightarrow\ S(\g)[\zeta]\ \rightarrow\ 0\ ,
$$
which induces an exact triangle in cohomology,
$$
\xymatrix {
H(U_{\zeta-\epsilon}(\g),d)  \ar[r]  & H(U_{\zeta}(\g),d) \ar[d] \\ 
& H(S(\g)[\zeta],d) \ar@/^/[lu]^{\delta} & 
}
$$
We claim that the connecting map $\delta$ is identically zero.
Let us recall its definition.
Given $w\in H(S(\g)[\zeta],d)$, let $\tilde{w}$ be a representative of $w$ in $S(\g)[\Delta]$,
and let $\tilde{\tilde{w}}$ be any representative of $\tilde w$ in $U_\zeta(\g)$.
Clearly $d\tilde w=0$, hence $d\tilde{\tilde w}\in U_{\zeta-\epsilon}(\g)$, 
and $\delta w$ is the corresponding class in $H(U_{\zeta-\epsilon}(\g),d)$.
On the other hand, by Lemma \ref{symlinear} we can choose $\tilde w$ to be a symmetric polynomial
in the closed elements of $\g$, and we let $\tilde{\tilde w}$ be the same polynomial, seen as an element
of $U_\zeta(\g)$.
But with this choice of $\tilde{\tilde w}$ we immediately have $d\tilde{\tilde w}=0$,
hence $\delta w=0$ as well.
We thus have exact sequences
$$
0\ \rightarrow\ H(U_{\zeta-\epsilon}(\g),d)\ \rightarrow\ H(U_\zeta(\g),d)\ \rightarrow\ H(S(\g)[\zeta],d)
\ \rightarrow\ 0\ .
$$
On the other hand, by Lemma \ref{symlinear}, we have
$H(S(\g)[\zeta],d)\simeq S(H(\g,d))[\zeta]$.
It then follows by induction that
$H(U_\zeta(\g),d)\simeq U_\zeta(H(\g,d))$, which completes the proof.
\end{proof}
\begin{remark}
Notice that, if $\g$ is a "linear" Lie algebra, then its cohomology $H(\g,d)$
is automatically a Lie algebra. 
In the non-linear case, though, it is not clear if $H(\g,d)$ is necessarily a non Lie algebra.
\end{remark}

The above argument can be repeated almost verbatim to prove the analogue statement in the
context of Lie conformal algebras. In other words we have the following
\begin{theorem}\label{Rlinear}
Let $R$ be a non-linear Lie conformal algebra,
together with a differential  $d:R\rightarrow R$, preserving the 
$\Gamma_+^\prime $-grading of $R$. 
Assume moreover that $H(R,d)$, with the induced $\lambda$-bracket, is again a non-linear 
Lie conformal algebra.
Then there is a canonical vertex algebra isomorphism
$$
H(V(R),d)\ \simeq\ V(H(R,d))\ .
$$
\end{theorem}

%%%
\subsection{Non-linear Lie algebras with a good almost linear differential}\label{secg}

Let $\g$ be a finite-dimensional vector superspace graded as follows,
\begin{equation}\label{grg}
\g\ =\ \bigoplus_{\substack{p,q\in\Gamma,\ p+q\in\Z_+\\ \Delta\in\Gamma_+^\prime }} \g^{pq}[\Delta]\ .
\end{equation}
The induced grading of the tensor algebra is
$\T(\g)=\bigoplus_{p,q\in\Gamma,\Delta\in\Gamma_+^\prime} \T^{pq}(\g)[\Delta]$.
We further assume $\g$ has a structure of non-linear Lie algebra (see Definition \ref{g})
with respect to the $\Gamma_+^\prime $-grading by subspaces
$\g[\Delta]=\bigoplus_{p,q}\g^{pq}[\Delta],\ \Delta\in\Gamma_+^\prime $.
In particular the non-linear Lie bracket satisfies the grading condition
\begin{equation}\label{gradcond1}
[\g[\Delta_1],\g[\Delta_2]]\ \subset\ \bigoplus_{\Delta<\Delta_1+\Delta_2}\T(\g)[\Delta]\ ,
\end{equation}
or, equivalently, $\Delta([a,b])<\Delta(a)+\Delta(b)$.
Moreover we assume that the corresponding $\Z_+$-grading of $\g$, given by
$\g^n=\bigoplus_{p+q=n,\, \Delta}\g^{pq}[\Delta],\ n\in\Z_+$, is a non-linear Lie algebra grading,
and the decreasing $\Gamma$-filtration of $\g$, given by
$F^p\g = \bigoplus_{p^\prime\geq p,q,\Delta}\g^{p^\prime q}[\Delta],\ p\in\Gamma$,
is a non-linear Lie algebra filtration.
The above assumptions can be combined by the following condition,
\begin{equation}\label{gradtotg}
[\g^{p_1q_1}[\Delta_1],\g^{p_2q_2}[\Delta_2]]\ \subset\
\bigoplus_{\substack{l\geq0\\
\Delta<\Delta_1+\Delta_2}}\T^{p_1+p_2+l,q_1+q_2-l}(\g)[\Delta]\ .
\end{equation}
The associated graded $\gr\,\g$ of the filtration $\{F^p\g,\ p\in\Gamma\}$ is the
space $\g$ with non-linear Lie bracket $[a,b]^\gr$, for $a\in\g^{p_1q_1},\
b\in\g^{p_2q_2}$, equal to the component of $[a,b]$ in
$\T^{p_1+p_2,q_1+q_2}(\g)$.
It is easy to check that, by definition, $\gr\,\g$ is again a non-linear Lie algebra, bigraded by $p,q$.

Let $U(\g)$ be the universal enveloping associative algebra of
the non-linear Lie algebra $\g$. It has the $\Z_+$-grading
$U(\g)=\bigoplus_{n\geq0}\ U(\g)^n$, where
$$
U(\g)^n\ =\ \Span_\C\Big\{a_1\cdots a_s\ \Big|\ a_i\in \g^{n_i},\
n_1+\cdots +n_s=n \Big\}\ ,
$$
the decreasing $\Gamma$-filtration by subspaces
$$ 
F^pU(\g)\ =\ \Span_\C\Big\{a_1\cdots
a_s\ \Big|\ a_i\in F^{p_i}\g,\ p_1+\cdots +p_s\geq p\Big\}\ ,
$$
and the increasing $\Gamma_+$-filtration by subspaces
$$
U_\Delta(\g)\ =\ \Span_\C\Big\{a_1\cdots a_s\ \Big|\ a_i\in
\g[\Delta_i],\ \Delta_1+\cdots+\Delta_s\leq\Delta\}\ .
$$
We let $F^pU(\g)^n=F^pU(\g)\cap U(\g)^n$ and
$F^pU_\Delta(\g)^n=F^pU(\g)\cap U(\g)^n\cap U_\Delta(\g)$. 
By construction, grading and filtrations of $U(\g)$ are compatible.
Notice that each filtered space $U_\Delta(\g)$ is finite-dimensional,
hence locally finite, i.e. for each $\Delta\in\Gamma_+$
and $n\geq0$ we have 
$F^pU_\Delta(\g)^n=0$ for $p \gg 0$.
Moreover, by the grading condition (\ref{gradcond1}), we have that
\begin{equation}\label{gradcond2}
[U_{\Delta_1}(\g),U_{\Delta_2}(\g)]\ \subset\ U_\Delta(\g)\ ,\quad
\text{ for some } \Delta<\Delta_1+\Delta_2\ .
\end{equation}
The associated graded algebra $\gr\, U(\g)$ is a $\Gamma$-bigraded
associative algebra, with a compatible $\Gamma_+$-filtration. 
Namely the $\Delta$-filtered subspace of bidegree $p,q\in\Gamma$ is
$$
\gr^{pq}U_\Delta(\g)\ =\
F^pU_\Delta(\g)^{p+q}/F^{p+\epsilon}U_\Delta(\g)^{p+q}\ .
$$
\begin{lemma}\label{ugr}
We have a canonical isomorphism $\gr\,U(\g)\simeq U(\gr\,\g)$
which preserves $\Gamma$-bigrading and $\Gamma_+$-filtration, i.e.
$$
\gr^{pq}U_\Delta(\g)\simeq\ U_\Delta(\gr\,\g)^{pq}\ .
$$
In particular the natural embedding $\g=\gr\,\g\subset \gr\,U(\g)$
is such that
$\g\cap\gr^{pq}U_\Delta(\g)=\bigoplus_{\Delta'\leq\Delta}\g^{pq}[\Delta']$.
\end{lemma}
\begin{proof}
Notice that $\g^{pq}\subset F^pU(\g)^{p+q}$ and $\g^{pq}\cap
F^{p+\epsilon}U(\g)^{p+q}=0$. We thus have a natural embedding
$\g^{pq}\subset\gr^{pq}\,U(\g)=F^{p}U(\g)^{p+q}/F^{p+\epsilon}U(\g)^{p+q}$,
which induces an injective map
$\gr\,\g\subset \gr\, U(\g)$. By universal property we then have
an associative algebra homomorphism
$U(\gr\,\g)\rightarrow\gr\,U(\g)$. This is a bijection since, by
the PBW Theorem, it maps a basis of $U(\gr\,\g)$ to a basis of
$\gr\,U(\g)$. The remaining part of the statement follows
immediately.
\end{proof}

\begin{definition}\label{defdnonlin}
Let $d:\ U(\g)\rightarrow U(\g)$ be a differential of $\Z_+$-degree 1, 
preserving both the $\Gamma$ and the $\Gamma_+$ filtrations:
\begin{equation}\label{graddg}
d(U(\g)^n)\subset U(\g)^{n+1}\ , \ \
d(F^pU(\g))\subset F^pU(\g)\ ,\ \ d(U_\Delta(\g))\subset
U_\Delta(\g)\ .
\end{equation}
Let $\grd:\ \gr\,U(\g)\rightarrow \gr\,U(\g)$ be the
corresponding associated graded differential.
We say that $d$ is an {\itshape almost linear differential} of $\g$
if $\grd$ preserves $\gr\,\g=\g$, it has
bidegree $(0,1)$, and it preserves the $\Gamma_+^\prime $-grading of
$\g$, namely
\begin{equation}\label{dnonlin}
\grd\big(\g^{pq}[\Delta]\big)\subset\g^{p,q+1}[\Delta]\ .
\end{equation}
In this case $\grd|_{\g}$ is a differential of the non-linear Lie algebra $\gr\,\g$,
and the action of $\grd$ on $\gr\,U(\g)\simeq U(\gr\,\g)$ is obtained
starting with $\grd|_{\g}$ and applying the Leibniz rule.
Moreover we say that $d$ is a {\itshape good} almost linear
differential of $\g$ if
\begin{equation}\label{ggood}
\Ker (\grd|_{\g^{pq}})\ =\ \im (\grd|_{\g^{p,q-1}})\ ,\quad \text{
for } p+q\neq0\ ,
\end{equation}
namely the cohomology $H^{pq}(\g,\grd)$ is concentrated in degree
$p+q=0$.
\end{definition}
From now on we let $d$ be a good almost linear differential of $\g$.
\begin{remark}
An equivalent way of writing condition (\ref{dnonlin}) in terms of the differential $d$ is
$$
d\big(\g^{pq}[\Delta]\big)\subset\g^{p,q+1}[\Delta]\oplus
F^{p+\epsilon}U_\Delta(\g)^{p+q+1}\ ,
$$
and $\grd|_{\g}$ corresponds to taking only the first component in this decomposition.
Moreover, an equivalent way of writing condition (\ref{ggood}) is
$$
\g^{pq}[\Delta]\cap d^{-1}\big(F^{p+\epsilon}U_\Delta(\g)^{p+q+1}\big)\ 
\subset\ d\g^{p,q-1}[\Delta]+F^{p+\epsilon}U_\Delta(\g)^{p+q} \ \ ,\ \text{ if } p+q\neq0\ .
$$
\end{remark}

By assumption (\ref{dnonlin}), $\gr\,\g$ is a bigraded filtered non-linear Lie algebra with differential $\grd$. Hence its cohomology $H(\g,\grd)$ is a space with induced bigrading and filtration,
\begin{eqnarray*}
&\displaystyle{
H(\g,\grd)\ =\!\!\! \bigoplus_{\substack{p,q\in\Gamma,\ p+q\in\Z_+\\ \Delta\in\Gamma_+^\prime }} H^{pq}(\g,\grd)[\Delta]\ , \quad\text{ where }
}\\
&\displaystyle{
H^{pq}(\g,\grd)[\Delta]\ =\ \big(\g^{pq}[\Delta]\cap\Ker\, \grd\big)\Big/ \grd\big(\g^{p,q-1}[\Delta]\big)\ .
}
\end{eqnarray*}
The assumption that $\grd$ is good is equivalent to say that $H^{pq}(\g,\grd)[\Delta]=0$ if $p+q\neq0$.
Hence the above decomposition reduces to
\begin{eqnarray}\label{diag}
&\displaystyle{
H(\g,\grd)\ =\! \bigoplus_{p\in\Gamma,\ \Delta\in\Gamma_+^\prime }H^{p,-p}(\g,\grd)[\Delta]\ , 
\ \ \text{ where } 
}\nonumber\\
&\displaystyle{\vphantom{\Big(} 
H^{p,-p}(\g,\grd)[\Delta]\ =\ \g^{p,-p}[\Delta]\cap\Ker \grd 
\ =\ \g^{p,-p}[\Delta]\cap d^{-1}\big(F^{p+\epsilon}U_\Delta(\g)^1\big)\ . \nonumber
}
\end{eqnarray}
In particular we have a natural embedding $H(\g,\grd)\subset \gr\,\g$.
It follows that $H(\g,\grd)$ has the induced structure of a non-linear Lie algebra.
Therefore the complex $(\gr\,\g,\grd)$ satisfies all assumptions of Theorem \ref{glinear}
and we have a canonical isomorphism
\begin{equation}\label{12set}
H(\gr\, U(\g),\grd)\ \simeq U(H(\gr\,\g,\grd))\ .
\end{equation}

We now fix $\Delta\in\Gamma_+$ and we consider the complex $(U_\Delta(\g),d)$. 
We want to show that it satisfies all the assumptions of Lemma \ref{homolalg}.
By the above observations, we only need to check that $(U_\Delta(\g),d)$ is good in the sense 
of Definition \ref{def-good}, namely $H^{pq}(\gr\,U_\Delta(\g),\grd)=0$ for $p+q\neq0$.
For this we need the following
\begin{lemma}\label{comp}
The differential $\grd$ is such that
$$
\grd\big(\gr^{pq}U(\g)\big)\cap\gr^{p,q+1}U_\Delta(\g)\ =\
\grd(\gr^{pq}U_\Delta(\g)\big)\ .
$$
Therefore there is a natural embedding
$$
H^{pq}(\gr\,U_\Delta(\g),\grd)\ \subset\ H^{pq}(\gr\,U(\g),\grd)\ .
$$
\end{lemma}
\begin{proof}

By assumption on $d$ we have $d\big(F^pU_\Delta(\g)^n\big)\subset
F^pU_\Delta(\g)^{n+1}$. It immediately follows that
$\grd\big(\gr^{pq}U_\Delta(\g)\big)\subset\gr^{p,q+1}U_\Delta(\g)$.
Hence we only have to show $\grd\big(\gr^{pq}U(\g)\big) \cap
\gr^{p,q+1}U_\Delta(\g) \subset \grd(\gr^{pq}U_\Delta(\g)\big)$.

Let $A\in U(\gr\,\g)^{pq}\simeq \gr^{pq}U(\g)$ be such that
$\grd(A)\in U_\Delta(\gr\,\g)^{p,q+1}$. We need to show that there
is $A'\in U_\Delta(\gr\,\g)^{pq}$ such that $\grd(A)=\grd(A')$. If
$A\in U_\Delta(\gr\,\g)^{pq}$, there is nothing to prove. Suppose
then $\Delta_1>\Delta$ is such that $A\in
U_{\Delta_1}(\gr\,\g)^{pq}$ and $A\notin
U_{\Delta_1-\epsilon}(\gr\,\g)^{pq}$.

By the grading condition (\ref{gradcond1}), the non-linear Lie algebra
$\gr\,\g$ is filtered by $\Gamma_+$, and the associated graded Lie
algebra, that we denote $\gr_\Delta\g$, is the space $\g$ with trivial
Lie bracket. Hence $\gr_\Delta U(\gr\,\g)\simeq U(\gr_\Delta\g) =
S(\g)$. We will denote by $X\mapsto\bar{X}$ the natural quotient
map $U_\Delta(\gr\,\g)^{pq}\twoheadrightarrow
S(\g)[\Delta]^{pq}\simeq U_\Delta(\gr\,\g)^{pq} /
U_{\Delta-\epsilon}(\gr\,\g)^{pq}$. By (\ref{dnonlin}) $\grd$ is a
differential of bidegree $(0,1)$ of the space $\g$ preserving the
$\Gamma_+^\prime $-grading. 
Hence $\grd$
extends, by the Leibniz rule, to a differential of
$U(\gr_\Delta\g)=S(\g)$ preserving the $\Gamma_+$-grading. It
then follows by Lemma \ref{symlinear} that
$H(S(\g),\grd)=S(H(\g,\grd))$ or, equivalently,
\begin{eqnarray}\label{nice}
& \Ker\Big(\grd: S(\g)[\Delta]^{pq}\rightarrow
S(\g)[\Delta]^{p,q+1}\Big) =
\im\Big(\grd: S(\g)[\Delta]^{p,q-1}\rightarrow S(\g)[\Delta]^{pq}\Big) \nonumber\\
& +\ S\Big(\Ker\big(\grd:\ \g\rightarrow \g\big)\Big)\cap
S(\g)[\Delta]^{pq}\ \ ,\ \ \ \forall\Delta\in\Gamma_+\ .
\end{eqnarray}

By assumption $A\in U_{\Delta_1}(\gr\,\g)^{pq},\ A\notin
U_{\Delta_1-\epsilon}(\gr\,\g)^{pq}$ and $\grd(A)\in
U_{\Delta_1-\epsilon}(\gr\,\g)$. 
This is equivalent to say that
$0\neq\bar{A}\in S(\g)[\Delta_1]^{pq}$ and $\grd(\bar{A})=0$. In
other words $\bar{A}\in \Ker\big(\grd:\ S(\g)[\Delta_1]^{pq}
\rightarrow S(\g)[\Delta_1]^{p,q+1}\big)$. By (\ref{nice}) we can
then find $\bar{B}\in\ S(\g)[\Delta_1]^{p,q-1}$ and $\bar{C}\in
S\big(\Ker\big(\grd:\ \g\rightarrow \g\big)\big)\cap
S(\g)[\Delta_1]^{pq}$ such that
$$
\bar{A}\ =\ \grd(\bar{B})+\bar{C}\ .
$$
Let then $B\in U_{\Delta_1}(\gr\,\g)^{p,q-1}$ be any preimage of
$\bar{B}$, and $C\in U_{\Delta_1}(\gr\,\g)^{pq}$ be a preimage
of $\bar{C}$ polynomial in the elements of $\Ker\big(\grd:\
\g\rightarrow \g\big)$, so that $\grd(C)=0$. We have
$$
A_1\ =\ A-\grd(B)-C\ \in\ U_{\Delta_1-\epsilon}(\gr\,\g)^{p,q}\ ,\
\ \grd(A_1)\ =\ \grd(A)\ .
$$
We can then repeat the same argument to conclude, by induction,
that there is an element $A'\in\ U_{\Delta}(\gr\,\g)^{p,q}$ such
that $\grd(A')=\grd(A)$, thus completing the proof. 
The second part of the lemma follows immediately from the first part.
\end{proof}

For every $p\in\Gamma,\ \Delta\in\Gamma_+^\prime $ we fix a basis
$\A_{p\Delta}=\{e_i,\ i\in\I_{p\Delta}\}$ of the space
$H^{p,-p}(\gr\,\g,\grd)[\Delta]=\g^{p,-p}[\Delta]\cap\Ker \grd$.

We also let $\A=\{e_i,\ i\in\I\}=\bigsqcup_{p\Delta}\A_{p\Delta}$, which is a basis of $H(\gr\,\g,\grd)$, 
and we fix a total ordering of the index set $\I$.
We assume such basis is compatible with the parity, and we denote $p(i)=p(e_i)\in\Z/2\Z$.
As in Section \ref{sec-nonlinlie} we denote
$$
\Sigma_{p\Delta}\ =\ \left\{e_{i_1}\dots e_{i_s}\ \left|
        \begin{array}{c}
        i_k\in\I_{p_k\Delta_k},\ i_1\leq\dots\leq i_s\\ 
        i_k<i_{k+1} \text{ if } p(i_k)=\bar{1}\\
        \sum_kp_k=p,\ \sum_k\Delta_k\leq\Delta
        \end{array}\right.\right\}\ ,
$$
moreover we let 
$\Sigma_p=\bigcup_\Delta\Sigma_{p\Delta},\ \Sigma_\Delta=\bigsqcup_p\Sigma_{p\Delta}$ 
and $\Sigma=\bigcup_{p,\Delta}\Sigma_{p\Delta}$.

\renewcommand{\theenumi}{\alph{enumi}}
\renewcommand{\labelenumi}{(\theenumi)}
\begin{lemma}\label{ggood2}
\begin{enumerate}
\item The complex $(U_\Delta(\g),d)$ is good in the sense of Definition \ref{def-good}.
\item For $p\in\Gamma$, $\Sigma_{p\Delta}$ is  a vector space basis 
of $H^{p,-p}(\gr\,U_\Delta(\g),\grd)$.
\end{enumerate}
\end{lemma}

\renewcommand{\theenumi}{\arabic{enumi}}
\renewcommand{\labelenumi}{\theenumi.}
\begin{proof}
Since, by (\ref{diag}), the non-linear Lie algebra $H(\gr\,\g,\grd)$ is concentrated 
on bidegrees $p,q$ with $p+q=0$,
so must be its universal enveloping algebra,
$$
U(H(\gr\,\g,\grd))\ =\ \bigoplus_{p\in\Gamma}U(H(\gr\,\g,\grd))^{p,-p}\ .
$$
By (\ref{12set}) we have a natural isomorphism $H(U(\gr\,\g),\grd)\simeq U(H(\gr\,\g,\grd))$. 
Hence, by looking separately at each bidegree $p,q\in\Gamma$, we get
\begin{eqnarray}\label{antonio}
H^{pq}(U(\gr\,\g),\grd) &=& 0\ ,\ \ \text{ if } p+q\neq0\ ,\nonumber\\
H^{p,-p}(U(\gr\,\g),\grd) &\simeq& U(H(\gr\,\g,\grd))^{p,-p}\ .
\end{eqnarray}
Thanks to Lemma \ref{comp}, there is a natural embedding 
$H^{pq}(\gr\,U_\Delta(\g),\grd)\subset H^{pq}(\gr\,U(\g),\grd)$, so that part (a) follows immediately from the first equation above.
By comparing the $\Gamma_+$-filtration in both sides of (\ref{antonio}) we get
$$
H^{p,-p}(\gr\,U_\Delta(\g),\grd)\ =\ U_\Delta(\gr\,\g)^{p,-p}\cap\Ker \grd
\ =\ U_\Delta(H(\gr\,\g,\grd))^{p,-p}\ .
$$
The second part of the lemma follows by the above identity and the PBW Theorem for (filtered)
non-linear Lie algebras.
\end{proof}

So far we showed that the complex $(U_\Delta(\g),d)$ satisfies the assumptions of 
Lemma \ref{homolalg}. We thus have, from (\ref{hom1}) and (\ref{hom2}),
\begin{eqnarray}
\vphantom{\Big(}  && F^pU_\Delta(\g)^n\cap\Ker d =
F^pU_\Delta(\g)^n\cap\im d\ ,\ \
\text{ for } n\neq0\ , \label{fond1}\\
\vphantom{\Big(} && F^pU_\Delta\!(\g)^0\!\cap\! d^{-1}(F^{p+\epsilon}\!U_\Delta(\g)^1\!) =
F^{p+\epsilon}\!U_\Delta(\g)^0\!+\!F^pU_\Delta(\g)^0\!\cap\!\Ker d\ .\label{fond2}
\end{eqnarray}
By assumption 
$e_i\in\g^{p,-p}[\Delta]\cap\Ker \grd\subset F^pU_\Delta(\g)^0\cap d^{-1}(F^{p+\epsilon}U_\Delta(\g)^1)$,
for $i\in\I_{p\Delta}$. Thanks to (\ref{fond2}) we can then find elements 
$\epsilon_i\in F^{p+\epsilon}U_\Delta(\g)^0$ and $E_i\in F^pU_\Delta(\g)^0\cap\Ker d$ such that
$E_i=e_i+\epsilon_i$. We will assume $E_i$ (hence $\epsilon_i$) has the same parity as $e_i$.
Recall that $H^0(U(\g),d)=U(\g)^0\cap\Ker d$, so that $E_i\in H^0(U(\g),d),\ \forall i\in\I$.

We then let $\B_{p\Delta}=\{E_i,\ i\in\I_{p\Delta}\}$ for every $p\in\Gamma,\ \Delta\in\Gamma_+^\prime $, 
and $\B=\{E_i,\ i\in\I\}=\bigsqcup_{p\Delta}\B_{p\Delta}$.
Moreover we define $H^{p,-p}(\g,d)[\Delta]$ and $H(\g,d)$ as the subspace of $H^0(U(\g),d)$ 
with basis $\B_{p\Delta}$ and $\B$ respectively:
\begin{equation}\label{h}
H(\g,d)\ =\ \bigoplus_{p,\Delta}H^{p,-p}(\g,d)[\Delta]\ \ ,\ \ \ 
H^{p,-p}(\g,d)[\Delta]\ =\ \Span_\C\ \B_{p\Delta}\ .
\end{equation}
Finally we introduce, as above, the collection of ordered monomials in the elements $E_i$,
$$
S_{p\Delta}\ =\ \left\{E_{i_1}\dots E_{i_s}\ \left|
        \begin{array}{c}
        i_k\in\I_{p_k\Delta_k},\ i_1\leq\dots\leq i_s\\ 
        i_k<i_{k+1} \text{ if } p(i_k)=\bar{1}\\
        \sum_kp_k=p,\ \sum_k\Delta_k\leq\Delta
        \end{array}\right.\right\}\ ,
$$
and we let 
$S_p=\bigcup_\Delta S_{p\Delta},\ S_\Delta=\bigsqcup_p S_{p\Delta}$ 
and $S=\bigcup_{p,\Delta}S_{p\Delta}$.

\renewcommand{\theenumi}{\alph{enumi}}
\renewcommand{\labelenumi}{(\theenumi)}
\begin{lemma}\label{free-basis}
\begin{enumerate}
\item The cohomology of the complex $(U(\g),d)$ is concentrated in degree 0:
$H^n(U(\g),d)=0$, for $n\neq0$.
\item The 0\st{th} cohomology $H^0(U(\g),d)$ is PBW generated by $H(\g,d)$. In particular
$S$ is a basis of $H^0(U(\g),d)$ compatible with the $\Gamma$ and $\Gamma_+$ filtrations of 
$H^0(U(\g),d)$.
\end{enumerate}
\end{lemma}
\renewcommand{\theenumi}{\arabic{enumi}}
\renewcommand{\labelenumi}{\theenumi.}
\begin{proof}
Let $A\in U(\g)^n\cap\Ker d,\ n\neq0$. For some $p\in\Gamma$ and $\Delta\in\Gamma_+$ we have
$A\in F^pU_\Delta(\g)^n$, hence, by (\ref{fond1}), it follows that $A\in F^pU_\Delta(\g)^n\cap\im d$,
namely $\bar A=0$ in $H^n(U(\g),d)=0$. This proves the first part of the lemma.

For the second part, we first show that the ordered monomials $E_{i_1}\dots E_{i_s}\in S_{p\Delta}$
span the space $F^pU_\Delta(\g)^0\cap\Ker d$.
For $A\in F^pU_\Delta(\g)^0\cap\Ker d$ we let $[A]$ be the corresponding element in
$\gr^{p,-p}U_\Delta(\g)=F^pU_\Delta(\g)^0/F^{p+\epsilon}U_\Delta(\g)^0$.
Notice that $\grd[A]=[dA]=0$, so that 
$[A]\in \gr^{p,-p}U_\Delta(\g)\cap\Ker \grd=H^{p,-p}(\gr\,U_\Delta(\g),\grd)$.
It then follows by the second part of Lemma \ref{ggood2} that
$$
A=A^\prime+\sum_{\vec{\text{\i}}}c_{\vec{\text{\i}}}e_{i_1}\dots e_{i_s}\ ,
$$
where $A^\prime\in F^{p+\epsilon}U_\Delta(\g)^0$ and the sum in the right hand side is a linear combination
of ordered monomials $e_{i_1}\dots e_{i_s}\in\Sigma_{p\Delta}$.
Recall that, by assumption, for $i\in\I_{p\Delta}$, we have 
$E_i=e_i+\epsilon_i\in F^pU_\Delta(\g)^0\cap\Ker d$ with $\epsilon_i\in F^{p+\epsilon}U_\Delta(\g)^0$.
We thus have
$$
A_1\ =\ A-\sum_{\vec{\text{\i}}}c_{\vec{\text{\i}}}E_{i_1}\dots E_{i_s}\ \in\ F^{p+\epsilon}U_\Delta(\g)^0\cap\Ker d\ .
$$
After repeating the same argument $N$ times we get, by induction, a sequence of elements
$$
A_N\ =\ A-\sum_{\vec{\text{\i}}}c_{\vec{\text{\i}}}E_{i_1}\dots E_{i_s}\ \in\ F^{p+N\epsilon}U_\Delta(\g)^0\cap\Ker d\ ,
$$
where now the indices $i_k\in\I_{p_k,\Delta_k}$ are such that 
$p\leq p_1+\dots+p_s\leq p+N\epsilon,\ \Delta_1+\dots+\Delta_s\leq\Delta$.
Hence $A_N=0$ by locally finiteness.

We are left to show that the ordered monomials $E_{i_1}\dots E_{i_s}\in S$ are linearly independent.
For this, suppose we have a linear combination which is zero,
\begin{equation}\label{indep}
A\ =\ \sum_{\vec{\text{\i}}}c_{\vec{\text{\i}}} E_{i_1}\dots E_{i_s}\ =\ 0\ ,
\end{equation}
where the sum is over "ordered" indices $\vec{\text{\i}}=(i_1,\dots,i_s),\ i_k\in\I_{p_k\Delta_k}$.
We can assume the coefficients $c_{\vec{\text{\i}}}$ are not all zero, and we let
$$
p\ =\ \min_{\vec{\text{\i}}\ |\ c_{\vec{\text{\i}}}\neq0} \Big\{p_1+\dots+p_s\Big\}\ .
$$
We then get by (\ref{indep}) 
$$
A\ =\ \sum_{\vec{\text{\i}}\ |\ p_1+\dots+p_s=p}c_{\vec{\text{\i}}} e_{i_1}\dots e_{i_s}+R\ =\ 0\ ,
$$
with $R\in F^{p+\epsilon}U(\g)^0$. We can now look at the image $[A]$ of $A$ in
$\gr^{p,-p}U(\g)^0=F^{p}U(\g)^0/F^{p+\epsilon}U(\g)^0$. We have
\begin{equation}\label{indep2}
[A]\ =\ \sum_{\vec{\text{\i}}\ |\ p_1+\dots+p_s=p}c_{\vec{\text{\i}}} e_{i_1}\dots e_{i_s}\ =\ 0\ .
\end{equation}
By the second part of Lemma \ref{ggood2}, the ordered monomials $e_{i_1}\dots e_{i_s}$ are
linearly independent in $\gr^{p,-p}U(\g)$. Hence (\ref{indep2}) implies $c_{\vec{\text{\i}}}=0$ for all
$\vec{\text{\i}}$ such that $p_1+\dots+p_s=p$, which is a contradiction.
\end{proof}

By Lemma \ref{free-basis} the associative algebra $H=H^0(U\g),d)$ is PBW generated
by the space $\h=H(\g,d)$. 
By definition (\ref{h}), $\h$ is $\Gamma_+^\prime $-graded with 
$\h[\Delta]=\Span_\C S_\Delta$. Moreover, by the secon part of Lemma \ref{free-basis}, 
the natural $\Gamma_+$-filtration $U_\Delta(\g)^0\cap\Ker d\subset H$ coincides
with the $\Gamma_+$-filtration $H_\Delta$ induced by the generating space
$\h$ as in (\ref{35}). By the grading condition (\ref{gradcond2}), we have
$[\h[\Delta_1],\h[\Delta_2]]\subset H_\Delta$, for some $\Delta<\Delta_1+\Delta_2$,
namely $H$ is "pregraded" by $\Gamma_+$, in the sense of Definition \ref{predef}.
In conclusion, all the assumptions of Theorem \ref{gmain2} hold, so that we have
\begin{theorem}\label{gnonlin}
The space $H(\g,d)$ admits a structure of non-linear Lie superalgebra
such that its universal enveloping associative algebra is canonically isomorphic 
to $H(U(\g),d)$:
$$
H(U(\g),d)\ \simeq\ U(H(\g,d))\ .
$$
\end{theorem}

%%%
\subsection{Non-linear Lie conformal algebras with a good almost linear differential}
\label{sec-nonlin-conf}

In this section we want to derive the same results as in the previous section in the context of
Lie conformal superalgberas. 

Let $R$ be a non-linear Lie conformal superalgebra, freely generated as a $\C[T]$-module
by a finite-dimensional subspace $\g$:
$R=\C[\ttt]\otimes\g$. We shall assume that $\g$ is graded as in (\ref{grg}).
This grading induces a vector space grading of $R$, by assuming $\ttt$
to have $\Gamma$-bidegree $(0,0)$ and $\Gamma_+$-degree $1$:
\begin{equation}\label{grR}
R\ =\ \bigoplus_{\substack{p,q\in\Gamma,\ p+q\in\Z_+\\ \Delta\in\Gamma_+^\prime }} R^{pq}[\Delta]\ \ ,\ \ \  
R^{pq}[\Delta]\ =\ \bigoplus_{n\geq0}\ttt^n\g^{pq}[\Delta-n]\ ,
\end{equation}
and of the tensor algebra $\T(R)$.
We will assume that $R$, as a (non-linear) Lie conformal algebra, is $\Z_+$-graded by
$\C[T]$submodules $R^n=\bigoplus_{\substack{p,q,\Delta\\ p+q=n}}R^{pq}[\Delta],\ n\in\Z_+$,
it is $\Gamma$-filtered by $F^pR=\bigoplus_{p^\prime\geq p,q,\Delta}R^{p^\prime,q}[\Delta],
\ p\in\Gamma$, and moreover we assume that the $\Gamma_+^\prime $-grading 
by subspaces $R[\Delta]=\bigoplus_{p,q}R^{pq}[\Delta],\ \Delta\in\Gamma_+^\prime $, 
is a conformal weight of $R$.
All these conditions mean the following relation
\begin{equation}\label{gradR}
R^{p_1,q_1}[\Delta_1]_{(n)}R^{p_2,q_2}[\Delta_2]\ \subset\ 
\bigoplus_{l\geq0}\T^{p_1+p_2+l,q_1+q_2-l}(R)[\Delta_1+\Delta_2-n-1]\ ,\ \ n\in\Z_+\ .
\end{equation}
The associated graded non-linear Lie conformal algebra $\gr\,R$ is defined as the 
$\C[T]$-module $R$ with 
$\lambda$-bracket $[a\ _\lambda\ b]^\gr$, for $a\in R^{p_1,q_1},\ b\in R^{p_2,q_2}$,
equal to the component of $[a\ _\lambda\ b]$ in $R^{p_1+q_1,p_2+q_2}[\lambda]$.
It is not hard to check that  $\gr\,R$ is a $\Gamma$-bigraded non-linear Lie conformal algebra, 
with a conformal weight $\Delta$.

Let $V(R)$ be the universal enveloping vertex algebra of $R$. 
Recall that on the vertex algebra $V(R)=\T(R)/\M(R)$ 
the $\lambda$-bracket $[a\ _\lambda\ b]$ of elements $a,b\in R$ 
is induced by the non-linear $\lambda$-bracket in $R$, and the normally
ordered product $:aB:$ with $a\in R,\ B\in V(R)$ is induced by the associative product in 
$\T(R)$. 
These two conditions, together with the axioms of vertex algebra,
uniquely define a vertex algebra structure on $V(R)$.
Since all axioms of vertex algebra are "homogeneus" with respect to the gradings 
and the filtration of $R$,
it is not hard to check that the vertex algebra $V(R)$ has induced $\Z_+$-grading 
$V(R)=\bigoplus_{n\geq0}V(R)^n$, $\Gamma$-filtration $F^pV(R)\subset V(R)$,
and $\Gamma_+$-grading $V(R)=\bigoplus_{\Delta\in\Gamma_+}V(R)[\Delta]$,
and all are additive with respect to the normally ordered product. Explicitly:
\begin{eqnarray*}
V(R)^n &=& \Span_\C\Big\{:a_{i_1}\cdots a_{i_s}:\ \Big|\ a_{i_k}\in R^{p_k,q_k}[\Delta_k] ,\
\sum_k(p_k+q_k)=n\Big\}\ ,\\
F^pV(R) &=& \Span_\C\Big\{:a_{i_1}\cdots a_{i_s}:\ \Big|\ a_{i_k}\in R^{p_k,q_k}[\Delta_k] ,\
\sum_k p_k\geq p\Big\}\ , \\
V(R)[\Delta] &=& \Span_\C\Big\{:a_{i_1}\cdots a_{i_s}:\ \Big|\ a_{i_k}\in R^{p_k,q_k}[\Delta_k] ,\
\sum_k\Delta_k=\Delta\Big\}\ .
\end{eqnarray*}
The vertex algebra structure is compatible with the above gradings and filtrations,
in the sense that, for each $n\in\Z$, we have:
\begin{equation}\label{gradV}
F^{p_1}V(R)^{n_1}[\Delta_1]_{(n)}F^{p_2}V(R)^{n_2}[\Delta_2]\ \subset\ F^{p_1+p_2}V(R)^{n_1+n_2}[\Delta_1+\Delta_2-n-1]\ .
\end{equation}
Notice that, since $\Delta(\ttt)=1$, each space $V(R)^n[\Delta]$ is finite dimensional, hence 
the corresponding filtration $F^pV(R)^n[\Delta]$ is locally finite.
We have the associated graded vertex algebra, defined by
$$
\gr\,V(R)=\bigoplus_{p,q,\Delta}\gr^{pq}V(R)[\Delta]\ ,\ \ 
\gr^{pq}V(R)[\Delta]\ =\ F^pV(R)^n[\Delta]/F^{p+\epsilon}V(R)^n[\Delta]\ ,
$$
with induced vertex algebra structure, which is a $\Gamma$-bigraded vertex algebra and it has a 
$\Gamma_+$-valued conformal weight.

The analogue of Lemma \ref{ugr} holds also in this context, namely we have
\begin{lemma}\label{vgr}
There is a canonical isomorphism $\gr\,V(R)\simeq V(\gr\,R)$
which preserves all gradings, namely such that
$$
\gr^{pq}V(R)[\Delta]\simeq\ V^{pq}(\gr\,R)[\Delta]\ .
$$
In particular the natural embedding $R=\gr\,R\subset \gr\,V(R)$
is such that
$R\cap\gr^{pq}V(R)[\Delta]=R^{pq}[\Delta]$.
\end{lemma}

\begin{definition}
Let $d:\ V(R)\rightarrow V(R)$ be a differential of 
the vertex algebra $V(R)$.
We assume that $d$ has $\Z_+$-degree $+1$, that it preserves the $\Gamma$-filtration
and the conformal weight $\Delta$. In other words
\begin{equation}\label{gradd}
d\big(F^pV(R)^n[\Delta]\big)\ \subset F^pV(R)^{n+1}[\Delta]\ .
\end{equation}
As before, we denote by $\grd:\ \gr\,V(R)\rightarrow \gr\,V(R)$ the
induced differential of the associated graded vertex algebra.
It clearly has bidegree $(0,1)$, it preserves the conformal weight $\Delta$, and it is
an odd derivation of all $n$\st{th} products of $\gr\,V(R)$.
We say that $d$ is an {\itshape almost linear differential} of the non-linear 
Lie conformal algebra $R$ if $\grd$ preserves $\g\subset \gr\,V(R)$, namely
$\grd(\g^{pq}[\Delta])\subset\g^{p,q+1}[\Delta]$;
or equivalenlty, in terms of $d$, 
\begin{equation}\label{dRnonlin}
d\big(\g^{pq}[\Delta]\big)\subset\g^{p,q+1}[\Delta]\oplus
F^{p+\epsilon}V(R)^{p+q+1}[\Delta]\ .
\end{equation}
In particular $\grd|_R$ is a differential of the non-linear Lie conformal algebra $\gr\,R=\C[\ttt]\g$,
and $\grd$ is the induced differential of the universal enveloping vertex algebra $V(\gr\,R)\simeq\gr\,V(R)$.

The corresponding cohomology is
$H(\gr\,R,\grd)
=\C[\ttt]H(\g,\grd)$, with $H(\g,\grd)$ $=\bigoplus_{p,q,\Delta}H^{pq}(\g,\grd)[\Delta]
=\bigoplus_{p,q,\Delta}\Ker\big(\grd\big|_{\g^{pq}[\Delta]}\big)/\im\big(\grd\big|_{\g^{p,q-1}[\Delta]}\big)$.
We say that $d$ is a {\itshape good} almost linear differential of $R$ if the cohomology $H(R,\grd)$
(or equivalently $H(\g,\grd)$) is concentrated at $\Z_+$-degree zero, namely
$H^{pq}(\g,\grd)=0$ if $p+q\neq0$, 
or equivalently, in terms of $d$,
\begin{equation}\label{dRgood2}
\g^{pq}[\Delta]\cap d^{-1}\big(F^{p+\epsilon}V(R)^{p+q+1}[\Delta]\big)\ 
\subset\ d\big(\g^{p,q-1}[\Delta]\big)+F^{p+\epsilon}V(R)^{p+q}[\Delta]\ .
\end{equation}
It follows that  $H(\gr\,R,\grd)$ has the decomposition
\begin{eqnarray}\label{dRgood}
\vphantom{\Big(}
&\displaystyle{
H(\gr\,R,\grd)=\bigoplus_{p,\Delta}H^{p,-p}(\gr\,R,\grd)[\Delta]\ ,\ \text{ where}}\\ 
\vphantom{\Big(}
&\displaystyle{
H^{p,-p}(\gr\,R,\grd)[\Delta]\ =\ R^{p,-p}[\Delta]\cap\Ker \grd
\ ,}\nonumber
\end{eqnarray}
in particular it is embedded in $\gr\,R$, hence it has the induced structure of a 
non-linear Lie conformal algebra.
We can thus apply Theorem \ref{Rlinear} to conclude
\begin{equation}\label{grRlinear}
H(\gr\,V(R),\grd)\ \simeq\ V(H(\gr\,R,\grd))\ .
\end{equation}
\end{definition}
From now on we let $d$ be a good almost linear differential of $R$.

\medskip

As we did before, for every $p\in\Gamma,\ \Delta\in\Gamma_+^\prime $, we fix a basis
$\A_{p\Delta}=\{e_i,\ i\in\I_{p\Delta}\}$ of $\g^{p,-p}[\Delta]\cap\Ker\grd$,
and we let $\A=\{e_i,\ i\in\I\}=\bigsqcup_{p\Delta}\A_{p\Delta}$.
We extend these bases to $\tA_{p\Delta}=\{e_{\ti},\ \ti\in\tI_{p\Delta}\}$, a basis of 
$H^{p,-p}(\gr\,R,\grd)[\Delta]=R^{p,-p}[\Delta]\cap\Ker\grd$,
and $\tA=\{e_\ti,\ \ti\in\tI\}=\bigsqcup_{p\Delta}\tA_{p\Delta}$, a basis of $H(\gr\,R,\grd)$.
For example we can take $\tI_{p\Delta}=\bigsqcup_{k\geq0}(\I_{p,\Delta-k},k)$
and $e_\ti=\ttt^{(k)}e_i$ for $\ti=(i,k)\in(\I_{p,\Delta-k},k)\subset\tI_{p,\Delta}$.
We then assume $\tI$ is totally ordered and we denote
$$
\tSigma_{p\Delta}\ =\ \left\{:e_{\ti_1}\dots e_{\ti_s}:\ \left|
        \begin{array}{c}
        \ti_k\in\tI_{p_k\Delta_k},\ \ti_1\leq\dots\leq \ti_s\\ 
        \ti_k<\ti_{k+1} \text{ if } p(\ti_k)=\bar{1}\\
        \sum_kp_k=p,\ \sum_k\Delta_k=\Delta
        \end{array}\right.\right\}\ .
$$
We also let 
$\tSigma_p=\bigsqcup_\Delta\tSigma_{p\Delta},\ \tSigma_\Delta=\bigsqcup_p\tSigma_{p\Delta}$ 
and $\tSigma=\bigsqcup_{p,\Delta}\tSigma_{p\Delta}$.

\renewcommand{\theenumi}{\alph{enumi}}
\renewcommand{\labelenumi}{(\theenumi)}
\begin{lemma}\label{Rgood2}
\begin{enumerate}
\item For $p,q\in\Gamma$ such that $p+q\neq0$ we have
$H^{pq}(\gr\,V(R),\grd)$ $=0$.
Hence, for every $\Delta\in\Gamma_+$, the complex $(V(R)[\Delta],d)$ is good.
\item For $p\in\Gamma,\ \Delta\in\Gamma_+$, the set $\tSigma_{p\Delta}$ is  a basis 
of $H^{p,-p}(\gr\,V(R),\grd)[\Delta]$.
\end{enumerate}
\end{lemma}

\renewcommand{\theenumi}{\arabic{enumi}}
\renewcommand{\labelenumi}{\theenumi.}
\begin{proof}
By Theorem \ref{Rlinear} and Lemma \ref{vgr} we have an isomorphism
$H(\gr\,V(R),\grd)$ $\simeq V(H(\gr\,R,\grd))$. Moreover by assumption (\ref{dRgood}),
$H(\gr\,R,\grd)$ is concentrated in zero $\Z_+$-degree, so must be its enveloping
vertex algebra $V(H(\gr\,R,\grd))$.
We thus have $H^{pq}(\gr\,V(R),\grd)=0$ if $p+q\neq0$, and 
$H^{p,-p}(\gr\,V(R),\grd)\simeq V^{p,-p}(H(\gr\,R,\grd))$.
Part (a) of the lemma follows immediately by the fact that $\grd$ preserves 
the conformal weight $\Delta$. 
Part (b) follows by (\ref{grRlinear}) and the PBW Theorem.
\end{proof}

By the first part of Lemma \ref{Rgood2}, the complex $(V(R)[\Delta],d)$ (which is finite dimensional, 
hence locally finite) satisfies all the assumptions of Lemma \ref{homolalg}.
We thus have by (\ref{hom1}) and (\ref{hom2})
\begin{eqnarray}
\vphantom{\Big(}  && F^pV(R)^n[\Delta]\cap\Ker d =
F^pV(R)^n[\Delta]\cap\im d\ ,\ \
\text{ for } n\neq0\ , \label{Rfond1}\\
\vphantom{\Big(} && F^pV(R)^0[\Delta]\cap d^{-1}(F^{p+\epsilon}V(R)^1[\Delta]) =
F^{p+\epsilon}V(R)^0[\Delta] \label{Rfond2}\\
\vphantom{\Big(} &&\qquad\qquad\qquad + F^pV(R)^0[\Delta]\cap\Ker d\ .\nonumber
\end{eqnarray}
Thanks to (\ref{Rfond2}), for each $i\in\I_{p\Delta}$ we can write 
$e_i=E_i-\epsilon_i\in\g^{p,-p}[\Delta]\cap\Ker\grd$, with
$\epsilon_i\in F^{p+\epsilon}V(R)^0[\Delta]$ and $E_i\in F^pV(R)^0[\Delta]\cap\Ker d$.
Moreover, for $\ti=(i,k)\in\tI_{p\Delta}$, we let $E_\ti=\ttt^{(k)}E_i$ and 
$\epsilon_\ti=\ttt^{(k)}\epsilon_i$, so that again $e_\ti=E_\ti-\epsilon_\ti$.
We then let
\begin{eqnarray*}
\vphantom{\Big(}
H^{p,-p}(\g,d)[\Delta]=\Span_\C\{E_i,\ i\in\I_{p\Delta}\} &\subset& 
H(\g,d)=\bigoplus_{p,\Delta}H^{p,-p}(\g,d)[\Delta]\ ,\\
\vphantom{\Big(}
H^{p,-p}(R,d)[\Delta]=\Span_\C\{E_\ti,\ \ti\in\tI_{p\Delta}\} &\subset& 
H(R,d)=\bigoplus_{p,\Delta}H^{p,-p}(R,d)[\Delta]\ .
\end{eqnarray*}
Finally we define the collection of ordered monomials in the elements $E_\ti$,
$$
\tS_{p\Delta}\ =\ \left\{:E_{\ti_1}\dots E_{\ti_s}:\ \left|
        \begin{array}{c}
        \ti_k\in\tI_{p_k\Delta_k},\ \ti_1\leq\dots\leq \ti_s\\ 
        \ti_k<\ti_{k+1} \text{ if } p(\ti_k)=\bar{1}\\
        \sum_kp_k=p,\ \sum_k\Delta_k=\Delta
        \end{array}\right.\right\}\ ,
$$
and we let 
$\tS_p=\bigsqcup_\Delta \tS_{p\Delta},\ \tS_\Delta=\bigsqcup_p \tS_{p\Delta}$ 
and $\tS=\bigsqcup_{p,\Delta}\tS_{p\Delta}$.
The analogue of Lemma \ref{free-basis} holds also in the vertex algebra set up, with the same proof
(using Lemma \ref{Rgood2} instead of Lemma \ref{ggood2}).
We thus have the following
\renewcommand{\theenumi}{\alph{enumi}}
\renewcommand{\labelenumi}{(\theenumi)}
\begin{lemma}\label{R-free-basis}
\begin{enumerate}
\item The cohomology of the complex $(V(R),d)$ is concentrated in degree 0:
$H^n(V(R),d)=0$, for $n\neq0$.
\item $H^0(V(R),d)$ is freely generated by $H(R,d)=\C[\ttt]H(\g,d)$. 
In particular
$\tS$ is a basis of $H^0(V(R),d)$ compatible with the $\Gamma$-filtration and 
the conformal weight.
\end{enumerate}
\end{lemma}

\begin{theorem}\label{Rnonlin}
\begin{enumerate}
\item The space $H(R,d)=\C[\ttt]H(\g,d)$ admits a structure of non-linear 
Lie conformal algebra
whose enveloping vertex algebra is canonically isomorphic to $H(V(R),d)$:
$$
H(V(R),d)\ \simeq\ V(H(R,d))\ .
$$
\item The decomposition $H(\g,d)=\bigoplus_{p,\Delta}H^{p,-p}(\g,d)[\Delta]$,
given by the subspaces $H^{p,-p}(\g,d)[\Delta]=\Span_\C\big\{E_i,\ i\in\I_{p\Delta}\big\}$,
induces a vertex algebra $\Gamma$-filtration and a $\Gamma_+$-valued
conformal weight $\Delta$ on $H(V(R),d)$, given by the subspaces
$F^pH(V(R),d)[\Delta]=F^pV(R)^0[\Delta]\cap\Ker d$.
In other words we have
\begin{eqnarray*}
F^{p_1}H(V(R),d)[\Delta_1]_{(n)}F^{p_2}H(V(R),d)[\Delta_2]\\
\subset\ F^{p_1+p_2}H(V(R),d)[\Delta_1+\Delta_2-n-1]\ .
\end{eqnarray*}
\end{enumerate}
\end{theorem}
\renewcommand{\theenumi}{\arabic{enumi}}
\renewcommand{\labelenumi}{\theenumi.}
\begin{proof}
By Lemma \ref{R-free-basis},
the vertex algebra $H(V(R),d)$ is freely generated by 
the free $\C[\ttt]$-module $H(R,d)=\C[\ttt]H(\g,d)$.
Moreover, the $\Gamma_+^\prime $-grading of $H(\g,d)$ induces
a conformal weight $\Delta$ on $H(V(R),d)$. Hence, by the observations in Example \ref{pregr_ex},
$H(V(R),d)$ in pregraded in the sense of Definition \ref{pregr}.
Part (a) is now an immediate corollary of Theorem \ref{Rmain2}. Part (b) is obvious.
\end{proof}

%%%
\subsection{Zhu algebra and cohomology}
\label{sec-zhu-cohom}

We make here the same assumptions we made in the previous section.
More precisely, we let $\g$ be a finite dimensional vector superspace graded as in (\ref{grg}),
we assume the free $\C[\ttt]$-module $R=\C[\ttt]\otimes\g$, graded according to (\ref{grR}),
is a non-linear Lie conformal algebra, and its $\lambda$-bracket satisfies the grading 
conditions (\ref{gradR}).
Therefore the universal enveloping vertex algebra $V(R)$ has induced $\Z_+$-grading $V(R)^n$,
$\Gamma$-filtration $F^pV(R)$, and $\Gamma_+$-valued conformal weight $\Delta$,
defined by a Hamiltonian operator $H$,
satisfying the grading conditions (\ref{gradV}).
We also fix $d:\ V(R)\rightarrow V(R)$, a good almost linear differential of $R$,
so that equations (\ref{gradd}), (\ref{dRnonlin}) and (\ref{dRgood2}) hold.
\begin{theorem}\label{main}
There is a canonical associative algebra isomorphism
$$
\phi\ :\  Zhu_H H(V(R),d)\ \stackrel{\sim}{\longrightarrow}\ H(Zhu_H V(R),\dz)\ ,
$$
where $\dz$ is a differential of $Zhu_H V(R)$ induced by $d:\ V(R)\rightarrow V(R)$.
\end{theorem}
The remaining part of this section will be devoted to the proof of Theorem \ref{main}.
First, let us fix some notation. We denote by $\pi_H$ the natural quotient map
$$
\pi_H\ :\ \ \Ker d\ \twoheadrightarrow \Ker d/\im d\ =\ H(V(R),d)\ ,
$$
and, as before, by $\pi_Z$ the natural quotient map
$$
\pi_Z\ :\ \ V(R)\ \twoheadrightarrow V(R)/J_{\hbar=1}(V(R))\ =\ Zhu_H V(R)\ .
$$
The vertex algebra $H(V(R),d)$ has the induced Hamiltonian operator $H$, 
so that it makes sense
to consider its Zhu algebra. We then let $\pi^H_Z$ be the natural quotient map
$$
\pi^H_Z\ :\ \ H(V(R),d)\ \twoheadrightarrow H(V(R),d)/J_{\hbar=1}(H(V(R),d))\ =\ Zhu_H H(V(R),d)\ .
$$
By assumption the differential $d$ is an odd derivation of all $n$\st{th} products of $V(R)$,
and it preserves the conformal weight $\Delta$. Hence it is a derivation of all products
$a_{(n,\hbar)}b$. In particular $d$ preserves the subspace $J_{\hbar=1}(V(R))\subset V(R)$,
and there is an induced differential $\dz$ of the quotient space $Zhu_H V(R)=V(R)/J_{\hbar=1}(V(R))$,
given by $\dz(\pi_Z(A))=\pi_Z(dA)$.
We can then consider the cohomology $H(Zhu_H V(R),\dz)$, and we let $\pi^Z_H$ be the natural
quotient map
$$
\pi^Z_H\ :\ \ \Ker \dz\ \twoheadrightarrow\ \Ker \dz/\im \dz\ =\ H(Zhu_H V(R),\dz)\ .
$$
An equivalent way to state Theorem \ref{main} is to say that 
the map 
\begin{equation}\label{532}
\phi:\ Zhu_H H(V(R),d) \rightarrow H(Zhu_H V(R),\dz)\ ,
\end{equation}
given by $\phi(\pi^H_Z\pi_H(A))=\pi^Z_H\pi_Z(A)$, is a well defined associative algebra 
isomorphism.
\medskip

% step1

As in the previous section, for each $p\in\Gamma,\ \Delta\in\Gamma_+^\prime $,
we fix a basis 
$\{e_i,\ \I_{p\Delta}\}$ of $\g^{p,-p}[\Delta]\cap d^{-1}\big(F^{p+\epsilon}V(R)^1[\Delta]\big)$,
and elements $\{E_i,\ i\in\I_{p\Delta}\}$ such that $E_i\in F^pV(R)^0[\Delta]\cap\Ker d$
and $E_i-e_i\in F^{p+\epsilon}V(R)^0[\Delta]$.
We then let
\begin{eqnarray*}
\vphantom{\Big(}
& \displaystyle{
H(\g,d)\ =\ \bigoplus_{p,\Delta}H^{p,-p}(\g,d)[\Delta]\ ,\ \ 
H^{p,-p}(\g,d)[\Delta]\ =\ \Span_\C\big\{\pi_H(E_i),\ i\in\I_{p\Delta}\big\}\ , }\\
\vphantom{\Big(}
& H(R,d)\ =\ \C[\ttt]H(\g,d)\ \subset\ H(V(R),d)\ .
\end{eqnarray*}
By Theorem \ref{Rnonlin}, $H(R,d)$ admits a structure of non-linear Lie conformal algebra
such that $H(V(R),d)\simeq V(H(R,d))$.
In particular $H(V(R),d)$ is PBW generated by the set
$\big\{\pi_H(\ttt^kE_i)\,|\, k\geq0,i\in\I=\bigsqcup_{p\Delta}\I_{p\Delta}\big\}$.
Moreover the natural $\Gamma_+^\prime $-grading of $H(\g,d)$ induces a conformal
weight $\Delta$ of the universal enveloping vertex algebra $H(V(R),d)$.
Hence the vertex algebra $H(V(R),d)$ satisfies all assumptions of Theorem \ref{zhuenvel}.
We thus get that the space $Zhu_H H(R,d)=\pi^H_Z H(\g,d)\simeq H(\g,d)$ is naturally endowed 
with a structure  of non-linear Lie superalgebra and $Zhu_H H(V(R),d)\simeq U(\pi^H_Z H(\g,d))$.
Hence $Zhu_H H(V(R),d)$ is PBW generated as associative algebra 
by the space $\pi^H_Z(H(\g,d))$.
We thus proved the following
\begin{lemma}\label{main1}
The associative superalgebra $Zhu_H H(V(R),d)$ is PBW generated by the set
$\big\{\pi^H_Z\pi_H(E_i)\,|\, i\in\I\big\}$.
\end{lemma}
\medskip

% step2

Consider now the associative algebra $Zhu_H V(R)$. By Theorem \ref{zhuenvel} it is isomorphic
to $U(Zhu_H R)$, where $Zhu_H R=\pi_Z R\simeq\g$ is a Lie superalgebra with Lie bracket
$[\pi_Z(a),\pi_Z(b)]=\pi_Z([a,b]_{\hbar=1})$, for $a,b\in R$.
By our assumptions the space $\pi_Z\g\simeq\g$ has a decomposition 
$$
\pi_Z\g\ =\ \bigoplus_{\substack{p,q\in\Gamma,\ p+q\in\Z_+\\ \Delta\in\Gamma_+^\prime }}\pi_Z\g^{pq}[\Delta]\ ,\ \ \pi_Z\g^{pq}[\Delta]\simeq\g^{pq}[\Delta]\ .
$$
\renewcommand{\theenumi}{\alph{enumi}}
\renewcommand{\labelenumi}{(\theenumi)}
\begin{lemma}\label{obv1}
\begin{enumerate}
\item We have the following inclusion
$$
\pi_Z R^{pq}[\Delta]\ \subset\ \bigoplus_{\Delta^\prime\leq\Delta}\pi_Z\g^{pq}[\Delta]\ .
$$
\item The non-linear Lie bracket on $\pi_Z\g$ satisfies the grading condition (\ref{gradtotg}), namely
$$
[\pi_Z\g^{p_1q_1}[\Delta_1],\pi_Z\g^{p_2q_2}[\Delta_2]]\ \subset\
\bigoplus_{\substack{l\geq0\\
\Delta<\Delta_1+\Delta_2}}\T^{p_1+p_2+l,q_1+q_2-l}(\pi_Z\g)[\Delta]\ .
$$
\end{enumerate}
\end{lemma}
\renewcommand{\theenumi}{\arabic{enumi}}
\renewcommand{\labelenumi}{\theenumi.}
\begin{proof}
Since $\ttt$ increases the conformal weight $\Delta$ by 1, 
$R^{pq}[\Delta]=\bigoplus_{k\geq0}\ttt^k\g^{pq}[\Delta]$.
On the other hand we have $\pi_Z(\ttt^k a)=\binom{-\Delta_a}{k}\pi_Z(a)$.
The first part of the lemma follows immediately. For part (b) we have,
by the definition of the product $[\ ,\ ]_{\hbar=1}$ and the assumption (\ref{gradR}),
\begin{eqnarray*}
[\pi_Z\g^{p_1q_1}[\Delta_1],\pi_Z\g^{p_2q_2}[\Delta_2]]
&=& \pi_Z\Big([\g^{p_1q_1}[\Delta_1],\g^{p_2q_2}[\Delta_2]]_{\hbar=1}\Big)\\
&\subset& 
\pi_Z\Big(\bigoplus_{l\geq0,\Delta<\Delta_1+\Delta_2}\T^{p_1+p_2+l,q_1+q_2-l}(R)[\Delta]\Big)\\
&\subset& 
\bigoplus_{l\geq0,\Delta<\Delta_1+\Delta_2}\T^{p_1+p_2+l,q_1+q_2-l}(\pi_Z\g)[\Delta]\ .
\end{eqnarray*}
\end{proof}

\begin{lemma}\label{obv2}
Let
$$
F^pU_\Delta(\pi_Z\g)^n\ =\ \Span_\C\left\{\pi_Z(a_1)\cdots
\pi_Z(a_s)\ \left|\ \begin{array}{c}
a_i\in \g^{p_i,q_i}[\Delta_i]\\ \sum_i(p_i+q_i)=n\\ \sum_ip_i\geq p,\ \sum_i\Delta_i\leq\Delta
\end{array}\right.\right\}\ ,
$$
Then we have the inclusion
$$
\pi_Z\big(F^pV(R)^n[\Delta]\big)\ \subset\ F^pU_\Delta(\pi_Z\g)^n\ .
$$
\end{lemma}
\begin{proof}
It follows by Lemma \ref{lem1} that every element in $F^pV(R)^n[\Delta]$ is a linear combination
of monomials $A=(\ttt^{k_1}a_1)_{(-1,\hbar=1)}\cdots_{(-1,\hbar=1)}(\ttt^{k_s}a_s)$,
with $a_i\in\g^{p_i,q_i}[\Delta_i]$ 
and $\sum_i(p_i+q_i)=n,\ \sum_ip_i\geq p,\ \sum_i(k_i+\Delta_i)=\Delta$.
On the other hand $\pi_Z(A)$ is, up to a constant coefficient, equal to $\pi_Z(a_1)\cdots\pi_Z(a_s)$,
which belongs to $F^pU_\Delta(\pi_Z\g)^n$.
\end{proof}
As noticed above, $d:\ V(R)\rightarrow V(R)$ is a derivation of all $(n,\hbar)$ products. 
Hence the induced map
$$
\dz\ :\ \ U(\pi_Z\g)\simeq Zhu_H V(R)\ \longrightarrow\ U(\pi_Z\g)\ ,
$$
given by $\dz(\pi_Z(A))=\pi_Z(dA)$
is a differential of the associative algebra $U(\pi_Z\g)\simeq Zhu_H V(R)$.
\begin{lemma}
We have
$$
\dz\big(F^pU_\Delta(\pi_Z\g)^n\big)\ \subset\ F^pU_\Delta(\pi_Z\g)^{n+1}\ .
$$
\end{lemma}
\begin{proof}
Let $\bar{A}=\pi_Z(a_1)\cdots\pi_Z(a_s)\in F^pU_\Delta(\pi_Z\g)^n$, with
$a_i\in\g^{p_i,q_i}[\Delta_i]$, and $\sum_i(p_i+q_i)=n,\ \sum_ip_i\geq p,\ \sum_i\Delta_i=\Delta$.
We have $\bar{A}=\pi_Z(A)$, with
$A=a^1_{(-1,\hbar=1)}\cdots _{(-1,\hbar=1)}a^s$, and it follows by Lemma \ref{lem1}
that $A\in F^pV(R)^n[\Delta]$.
By assumption (\ref{gradd}) $dA\in F^pV(R)^{n+1}[\Delta]$,
hence by Lemma \ref{obv2} above, we conclude 
that $\dz(\bar{A})=\pi_Z(dA)\in F^pU_\Delta(\g)^{n+1}$,
as we wanted.
\end{proof}
So far we proved that $\dz:\ U(\pi_Z\g)\rightarrow U(\pi_Z\g)$ satisfies the conditions (\ref{graddg}).
We want to prove that $\dz$ is a good almost linear differential of the non-linear Lie algebra $\pi_Z\g$.
For this, we will need the following
\begin{lemma}\label{obv3}
We have
$$
\pi_Z\g^{pq}[\Delta]\cap\left.\dz\right.^{-1}\!\big(\!F^{p+\epsilon}\!U_\Delta(\pi_Z\g)^{p+q+1}\!\big)
=\pi_Z\!\Big(\\g^{pq}[\Delta]\cap d^{-1}\!\big(\!F^{p+\epsilon}V(R)^{p+q+1}[\Delta]\big)\!\Big)\ .
$$
\end{lemma}
\begin{proof}
The inclusion $\supset$ follows immediately by Lemma \ref{obv2} above, so we only need to prove 
the opposite inclusion.
Let $a\in\g^{pq}[\Delta]$ be such 
that $\dz(\pi_Z(a))=\pi_Z(da)\in F^{p+\epsilon}U_\Delta(\pi_Z\g)^{p+q+1}$.
By assumption (\ref{dRnonlin}) on $d$, we have
$da=\alpha+A\in\g^{p,q+1}[\Delta]\oplus F^{p+\epsilon}V(R)^{p+q+1}[\Delta]$.
We then have
$\pi_Z(\alpha)=\pi_Z(da)-\pi_Z(A)\in F^{p+\epsilon}U_\Delta(\pi_Z\g)^{p+q+1}$.
But $\pi_Z(\alpha)\in \pi_Z\g^{p,q+1}[\Delta]$ 
and $\pi_Z\g^{p,q+1}[\Delta]\cap F^{p+\epsilon}U_\Delta(\pi_Z\g)^{p+q+1}=0$. 
Hence $\pi_Z(\alpha)=0$, and $\alpha=0$.
In conclusion $da\in F^{p+\epsilon}V(R)^{p+q+1}[\Delta]$, which is what we wanted.
\end{proof}
We now use assumption (\ref{dRnonlin}) and Lemma \ref{obv2} to get
\begin{eqnarray*}
\vphantom{\Big(}
\dz\big(\pi_Z\g^{pq}[\Delta]\big) &=& \pi_Z\big(d\g^{pq}[\Delta]\big)\ 
\subset\ \pi_Z\big(\g^{p,q+1}[\Delta]\oplus F^{p+\epsilon}V(R)^{p+q+1}[\Delta]\big)\\ 
\vphantom{\Big(}
&\subset& \pi_Z\g^{p,q+1}[\Delta]\oplus F^{p+\epsilon}U_\Delta(\pi_Z\g)^{p+q+1}\ ,
\end{eqnarray*}
hence $\dz$ is an almost linear differential of $\pi_Z\g$.
Moreover, by assumption (\ref{dRgood2}), Lemma \ref{obv2} and Lemma \ref{obv3},
we have, for $p+q\neq0$,
$$
\pi_Z\g^{pq}[\Delta]\cap\left.\dz\right.^{-1}\big(F^{p+\epsilon}U_\Delta(\pi_Z\g)^{p+q+1}\big)\ 
\subset\ \dz\big(\pi_Z\g^{p,q-1}[\Delta]\big)+F^{p+\epsilon}U_\Delta(\pi_Z\g)^{p+q}\ ,
$$
hence the differential $\dz$ is good.
We showed so far that $\pi_Z,\dz$ satisfy the conditions of Theorem \ref{gnonlin}.
Let $e_i,E_i$ be as above; namely $\{e_i,\ i\in\I_{p\Delta}\}$ is a basis of
$\g^{p,-p}[\Delta]\cap d^{-1}\big(F^{p+\epsilon}V(R)^1[\Delta]\big)$ 
and $E_i\in F^pV(R)^0[\Delta]\cap\Ker d$ is such that $E_i-e_i\in F^{p+\epsilon}V(R)^0[\Delta]$.
By Lemma \ref{obv3}, $\{\pi_Z(e_i)\,|\, i\in\I_{p\Delta}\}$ is then a basis 
of $\pi_Z\g^{p,-p}[\Delta]\cap \left.\dz\right.^{-1}\big(F^{p+\epsilon}U_\Delta(\pi_Z\g)^1\big)$,
and, for $i\in\I_{p\Delta}$, $\pi_Z(E_i)\in F^pU_\Delta(\pi_Z\g)^0\cap\Ker \dz$
is such that $\pi_Z(E_i)-\pi_Z(e_i)\in F^{p+\epsilon}U_\Delta(\pi_Z\g)^0$.
Notice that by Lemma \ref{free-basis}, the quotient map $\pi^Z_H$ restricts to an isomorphism
$U(\pi_Z\g)^0\cap\Ker \dz \stackrel{\sim}{\rightarrow}H(Zhu_H V(R),\dz)$.
As in (\ref{h}) we then let
\begin{eqnarray*}
\vphantom{\Big(}
&& H^{p,-p}(\pi_Z\g,\dz)[\Delta]\ =\ \Span_\C\big\{\pi^Z_H\pi_Z(E_i),\ i\in\I_{p\Delta}\big\}\ , \\
\vphantom{\Big(}
&& H(\pi_Z\g,\dz)\ =\ \bigoplus_{p,\Delta}H^{p,-p}(\pi_Z\g,\dz)[\Delta]\ \subset\ H(Zhu_H V(R),\dz)\ . 
\end{eqnarray*}
By Theorem \ref{gnonlin}, $H(\pi_Z\g,\dz)$ is a non-linear Lie algebra and
$H(Zhu_H V(R),\dz)$ $\simeq U(H(\pi_Z\g,\dz))$.
The following lemma follows immediately by the above observations.
\begin{lemma}\label{main2}
The associative algebra $H(Zhu_H V(R),\dz)$ is PBW generated
by the set $\big\{\pi^Z_H\pi_Z(E_i)\,|\, i\in\I\big\}$.
\end{lemma}
\medskip

% conclusion

\begin{proof}[Proof of Theorem \ref{main}]
First, we need to show that the map $\phi$, given by (\ref{532}), is well defined.
The natural quotient map $\pi_Z:\ V(R)\rightarrow Zhu_H V(R)$
is such that
$\pi_Z(\Ker d)\subset\Ker \dz$
and $\pi_Z(d(V(R)))= \dz(Zhu_H V(R))$.
Hence it induces a canonical map
\begin{equation}\label{interm}
H(V(R),d)\ \longrightarrow H(Zhu_H V(R),\dz)\ ,
\end{equation}
given by $\pi_H(A)\mapsto \pi^Z_H\pi_Z(A)$.
Recall that $\pi_H:V(R)\rightarrow H(V(R),d)$ is a vertex algebra homomorphism 
which preserves the conformal weight.
Hence $\pi_H(A)_{(-2,\hbar)}\pi_H(B)=\pi_H(A_{-2,\hbar}B)$.
And obviously $\pi^Z_H\pi_Z(A_{(-2,\hbar=1)}B)=0$.
Therefore $J_{\hbar=1}(H(V(R),d))=H(V(R),d)_{(-2,\hbar=1)}H(V(R),d)$
is in the kernel of (\ref{interm}),
which of course guarantees that $\phi:\ \pi^H_Z\pi_H(A)\mapsto \pi^Z_H\pi_Z(A)$ is well defined.
The fact that $\phi$ is an algebra homomorphism is immediate:
\begin{eqnarray*}
\phi(\pi^H_Z\pi_H(A)\cdot \pi^H_Z\pi_H(B))
&=& \phi(\pi^H_Z(\pi_H(A_{-1,\hbar=1}B))) 
\ =\ \pi^Z_H(\pi_Z(A_{-1,\hbar=1}B)) \\
&=& \pi^Z_H(\pi_Z(A))\cdot\pi^Z_H(\pi_Z(B))
\ =\ \phi(A)\cdot\phi(B)\ .
\end{eqnarray*}
Finally, $\phi$ is an isomorphism since, by Lemmas \ref{main1} and \ref{main2},
it maps a PBW generating set of $Zhu_H H(V(R),d)$
to a PBW generating set of $H(Zhu_H V(R),\dz)$.
\end{proof}

%\newpage
%%%%%%%% SECTION %%%%%%%%%%%%%%%%%%%%%%%%%%%%%

\section{Finite and affine W-algebras}\label{sec:6}
\setcounter{equation}{0}

%%%

\subsection{Definition of the affine W-algebra $W_k(\g,x,f)$}
\label{sec-aff-W}

% set up

Let $\g$ is a simple finite dimensional Lie superalgebra, with a 
non-degenerate even supersymmetric invariant bilinear form $(\,. |\,. )$, 
and let $x,f\in\g$ be a pair of even elements such that 
$\ad\,x$ is diagonalizable on $\g$ with half-integer eigenvalues 
and $[x,f]=-f$. Let
\begin{equation}\label{halfgrad}
  \g=\oplus_{j\in\frac{1}{2}\ZZ} \g_j 
\end{equation}
be the eigenspace decomposition with respect to $\ad\,x$.
Let $\g_+$, $\g_-$, $\g_{\geq}$  and 
$\g_{\leq}$ 
denote the sums of all eigenspaces $\g_j$
with $j$ positive, negative, non-negative, and non-positive, respectively,
and let 
$\pi_+$, $\pi_-$, $\pi_{\geq}$, $\pi_{\leq}$, 
denote the corresponding
projections of $\g$ on $\g_+$, $\g_-$, etc.
We shall assume that the pair $x,f$ is ``good'', i.e.
the centraliser $\g^f$ of $f$ in $\g$ lies in $\g_{\leq}$. 
In this case the grading (\ref{halfgrad}) is also called good.
It is easy to see that for a given $x$, all $f$ such that the pair $x,f$
is good, form a single $exp(\g_0)_{\bar{0}}$-orbit, which is Zariski
dense in $(\g_{-1})_{\bar{0}}$. 
A case of special interest is when the pair $x,f$ is a part of an $\ssl_2$-triple 
$\ssl_2\simeq\Span_\C\{e,h=2x,f\}\subset\g$. In this case the spectrum of 
$\ad\,x$ 
is automatically in $\frac12 \Z$, and the pair $x,f$ is good by representation 
theory of $\ssl_2$. 
One can find a detailed study and classification of good gradings of simple 
Lie algebras in \cite{EK}.

Following \cite{KRW,KW1}, we recall here the construction of the family of
vertex algebras $W_k(\g,x,f)$, depending on a parameter $k\in\C$.
We define a skew-supersymmetric bilinear form $\langle\,. |.\, \rangle_{ne}$ 
on $\g_{\frac12}$ by (see \ref{eq:0.6})
$$
\langle a|b\rangle_{ne}\ =\ (f,[a,b])\ .
$$
This  form is non-degenerate since the map $\ad \,f : \g_{-1/2} \rightarrow 
\g_{1/2}$ is a vector space isomprphism.

For each $j\in\frac{1}{2}\Z$ we fix a basis $\{u_\alpha\,|\,\alpha\in S_j\}$ of 
$\g_j$ compatible with the parity of $\g$.
We denote $p(\alpha)=p(u_\alpha)$, the parity of $u_\alpha$, let $s(\alpha)=s(u_{\alpha})=(-1)^{p(\alpha)}$, and let 
$j_\alpha=j$ if $u_\alpha\in\g_j$. 
Furthermore, we let $S_+=\bigsqcup_{j>0}S_j$ and 
$S_\leq=\bigsqcup_{j\leq0}S_j$,
so that $\{u_\alpha\,|\, \alpha\in S_+\}$ is a basis of 
$\g_+$
and $\{u_\alpha\,|\, \alpha\in S_\leq\}$ is a basis of 
$\g_\leq$
% 5.
For $j\in\frac12 \Z$, we also let $\{u^\alpha\,|\, \alpha\in S_j\}$ be the basis 
of $\g_{-j}$,
dual to $\{u_\alpha\,|\, \alpha\in S_j\}$ with respect to the bilinear 
form $(\ . |.\ )$, i.e.
$$
(u_\alpha|u^\beta)\ =\ s(\alpha)(u^\beta|u_\alpha)\ =\ \delta_{\alpha}^{\beta}
\  ,\, \alpha,\beta\in S_j\ .
$$
In particular $\{u^{\alpha}\}_{\alpha\in S_+}$ is a basis of 
$\g_-$ dual to the basis $\{u_\alpha\}_{\alpha\in S_+}$ of $\g_+$.
We shall identify $\g_-$ and $\g_+^*$ via the bilinear form $(\,. |\,. )$.
% 7.
Finally, we let $\{u_\alpha\}$ and $\{v_\alpha\}$, $\alpha\in S_{\frac12}$, 
be a 
pair of dual bases of $\g_{\frac12}$ with respect to the skew-supersymmetric 
form $\langle\,. |.\, \rangle_{ne}$, i.e.
$$
\langle u_\alpha|v_\beta\rangle_{ne}\ =\ (f|[u_\alpha,v_\beta])\ =
\ \delta_{\alpha,\beta},
\ \ \, \alpha,\beta\in S_{1/2}\ .
$$
It is immediate to check that, with the above notation, we have 
\begin{equation}\label{change}
u^\alpha\ =\ [v_\alpha,f]\ ,\,\alpha\in S_{1/2}\ .
\end{equation}

% def of R_k(g)
% a.
The building blocks of the complex that defines the affine $W$-algebra
are the universal enveloping vertex algebras of 
current and fermionic Lie conformal algebras,
described in Subsection \ref{sec:1.10}. However, we shall use a slighly 
more convenient 
language of non-linear Lie conformal algebras (cf. Example \ref{newex}). 
First, the {\itshape current Lie conformal algebra} $Cur_k\g$ of level $k$ 
is, by definition,
the free $\C[\ttt]$-module $\C[\ttt]\otimes\g$, together with the non-linear
$\lambda$-bracket
$$
[a\ _\lambda\ b]\ =\ [a,b]+\lambda k(a|b)\ ,\ \ a,b\in\g\ .
$$
% b.

Second, let $\varphi_{\g_+}=\{\varphi_a\,|\,a\in\g_+\}$ be a vector superspace 
isomorphic to $\Pi\g_+$, where $\Pi$ denotes parity-reversing, 
via $a\mapsto\varphi_a$,
and we let $\varphi^{\g_-}=\{\varphi^a\,|\,a\in\g_-\}$ be a vector superspace 
isomorphic to $\Pi\g_-$, via $a\mapsto\varphi^a$.
The projections $\pi_+$ and $\pi_-$
allow us to extend the notation
$\varphi_a,\ \varphi^a$ to all $a\in\g$, by letting
$$
\varphi_a\ :=\ \varphi_{\pi_+a}\ ,\ \ \varphi^a\ :=\ \varphi^{\pi_-a}\ ,\ \, 
a\in\g\ .
$$
The superspace $\varphi_{\g_+}\oplus\varphi^{\g_-}$ is endowed
with the following non-degenerate skew-supersymmetric bilinear form
 ($a,b\in\g$):
$$
\langle\varphi_a|\varphi_b\rangle_{ch}=\langle\varphi^a|\varphi^b\rangle_{ch}=
0\ ,\ \ \langle\varphi_a|\varphi^b\rangle_{ch}\ =\ s(a)\langle\varphi^b|
\varphi_a\rangle_{ch} \ =\ (\pi_+a|\pi_-b)\ .
$$
Here, as before, $s(a)=(-1)^{p(a)}$, where $p(a)$ denotes the parity of $a\in\g$.
We will denote $\varphi_\alpha=\varphi_{u_\alpha},\ 
\varphi^\alpha=\varphi^{u^\alpha},\ \alpha\in S_+$.
Then $\{\varphi_\alpha,\, \varphi^\alpha \,|\, \alpha\in S_+\}$ is a basis 
of $\varphi_{\g_+}\oplus\varphi^{\g_-}$, such that
$\langle\varphi_\alpha|\varphi_\beta\rangle_{ch}=
\langle\varphi^\alpha|\varphi^\beta\rangle_{ch}=0,\
\langle\varphi_\alpha|\varphi^\beta\rangle_{ch}=\delta_\alpha^\beta$.
The corresponding {\itshape charged free fermion} non-linear Lie conformal 
algebra $R_{ch}$
is, by definition, the free 
$\C[\ttt]$-module $\C[\ttt]\otimes (\varphi_{\g_+}\oplus\varphi^{\g_-})$,
with the $\lambda$-bracket
$$
[\phi\ _\lambda\ \psi]\ =\ \langle\phi|\psi\rangle_{ch}\ ,\ \ 
\phi,\psi\in\varphi_{\g_+}\oplus\varphi^{\g_-}\ .
$$
% c.

Finally, let $\Phi_{\g_{1/2}}=\{\Phi_a\,|\, a\in\g_{1/2}\}$ be a vector superspace 
isomorphic to $\g_{1/2}$, via $a\mapsto\Phi_a$,
together with the skew-supersymmetric bilinear form $\langle\ |\ \rangle_{ne}$ 
defined above.
We will denote $\Phi_\alpha=\Phi_{u_\alpha},\ \Phi^\alpha=\Phi_{v_\alpha},\ 
\alpha\in S_{1/2}$, so that
$\langle\Phi_\alpha|\Phi^\beta\rangle_{ne}
=-s(\alpha)\langle\Phi^\beta|\Phi_\alpha\rangle_{ne}=\delta^\beta_\alpha$.
As before, the projection $\pi_{1/2}$: $\g\twoheadrightarrow\g_{1/2}$
allows us to extend the notation $\Phi_a$ to all $a\in\g$ by letting
$$
\Phi_a\ =\ \Phi_{\pi_{1/2}a}\ ,\ \, a\in\g\ .
$$
The corresponding {\itshape neutral free fermion} non-linear Lie conformal 
algebra $R_{ne}$
is, by definition, the free $\C[\ttt]$-module $\C[\ttt]\otimes \Phi_{\g_{1/2}}$,
with the $\lambda$-bracket
$$
[\Phi_a\ _\lambda\ \Phi_b]\ =\ \langle\Phi_a|\Phi_b\rangle_{ne}
\ ,\ \ a,b\in \g_{1/2}\ .
$$

% vertex algebra C_k(g)

Now we define the (non-linear) Lie conformal algebra $R_k(\g,x)$ 
as the direct sum of the non-linear Lie conformal algebras $Cur_k\g$, $R_{ch}$ 
and $R_{ne}$: 
$$
R_k(\g,x)\, =\, \C[T]r(\g,x)\,,\hbox{where}\, \ r(\g,x)\, 
=\, \g\oplus\varphi_{\g_+}\oplus\varphi^{\g_-} \oplus\Phi_{\g_{1/2}}\ ,
$$
and we let $C_k(\g,x)= V(R_k(\g,x))$ be the corresponding universal enveloping 
vertex algebra.
(Note that we are taking the complex, described in Subsection \ref{sec:0.4},
for the choice $\fs = \fg_{1/2}$.)

% completeness rel's

The following relations are obvious and they are useful for many
computations ($a\in\g$),
\begin{eqnarray*}
a &=& \sum_{\alpha\in S}(a|u^\alpha)u_\alpha\ =\ \sum_{\alpha\in S}(u_\alpha|a)u^\alpha\ , \\
\varphi_a &=& \sum_{\alpha\in S_+}(a|u^\alpha)\varphi_\alpha
\ ,\ \ \varphi^a\ =\ \sum_{\alpha\in S_+}(u_\alpha|a)\varphi^\alpha\ , \\
\Phi_a &=& \sum_{\alpha\in S_{1/2}} (f|[a,v_\alpha]) \Phi_\alpha
\ =\ \sum_{\alpha\in S_{1/2}} (f|[u_\alpha,a]) \Phi^\alpha\ .
\end{eqnarray*}

% gradings

% a. 
We define the {\itshape energy momentum} element $L\in C_k(\g,x)$ by 
$$
L\ =\ L^\g+Tx+L^{ne}+L^{ch}\ \in\ C_k(\g,x)\ ,
$$
where $L^{\g}$ is defined by (\ref{eq:1.57}), $L^{ne}=L^A$
is defined by (\ref{eq:1.59}), where
$A=\Phi_{\g_{1/2}}$, 
and, finally, $L^{ch}=L^{A,\vec{m}}$ is defined by (\ref{eq:1.61}), 
where $A=\varphi_{\g_+}\oplus\varphi^{\g_-}$
and $\vec{m}=(j_{\alpha})$.
It follows that 
$L$ is a Virasoro element, i.e. 
$$
[L\ _\lambda\ L]\ =\ (T+2\lambda)L+\frac{1}{12}c_k(\g,x)\lambda^3\ ,
$$
where, due to (\ref{eq:1.58}),(\ref{eq:1.60}), and (\ref{eq:1.62}),
the central charge $c_k(\g,x)$ is given by
$$
c_k(\g,x)\ =\ \frac{k\,\sdim\,\g}{k+h^{\vee}}-12k(x|x)
-\sum_{\alpha\in S_+}s(\alpha)(12j_\alpha^2-12j_\alpha+2)-
\frac12\sdim\,\g_{1/2}\ .
$$
It is straightforward to compute the $\lambda$-bracket of $L$ with the 
generators of $C_k(\g,x)$:
\begin{eqnarray*}
&\displaystyle{
{[L\ _\lambda\ a]}\ =\ Ta+\lambda(a-[x,a])-k\lambda^2(x|a)
\ ,\ \  {[L\ _\lambda\ \Phi_a]}\ =\ (T+\frac12 \lambda)\Phi_a\ ,
}\\
&\displaystyle{
{[L\ _\lambda\ \varphi_a]}\ =\ T\varphi_a+\lambda\varphi_{a-[x,a]}
\ ,\ \ {[L\ _\lambda\ \varphi^a]}\ =\ T\varphi^a+\lambda\varphi^{-[x,a]}\ .
}
\end{eqnarray*}
In other words, all generators 
$u_\alpha (\alpha\in S),\ \varphi_\alpha,\varphi^\alpha (\alpha\in S_+)$,
and $\Phi_\alpha (\alpha\in S_{1/2})$,
are eigenvectors of the Hamiltonian operator $H=L_0$, and are primary 
elements, except for 
$u_\alpha (\alpha\in S_0)$ such that 
$(u_\alpha|x)\neq0$,
with the following conformal weights:
\begin{eqnarray}\label{cw}
\Delta{(u_\alpha)} &=& 1-j_\alpha\  \ (\alpha\in S)
\ ,\ \ \Delta{(\Phi_\alpha)}\ =\ 1/2\  \ (\alpha\in S_{1/2})\ ,\nonumber\\
\Delta{(\varphi_\alpha)} &=& 1-j_\alpha\  \ (\alpha\in S_+)
\ ,\ \ \Delta{(\varphi^\alpha)}\ =\ j_\alpha\  \ (\alpha\in S_+)\ .
\end{eqnarray}
%
% b.
Next we define the {\itshape charge decomposition} 
$C_k(\g,x)=\bigoplus_{n\in\Z}C_k(\g,x)^n$
by letting
\begin{eqnarray}
\label{chdeg}
& \charge \,a\ =\ \charge\,\Phi\ =\ 0\ ,\, a\in\g,\ \Phi\in\Phi_{\g_{1/2}}\ ,\ 
\ \charge\,\ttt\ =\ 0\ ,\nonumber\\
& \charge\,\varphi_a\ =\ -1\ , \ \,  a\in\g_+\ ,
\qquad \charge\,\varphi^a\ =\  1\ ,\ \, a\in\g_-\ .
\end{eqnarray}
This makes $C_k(\g,x)$ a $\Z$-graded vertex algebra.
Notice that we are choosing the sign in (\ref{chdeg}) opposite to that of \cite{KW1}.

% differential

Now we introduce the following element of $C_k(\g,x)$ (cf. Section \ref{sec:0.4}):
\begin{eqnarray}\label{differ}
d &=& \sum_{\alpha\in S_+} s(\alpha):\varphi^\alpha u_\alpha:
\ +\ \sum_{\alpha\in S_{1/2}}:\varphi^\alpha\Phi_\alpha:\ +\ \varphi^f \nonumber\\
&+& \frac{1}{2}\sum_{\alpha,\beta\in S_+}s(\alpha)
        :\varphi^\alpha\varphi^\beta\varphi_{[u_\beta,u_\alpha]}:\ . 
\end{eqnarray}
\renewcommand{\theenumi}{\alph{enumi}}
\renewcommand{\labelenumi}{(\theenumi)}
\begin{lemma}\label{lemdiff}
\begin{enumerate}
\item The element $d\in C_k(\g,x)$ has parity $\bar{1}$ and charge $1$.
Moreover $d$ is a primary element of conformal weight $1$, i.e.
$[L _\lambda d] = (T+\lambda)d$, or $[d_\lambda L] =\lambda L$. In
particular, $d_{(0)}L = 0$.
\item The element $d$ satisfies the following $\lambda$-bracket relations 
with the generators of $C_k(\g,x)$ ($a\in\g$):
\begin{eqnarray}\label{dgen-old}
{[d\ _\lambda\ a]} 
&=& \sum_{\alpha\in S_+} s(\alpha):\varphi^\alpha [u_\alpha,a]: 
\ +\ k\,s(a)(\ttt+\lambda)\varphi^a\ , \nonumber\\
{[d\ _\lambda\ \varphi_a]}
&=& \pi_+ a + (a|f) + s(a)\Phi_a + 
\sum_{\alpha\in S_+}:\varphi^\alpha\varphi_{[u_\alpha,\pi_+ a]}:\ , \\
{[d\ _\lambda\ \varphi^a]}
&=& \frac{1}{2}\sum_{\alpha\in S_+} s(\alpha) :\varphi^\alpha\varphi^
{[u_\alpha,a]}:
\ ,\ \ {[d\, _\lambda\, \Phi_a]} \ =\ \varphi^{[\pi_{1/2}a,f]}\ .\nonumber
\end{eqnarray}
Using (\ref{change}) we can rewrite the last equation 
as $[d\ _\lambda\ \Phi^\alpha]=\varphi^\alpha$, $\alpha\in S_{1/2}$.
\item The $\lambda$-bracket of $d$ with itself is trivial:
$$
[d\ _\lambda\ d]\ =\ 0\ .
$$
\item $d_{(0)}$ is an odd derivation of all $n$\st{th}-products in 
$C_k(\g,x)$,  it commutes with $\ttt$ and $H$, 
it increases the charge by 1,
and $d_{(0)}^2=0$,
hence it is a differential of the vertex algebra $C_k(\g,x)$.
\end{enumerate}
\end{lemma}
\renewcommand{\theenumi}{\arabic{enumi}}
\renewcommand{\labelenumi}{\theenumi.}
\begin{proof}
(a) and (b) follow by a straightforward computation. 
(c) follows from (b), after another long computation.
(d) is immediate by Proposition \ref{prop:1.8}(a) and, in view of (c), 
by the commutator formula (\ref{eq:1.19}).
See \cite{KRW,KW1} for the details of the proof.
\end{proof}

\begin{definition}
We define the {\itshape affine $W$-algebra} $W_k(\g,x,f)$ to be the 
vector superspace
$$
W_k(\g,x,f)\ =\ H(C_k(\g,x),d_{(0)})\ ,
$$
together with the vertex algebra structure induced from $C_k(\g,x)$
(see Proposition \ref{prop:1.8}(a)), and the 
Hamiltonian operator $H$, defined by (\ref{cw}).
\end{definition}

\begin{remark} 
\label{rem:6.3}
Since, given $x$, all elements $f$, such that the pair $x,f$
is good, are conjugate by $exp (\g_0)_{\bar0}$, the $W$-algebra $W_k(\g,x,f)$
is independent, up to isomorphism, of the choice of $f$, hence we shall drop 
$f$ from its notation. 
Of course, $W_k(\g,x)$ depends, up to isomorphism, only on the conjugacy class 
of $x$ in $\g$.
\end{remark}
\begin{remark}\label{rem-oct30}
After the present paper was completed, the paper \cite{BrG} came to our attention,
where the authors study the finite $W$-algebras in their Whittaker model definition,
attached to more general good $\R$-gradings:
$\g=\bigoplus_{j\in\R}\g_j$, defined by the property that there exists $f\in\g_{-1}$
for which $\g^f\subset\g_{<1/2}$.
Using this, they prove that all the corresponding finite $W$-algebras with the same $f$ are
isomorphic.
It is easy to see that if we let $\g_+=\g_{\geq1/2},\,\g_-=\g_{\leq-1/2}$,
then the definitions of the affine and finite $W$-algebras $W_k(\g,x,f)$ and $W^\tf(\g,x,f)$
in Sections \ref{sec-aff-W} and \ref{sec-Wfin} are still valid.
Moreover, using the results of \cite{BrG} on good $\R$-gradings,
it is not difficult to show that affine $W$-algebras $W_k(\g,x,f)$
with given $f$ are isomorphic.
However, it is still useful to vary $x$, which defines a good $\R$-grading with given $f$,
since this produces different energy-momentum fields,
hence different Hamiltonians $H$ and different categories of positive-energy
$W_k(\g,x,f)$-modules.
One such example is treated in detail in \cite{KRW}.
\end{remark}

%%%

\subsection{Definition of the finite W-algebra $W^\tf(\g,x)$}\label{sec-Wfin}

In order to define the finite W-algebra $W^\tf(\g,x)$ we look at the $H$-
twisted Zhu algebra 
of the vertex algebra $C_k(\g,x)$.
According to Definition \ref{zhulie}, the $H$-twisted Zhu non-linear 
Lie algebra associated to $R_k(\g,x)$ is the vector superspace
$$
Zhu_H R_k(\g)\ =\ \overline{r}(\g,x)
\ \simeq\ \overline{\g}\oplus\overline{\varphi}_{\g_+}
\oplus\overline{\varphi}^{\g_-}\oplus\overline{\Phi}_{\g_{1/2}}\ ,
$$
endowed with the non-linear Lie bracket $[\ ,\ ]$ induced by the 
bracket $[\,_*\,]$
on $R_k(\g,x)$, namely
\begin{eqnarray*}
{[\overline a,\overline b]} &=& \overline{[a,b]}\ -\ k([x,a]|b)
\ ,\ \ \text{ for } \overline a,\overline b\in\overline \g\ ,\\
{[\overline \phi,\overline \psi]} &=& \langle\overline\phi|\overline\psi\rangle_{ch}
\ ,\ \ \text{ for } \overline\phi,\overline\psi\in\overline{\varphi}_{\g_+}\oplus\overline{\varphi}^{\g_-}\ ,\\
{[\overline{\Phi},\overline{\Phi}^\prime]} &=& \langle\overline{\Phi}|\overline{\Phi}^\prime\rangle_{ne}
\ ,\ \ \text{ for } \overline\Phi,\overline\Phi^\prime \in \overline\Phi_{\g_{1/2}}\ ,
\end{eqnarray*}
and all other brackets equal zero. Indeed, since
 $[a_*b] =\sum_{j \in \Z_+} \binom{\Delta_a-1}{j}a_{(j)}b$,
we have ${[\overline a,\overline b]} = \overline{[a,b]}\ +\ k(\Delta_a -1)
(a|b)$, which gives the first formula; all other formulas are obvious.
Here and further we denote elements of $Zhu_H C_k(\g,x)$ with an overbar, 
in order to distinguish them from the same elements in $C_k(\g,x)$.
By Corollary \ref{corzhuenv}, the $H$-twisted Zhu algebra of $C_k(\g,x)$,
which we denote $C^{\tf}_k(\g,x)$, is isomorphic to 
the universal enveloping associative algebra of $\overline r(\g,x)$,
$$
C^{\tf}_k(\g,x)\ :=\ Zhu_H C_k(\g,x)\ \simeq\ U(\overline r(\g,x))\ .
$$
\begin{remark}\label{16ott05}
Notice that the associative algebras $C_k^\tf(\g,x)$, with $k\in\C$, are all isomorphic.
Indeed we have an explicit isomorphism 
$\sigma:\ C_k^\tf(\g,x)\stackrel{\sim}{\rightarrow}C_{k=0}^\tf(\g,x)$,
given by $\sigma(a)=a+k(x|a)$, for $a\in\g$, with all other generators unchanged:
$\sigma(\varphi_a)=\varphi_a,\ \sigma(\varphi^a)=\varphi^a,\ \sigma(\Phi_a)=\Phi_a$. 
Hence we shall denote, from now on, $C^\tf(\g,x)$ without subscript $k$.
\end{remark}

% differential

Let $\pi_Z$ be the natural quotient map from $C_k(\g,x)$ to $C^\tf(\g,x)$. 
It is immediate to check, using Remark 4.12(b), that
the element $\overline{d}=\pi_Z(d)\in C^\tf(\g,x)$, corresponding to $d$ 
in (\ref{differ}), 
is given by
\begin{eqnarray}\label{dba}
\overline d &=& \sum_{\alpha\in S_+}s(\alpha) \overline{\varphi}^\alpha \overline
{u}_\alpha
\ +\ \sum_{\alpha\in S_{1/2}} \overline{\varphi}^\alpha \overline{\Phi}_\alpha\ +\ \overline{\varphi}^f \nonumber\\
&+& \frac{1}{2}\sum_{\alpha,\beta\in S_+}s(\alpha)
\overline{\varphi}^\alpha\overline{\varphi}^\beta \overline{\varphi}_{[u_\beta,u_\alpha]}\ . 
\end{eqnarray}
Moreover, since $\Delta(d)=1$, we have $[d_*A]=d_{(0)}A$ for every $A\in C_k(\g,x)$,
hence $[\overline d,\pi_Z A]\ =\ \pi_Z(d_{(0)}A)$.
In other words, the action on $C^\tf(\g,x)$ induced by $d_{(0)}$ is given by 
the adjoint action 
of $\overline d$:
$$
ad\,\overline d:\ C^{\tf}(\g,x)\rightarrow C^{\tf}(\g,x)\ .
$$
It follows that $ad\,\overline d$ is an odd derivation of the 
associative superalgebra $C^{\tf}(\g,x)$, and 
$(ad\,\overline d)^2=0$.

% definition of the finite W-algebra

\begin{definition}\label{def-finW}
The {\itshape finite $W$-algebra} $W^{\tf}(\g,x)$ is, by definition, 
the associative superalgebra
$$
W^{\tf}(\g,x)\ =\ H(C^{\tf}(\g,x),ad\,\overline d)\ .
$$
\end{definition}
By Remark \ref{16ott05}, the finite $W$-algebra is, up to isomorphism, independent of $k$, 
namely the isomorphism of 
$W^{\tf}_k$ with $W^{\tf}_0$ is given by the change of generators,
described in the Remark. 
Here we are using the fact that the isomorphism $\sigma$ leaves $\overline d$ unchanged.

%%%

\subsection{Structure of the affine W-algebra $W_k(\g,x)$}

% building blocks

The following vertex operators of the vertex algebra $C_k(\g,x)$, will be
important ``building blocks'' for the affine $W$-algebra (cf. the second formula
in (\ref{dgen-old})):
\begin{equation}\label{bb}
J_a\ =\ a+\sum_{\alpha\in S_+}:\varphi^\alpha \varphi_{[u_\alpha,a]}:\ ,\ \ a\in\g\ .
\end{equation}
In other words, we define an injective linear map $\g\stackrel{\sim}{\rightarrow}
J_\g\subset C_k(\g)$,
given by $a\mapsto J_a$.
We will also denote $J_\alpha=J_{u_\alpha},\ \alpha\in S$. 
Notice that $J_a$ has the same
parity, the same charge degree, and the same conformal weight as $a$.
It is clear from the definitions that the space 
$J_\g\oplus \varphi_{\g_+}\oplus \varphi^{\g_-}\oplus \Phi_{\g_{1/2}}$ is a 
freely generating
subspace of the vertex algebra $C_k(\g,x)$.
We can then rewrite all formulas in Section \ref{sec-aff-W} in terms of this 
new set of generators.

% formulas

First, we compute the $\lambda$-bracket of $J_a,\ a\in\g$, with the 
generators of $C_k(\g,x)$.
We immediately get for $a,b\in\g$,
\begin{equation}\label{lnew}
{[J_a\ _\lambda\ \varphi_b]} = s(a)\varphi_{[a,\pi_+ b]}\ ,\ \
{[J_a\ _\lambda\ \varphi^b]} = \varphi^{[a,\pi_- b]}\ ,\ \
{[J_a\ _\lambda\ \Phi_b]} = 0\ .
\end{equation}
It follows by a straightforward but rather lengthy computation, that
\begin{eqnarray}\label{jj}
{[J_a\ _\lambda\ J_b]} &=& J_{[a,b]}
+\lambda (k(a|b)+\str_{\g_+}(\pi_+ (ad\,a) \pi_+ (ad\,b)))\\
&+& \sum_{\alpha\in S_+}:\varphi^\alpha
(\varphi_{[\pi_\leq [u_\alpha,a],b]}-p(a,b)\varphi_{[\pi_\leq [u_\alpha,b],a]}):\ . \nonumber
\end{eqnarray}
Notice that the last term in the above equation is identically zero if $a$ 
and $b$ are both in $\g_\geq$
or both in $\g_\leq$.
We then denote
\begin{equation}\label{psi}
\psi_k(a|b)\ =\ k(a|b)+\str_{\g_+}\big((\pi_+ ad\,a)(\pi_+ ad\,b)\big)\ ,
\end{equation}
so that equation (\ref{jj}) implies, for $a,b\in\g_\leq$ or $a,b\in\g_\geq$:
\begin{equation}\label{jj2}
{[J_a\ _\lambda\ J_b]}\ =\ J_{[a,b]}+\psi_k(a|b)\ .
\end{equation}

We are also interested in the $\lambda$-bracket of $d$ with the generators of 
$C_k(\g,x)$.
It immediately follows from (\ref{dgen-old}) that
\begin{equation}\label{dgen}
{[d\ _\lambda\ \varphi_a]}\ =\ J_{\pi_+ a}+(a|f)+s(a)\Phi_a\ .
\end{equation}
We are left to compute $[d\ _\lambda\ J_a]$, $a\in\g$.
This is obtained by another somewhat lengthy computation:
\begin{eqnarray}\label{dj}
{[d\ _\lambda\ J_a]} &=& \sum_{\alpha\in S_+}s(\alpha):\varphi^\alpha 
J_{\pi_\leq[u_\alpha,a]}:
-s(a)\sum_{\alpha\in S_+}:\varphi^\alpha \Phi_{[u_\alpha,a]}: \nonumber\\
&+& (T+\lambda)\sum_{\alpha\in S_+}\psi_k(a|u_\alpha)\varphi^\alpha
+s(a)\varphi^{[f,a]}\ .
\end{eqnarray}
Notice that the first three terms in the RHS are identically zero 
for $a\in\g_+$.
In other words, we have
$$
{[d\ _\lambda\ J_a]}\ =\ s(a)\varphi^{[f,a]}\ =
\ -{[d\ _\lambda\ s(a)\Phi_a]}\ , \,   a\in\g_+\ .
$$

% R+ and R

We now let $R_+$ and $R$ be the following free $\C[T]$-submodules of 
$C_k(\g,x)$,
\begin{eqnarray}\label{R+R}
R_+ &=& \C[T]r_+\ ,\ \ r_+=\varphi_{\g_+}\oplus d_{(0)}\varphi_{\g_+}\ , \\
R &=& \C[T]r\ ,\ \ r= J_{\g_\leq}\oplus \varphi^{\g_-}\oplus \Phi_{\g_{1/2}}\ ,\nonumber
\end{eqnarray}
where $J_{\g_\leq}=\{J_a, \ a\in\g_\leq\}$.
Recall that (cf. (\ref{dgen-old}))
$$
d_{(0)}\varphi_a\ =\ J_a+(a|f)+s(a)\Phi_a\ ,\,  a\in\g_+\ .
$$
Hence $R_+\cap R=0$, and $R_+\oplus R$ is a freely generating subspace for 
the vertex algebra $C_k(\g,x)$.
Moreover, it follows from the $\lambda$-bracket relations (\ref{lnew}) and 
(\ref{jj2}) that
both $R_+$ and $R\oplus\C$ 
are closed under the $\lambda$-bracket.
More precisely, $R_+$ has the following structure of a Lie conformal algebra 
($a,b\in\g_+$):
\begin{eqnarray}\label{lR+}
\vphantom{\Big(} & {[\varphi_a\ _\lambda\ \varphi_b]}\ =\ 0\ ,\ \ 
{[d_{(0)}\varphi_a\ _\lambda\ \varphi_b]}\ =\ s(a)\varphi_{[a,b]}\ ,\nonumber\\ 
\vphantom{\Big(} & {[d_{(0)}\varphi_a\ _\lambda\ d_{(0)}\varphi_b]}\ =\ d_{(0)}\varphi_{[a,b]}\ .
\end{eqnarray}
The third equation is obtained by applying $d_{(0)}$ to both sides of the second equation
and recalling that $d_{(0)}$ is an odd derivation of the $\lambda$-bracket and $d_{(0)}^2=0$.
Similarly, the $\lambda$-bracket of $C_k(\g,x)$ restricted to $R$ gives the 
following structure
of a non-linear Lie conformal algebra on $R$ 
(for $J_a,J_b\in J_{\g_\leq},\ \varphi^a,\varphi^b\in\varphi^{\g_-},\ \Phi_a,
\Phi_b\in\Phi_{\g_{1/2}}$):
\begin{eqnarray}\label{lR}
\vphantom{\Big(} &  {[J_a\ _\lambda\ J_b]}\ =\ J_{[a,b]}+\lambda\psi_k(a|b)\ ,\ \ 
{[J_a\ _\lambda\ \varphi^b]}\ =\ \varphi^{[a,b]}\ ,\\ 
\vphantom{\Big(} & {[\Phi_a\ _\lambda\ \Phi_b]}\ =\ (f|[a,b])\ ,\ \ 
{[J_a\, _\lambda\, \Phi^b]} = 0 = {[\varphi^a\, _\lambda\, \varphi^b]} = {[\varphi^a\, _\lambda\, \Phi_b]}
\ .\nonumber
\end{eqnarray}
In conclusion, the universal enveloping vertex algebras $V(R_+)$ and $V(R)$ 
are naturally
vertex subalgebras of $C_k(\g,x)$, and, since $C_k(\g,x)$ is freely generated 
by $R_+\oplus R$,
we have a vector space isomorphism
\begin{equation}\label{tensor}
C_k(\g,x)\ \simeq\ V(R_+)\otimes V(R)\ ,
\end{equation}
which extends the natural embeddings of $V(R_+)$ and $V(R)$ in $C_k(\g,x)$.
Notice that  (\ref{tensor}) is not a vertex algebra isomorphism since, for 
example, 
the $\lambda$-bracket of $J_a,\ a\in\g_\leq$ and $\varphi_b,\ b\in\g_+$, is 
not necessarily zero
(see (\ref{lnew})).

% action of d

We further observe that the action of the differential $d_{(0)}$ preserves 
both vertex subalgebras
$V(R_+)$ and $V(R)$. 
More precisely, $d_{(0)}$ obviously restricts to a differential, which we
denote by $d_+$, of the Lie conformal 
algebra $R_+=\C[T](\varphi_{\g_+}\oplus d_{(0)}\varphi_{\g_+})=\C[T]r_+$, 
commuting with $T$, and given by:
\begin{displaymath}
d_+(\varphi_a) = d_{(0)}\varphi_a ,\,  d_+(d_{(0)}\varphi_a) = 0.
\end{displaymath}
The corresponding cohomology vanishes, since $\Ker\, d_+=\im\, d_+=\C[T]d_{(0)}
\varphi_{\g_+}$.
On the other hand, $d_{(0)}$ restricts to a vertex algebra differential,
which we denote by $d$, of $V(R)\subset C_k(\g,x)$,
but it does not preserve $R=\C[T]r$. 
Namely the action of $d$ on $r=J_{\g_\leq}\oplus \varphi^{\g_-}\oplus \Phi_{\g_{1/2}}$ is
"non-linear", and it is given explicitly by
\begin{eqnarray}\label{dR}
d(J_a) &=& \sum_{\alpha\in S_+}s(\alpha):\varphi^\alpha J_{\pi_\leq[u_\alpha,a]}:
-s(a)\sum_{\alpha\in S_+}:\varphi^\alpha \Phi_{[u_\alpha,a]}: \nonumber\\
&+& \sum_{\alpha\in S_+}\psi_k(a|u_\alpha)T\varphi^\alpha
-s(a)\varphi^{[a,f]}\ , \,  a\in\g_\leq , \nonumber\\
d(\varphi^a) &=& \frac12 \sum_{\alpha\in S_+}s(\alpha):\varphi^\alpha 
\varphi^{[u_\alpha,a]}:
\ , \, a\in\g_- ,\\
d(\Phi_a) &=& \varphi^{[a,f]}\ ,\ \, a\in\g_{1/2}.\nonumber
\end{eqnarray}
Recall that the affine W-algebra is defined as $W_k(\g,x)=H(C_k(\g,x),d_{(0)})$. We thus get, by the above observations,
\begin{eqnarray}\label{isoms}
W_k(\g,x) &=& H(C_k(\g,x),d_{(0)})
\ \simeq\ H(V(R_+)\otimes V(R),d_+\otimes 1+1\otimes d) \nonumber\\
&\simeq& H(V(R_+),d_+)\otimes H(V(R),d)
\ \simeq\ H(V(R),d) \ .
\end{eqnarray}
In the third equality we used the Kunneth Lemma, and for the last equality just notice 
that, by Theorem \ref{Rlinear}, $H(V(R_+),d_+)\simeq V(H(R_+,d_+))=\C$.

% cw

We finally observe that the decomposition (\ref{tensor}) is compatible
with the $H$-grading (\ref{cw}) of $C_k(\g,x)$.
In particular $R=\C[T]r$ has the induced $\frac12 \NN$-grading, 
given by
\begin{equation}\label{cwr}
\Delta(J_\alpha)\ =\ 1-j_\alpha\ ,\ \ 
\Delta(\varphi^\alpha)\ =\ j_\alpha\ ,\ \ 
\Delta(\Phi_\alpha)\ =\ 1/2\ ,
\end{equation}
which induces a $\frac12 \Z_+$-valued $H$-grading of the universal
enveloping vertex algebra $V(R)$.
Moreover, since $H$ commutes with $d$,
we have an induced $H$-grading of the affine $W$-algebra $W_k(\g,x)\simeq 
H(V(R),d)$.

% concl

We can summarize all the results obtained so far in the following
\begin{theorem}\label{str-aff-W}
The vertex algebra $W_k(\g,x)$ is isomorphic to the vertex algebra
$H(V(R),d)$,
where $R=\C[T]r,\ r=J_{\g_\leq}\oplus \varphi^{\g_-}\oplus \Phi_{\g_{1/2}}$, 
is the non-linear
Lie conformal algebra with $\lambda$-bracket given by (\ref{lR}),
$V(R)$ is the universal enveloping vertex algebra of $R$, 
and $d$ is a differential of $V(R)$, 
given by (\ref{dR}).

Moreover, the $H$-grading on $W_k(\g,x)$ 
is induced by the $\frac12 \NN$-grading of $r$ given by 
(\ref{cwr}).
\end{theorem}

%%%

\subsection{Structure of the finite W-algebra $W^\tf(\g,x)$}

% intro & notation

We want to apply the arguments from the last section to the finite 
W-algebra $W^\tf(\g,x)$.
As before, we denote by $\pi_Z$ the natural quotient map from
the vertex algebra $C_k(\g,x)$ to its $H$-twisted Zhu algebra $C^\tf(\g,x)$.
In particular, using notation introduced in Section \ref{sec-Wfin}, 
we have ($a\in\g$)
$$
\pi_Z(a)\ =\ \overline a\ ,\ \ \pi_Z(\varphi_a)\ =\ \overline{\varphi}_a
\ ,\ \ \pi_Z(\varphi^a)\ =\ \overline{\varphi}^a\ ,\ \ \pi_Z(\Phi_a)\ =
\ \overline{\Phi}_a\ .
$$
We also denote by $\overline{J}_a\in C^\tf(\g,x)$ the image via $\pi_Z$ 
of $J_a\in C_k(\g,x)$,
defined in (\ref{bb}).
We have:
\begin{equation}\label{bbfin}
\overline{J}_a\ =\ \overline a 
+ \sum_{\alpha\in S_+}\overline{\varphi}^\alpha \overline{\varphi}_{[u_\alpha,a]}
+ \str_{\g_+}((ad\,x)(ad\,a))\ ,\ \ a\in\g\ .
\end{equation}
% prelim
The following identities can be used to prove equation (\ref{bbfin}),
and they will be useful to translate all other results from the last section from the affine to the finite case.
We have, for $a,b,c\in\g$:
\begin{eqnarray}\label{prelim-eq}
\pi_Z(:\varphi^a \varphi_b:) &=& \overline{\varphi}^a \overline{\varphi}_b + (x|[\pi_- a,\pi_+ b])
\ ,\nonumber\\
\pi_Z(:\varphi_a \varphi^b:) &=& \overline{\varphi}_a \overline{\varphi}^b 
+ (x|[\pi_+ a,\pi_- b]) - (a|b)\ ,\nonumber\\
\pi_Z(:\varphi^a \varphi^b \varphi_c:) &=& \overline{\varphi}^a \overline{\varphi}^b \overline{\varphi}_c  
+ (x|[\pi_- b,\pi_+ c]) \overline{\varphi}^a \\
&-& s(a)s(b)p(a,b)(x|[\pi_- a,\pi_+ c]) \overline{\varphi}^b\ ,\nonumber\\
\pi_Z(:\varphi^a J_b:) &=& \overline{\varphi}^a \overline{J}_b 
+ s(b) \overline{\varphi}^{[[x,\pi_- a],b]}\ . \nonumber
\end{eqnarray}
We explain how to derive the first relation, all the others are obtained in a similar way.
By definition
$\overline{\varphi}^a \overline{\varphi}_b=\pi_Z(\varphi^a *_{-1}\varphi_b)$.
Hence the first equation in (\ref{prelim-eq}) follows by the identity
$$
\varphi^a*_{-1}\varphi_b\ =\ :\varphi^a\varphi_b:-(x|[\pi_-a,\pi_+b])\ ,
$$
which is easy to check by the definition (\ref{star-pr-br}) of the $*_{-1}$-product.

Notice that the space 
$\overline{J}_\g\oplus \overline{\varphi}_{\g_+}
\oplus \overline{\varphi}^{\g_-}\oplus \overline{\Phi}_{\g_{1/2}}$
PBW generates $C^\tf(\g,x)$.
The commutation relations among these generators immediately follow 
from equations (\ref{lnew}) and (\ref{jj}),
via the condition
\begin{equation}\label{commnew}
[\pi_Z A,\pi_Z B]\ =\ \pi_Z [A_*B]\ ,\,\,  A,B\in C_k(\g,x)\ .
\end{equation}
Moreover, recall that the adjoint action of $\overline{d}$ on $C^\tf(\g,x)$ 
is induced, via $\pi_Z$,
by the action of $d_{(0)}$ on $C_k(\g,x)$.
Namely
\begin{equation}\label{commdgen}
[\overline{d},\pi_Z A]\ =\ \pi_Z(d_{(0)}A)\ ,\ \, A\in C_k(\g,x)\ .
\end{equation}
% r+ and r
As in the previous section, we then define the spaces
$$
\overline{r}_+\ =\ \overline{\varphi}_{\g_+}\oplus [\overline{d},\overline{\varphi}_{\g_+}]\ , \ \ 
\overline{r}\ =\ \overline{J}_{\g_\leq}\oplus \overline{\varphi}^{\g_-}\oplus \overline{\Phi}_{\g_{1/2}}
\ ,
$$
which are non-linear Lie subalgebras of $C^\tf(\g,x)$,
with the following commutation relations.
In $\overline{r}_+$ we have, for $a,b\in\g_+$:
\begin{eqnarray}\label{commr+}
\vphantom{\Big(} & {[\overline{\varphi}_a\ _\lambda\ \overline{\varphi}_b]}\ =\ 0\ ,\ \ 
{[[\overline{d},\overline{\varphi}_a],\overline{\varphi}_b]}
\ =\ s(a)\overline{\varphi}_{[a,b]}\ ,\nonumber\\ 
\vphantom{\Big(} & {[[\overline{d},\overline{\varphi}_a],[\overline{d},\overline{\varphi}_b]]}
\ =\ [\overline{d},\overline{\varphi}_{[a,b]}]\ .
\end{eqnarray}
The second equation follows by (\ref{commnew}) and the identity
$$
[\overline{d},\overline{\varphi}_a]\ =\ \overline{J}_{\pi_+ a}+(a|f)+s(a) 
\overline{\Phi}_a\ .
$$
The last equation follows by the fact that $(ad\,\overline{d})^2=0$.
In $\overline{r}$ we have, again from (\ref{commnew}),
(for $\overline{J}_a,\overline{J}_b\in \overline{J}_{\g_\leq}
,\ \overline{\varphi}^a,\overline{\varphi}^b\in\overline{\varphi}^{\g_-}
,\ \overline{\Phi}_a,\overline{\Phi}_b\in\overline{\Phi}_{\g_{1/2}}$),
\begin{eqnarray}\label{commr}
\vphantom{\Big(} &  {[\overline{J}_a,\overline{J}_b]}
\ =\ \overline{J}_{[a,b]}-\psi_k([x,a]|b)\ ,\ \ 
{[\overline{J}_a,\overline{\varphi}^b]}
\ =\ \overline{\varphi}^{[a,b]}\ ,\\ 
\vphantom{\Big(} & {[\overline{\Phi}_a,\overline{\Phi}_b]}\ =\ (f|[a,b])\ ,\ \ 
{[\overline{J}_a,\overline{\Phi}^b]} = 0 
= {[\overline{\varphi}^a,\overline{\varphi}^b]} = {[\overline{\varphi}^a,\overline{\Phi}_b]}
\ .\nonumber
\end{eqnarray}
The universal enveloping algebras $U(\overline{r}_+)$ and $U(\overline{r})$ are naturally
subalgebras of $C^\tf(\g,x)$ and, by the PBW Theorem,
we have a vector space isomorphism
$$
C^\tf(\g,x)\ \simeq\ U(\overline{r}_+)\otimes U(\overline{r})\ .
$$
% action of d
Moreover the adjoint action of $\overline{d}$ restricts 
to a differential $\overline{d}_+$ of $\overline{r}_+$, hence of $U(\overline{r}_+)$,
is given by
$$
\overline{d}_+(\overline{\varphi}_a)\ =\ [\overline{d},\overline{\varphi}_a]
\ ,\ \  \overline{d}_+([\overline{d},\overline{\varphi}_a])\ =\ 0\ ,
$$
and it restricts to a "non-linear" differential of $U(\overline{r})$,
an odd derivation of the associative product, which we denote again by 
$\overline{d}$.
It can be computed explicitly using (\ref{commdgen}) and it is given by
\begin{eqnarray}\label{dr}
\overline{d}(\overline{J}_a) 
&=& \sum_{\alpha\in S_+}s(\alpha) \overline{\varphi}^\alpha \overline{J}_{\pi_\leq[u_\alpha,a]}
+s(a)\overline{\varphi}^{\sum_{S_+}[[x,u^\alpha],\pi_\leq[u_\alpha,a]]} \nonumber\\
&& \!\!\!\!\!\!\!\!\!\!\!\!\!\!\!\!\!\!\!\!\!\!\!\!\!\!\!\!\!\!
-s(a)\!\!\! \sum_{\alpha\in S_+}\!\! \overline{\varphi}^\alpha \overline{\Phi}_{[u_\alpha,a]} 
+\!\!\! \sum_{\alpha\in S_+}\!\! \psi_k([x,a]|u_\alpha) \overline{\varphi}^\alpha
-s(a)\overline{\varphi}^{[a,f]}\ , \, a\in\g_\leq ,\,\nonumber\\
\overline{d}(\overline{\varphi}^a) 
&=& \frac12 \sum_{\alpha\in S_+}s(\alpha) \overline{\varphi}^\alpha \overline{\varphi}^{[u_\alpha,a]}
\ ,\, a\in\g_-\ ,\\ 
\overline{d}(\overline{\Phi}_a) &=& \overline{\varphi}^{[a,f]}\ ,\ \, 
a\in\g_{1/2}\,.\nonumber
\end{eqnarray}
Recalling Definition \ref{def-finW} of the finite W-algebra,
we can repeat the same arguments from last section, equation (\ref{isoms}), to conclude that
$$
W^\tf(\g,x) = H(C^\tf(\g,x),ad\,\overline{d})
\ \simeq\ H(U(\overline{r}),\overline{d}) \ .
$$
Moreover, it is clear from the above definitions that the non-linear Lie algebra $\overline{r}$
coincides with the $H$-twisted Zhu algebra of the non-linear Lie conformal algebra
$R$, defined in (\ref{R+R}).
Hence, by Theorem \ref{zhuenvel} we have
$$
Zhu_H V(R)\ \simeq\ U(Zhu_H R)\ \simeq\ U(\overline{r})\ .
$$
Finally, the action of $\overline{d}$ on $U(\overline{r})$ is induced by the action
of $d$ on $V(R)$, defined in (\ref{dR}), namely
$$
\overline{d}(\pi_Z A)\ =\ \pi_Z(d A)\ ,\ \, A\in V(R)\ ,
$$
where, as before, $\pi_Z$ denotes the quotient map from the vertex algebra 
$V(R)$ to its
$H$-twisted Zhu algebra $U(\overline{r})$.

We can summarize all the above results in the following
\begin{theorem}\label{str-fin-W}
The finite W-algebra $W^\tf(\g,x)$ 
is isomorphic to the associative algebra
$ H(U(\overline{r}),\overline{d})$,
where $\overline{r}
=\overline{J}_{\g_\leq}\oplus \overline{\varphi}^{\g_-}\oplus 
\overline{\Phi}_{\g_{1/2}}
\simeq Zhu_H R$
is the $H$-twisted Zhu algebra of the (non-linear) Lie conformal algebra $R$
(defined in Theorem \ref{str-aff-W}), with the Lie bracket given explicitly 
by equations (\ref{commr}),
and $\overline{d}$ is the differential on $U(\overline{r})\simeq 
Zhu_H V(R)$
induced by $d$ on $V(R)$ (again defined in Theorem \ref{str-aff-W}),
given explicitly by equations (\ref{dr}).
\end{theorem}

%%%

\subsection{Relation between finite and affine W-algebras}

Recall that, by Theorem \ref{str-aff-W}, the affine W-algebra can be defined as 
$W_k(\g,x)=H(V(R),d)$, where $R=\C[T]r,\ r=J_{\g_\leq}\oplus \varphi^{\g_-}\oplus \Phi_{\g_{1/2}}$,
is the non-linear Lie conformal algebra defined by (\ref{lR}), and 
the differential $d:\ V(R)\rightarrow V(R)$ is given by (\ref{dR}).
We now want to show that the pair $(R,d)$ satisfies all assumptions of Theorem \ref{Rnonlin}
and Theorem \ref{main}.
For this we need to decompose the generating space $r$ as in (\ref{grg}):
$$
r\ =\ \bigoplus_{\substack{p,q\in\frac12 \Z\\ 
\Delta\in\frac12 \Z_+}} r^{pq}[\Delta]\ .
$$
This is obtained by assigning the following degrees
\begin{center}
\begin{tabular}{c|ccc}
& $\vphantom{\Big()}p$ & $q$ & $\Delta$ \\
\hline
$\vphantom{\Big()} J_\alpha,\ \alpha\in S_\leq$ 
 & $j_\alpha-\frac12$ & $-j_\alpha+\frac12$ & $1-j_\alpha$ \\
$\vphantom{\Big()} \varphi^\alpha,\ \alpha\in S_+$ 
 & $-j_\alpha+\frac12$ & $j_\alpha+\frac12$ & $j_\alpha$ \\
$\vphantom{\Big()}\Phi_\alpha,\ \alpha\in S_{1/2}$ & $0$ & $0$ & $\frac12$ 
\end{tabular}
\end{center}
Notice that $\Delta$ coincides with the conformal weight (\ref{cw}), and $p+q$ coincides
with the charge (\ref{chdeg}).
It is immediate to check that the Lie conformal algebra structure (\ref{lR}) of $R$ satisfies the grading 
conditions (\ref{gradR}), namely
$$
r^{p_1 q_1}[\Delta_1]_{(n)}r^{p_2 q_2}[\Delta_2]\ \subset\ 
\bigoplus_{l\geq0}\T^{p_1+p_2+l,q_1+q_2-l}(R)[\Delta_1+\Delta_2-n-1]\ .
$$
In particular, the universal enveloping vertex algebra $V(R)$ has the induced 
$\Z$-charge grading,
the increasing $\frac12 \Z$-filtration given by $p$, and $\frac12 \Z_+$-valued conformal weight $\Delta$.
We finally need to show that $d:\ V(R)\rightarrow V(R)$ is a good almost linear differential of $R$,
according to Definition \ref{defdnonlin}.
Indeed, $d$ leaves unchanged the conformal weight and changes the charge degree by +1.
It is also immediate to check that the action of $d$ preserves the increasing filtration given by $p$.
In other words condition ({\ref{gradd}) holds.
The associated graded differential is easy to write down:
$$
\grd(J_a)\ =\ -s(a)\varphi^{[a,f]}
\ ,\ \ \grd(\varphi^a)\ =\ 0
\ ,\ \ \grd(\Phi_a)\ =\ \varphi^{[a,f]}\ .
$$
In particular, it is a "linear" differential of $r$ of bidegree $(0,1)$, with 
$\Ker(\grd|_r)=J_{\g^f}\oplus \varphi^{\g_-}$ and $\im(\grd|_r)=\varphi^{\g_-}$.
Hence the corresponding cohomology is concentrated at the zero charge:
$$
H(r,\grd)\ \simeq\ \Ker\big(\grd|_{r^0}\big)\ =\ J_{\g^f}\ .
$$
In other words, conditions (\ref{dRnonlin}) and (\ref{dRgood2}) hold.

The following result is now an immediate consequence of Theorem \ref{Rnonlin}
(see also \cite{KW1}):
\begin{theorem}\label{appl1}
Fix bases $\{u_i\,|\, i\in I_j\}$ of $\g^f_j,\ j\leq0$, and let $J_i=J_{u_i},\ i\in I=\sqcup_j I_j$.
For every $i\in I_j$ we can find an element $E_i\in F^{j-1/2}V(R)$
of zero charge and of conformal weight $1-j$, such that $d(E_i)=0$ and
$E_i-J_i\in F^jV(R)$.
Moreover the space
\begin{equation}\label{gener-ker}
H(R,d)\ =\ \C[T]H(r,d)\ ,\ \hbox{where }\ H(r,d)\ =\ \Span_\C\big\{E_i,\ i\in I
\big\},
\end{equation}
admits a structure of a non-linear Lie conformal algebra such that
$W_k(\g,x)\ \simeq\ V(H(R,d))$.
In particular $W_k(\g,x)=H^0(V(R),d)$,
i.e. the cohomology of the complex $(C_k(\g,x),d_{(0)})$ is concentrated at zero charge.
\end{theorem}

We now recall that, by Theorem \ref{str-fin-W}, the finite W-algebra is isomorphic as associative algebra
to $H(Zhu_H V(R),\dz)$. We thus immediately get from Theorem \ref{main} 
and Theorem \ref{str-aff-W} the main result of this section:
\begin{theorem}\label{fine}
There is a canonical associative algebra isomorphism
$$
Zhu_H W_k(\g,x)\ \stackrel{\sim}{\longrightarrow} W^\tf(\g,x)\ \simeq\ U(H(r,d))\ ,
$$
where $H(r,d)$ is the space in (\ref{gener-ker}), with the structure of 
a non-linear Lie algebra
induced by the 
bracket $[\,_*\,]$ on $W_k(\g,x)$.
\end{theorem}

\begin{remark}\label{formal}
Let $K_0$ denote the subspace of zero charge $d$-closed elements of $V(R)$.
Then, obviously, the canonical map $K_0\rightarrow W_k(\g,x)$
is an isomorphism of vertex algebras.
Thus we have the "formality" property, i.e. the complex $(W_k(\g,x),d=0)$ 
embeds in the complex $(C_k(\g,x),d_{(0)})$ as the subcomplex consisting
of zero-charge $d_{(0)}$-closed elements of the subcomplex $(V(R),d)$.
Similar "formality" property holds for the finite $W$-algebras
\end{remark}

%%%

\section{Quasiclassical limit: quantum and classical $W$-algebras}\label{sec:pva}

If we remove the integral terms ("quantum corrections") in formulas (\ref{eq:1.36}),
(\ref{eq:1.37}), (\ref{eq:1.38}) and (\ref{eq:1.39}) of the fourth definition of a vertex algebra,
we arrive at the definition of a Poisson vertex algebra (cf. \cite{FB}).
\begin{definition}\label{pva}
A {\itshape Poisson vertex algebra} is a quintuple
$(\VPA, \vac ,T ,\{ \, . \, {}_{\lambda} \, . \, \}, \cdot)$, where
\romanparenlist
\begin{enumerate}
\item %%i
$(\VPA,T,\{ \, . \, _{\lambda} \, . \, \})$ is a Lie conformal superalgebra,
\item %%ii
$(\VPA,\vac ,T , \cdot)$ is a unital associative commutative differential superalgebra,
\item %%iii
the operations 
$\{ \, . \, {}_{\lambda} \, . \, \}$ 
and $\cdot$ are
related by the Leibniz rules ($=$ commutative Wick formulas): 
\begin{eqnarray}
\vphantom{\Big(}
\{a_{\lambda} bc \} &=& \{ a_{\lambda} b \} c+ p(a,b) b \{
a_{\lambda} c \}\ ,\label{xxx1}\\ 
\vphantom{\Big(}
\{ ab_{\lambda} c\} &=&
(e^{T\partial_{\lambda}}a)\{ b_{\lambda}c \} + p(a,b)
(e^{T\partial_{\lambda}}b) \{ a_{\lambda}c\} \label{xxx2}
\end{eqnarray}
$\big(= p(a,b)p(a,c)\{b_{\lambda+T} c\}_{\rightarrow}a+p(b,c)\{a_{\lambda+T} c\}_{\rightarrow} b$,
where the arrow means that $\lambda+T$ should be moved to the right$\big)$.
\end{enumerate}
\end{definition}
As for vertex algebras, given a Poisson vertex algebra $\VPA$ we can define $n\st{th}$ products
for every $n\in\Z$ as follows:
$$
a_{(-n-1)}b\ =\ \big(T^{(n)}a\big)b\ ,\ \ 
a_{(n)}b\ =\ \frac{d^n}{d\lambda^n}\{a\ _\lambda\ b\}\big|_{\lambda=0}\ ,\ \ n\in\Z_+\ .
$$
\begin{example}\label{gfz}
Let $\VPA=\C[u_i^{(n)}\ |\ i\in I,\, n\in\Z_+]$, be the algebra of polynomials in the even indeterminates
$u_i^{(n)}$, let $T$ be the derivation of $\VPA$ defined by $Tu_i^{(n)}=u_i^{(n+1)}$,
and let $\vac=1$.
Define a $\lambda$-bracket on $\VPA$ by
$$
\{P\ _\lambda\ Q\}\ =\ \sum_{i,j\in I,\,p,q\in\Z_+} \frac{\partial Q}{\partial u_j^{(q)}}(\lambda+T)^q
\{u_i\ _{\lambda+T}\ u_j\}_\rightarrow (-\lambda-T)^p \frac{\partial P}{\partial u_i^{(p)}}\ ,
$$
where the $\lambda$-bracket between the $u_i=u_i^{(0)}$ is defined such that
skewsymmetry (resp. Jacobi identity) holds for any pair $u_i,\,u_j$ (resp.
triple $u_i,\ u_j,\ u_k$).
As before, $\{a_{\lambda+T}b\}_\rightarrow$  means that the powers of $\lambda+T$ are placed 
on the right.
Then it is not difficult to check that
$\VPA$ is a Poisson vertex algebra.
For example we can take $\{u_i\ _\lambda\ u_j\}=P_{i,j}(\lambda)$
be arbitrary polynomials such that $P_{i,j}(\lambda)=-P_{j,i}(-\lambda)$
for every $i,j\in I$.
The simplest example of this is $\VPA=\C[u^{(n)}|n\in\Z_+]$ where $\{u_\lambda u\}=\lambda$,
called the Gardner--Faddeev--Zakharov (GFZ) bracket.
\end{example}

\begin{remark}
Lie conformal algebras provide a very convenient framework for both classical
and quantum Hamiltonian systems.
This is based on the obvious observation that, given a Lie conformal algebra 
$(\VPA,\{\cdot \,_\lambda\,\cdot\})$, the bracket $\{a\,,\,b\}=\{a\,_\lambda\,b\}|_{\lambda=0}$
on $\VPA$ satisfies the Jacobi identity for Lie algebras,
it is such that $\{T\VPA\,,\,\VPA\}=0$, and the subspace $T\VPA$ is a two-sided ideal
with respect to this bracket,
so that $(\VPA/T\VPA,\{\cdot\,,\,\cdot\})$ is a Lie algebra.
A "local functional" is an element of $\VPA/T\VPA$.
Given $h\in \VPA/T\VPA$, the corresponding "Hamiltonian equation" is,
by definition, the equation
$$
\dot{u}\ =\ \{h\,,\,u\}\ ,\ \ u\in \VPA\ .
$$
A local functional $h_1$ is said to be an "integral of motion" for this equation if $\dot{h_1}=0$,
or, equivalently, if $h$ and $h_1$ are "in involution", i.e. 
$\{h\,,\,h_1\}=0$.
If $\VPA$ is a Poisson vertex algebra, we obtain in this way classical Hamiltonian equations.
For example, if $\VPA=\C[u,u^\prime,u^{\prime\prime},\dots]$ is the Poisson vertex algebra
with the GFZ bracket $\{u\,_\lambda\,u\}=\lambda$,
then the local functionals $h_0=\int u,\, h_1=\int u^2,\, h_2=\frac12\int(u^3-{u^\prime}^2),\,\dots$
are in involution (here the sign of the integral simply means taking the image in $\VPA/T\VPA$),
and the Hamiltonian equation $\dot u=\{h_2\,,\, u\}$ is the classical KdV equation
$\dot{u}=3uu^\prime+u^{\prime\prime\prime}$.
In exactly the same way, given a vertex algebra $V$, one defines a "local functional"
$h\in V/TV$ and the corresponding "quantum Hamiltonian equation"
$\dot{u}=[h,u],\,u\in V$.
In particular, taking the vertex algebra $V^k(\g)$, where $\g$ is the 1-dimensional (even) 
Lie algebra with a non-degenerate bilinear form
(its quasi-classical limit, discussed below, is the GFZ Poisson vertex algebra),
one can in the same way construct and study quantum Hamiltonian equations
and their integrals of motion (see \cite{FF3}).
\end{remark}

We next want to define the $(H,\Gamma)$-twisted Zhu algebra of a Poisson vertex algebra
(cf. Section \ref{sec:2.2}).
A {\itshape Hamiltonian operator} $H$ on a Poisson vertex algebra $\VPA$ is a diagonalizable
operator on $\VPA$ such that (cf. (\ref{eq:0.2}))
$$
H(a_{(n)}b)\ =\ (Ha)_{(n)}b+a_{(n)}(Hb)-(n+1)a_{(n)}b\ ,\ \ a,b\in \VPA,\ n\in\Z\ .
$$
As usual, if $a$ is an eigenvector of $H$, its eigenvalue is denoted by $\Delta_a$.
Furthermore, let $\Gamma$ be an additive subgroup of $\R$ containing $\Z$.
A Poisson vertex algebra $\VPA$ is $\Gamma/\Z$-{\itshape graded} if we have
a decomposition $\VPA=\bigoplus_{\bar{\gamma} \in \Gamma /\Z}{}^{\bar{\gamma}}\VPA$
by $H$-invariant subspaces, such that
$$
{}^{\bar{\alpha}} \VPA {}^{\bar{\beta}}\VPA
\subset {}^{\bar{\alpha} +\bar{\beta}} \VPA\ ,\ \ 
\big\{{}^{\bar{\alpha}} \VPA\ _\lambda\ {}^{\bar{\beta}}\VPA\big\}
\subset \C[\lambda]\otimes {}^{\bar{\alpha} +\bar{\beta}} \VPA\ .
$$
We will use the notation introduced in Section \ref{sec:2.2}:
given $a\in {}^{\bar{\gamma_a}} \VPA[\Delta_a]$,
we let $\epsilon_a$ be the maximal non-positive real number in $\bar{\gamma}_a-\bar{\Delta}_a$,
we let $\gamma_a=\Delta_a+\epsilon_a$,
and for homogeneus $a,b\in \VPA$ we let $\chi(a,b)=1$ or $0$
depending on whether $\epsilon_a+\epsilon_b\leq-1$ or not.
We also let $H^\prime$ be the diagonalizable operator of $\VPA$
such that $H^\prime(a)=\gamma_a a$.
Consider the space $\VPA[\hbar]$, with the $\C[\hbar]$-bilinear
commutative associative product induced by the product $ab$ on $\VPA$,
and define the following $\C[\hbar]$-bilinear $\hbar$-{\itshape bracket} on $\VPA[\hbar]$:
\begin{equation}\label{P-hbr}
\{a,b\}_\hbar\ =\ \sum_{j\in\Z_+}\binom{\gamma_a-1}{j}\hbar^j a_{(j)}b\ ,\ \ a,b\in \VPA\ .
\end{equation}
Let $\VPA_\Gamma$  (resp. $\VPA_{\hbar ,\Gamma}$)
be the $\C$-span (resp. $\C [\hbar]$-span in $\VPA[\hbar]$) of all elements
$a \in \VPA$ such that $\epsilon_a =0$.
Moreover let $J_{\hbar,\Gamma}\subset \VPA_{\hbar,\Gamma}$ 
be the $\C [\hbar]$-submodule generated by all the elements 
$a_{(-2+\chi (a,b),\hbar ,\Gamma)}b$, with $a,b \in \VPA$ such that 
$\epsilon_a+\epsilon_b \in \Z$, where we let
$$
a_{(-2,\hbar,\Gamma)}b\ =\ ((T+\hbar H^\prime)a)b\ ,\ \ 
a_{(-1,\hbar,\Gamma)}b\ =\ ab\ ,\ \ a,b\in \VPA\ .
$$
If we put $b=1$ in the above expression, we get that $(T+\hbar H^\prime)a\in J_{\hbar,\Gamma}$
for all $a\in \VPA_{\hbar,\Gamma}$.
Hence, by induction,
\begin{equation}\label{tn}
T^{(n)}a\ \equiv\ \hbar^n\binom{-\gamma_a}{n}a\ \ \mod J_{\hbar,\Gamma}\ \ ,
\quad \text{ for } \ a\in\VPA_{\hbar,\Gamma}\ .
\end{equation}
\begin{theorem}\label{P-2.7}
$J_{\hbar,\Gamma}$ is a two-sided ideal of $\VPA_{\hbar,\Gamma}$ with respect to both
the commutative associative product $\cdot$ and the $\hbar$-bracket $\{\cdot\,,\,\cdot\}_\hbar$.
Moreover, 
$$
Zhu_{\hbar,\Gamma}\VPA\ :=\ \VPA_{\hbar,\Gamma}/J_{\hbar,\Gamma}
$$ 
is a Poisson superalgebra
over $\C[\hbar]$, with commutative associative product induced by $\cdot$
and with Lie bracket induced by $\{\cdot\,,\,\cdot\}_\hbar$.
\end{theorem}
\begin{proof}
It is clear, by the definition, that $J_{\hbar,\Gamma}$ is a two-sided ideal with respect
to the commutative associative product of $\VPA_{\hbar,\Gamma}$.
Moreover, it immediately follows by the definition (\ref{P-hbr}) of the $\hbar$-bracket
and by the left Leibniz rule (\ref{xxx1}), that
\begin{equation}\label{P-leib1}
\{a,bc\}_\hbar\ =\ \{a,b\}_\hbar c+p(a,b)\{a,c\}_\hbar\ ,\ \ a,b\in \VPA\ .
\end{equation}
It is not hard to prove, by a direct computation, that the following identity holds
for all $a,b\in \VPA_{\Gamma}$ (cf. (\ref{2.10-eq}):
\begin{equation}\label{P-sesq}
\{a,(T+\hbar H)b\}_\hbar\ =\ (T+\hbar H)\{a,b\}_\hbar\ .
\end{equation}
If we then replace $b$ by $(T+\hbar H)b$ in (\ref{P-leib1}) and we use (\ref{P-sesq}), we get, 
for $a,b,c\in \VPA_\Gamma$,
\begin{equation}\label{P-leib2}
\{a,b_{(-2,\hbar,\Gamma)}c\}_\hbar\ =\ {\{a,b\}_\hbar}_{(-2,\hbar,\Gamma)} c
+p(a,b)b_{(-2,\hbar,\Gamma)}\{a,c\}_\hbar\ .
\end{equation}
Next, we want to prove that the $\hbar$-bracket is skewsymmetric in the quotient space
$\VPA_{\hbar,\Gamma}/J_{\hbar,\Gamma}$, namely
\begin{equation}\label{P-skew}
\{b,a\}_\hbar\ \equiv\ -p(a,b)\{a,b\}_\hbar\ \ \mod J_{\hbar,\Gamma}\ ,\ \ a,b\in \VPA_\Gamma\ .
\end{equation}
By definition of the $\hbar$-bracket and by the skewsymmetry os the $\lambda$-bracket,
or, equivalently, by (\ref{n-skew}), we have
\begin{eqnarray*}
\{b,a\}_\hbar &=& \sum_{i\in\Z_+}\binom{\Delta_b-1}{i}\hbar^i b_{(i)}a\\
&=& p(a,b)\sum_{i,j\in\Z_+}\binom{\Delta_b-1}{i}\hbar^i (-1)^{i+j+1} T^{(j)}\big( a_{(i+j)}b \big)\ .
\end{eqnarray*}
We then use (\ref{tn}) to get, for $a,b\in \VPA_\Gamma$,
\begin{equation}\label{prx}
\{b,a\}_\hbar
\equiv -p(a,b)\!\! \sum_{i,j\in\Z_+}\!\! \binom{\Delta_b-1}{i}\binom{-\Delta_a-\Delta_b+i+j+1}{j}
(-\hbar)^{i+j}  a_{(i+j)}b
\end{equation}
$\mod J_{\hbar,\Gamma}$.
Thanks to (\ref{eq:2.20}) with $x=\Delta_b-1,\, y=-\Delta_a-\Delta_b+n+1$,
the RHS of (\ref{prx}) is equal to
$$
-p(a,b)\sum_{n\in\Z_+}\binom{-\Delta_a+n}{n} (-\hbar)^{n}  a_{(n)}b\ ,
$$
which in turn is equal to $-p(a,b)\{a,b\}_\hbar$, thus proving (\ref{P-skew}).
It follows by (\ref{P-leib2}) and (\ref{P-skew}) that $J_{\hbar,\Gamma}\subset \VPA_{\hbar,\Gamma}$
is a two-sided ideal with respect to the $\hbar$-bracket.
We are then left to prove that the $\hbar$-bracket satisfies 
the Jacobi identity in $\VPA_{\hbar,\Gamma}/J_{\hbar,\Gamma}$.
In fact, we will prove more, that the Jacobi identity holds already in $\VPA_{\hbar,\Gamma}$,
namely that, for $a,b,c\in \VPA_\Gamma$,
\begin{equation}\label{P-jac}
\{a,\{b,c\}_\hbar\}_\hbar\,-\,p(a,b)\{b,\{a,c\}_\hbar\}_\hbar\ =\ \{\{a,b\}_\hbar ,c\}_\hbar\ .
\end{equation}
By the definition of the $\hbar$-bracket, the LHS of (\ref{P-jac}) is
\begin{equation}\label{pry}
\sum_{i,j\in\Z_+}\binom{\Delta_a-1}{i}\binom{\Delta_b-1}{j}\hbar^{i+j}
\Big(a_{(i)}\big(b_{(j)}c\big)-p(a,b)b_{(j)}\big(a_{(i)}c\big)\Big)\ .
\end{equation}
By the Jacobi identity for the $\lambda$-bracket, or, equivalently, using (\ref{n-jac}),
we get that (\ref{pry}) is equal to
$$
\sum_{i,j,k\in\Z_+}\binom{\Delta_a-1}{i}\binom{\Delta_b-1}{j}\binom{i}{k}\hbar^{i+j}
\big(a_{(k)}b\big)_{(i+j-k)}c\ .
$$
In the above expression we replace
$\binom{\Delta_a-1}{i}\binom{i}{k}$ by $\binom{\Delta_a-1}{k}\binom{\Delta_a-k-1}{i-k}$,
and we make the change of variables
$n=i+j-k,\, l=i-k$, to get
$$
\sum_{n,k\in\Z_+}\binom{\Delta_a-1}{k}
\sum_{l=0}^n \binom{\Delta_a-k-1}{l}\binom{\Delta_b-1}{n-l}
\hbar^{n+k} \big(a_{(k)}b\big)_{(n)}c\ .
$$
By (\ref{eq:2.20}) with $x=\Delta_a-k-1$ and $y=\Delta_b-1$,
we then get
$$
\sum_{n,k\in\Z_+}\binom{\Delta_a-1}{k}\binom{\Delta_a+\Delta_b-k-2}{n}
\hbar^{n+k} \big(a_{(k)}b\big)_{(n)}c\ ,
$$
which is in turn equal to $\{\{a,b\}_\hbar,c\}_\hbar$, thus proving (\ref{P-jac}).
\end{proof}

\begin{definition}\label{fam-ass}
A {\itshape family of associative algebras} $A_\epsilon$ is
an associative superalgebra over the algebra of polynomials $\C[\epsilon]$,
such that
\begin{equation}\label{xxx3}
[A_\epsilon\,,\, A_\epsilon]\ \subset\ \epsilon A_\epsilon\ .
\end{equation}
It is called {\itshape regular} if  the operator of multiplication by $\epsilon$ has zero kernel.
Then $A^\cl=A_\epsilon/\epsilon A_\epsilon$ is a commutative associative superalgebra,
with the Poisson bracket $\{\bar a,\bar b\}$
defined by taking preimages $a,b\in A_\epsilon$ of $\bar a,\bar b\in A^\cl$,
writing
\begin{equation}\label{xxx4}
[a,b]\ =\ \epsilon\{a,b\}\ \in\ \epsilon A_\epsilon\ ,
\end{equation}
and taking the image of $\{a,b\}$ in $A^\cl$.
It is easy to see that this bracket is well defined and it makes $A^\cl$
a Poisson superalgebra.
The Poisson superalgebra $A^\cl$ is called the {\itshape quasiclassical limit}
of the family $A_\epsilon$,
and we let $\pi_0:\, A_\epsilon\twoheadrightarrow A^\cl$ be the natural quotient map.
Similarly, $A=A_\epsilon/(\epsilon-1)A_\epsilon$ is an associative algebra over $\C$,
and we let $\pi_1:\, A_\epsilon\twoheadrightarrow A$ be the natural quotient map.
\end{definition}
\begin{remark}\label{remzhu}
The algebra $Zhu_{\hbar,\Gamma} V$ over $\C[\hbar]$ (defined in Section \ref{sec:2.2})
is a family of associative algberas, by Theorem \ref{th:2.12}(f).
However $Zhu_{\hbar,\Gamma} V$ is not, in general, a regular family of associative
algebras, in the sense of Definition \ref{fam-ass}, 
since the multiplication by $\hbar$ might have a non-zero kernel.
For example, if $V$ is an in Example \ref{stupid}, then
$$
Zhu_{\hbar,H} V\ \simeq\ \C[\hbar] V/\Big(\C\oplus\Big(\bigoplus_{n\geq1}\hbar V[n]\Big)\Big)
\ =\ \C[\hbar]\oplus \Big(\bigoplus_{n\geq1} V[n]\Big)\ ,
$$
and $\Ker\,\hbar \simeq \bigoplus_{n\geq1}V[n]\neq0$.
\end{remark}

However, by Proposition \ref{old:2.17}(a),
$Zhu_{\hbar,\Gamma}V/\hbar Zhu_{\hbar,\Gamma}V
\simeq V_\Gamma/(V_\Gamma\cap V_{(-2+\chi)}V)$ 
is still a Poisson algebra, that we will call, again, the {\itshape quasiclassical limit} 
of $Zhu_{\hbar,\Gamma}V$.
We then let $\psi_0$ be the natural quotient map
$\psi_0:\, Zhu_{\hbar,\Gamma}V\twoheadrightarrow V_\Gamma/(V_\Gamma\cap V_{(-2+\chi)}V)$.
More generally, recall (see Section \ref{sec:2.2}) that we can specialize
$Zhu_{\hbar,\Gamma}V$ at any value $\hbar=c\in\C$, to get the associative 
superalgebra
$Zhu_{\hbar,\Gamma}V/(\hbar-c) Zhu_{\hbar,\Gamma}V\simeq V_\Gamma/ J_{\hbar=c,\Gamma}$.
Denote by $\psi_c$ the quotient map
$Zhu_{\hbar,\Gamma}V\twoheadrightarrow V_\Gamma/J_{\hbar=c,\Gamma}$, and
recall the notation $Zhu_\Gamma V=V_\Gamma/ J_{\hbar=1,\Gamma}$
for the $(H,\Gamma)$-twisted Zhu algebra, from Section \ref{sec:2.2}.

We will use the same notation also for a Poisson vertex algebra $\VPA$.
Namely we let again $\psi_c$ be the quotient map
$Zhu_{\hbar,\Gamma}\VPA\twoheadrightarrow 
Zhu_{\hbar,\Gamma}\VPA/(\hbar-c)Zhu_{\hbar,\Gamma}\VPA\simeq
\VPA_\Gamma/J_{\hbar=c,\Gamma}$ for arbitrary $c\in \C$,
and we let $Zhu_\Gamma\VPA=\VPA_\Gamma/J_{\hbar=1,\Gamma}$.
Obviously $\VPA_\Gamma/J_{\hbar=c,\Gamma}$ is a Poisson superalgebra over $\C$
for every choice of $c\in\C$.

\begin{definition}\label{fam-ver}
A {\itshape family of vertex algebras} $V_\epsilon$ is a vertex
algebra over $\C[\epsilon]$ such that 
\begin{equation}\label{xxx5}
[V_\epsilon\ _\lambda\ V_\epsilon]\ \subset\ \C[\lambda]\otimes \epsilon V_\epsilon\ .
\end{equation}
As before, $V_\epsilon$ is called {\itshape regular} if the operator of
multiplication by $\epsilon$ has zero kernel.
Then the normally ordered product on $V_\epsilon$ induces the structure of a unital
commutative associative differential superalgebra on
$(V^\cl = V_\epsilon/\epsilon V_\epsilon,\vac,T,\cdot)$, endowed with the 
$\lambda$-bracket $\{\bar a\ _\lambda\ \bar b\}$, defined by taking preimages
$a,b\in V_\epsilon$ of $\bar a,\bar b\in V^\cl$, writing
\begin{equation}\label{xxx6}
[a\ _\lambda\ b]\ =\ \epsilon\{a\ _\lambda\ b\}\ ,
\end{equation}
and taking the image of $\{a\ _\lambda\ b\}$ in $\C[\lambda]\otimes V^\cl$.
It is easy to see, using the fourth definition of a vertex algebra,
that $(V^\cl,\vac,T,\{\cdot\,_\lambda\,\cdot\},\cdot)$
is a Poisson vertex algebra.
Likewise, this Poisson vertex algebra is called the 
{\itshape quasiclassical limit} of the family $V_\epsilon$,
and we let $\pi_0:\, V_\epsilon\twoheadrightarrow V^\cl$ be the natural quotient map.
Moreover, $V=V_\epsilon/(\epsilon-1)V_\epsilon$ is clearly a vertex algebra over $\C$,
and we let $\pi_1:\, V_\epsilon\twoheadrightarrow V$ be the corresponding quotient map.
\end{definition}

\begin{example}
A general example of a regular family of associative algebras $U(\g)_\epsilon$
is obtained by taking a Lie superalgebra $\g$, defining a new $\C[\epsilon]$-bilinear
Lie bracket on $\C[\epsilon]\otimes\g$ by the formula $[a,b]_\epsilon=\epsilon[a,b]$ for $a,b\in\g$,
and denoting by $U(\g)_\epsilon$ the universal enveloping superalgebra over $\C[\epsilon]$
of $\C[\epsilon]\otimes\g$.
It is clear that the quasiclassical limit of the family $U(\g)_\epsilon$
is the unital commutative associative superalgebra $S(\g)$
with the Poisson bracket which coincides with the Lie bracket on $\g$
and is extended to $S(\g)$ by the Leibniz rule.

In a similar fashion one constructs a regular family of vertex algebras $V(R)_\epsilon$,
starting from a Lie conformal superalgebra $R$, by defining the $\C[\epsilon]$-bilinear 
$\lambda$-bracket $[a\ _\lambda\ b]=\epsilon[a\ _\lambda\ b]$ on $\C[\epsilon]\otimes R$,
and denoting $V(R)_\epsilon$ the universal enveloping vertex algebra over $\C[\epsilon]$
of $\C[\epsilon]\otimes R$.
It is easy to see that the quasiclassical limit of the family $V(R)_\epsilon$
is the unital commutative associative differential superalgebra $S(R)$
with the Poisson $\lambda$-bracket which coincides with
the $\lambda$-bracket on $R$ and extends it to $S(R)$ by the Leibniz rules
(\ref{xxx1}) and (\ref{xxx2}).
\end{example}

\begin{example}\label{rem-516}
In exactly the same way one constructs a regular family of associative superalgebras 
(resp. vertex algebras)
in a slightly more general case of a non-linear Lie superalgebra (resp. Lie conformal
superalgebra), which is a central extension of a Lie superalgebra (resp. Lie conformal
superalgebra), as described in Example \ref{newex}.
Also, the quasiclassical limits are computed in the same way.
However this construction does not work for arbitrary non-linear Lie superalgebras
or Lie conformal superalgebras.
\end{example}
\begin{example}
The GFZ Poisson vertex algebra from Example \ref{gfz} is the quasiclassical limit of the family
$B_\epsilon=V(R)_\epsilon$, where $R=\C[T]u$ is a non-linear Lie conformal algebra
with $\lambda$-bracket $[u\ _\lambda\ u]=\lambda$.
Note that $B_{\epsilon=1}$ is the vertex algebra of a free boson (cf. \cite{K}).
\end{example}

Let $V_\epsilon$ be a regular family of vertex algebras over $\C[\epsilon]$,
with a Hamiltonian operator $H$ and a $\Gamma/\Z$-grading.
According to the above observations, starting from $V_\epsilon$ we can build the following 
commutative diagram of fundamental objects:
%%%%%% DIAGRAM %%%%%%%%%%%%%%%%%%%%%%%%%%
\begin{equation}\label{maxi}
\UseTips
\!\!\!\!\!\!\!
\xymatrix{
% 1
V^\cl \ar@{.>}[d]^{Zhu_{\hbar,\Gamma}} 
\!\!\!\!\!\!
& & \ar[l]_{\pi_0} 
& \!\!\!\!\!\!
V_\epsilon  \ar@{.>}[d]^{Zhu_{\hbar,\Gamma}} 
\!\!\!\!\!\!
& \ar[r]^{\pi_1} &
& \!\!\!\!\!\!
V \ar@{.>}[d]_{Zhu_{\hbar,\Gamma}} \\
% 2
Zhu_{\hbar,\Gamma}V^\cl \ar[d]^{\psi_1} \ar@/_2pc/[dd]_{\psi_0} 
\!\!\!\!\!\!
& & \ar[l]_{\pi_0}
& \!\!\!\!\!\!
Zhu_{\hbar,\Gamma}V_\epsilon  \ar[d]^{\psi_1} \ar@/_2pc/[dd]_{\psi_0} 
\!\!\!\!\!\!
& \ar[r]^{\pi_1} &
& \!\!\!\!\!\!
Zhu_{\hbar,\Gamma}V \ar[d]^{\psi_1} \ar@/_2pc/[dd]_{\psi_0}  \\
% 3
Zhu_{\Gamma}V^\cl 
\!\!\!\!\!\!
& & \ar[l]_{\pi_0} 
& \!\!\!\!\!\!
Zhu_{\Gamma}V_\epsilon 
\!\!\!\!\!\!
& \ar[r]^{\pi_1} &
& \!\!\!\!\!\!
Zhu_{\Gamma}V \\
% 4
\frac{V^\cl_\Gamma}{V^\cl_\Gamma\cap\big(V^\cl_{(-2+\chi)}V^\cl\big)}
\!\!\!\!\!\!\!\!\!\!\!\!
& & \ar[l]_{\pi_0} 
& \!\!\!\!\!\!\!\!\!\!\!\!
\frac{{V_{\epsilon}}_\Gamma}{{V_\epsilon}_\Gamma\cap\big({V_\epsilon}_{(-2+\chi)}{V_\epsilon}\big)}
\!\!\!\!\!\!\!\!\!\!\!\!
& \ar[r]^{\pi_1} &
& \!\!\!\!\!\!\!\!\!\!\!\!
\frac{V_\Gamma}{V_\Gamma\cap\big(V_{(-2+\chi)}V\big)}
}
\end{equation}
%%%%%%%%%%%%%%%%%%%%%%%%%%%%%%%%%%%%%%
In the above diagram, horizontal left arrows denote the canonical map $\pi_0$
from an $\epsilon$-family of algebras to its quasiclassical limit, described
in Definitions \ref{fam-ass} and \ref{fam-ver}.
Similarly, the horizontal right arrows denote the quotient map $\pi_1$ from an
$\epsilon$-family of algebras to its specialization at $\epsilon=1$,
described in Definitions \ref{fam-ass} and \ref{fam-ver}.
The dotted vertical arrows stand for taking the family of associative algebras
$Zhu_{\hbar,\Gamma}$ for
a vertex algebra or a Poisson vertex algebra.
The straight (respectively curved) vertical arrows denote the canonical
quotient maps $\psi_1$ (resp. $\psi_0$) 
to the specializations at $\hbar=1$ (resp. $\hbar=0$),
as described after Remark \ref{remzhu}.
\begin{example}\label{10nov}
Consider the special case of a family of vertex algebras $V(R)_\epsilon$ associated
to a Lie conformal algebra $R$, or to a non-linear Lie conformal algebra
of the type discussed in Example \ref{rem-516}.
Suppose $R$ has a Hamiltonian operator $H$, and take the $\Gamma/\Z$-grading of $V(R)_\epsilon$
induced by the action of $H$ (see Example \ref{ex-delta}).
Then we can build, starting from $V(R)_\epsilon$, the whole diagram (\ref{maxi}).
In particular, at the top right and left corners we have respectively $V=V(R)$,
the universal enveloping vertex algebra of $R$,
and $V^\cl=S(R)$, the symmetric algebra of $R$, with $\lambda$-bracket extending
the $\lambda$-bracket of $R$ by the Leibniz rules (\ref{xxx1}) and (\ref{xxx2}).
Moreover, at the bottom left and right corners we have isomorphic Poisson algebras:
\begin{equation}\label{LRB}
V^\cl/(V^\cl_{(-2)}V^\cl)\ \simeq\ S(R/TR)\ \simeq\ V/(V_{(-2)}V)\ ,
\end{equation}
where $S(R/TR)$ has Lie bracket induced by the $0\st{th}$ product of $R/TR$
and extended by the Leibniz rule.
To prove (\ref{LRB}), note that $S(R)_{(-2)}S(R)$ is the ideal of $S(R)$ generated by $TR$,
which implies the first isomorphism.
Similarly $V(R)_{(-2)}V(R)$ is the ideal of $V(R)$, with respect to the $-1\st{st}$ product,
generated by $TR$.
Hence the natural linear map $R\rightarrow V(R)/(V(R)_{(-2)}V(R))$
induces a surjective homomorphism of associative algebras
$\varphi:\, S(R/TR)\twoheadrightarrow V(R)/(V(R)_{(-2)}V(R))$.
To prove that this is an isomorphism, we construct the inverse map as follows.
The quotient map $R\twoheadrightarrow R/TR$ extends to a surjective linear map
$\T(R)\twoheadrightarrow S(R/TR)$. 
It is easy to see that the subspace $\M(R)\subset \T(R)$
(see (\ref{MR})) is in the kernel, hence we get a surjective linear map 
$V(R)\twoheadrightarrow S(R/TR)$.
It is also easy to see that $V(R)_{(-2)}V(R)$ is in the kernel of this map,
hence we get a map $V(R)/(V(R)_{(-2)}V(R))\rightarrow S(R/TR)$
which is the inverse of $\varphi$.
It is clear that $\varphi$ is a Poisson algebra isomorphism.
\end{example}
\begin{example}
We present here an example of a regular family of vertex algebras $V(R)_\epsilon$
for which the Poisson algebras at the bottom left and right corners of diagram (\ref{maxi})
are not isomorphic.
Let $U=\bigoplus_{n\in\Z_+}U[n]$ be a graded unital commutative associative algebra
with $U[0]=\C1$, and let $H$ be the diagonalizable operator on $U$ such that
$H|_{U[n]}=n I_{U[n]}$.
Then $V_\epsilon=\C[\epsilon]\otimes U$ is a regular family of vertex algebras with
$\vac=1,\, T=\epsilon H$ and $a_{(n)}b=\delta_{n,-1}ab$.
In this case $V=(U,\vac=1,T=H,Y(a,z)=L_a)$
and $V/V_{(-2)}V\simeq\C$,
while $V^\cl=(U,\vac=1,T=0,Y(a,z)=L_a)$ (cf. Example \ref{stupid})
and $V^\cl/(V^\cl_{(-2)}V^\cl)\simeq U$.
\end{example}

Returning to $W$-algebras, consider the family of $W$-algebras $W_k(\g,x)_\epsilon$
defined as follows.
Let $C_k(\g,x)_\epsilon=V(R(\g,x))_\epsilon$ be the regular family of vertex algebras
associated to the non-linear Lie conformal algebra $R_k(\g,x)$,
as explained in Remark \ref{rem-516}.
Let $d_\epsilon$ be the element of $C_k(\g,x)$ given by (\ref{differ}).
Then $(C_k(\g,x)_\epsilon,{d_\epsilon}_{(0)})$ is a regular family of vertex algebras
with a differential, and $W_k(\g,x)_\epsilon$ is, by definition,
the corresponding cohomology.
Notice that $W_k(\g,x)_\epsilon$ is, by the above construction, a vertex algebra over $\C[\epsilon]$.
Moreover, by formality (cf. Remark \ref{formal}), it is a vertex subalgebra of $C_k(\g,x)_\epsilon$,
hence it is a regular family of vertex algebras,
according to Definition \ref{fam-ver}.
By the observations in Section \ref{sec-aff-W}, $W_k(\g,x)_\epsilon$
has a Hamiltonian operator defined by (\ref{cw}),
and we take the $\Gamma/\Z$-grading of $W_k(\g,x)_\epsilon$,
induced by $H$ (see Example \ref{ex-delta}).

We then build, starting from $W_k(\g,x)_\epsilon$, the whole diagram (\ref{maxi})
of fundamental objects.
In particular, in the top right corner we have the usual affine $W$-algebra $W_k(\g,x)$,
while in the top left corner we have the {\itshape classical affine} $W$-{\itshape algebra}
$W_k^\cl(\g,x)$.
(We thus recover, in the case when $f$ is a principal nilpotent element, the so called 
{\itshape classical Drinfeld--Sokolov reduction} \cite{DS,FF1}.)
In the third row we have on the right the usual finite $W$-algebra $W^\tf(\g,x)$,
and on the left the {\itshape classical finite} $W$-{\itshape algebra} $W^{\tf,\cl}(\g,x)$.
Moreover, in the bottom right corner we have a Poisson algebra which,
by Proposition \ref{last}, is isomorphic to the associated graded of the finite $W$ algebra
$W^\tf(\g,x)$ with respect to the filtration (\ref{240}) induced by the conformal weight,
which coincides with the so called Kazhdan filtration.
We thus recover the Poisson algebra on the Slodowy slice
(see \cite{GG,P1,P2}).
Finally, we claim that the Poisson algebra at the bottom left corner is isomorphic
to the one at the bottom right corner.
To see this, first observe that, by Example \ref{10nov}, they are isomorphic at the level of
complexes, and then just notice that, since the differentials are derivations of all $n\st{th}$
products, taking cohomology commutes with taking quotients by $V_{(-2)}V$.

%%%%%%%%% APPENDIX %%%%%%%%%%%%%%%%%%%%%%%%%%%%

%%% addition 3

\appendice{
Three equivalent definitions\\ of finite $W$-algebras}

\begin{center}
{\normalsize
A. D'Andrea,\, C. De Concini,\, A. De Sole,\, R. Heluani and V. Kac
\par
}
\end{center}

\bigskip

We prove here that the definition of finite $W$-algebras
via the Whittaker models, which goes back to \cite{Ko} (see \cite{P1,GG}),
is equivalent to the definition presented in Section \ref{sec:6} of this paper
(in fact, to the slightly more general definition, indicated in Section \ref{sec:0.4},
with arbitrary coisotropic subspace $\mathfrak s\subset\g_{1/2}$).
For simplicity, we consider here the construction for a Lie algebra $\g$,
as the definition of finite $W$-algebras via the Whittaker models is known in the even case.
However most of our arguments can be generalized to the super case.

We will use the notation introduced in Section \ref{sec:6}:
$\g$ is a simple, finite dimensional Lie algebra, with a non-degenerate\
symmetric invariant bilinear form $(\cdot\,|\,\cdot)$,
and $x,f\in\g$ form a "good" pair of elements of $\g$,
i.e. $\ad x$ is diagonalizable on $\g$ with half integral eigenvalues
and eigenspace decomposition $\g=\bigoplus_{j\in\frac12 \Z}\g_j$,
and $f\in\g_{-1}$ is such that $\g^f=\{a\in\g\,|\,[f,a]=0\}\subset\bigoplus_{j\leq0}\g_j$.
We choose an isotropic subspace $\lag\subset\g_{1/2}$ with respect to the skewsymmetric
non-degenerate bilinear form $(f|[a,b])$ on $\g_{1/2}$,
and denote by $\lag^\perp(=\mathfrak s)\subset\g_{1/2}$ the orthogonal 
complement of $\lag$ with respect  to this bilinear form.
Obviously $\lag\subset\lag^\perp$, and $(f|[a,b])$ induces a non-degenerate skewsymmetric
bilinear form on $\lag^\perp/\lag$.
We then let $\m_\lag=\lag\oplus\g_{\geq1}$, and $\n_\lag=\lag^\perp\oplus\g_{\geq1}\subset\g$.
They are obviously subalgebras of $\g$, and $\m_\lag$ is an ideal of $\n_\lag$,
with $\n_\lag/\m_\lag\simeq\lag^\perp/\lag$.
(More generally, one can take a good $\R$-grading 
and let $\m_\lag=\lag+\g_{>1/2},\, \n_\lag=\lag^{\perp}+\g_{>1/2}$, see Remark \ref{rem-oct30}.)

%%%
\subsection*{The first definition, via quantum Hamiltonian reduction}

Recall the Definition \ref{def-finW} of the finite $W$-algebra $W^\tf(\g,x)$.
We can generalize it to an arbitrary choice of the isotropic subspace $\lag\subset\g_{1/2}$
as follows.
In analogy with the notation introduced in Section \ref{sec:6},
we let 
$\varphi_{\n_\lag}=\{\varphi_a\,|\,a\in\n_\lag\}$ be the vector space $\n_\lag$ with odd parity,
$\varphi^{\n^*_\lag}=\{\varphi^\phi\,|\,\phi\in\n^*_\lag\}$ be the vector space 
$\n^*_\lag$ with odd parity,
and $\Phi_{\lag^\perp/\lag}=\{\Phi_{[a]}\,|\,[a]\in\lag^\perp/\lag\}$ be a vector superspace 
isomorphic to $\lag^\perp/\lag$.
The usual pairing of $\n_\lag$ and $\n^*_\lag$ induces a non-degenerate skewsymmetric
bilinear form on $\varphi^{\n^*_\lag}\oplus\varphi_{\n_\lag}$,
which we denote by $\langle\cdot|\cdot\rangle_{ch}$,
and, as notice above, the inner product $(f|[a,b])$ on $\g_{1/2}$ induces a non-degenerate
skewsymmetric bilinear form on $\Phi_{\lag^\perp/\lag}$, which we denote 
by $\langle\cdot|\cdot\rangle_{ne}$.
We then define the non-linear Lie superalgebra
$$
r\ =\ \g\oplus\varphi^{\n^*_\lag}\oplus\varphi_{\n_\lag}\oplus\Phi_{\lag^\perp/\lag}\ ,
$$
with non-linear Lie bracket given by
\begin{eqnarray*}
&& [a,b]\,=\ \ \text{ Lie bracket of } \g\ , \ \ a,b\in\g \ ,\\
&& {[\varphi_a,\varphi^\phi]}\ =\ [\varphi^\phi,\varphi_a]\ =\ \phi(a)
\ =:\ \langle\varphi_a|\varphi^\phi\rangle_{ch}\ ,\ \ a\in\n_\lag,\,\phi\in\n^*_\lag \ ,\\
&& {[\Phi_{[a]},\Phi_{[b]}]}\ =\ (f|[a,b])\ =:\ \langle\Phi_{[a]}|\Phi_{[b]}\rangle_{ch}
\ ,\ \ [a],[b]\in\lag^\perp/\lag\ ,
\end{eqnarray*}
and all other Lie brackets among generators equal zero.
We denote by $C^{I}=U(r)$ the universal enveloping algebra of $r$,
and we let $d$ be the following odd element of $C^{I}$:
$$
d\ =\ \sum_{\alpha\in S} \varphi^\alpha
\big(u_\alpha+(f|u_\alpha)+\Phi_{[u_\alpha]}\big)
+\frac{1}{2}\sum_{\alpha,\beta\in S} \varphi^\alpha\varphi^\beta
\varphi_{[u_\beta,u_\alpha]}\ \in\ U(R)\ ,
$$
where $\{u_\alpha\},\,\{v^\alpha\},\,\alpha\in S$, are dual bases of $\n_\lag$ and $\n^*_\lag$,
namely $v^\beta(u_\alpha)=\delta_{\alpha,\beta}$,
$\varphi_\alpha=\varphi_{u_\alpha},\,\varphi^\alpha=\varphi^{v^\alpha}$ for $\alpha\in S$,
and $\Phi_{[a]}$, for $a\in\n_\lag$, is defined via the isomorphism 
$\n_\lag/\m_\lag\simeq\lag^\perp/\lag$.
It is not hard to check that $[d,d]=0$, so that $d^{I}=\ad(d)$ is a differential of the associative
superalgebra $C^{I}$.
\begin{definition}\label{first}
The finite $W$-algebra is
$W^{I}\ =\ H(C^{I},d^{I})$.
\end{definition}
It is clear that, in the special case $\lag=0$, Definition \ref{first} reduces to Definition \ref{def-finW}.
\begin{remark}\label{d-gener}
One can easily compute the action of $d^{I}$ on the generators of $C^{I}$:
\begin{eqnarray*}
\vphantom{\Big(}
d^{I}(a) &=& \sum_{\alpha}\varphi^\alpha[u_\alpha,a] \\
\vphantom{\Big(}
d^{I}(\varphi^\phi) &=& \frac{1}{2}\sum_{\alpha}\varphi^\alpha\varphi^{u_\alpha\cdot\phi} \\
\vphantom{\Big(}
d^{I}(\varphi_a) &=& a+(f|a)+\Phi_{[a]}
+\sum_{\alpha}\varphi^\alpha\varphi_{[u_\alpha,a]} \\
\vphantom{\Big(} 
d^{I}(\Phi_{[a]}) &=& \sum_{\alpha}\varphi^\alpha(f|[u_\alpha,a])
\end{eqnarray*}
In the second equation $a\cdot\phi$ denotes the coadjoint action of $\n_\lag$ on $\n_\lag^*$,
i.e. $(a\cdot\phi)(b)=\phi([b,a])$.
\end{remark}
\begin{remark}\label{xa}
For $a\in\n_\lag$, we let $X_a = a+(f|a)+\Phi_{[a]}$.
Then the map $a\mapsto X_a$ defines a Lie algebra isomorphism 
$\n_\lag\simeq X_{\n_\lag}\subset C^{I}$,
namely $X_{[a,b]}=[X_a,X_b],\,\forall a,b\in\n_\lag$.
Moreover, the action of $d^{I}$ on $X_{\n_\lag}$ is given by
$$
d^{I}(X_a)\ =\ \sum_{\alpha}\varphi^\alpha X_{[u_\alpha,a]}\ .
$$
\end{remark}

%%%
\subsection*{The second definition, via Whittaker models}

Let $\C_{\pm\chi}\simeq\C$ be the 1-dimensional representations of $\m_\lag$ given by
$m\cdot 1=\pm(f|m)$.
Let $M^0_\lag$ be the left $\g$-module induced by $\C_{-\chi}$:
\begin{equation}\label{ml0}
M^0_\lag\ =\ \Ind_{\m_\lag}^\g \C_{-\chi}\ =\
U(\g)\otimes_{U(\m_\lag)}\C_{-\chi}\ .
\end{equation}
$M^0_\lag$ is, by definition, a left $\g$--module via left multiplication.
However it is easy to check that $M^0_\lag$ is also a left $\n_\lag$-module
with respect to the adjoint action of $\n_\lag$ on $U(\g)$.
\begin{theorem}\label{ggthm}
\cite{GG} The Lie algebra cohomology of $\n_\lag$ with coefficients in $M^0_\lag$
is concentrated at degree 0:
$$
H(\n_\lag,M^0_\lag)\ =\ (M^0_\lag)^{\n_\lag}\ ,
$$
where the space of $\ad\n_\lag$-invariants $(M^0_\lag)^{\n_\lag}$ is an associative algebra with
product induced from $U(\g)$.
Moreover this algebra is independent of the choice of $\lag$, up to isomorphism.
\end{theorem}
\begin{definition}\label{second}
The finite $W$-algebra is $W^{II}=(M^0_\lag)^{\n_\lag}$. 
\end{definition}
By Theorem \ref{ggthm}, $W^{II}$
is the cohomology of the complex
$$
C^{II}\ =\ \Lambda (\n_\lag^*) \otimes (U(\g)\otimes_{U(\m_\lag)}\C_{-\chi})\ , 
$$
with differential
\begin{eqnarray*}
d^{II}\big(\Psi \otimes (A\otimes_{U(\m_\lag)}1)\big) &=&
 \frac{1}{2}\sum_\alpha v^\alpha\wedge u_\alpha\cdot\Psi\otimes (A\otimes_{U(\m_\lag)}1) \\
&& +\sum_\alpha v^\alpha\wedge\Psi \otimes (\ad u_\alpha(A)\otimes_{U(\m_\lag)}1)\ .
\end{eqnarray*}
As before, $a\cdot\Psi$ denotes the coadjoint action of $\n_\lag$ on $\n_\lag^*$,
extended to $\Lambda(\n_\lag^*)$.

In the rest of the Appendix we will prove the following
\begin{theorem}\label{main-app}
There is an associative algebra isomorphism $W^{I}\simeq W^{II}$.
\end{theorem}

%%%
\subsection*{The third definition of finite $W$-algebras}

We present here an equivalent definition of the complex $(C^{I},d^{I})$ defining the algebra $W^{I}$,
which better enlightens the relation to the complex $(C^{II},d^{II})$ and the corresponding 
cohomology $W^{II}$.
Let $U(\n_\lag)_{-\chi}$ be the space $U(\n_\lag)$ with left action of $\m_\lag$
given by $m\cdot A=(m-(f|m))A$,
namely, as $\m_\lag$-module, $U(\n_\lag)_{-\chi}\simeq U(\n_\lag)\otimes\C_{-\chi}$.
Consider the induced $\g$-module,
$$
M_\lag\ =\ \Ind_{\m_\lag}^\g U(\n_\lag)_{-\chi}\ =\
U(\g)\otimes_{U(\m_\lag)} U(\n_\lag)_{-\chi}\ .
$$
$M_\lag$ is clearly a left $\g$--module by left multiplication.
Moreover,  a simple computation shows that $M_\lag$ has also the structure of left $\n_\lag$-module,
induced by the adjoint action of $\n_\lag$ on $U(\g)$,
and of right $\n_\lag$-module, given by the right multiplication of $\n_\lag$ on $U(\n_\lag)$.
Notice that
\begin{equation}\label{switch1}
M_\lag\ \simeq\ U(\g)_\chi\otimes_{U(\m_\lag)}U(\n_\lag)\ ,
\end{equation}
where $\m_\lag$ acts on $U(\n_\lag)$ by left multiplication,
and on $U(\g)_\chi\simeq U(\g)\otimes\C_\chi$ by right action.
Similarly
\begin{equation}\label{switch2}
M^0_\lag\ \simeq\ U(\g)_\chi\otimes_{U(\m_\lag)}\C\ .
\end{equation}

We then define, starting from $M_\lag$, a mixed Lie algebra cohomology and homology 
complex as follows.
We first define the complex $(C^h,d^h)$ for the Lie algebra homology of $\n_\lag$ with
coefficients on the right module $M_\lag$, namely
$$
C^h = M_\lag \otimes \Lambda(\n_\lag)\ , 
$$
and $d^h$ is the following differential on $C^h$:
\begin{eqnarray}\label{dh}
d^h(m\otimes c_1\wedge\cdots\wedge c_s) &=&
\sum_{1\leq k\leq s} (-1)^k m\cdot c_k\otimes c_1\wedge\stackrel{k}{\check{\cdots}}\wedge c_s \\
&+&\!\!\!\!\!\!\!\!\!
\sum_{1\leq j<k\leq s}(-1)^{j+k+1}m\otimes [c_j,c_k]\wedge
c_1\stackrel{j}{\check{\cdots}}\stackrel{k}{\check{\cdots}}\wedge c_s\ .\nonumber
\end{eqnarray}
Clearly $C^h$ is again a left $\n_\lag$-module, with respect to the adjoint action of $\n_\lag$,
hence we can talk about the Lie algebra cohomology of $\n_\lag$ with coefficients on $C^h$.
We thus get a new complex $(C^c,d^c)$, where
$$
C^c\ =\  \Lambda (\n^*_\lag) \otimes M_\lag \otimes \Lambda(\n_\lag)\ , 
$$
and $d^c$ is the usual Lie algebra cohomology differential:
\begin{eqnarray}\label{dc}
d^c(\Psi \otimes m\otimes C) &=&
 \frac{1}{2}\sum_\alpha v^\alpha\wedge u_\alpha\cdot\Psi\otimes m\otimes C \nonumber\\
&& +\sum_\alpha v^\alpha\wedge\Psi \otimes \ad u_\alpha(m\otimes C)\ .
\end{eqnarray}

We finally define a new complex $(C^{III},d^{III})$, where $C^{III}$ is the same as above,
$$
C^{III}\ =\  \Lambda(\n^*_\lag) \otimes \big(U(\g)\otimes_{U(\m_\lag)}U(\n_\lag)_{-\chi}\big) 
\otimes \Lambda(\n_\lag)\ , 
$$
and $d^{III}$ is a mixed cohomology-homology differential:
$$
d^{III}\ =\ d^c+(-1)^{\delta-1}\otimes d^h\ ,
$$
with $d^c$ as in (\ref{dc}), $d^h$ as in (\ref{dh}), and $\delta$ denotes the usual grading  of 
$\Lambda(\n^*_\lag)$.
\begin{proposition}\label{p12}
There is an isomorphism of complexes $(C^{III},d^{III})\rightarrow (C^{I},d^{I})$
given by the following linear map
\begin{eqnarray*}
i &:& (\phi^1\wedge\cdots\wedge\phi^i) \otimes 
(a_1\dots a_s \otimes_{U(\m_\lag)}b_1\dots b_t)
\otimes (c_1\wedge\cdots\wedge c_j) \\
&\mapsto& \varphi^{\phi^1}\cdots\varphi^{\phi^i} 
a_1\dots a_s X_{b_1}\dots X_{b_t} 
\varphi_{c_1}\cdots\varphi_{c_j}
\end{eqnarray*}
\end{proposition}
\begin{proof}
It is immediate to check that $i$ is well defined. It is obviously surjective, and it is injective
thanks to the PBW Theorem for $U(r)$.
We are left to prove that $i\circ d^{III}=d^{I}\circ i$.
This is a straightforward computation, which can be done 
using the formulas in Remark \ref{d-gener}.
\end{proof}

\bigskip

%%%
%\subsection*{Proof of Theorem \ref{main-app}}

\noindent {\large{\bfseries {\itshape Proof of Theorem \ref{main-app}.}}}
Thanks to Proposition \ref{p12}, in order to prove Theorem \ref{main-app}
we need to show that $(C^{III},d^{III})$ and $(C^{II},d^{II})$ are quasi-isomorphic complexes,
and the induced isomorphism of cohomologies $W^{III}\rightarrow W^{II}$
is an isomorphism of associative algebras.
By definition, $(C^{III},d^{III})$ is a bicomplex:
$$
C^{III}\ =\ 
\bigoplus_{p\geq0,q\leq0}C^{p,q}\ ,\ \
d^{III}\ =\ d_1+d_2\ ,
$$
where
$$
C^{p,q}\ =\ \Lambda^p (\n_\lag^*)
\otimes (U(\g)\otimes_{U(\m_\lag)} U(\n_\lag)_{-\chi}) \otimes \Lambda^{-q}(\n_\lag)\ , 
$$
and 
$d_1=d^c,\,d_2\ =\ (-1)^{\delta-1}\otimes d^h$ so that
$$
d_1^2\,=\,d_2^2\,=\,d_1d_2+d_2d_1\,=\,0\ ,
$$
and
$d_1:\ C^{p,q}\rightarrow C^{p+1,q},\, d_2:\ C^{p,q}\rightarrow C^{p,q+1}$.
Notice that $d^c$ and $d^h$ commute, since $d^c$ is defined in terms of the left action 
of $\n_\lag$ on $M_\lag$, while $d^h$ is defined in terms of the right action 
of $\n_\lag$ on $M_\lag$.
We can then study the corresponding spectral sequence $(E_r,d^r;\ r\geq1)$.
The first term of the spectral sequence is
\begin{eqnarray*}
\vphantom{\Big(}
E_1^{p,q} &=& H^{p,q}(C^{III},d^c) \\
\vphantom{\Big(}
&=& \Lambda^p (\n_\lag^*)\otimes 
H^{q}((U(\g) \otimes_{U(\m_\lag)}  U(\n_\lag)_{-\chi}) \otimes \Lambda(\n_\lag)\ ,d^h) \\
\vphantom{\Big(}
&=& \Lambda^p (\n_\lag^*\otimes) 
H^{q}((U(\g)_\chi \otimes_{U(\m_\lag)}  U(\n_\lag)) \otimes \Lambda(\n_\lag)\ ,d^h) \\
\vphantom{\Big(}
&=& \Lambda^p (\n_\lag^*)\otimes (U(\g)_\chi \otimes_{U(\m_\lag)} 
H^{q}( U(\n_\lag) \otimes \Lambda(\n_\lag)\ ,d^h)) \\
\vphantom{\Big(}
&=& \delta_{q,0} \Lambda^p (\n_\lag^*) \otimes (U(\g)_\chi \otimes_{U(\m_\lag)} \C) \\
\vphantom{\Big(}
&=& \delta_{q,0} \Lambda^p (\n_\lag^*) \otimes (U(\g)
\otimes_{U(\m_\lag)} \C_{-\chi}) \ =\ \delta_{q,0}C^{II,p}\ .
\end{eqnarray*}
In the second and fifth equalities above we used (\ref{switch1}) and (\ref{switch2}) respectively.
The third equality is obvious, since $d^h$ is defined in terms of the right action of $\n_\lag$
on ${U(\g)_\chi}\otimes_{U(\m_\lag)}U(\n_\lag)$ given by right multiplication on $U(\n_\lag)$.
Finally, the fourth equality follows by the isomorphism $H_q(\n,U(\n))\simeq\delta_{q,0}\C$.
The corresponding differential on $E_1^{pq}=\delta_{q,0}C^{II,p}$ is $d_1=d^c=d^{II}$.

Since the complex $E_1$ is concentrated at $q=0$, it immediately follows that the spectral 
sequence stabilizes at $r=2$, namely $E_\infty=E_2$.
Moreover, since the bigrading of $C^{III}$ is bounded, we also know that
$$
E_{\infty}\,\simeq\,\text{gr}\,H(C^{III},d^{III})\,\simeq\,H(C^{III},d^{III})\ .
$$
In conclusion, we have
$H(C^{III},d^{III})\,\simeq\,E_2\,\simeq\,H(C^{II},d^{II})$.

The above arguments, together with Proposition \ref{p12}, show that the complexes
$(C^I,d^I)$ and $(C^{II},d^{II})$ are quasi-isomorphic.
In fact, we can write down an explicit quasi-isomorphism
$\pi:\,(C^I,d^I)\rightarrow(C^{II},d^{II})$ as follows:
\begin{eqnarray*}
&& \pi(\varphi^{\phi^1}\cdots\varphi^{\phi^i}a_1\cdots a_sX_{b_1}\cdots X_{b_t}
\varphi_{c_1}\cdots\varphi_{c_j})  \\
&& \ \ \ =\ \delta_{t,0}\delta_{j,0}
\phi^1\wedge\cdots\wedge\phi^i\otimes(a_1\cdots a_s\otimes_{U(\m_\lag)}1)\ .
\end{eqnarray*}

We are left to prove that the corresponding isomorphism of cohomologies
$H(C^I,d^I)\simeq H(C^{II},d^{II})$ preserves the associative algebra structures.
Recall that in $H(C^{II},d^{II})$ the associative product is induced by $U(\g)$,
namely if 
$(A\otimes_{U(\m_\lag)}1),\,(B\otimes_{U(\m_\lag)}1)\in (M_\lag^0)^{\n_\lag}\simeq H(C^{II},d^{II})$,
then $(A\otimes_{U(\m_\lag)}1)\cdot(B\otimes_{U(\m_\lag)}1)=(AB)\otimes_{U(\m_\lag)}1$.
On the other hand, if we pick appropriate preimages in the complex $(C^I,d^I)$,
say $A$ and $B$, their product is $AB$.
The claim follows immediately.

\begin{proof}[]\end{proof}

%%%%%%%% BIBLIOGRAPHY %%%%%%%%%%%%%%%%%%%%%%%%%%

%%%%%%%%%%%%%%%%%%%%%%%%%%%%%%%%
%%%%%%%%%%%%%%%%%%%%%%%%%%%%%%%%

\renewcommand{\theenumi}{\alph{enumi}}
\renewcommand{\labelenumi}{(\theenumi)}

\end{document}